\documentclass[aps,pra,notitlepage,superscriptaddress,floatfix]{revtex4-1}

\usepackage{bm}
\usepackage[colorlinks=true, citecolor=blue, urlcolor=blue]{hyperref}\usepackage{epsf,graphics,graphicx}
\usepackage{amsmath}
\usepackage{latexsym}
\usepackage{amssymb}
\usepackage{graphicx}
\usepackage{color}
\usepackage{soul,xcolor}
\setstcolor{red}
\usepackage{ulem}
\usepackage{cancel}
\usepackage{tikz}
\usepackage[export]{adjustbox}
\usepackage{subcaption}
\usepackage{braket}
\usepackage{float}

\newcommand{\beq}{\begin{equation}}
	\newcommand{\eeq}{\end{equation}}

\def\bea{\begin{eqnarray}}
	\def\eea{\end{eqnarray}}
\def\ba{\begin{array}}
	\def\ea{\end{array}}

\usepackage{epsfig}

\input{epsf}

\begin{document}
\title{Uncovering the Fractal Nature of Water Vapor Distribution above the Surface of the Earth}
\author{Anirudha Menon}
\affiliation{Physics and Applied Mathematics Unit, Kolkata 700108, India} \email{amenon@ucdavis.edu}
\author{Banasri Basu}
\affiliation{Physics and Applied Mathematics Unit, Kolkata 700108, India} \email{sribbasu1@gmail.com}

\date{\today}

\begin{abstract}
Fractals have been at the heart of geophysical and geospatial studies in the recent past. We examine the emergent fractal character of water vapor distributions above the surface of the Earth as a function of both image resolution (number of pixels) and moisture content percentile. We calculate physically relevant quantities such as fractal dimension, number of clusters, and size of the largest cluster with varying vapor percentile using computational methods and algorithms. Our analysis unravels a potential multi-fractal character of the data which we construct using the box-counting method to calculate the generalized dimension. We examine the nature of the percolation that occurs as the vapor percentile is varied and comment on the universality class of the transition. We test the applicability of Korcak's law on our system and determine the quality of the fit using the Kolmogorov-Smirnoff statistic. We show that the fractal character of the distribution is exact as a function of image resolution and approximate in some regimes as a function of the vapor percentiles.
\end{abstract}

\maketitle

\section{Introduction}
Initiated by a series of papers of Mandelbrot and Wallis \cite{1,2,3}, the fractal analysis of geophysical variables, over a time period, has become an established tool for the investigation of the characteristics of the systems. This has been followed by applications of the technique to various geophysical phenomena \cite{4,5,6}.  The geophysical phenomena, particularly climate variability and trends, have significant environmental and socio-economic impacts. Understanding the emergent fractal structures embedded in certain geographic distributions of physical quantities will lead to further discoveries and characterizations of these quantities.

A fractal \cite{6a} is a beautiful and infinite pattern that is self-similar from the macroscopic to the microscopic. Such objects are  non-integer dimensional in the conventional sense. Fractal dimensional analysis can identify and quantify how the geometry of patterns repeat from one scale to another. Scaling laws are at the heart of fractal analyses. Many natural and social-economic phenomena display a scale distribution \cite{1,2,3,6a}, including coastline and mountain formations. These distributions are approximate fractals, i.e., that they appear scale-invariant in a certain range of scales. It is well understood that while ideal fractals exist as mathematical constructions, the fractal structures found in nature are approximate.

Numerous attempts \cite{7,8,9,10,11,12,13,14,15,16,17,18} that  elucidate the spatio-temporal patterns of land surface temperature (LST) and  reveal  the scaling properties in these distributions motivated us to study the fractal analysis of the distribution of water vapour data. Based on theory, fractal analysis can be used to decompose irregular data sets to many microscopic pictures on multiple scales and reveal their scaling properties in various geospatial distributions.

A generalization of fractals exist to the realm of multi-fractals, which first appeared in works \cite{101} by Mandelbrot involving multiplicative cascade models. Since then, they have found applications in the characterizations of many different physical systems such as power grid networks \cite{102}, payment netoworks \cite{102b}, and the quantum entangled states \cite{103}. Typically, the box counting method is used to determine the generalized dimension which depends on the moment of the multi-fractal measure. The variation of the generalized dimension as a function of the moments determine the underlying non-fractal, mono-fractal, or multi-fractal nature of the distribution.

In this paper, we concentrate on the analysis of Earth’s primary greenhouse gas, water vapour, which is a critical component of Earth’s climate system. Accurate quantification of water vapour data and the interpretation of its physical characteristics are crucial in understanding how global climate systems, such as global warming, are changing over time. Urban land use \cite{18b,18c}, urban infrastructure \cite{18d}, and network distributions \cite{18e,18f,18g} have been studied using fractal based analyses and techniques. Similar techniques have also been developed for atmospheric sciences \cite{18h}. We, however, seek to characterize water vapor distribution above the entire planet, and not restrict to cities or urban areas. We also analyze the applicability of Korcak's law to our system and provide substantial data driven evidence in the affirmative.  

The structure of this manuscript is as follows: In section II we discuss the methods of data acquisition, data extraction, and data processing. In section III, we examine the fractal and multi-fractal nature of the  distribution as a function of the resolution of the images obtained, namely at $180 \times 360$ pixels and $360 \times 720$ pixels. We also seek to identify a possible percolation transition and characterize it. In section IV we examine the fractal and multi-fractal character of the vapor distribution as a function of the vapor percentile. The above mentioned percolation transition is also examined in detail in this case. We discuss our observations in both sections III and IV. In section V we summarize our findings and conclude. 

\section{Data Extraction and Processing}

\begin{figure}
	\begin{subfigure}[t]{0.45\textwidth}
		\centering
		\includegraphics[width=\linewidth]{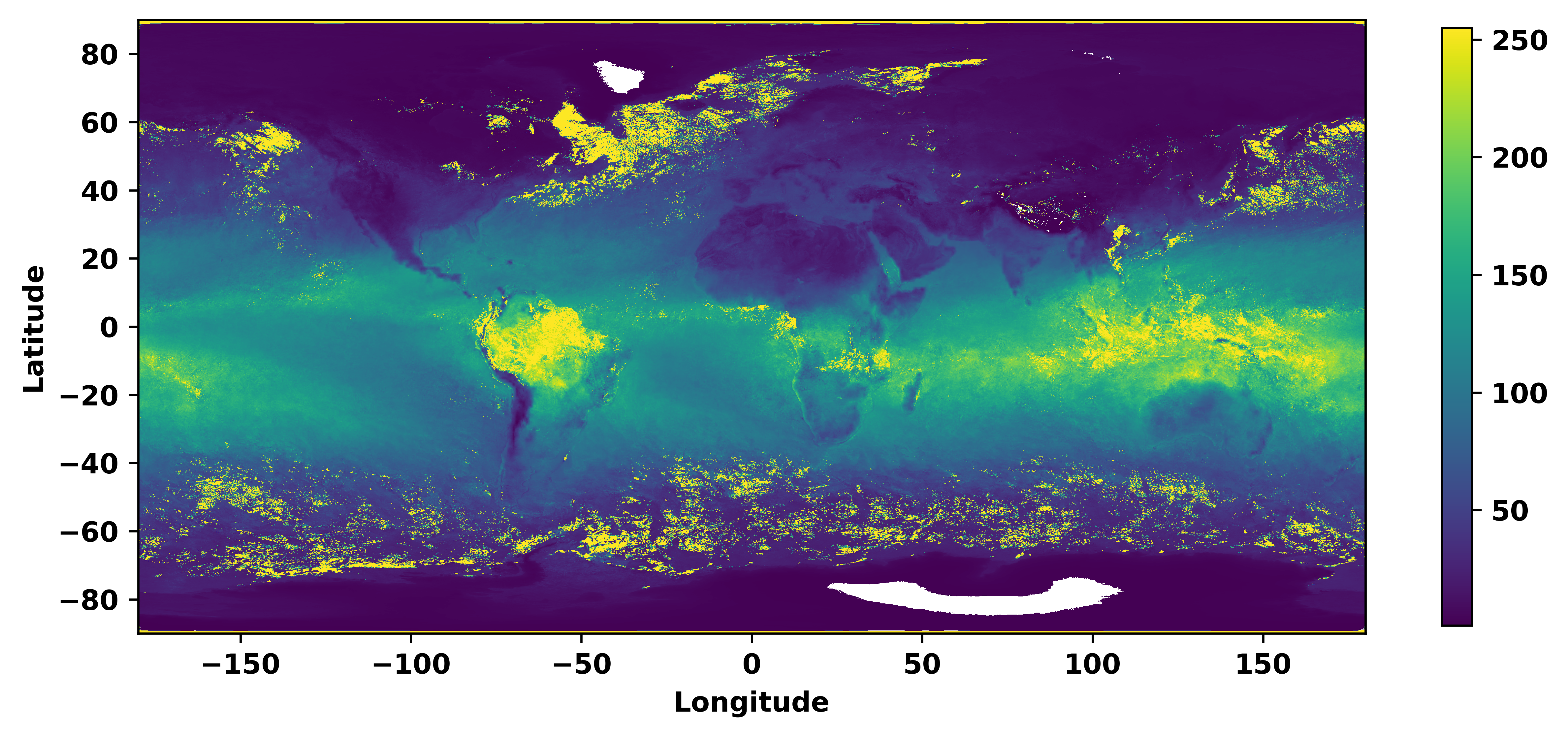} 
	\end{subfigure}
	\hfill
	\begin{subfigure}[t]{0.45\textwidth}
		\centering
		\includegraphics[width=\linewidth]{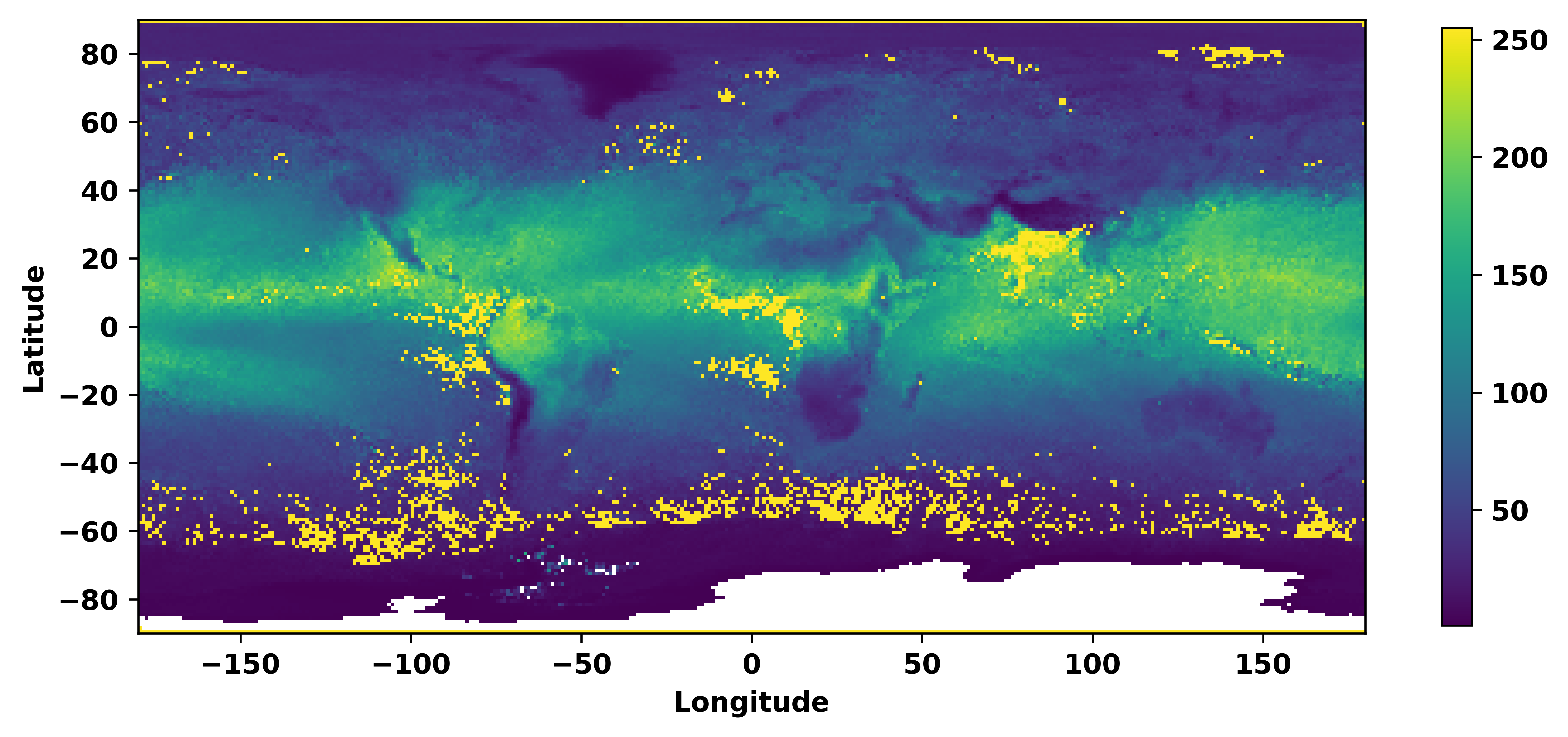} 
	\end{subfigure}
	\begin{subfigure}[t]{0.45\textwidth}
		\centering
		\includegraphics[width=\linewidth]{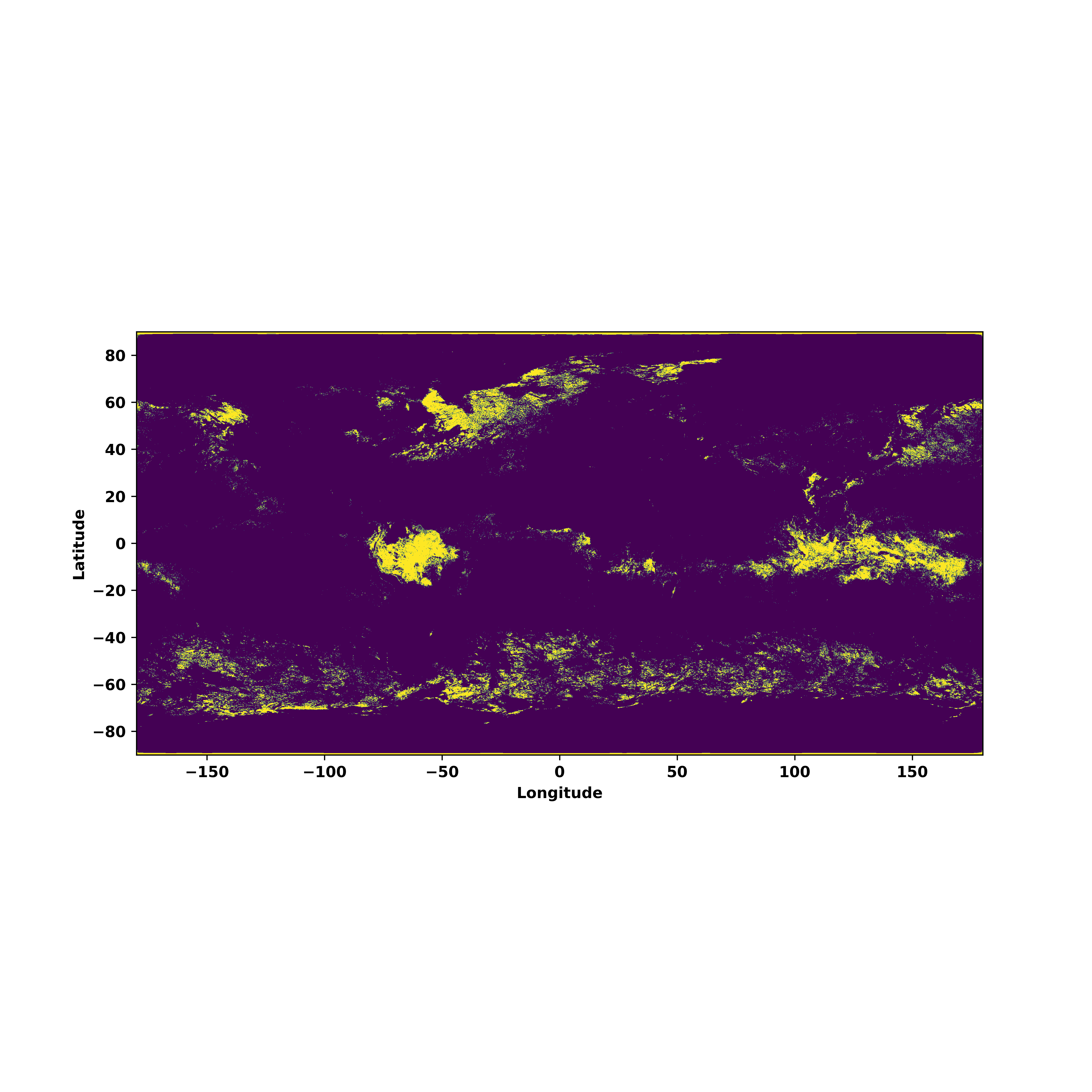} 
	\end{subfigure}
	\hfill
	\begin{subfigure}[t]{0.45\textwidth}
		\centering
		\includegraphics[width=\linewidth]{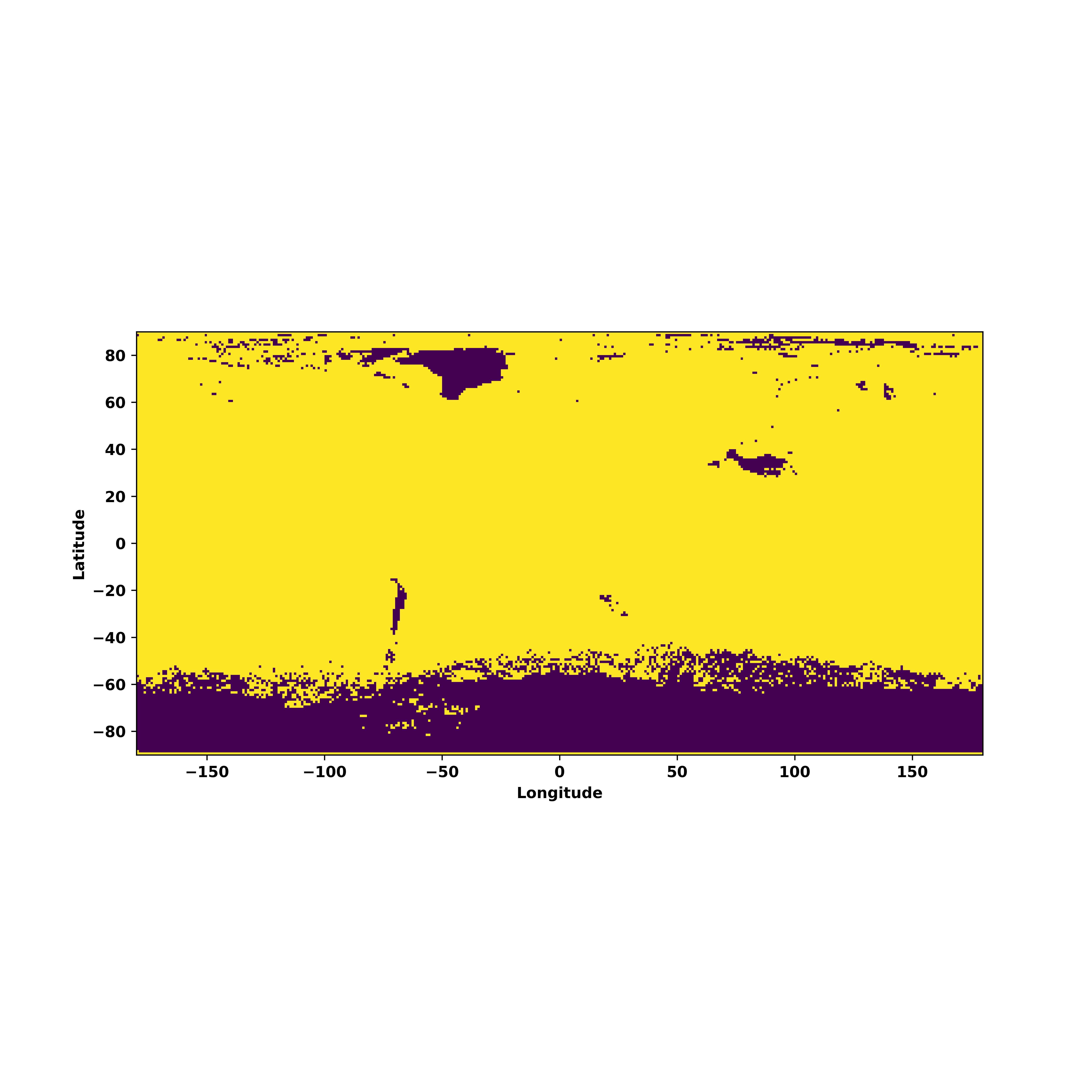} 
	\end{subfigure}
	\caption{Visualization of the MODIS water vapor data: The top panel shows the distributions at two different resolutions, namely $180\times 360$ pixels and $360\times 720$ pixels. The bottom panel shows the same data after the binning procedure described above. The left plot is at the 80th percentile while the right plot is at the 20th percentile.}
	\label{0}
\end{figure}

Data was obtained in the Geotiff format for the monthly global distribution of water vapor from the TERRA MODIS (Moderate Resolution Imaging Spectroradiometer) dataset \cite{18a} for 2021 at $360\times 720$ pixels and for the years 2012-2021 at $180\times 360$ pixels. The MODIS atmospheric water vapor data estimates the total column water vapor made from integrated MODIS infrared retrievals of atmospheric moisture profiles in clear scenes over land and ocean. Note that the total column water vapor can only be obtained in clear-sky conditions. Representative datafiles are plotted in Fig. \ref{0}, where the data in the top row is for two different months at the resolutions of $180\times 360$ and $360\times 720$ pixels, respectively, for the left and right plots. This data was then extracted using the Rasterio module in python v3.9 and converted to the numpy array format. Different values of vapor percentiles were chosen from $0$ to $100$ and the numpy arrays were binned binarily based on whether the value in a given pixel was greater or lesser than the given vapor percentile. The bottom row in Fig. \ref{0} represents the binned data at different vapor percentiles. Once this was completed for all data files, a clustering algorithm was used to determine the clusters in the binarily binned arrays. This algorithm was recursive and identified all the pixels in the Moore neighborhood of a given pixel. Two pixels are said to be in the same cluster if they both have value 1 and are in each others Moore neigborhood. The pixels were added to a list if they satisfy the above criteria, and the neighboring cells of these added cells were identified. The whole process was repeated until the entire grid was covered. A separate algorithm was used to calculate the perimeters and areas of the clusters. While the areas of clusters were trivial to calculate, since they are simply the number of cells in a cluster, the perimeters were harder to determine and required a caching of the neighbors of each cell. The data produced from these computations was then stored on disk for further analysis.

\section{Resolution based Numerical Analysis}

The self-similar property underlying the original fractal model assumes that the form or pattern of the spatial phenomenon remains unchanged throughout all scales. Closely related to scale is the concept of resolution. Resolution refers to the smallest distinguishable parts in an object or a sequence and is often determined by the least count of the instrument or the sampling interval used in a study. Moreover, fractals may be useful in identifying the breakpoints in a pattern or distribution at a particular scale where the processes contributing to these patterns become unstable. In this regard, it  is important to have a clear understanding of the effect of resolution in fractal dimensional analysis. 

We considered the binarily binned data with the clusters being formed by the aggregation of the cells with value $1$. We examined data at two different resolutions of $180 \times 360$ pixels and $360 \times 720$ pixels in order to see how the nature of the distribution varied as a function of resolution. Alternatively, we could also have binned the cells with value $0$ and this would give us a complementary distribution to the one we are currently studying. An analysis of this complimentary distribution, while potentially interesting, is relegated to a future work. 

\subsection{Area-Perimeter Relation and Fractal Dimension}

In order to quantify the fractal nature of the water vapor distribution, we consider the aggregated area-perimeter fractal dimension \cite{6a,18} of the clusters at each vapor percentile $V$ using the following equation

\begin{eqnarray}
	\sum_{c \subset C} P_c = k\big{[}\sum_{c \subset C} A_c\big{]}^{D/2},
\end{eqnarray}

where $c$ represtens each cluster in the total set of clusters $C$, $A_c$ and $P_c$ are the area and the perimeter of the cluster $c$, $k=2\sqrt{\pi}$, and $D$ is the fractal dimension. The value of $k$ is determined based on the limiting case of a circle. When the shapes of the clusters are more elongated or convoluted, we approach the linear limit of $P=A$ and $D=2$. 

The distributions of clusters are obtained at different values of the water vapor percentile ranging from $0$ to $100$. We present below a log-log plot of the distribution of the cumulative perimeter vs cumulative area at different percentiles for the year 2021. All 12 months of the year are considered in the analysis at two different resolutions of $180 \times 360$ pixels and $360 \times 720$ pixels. The slope of the log-log plots represents the fractal dimension up to an additive constant and is only shown for the month of February to maintain clarity.  

\begin{figure}[H]
	\centering
	\begin{subfigure}[t]{0.45\textwidth}
		\centering
		\includegraphics[width=\linewidth]{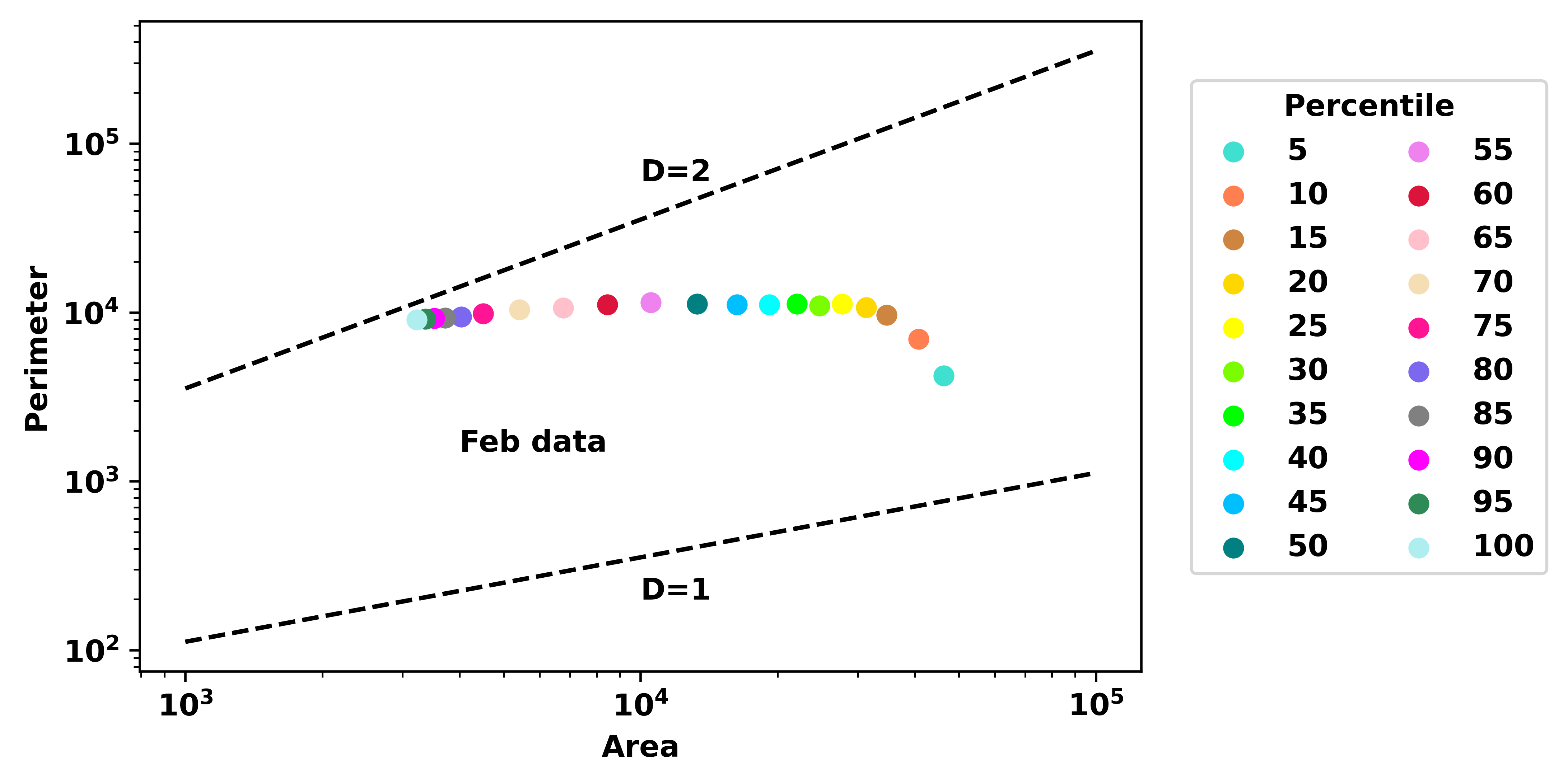} 
	\end{subfigure}
	\begin{subfigure}[t]{0.45\textwidth}
		\centering
		\includegraphics[width=\linewidth]{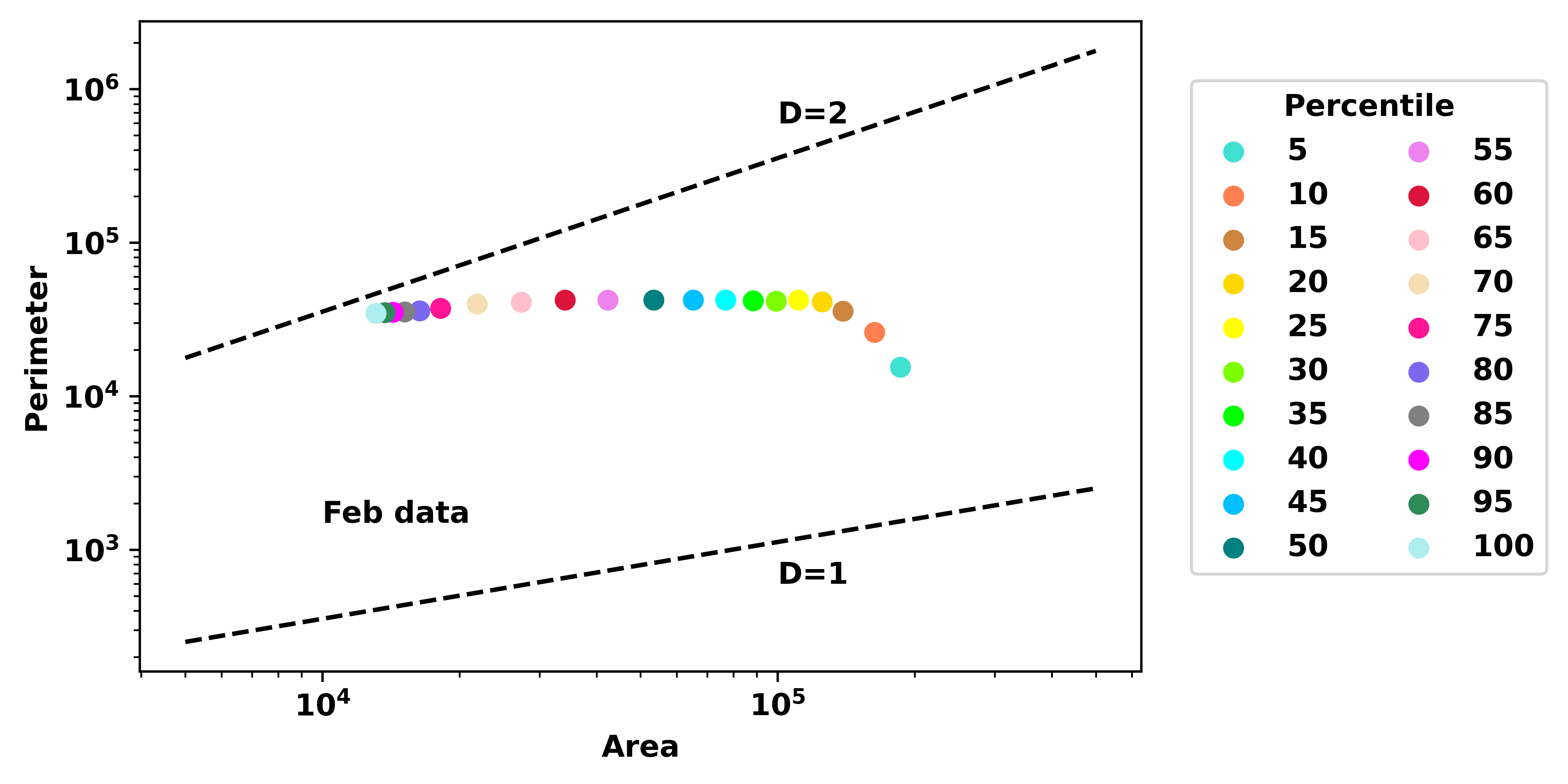} 
	\end{subfigure}
	\caption{Aggregate area-perimeter log-log plot at 180x360 pixels and 360x720 pixels on the left and right, respectively. The fact that the distributions look almost identical at these two resolutions leads us to explore further.}
	\label{22}
\end{figure}

As a very interesting observation, we note that the distributions (and in particular, the slopes) look visually identical at both resolutions, indicating that there maybe some universality in the data as a function of resolution. Thus, we find motivation to explore this aspect further. 

\begin{figure}[H]
	\centering
	\begin{subfigure}[t]{0.3\textwidth}
		\centering
		\includegraphics[width=\linewidth]{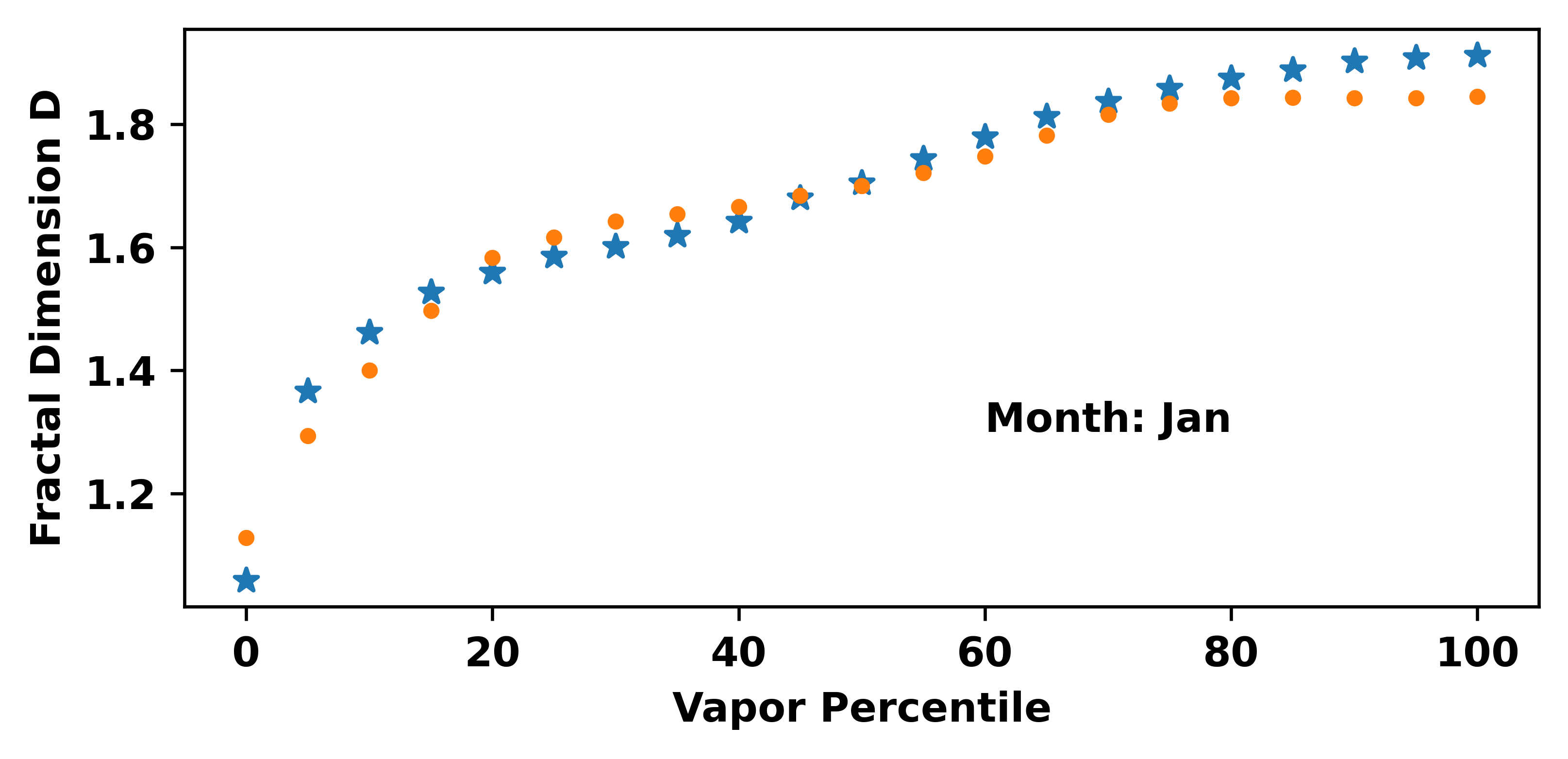} 
	\end{subfigure}
	\begin{subfigure}[t]{0.3\textwidth}
		\centering
		\includegraphics[width=\linewidth]{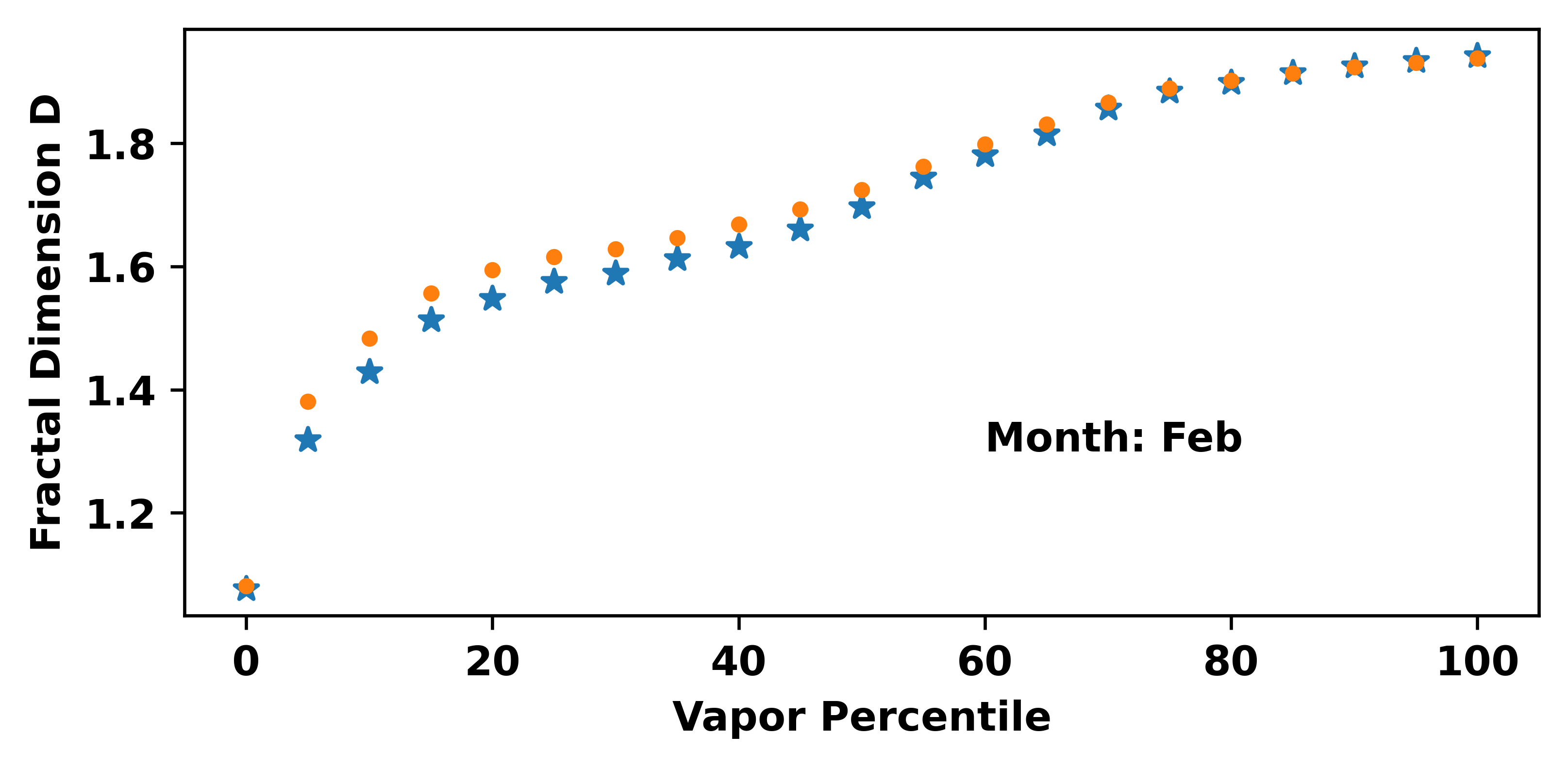} 
	\end{subfigure}
	\begin{subfigure}[t]{0.3\textwidth}
		\centering
		\includegraphics[width=\linewidth]{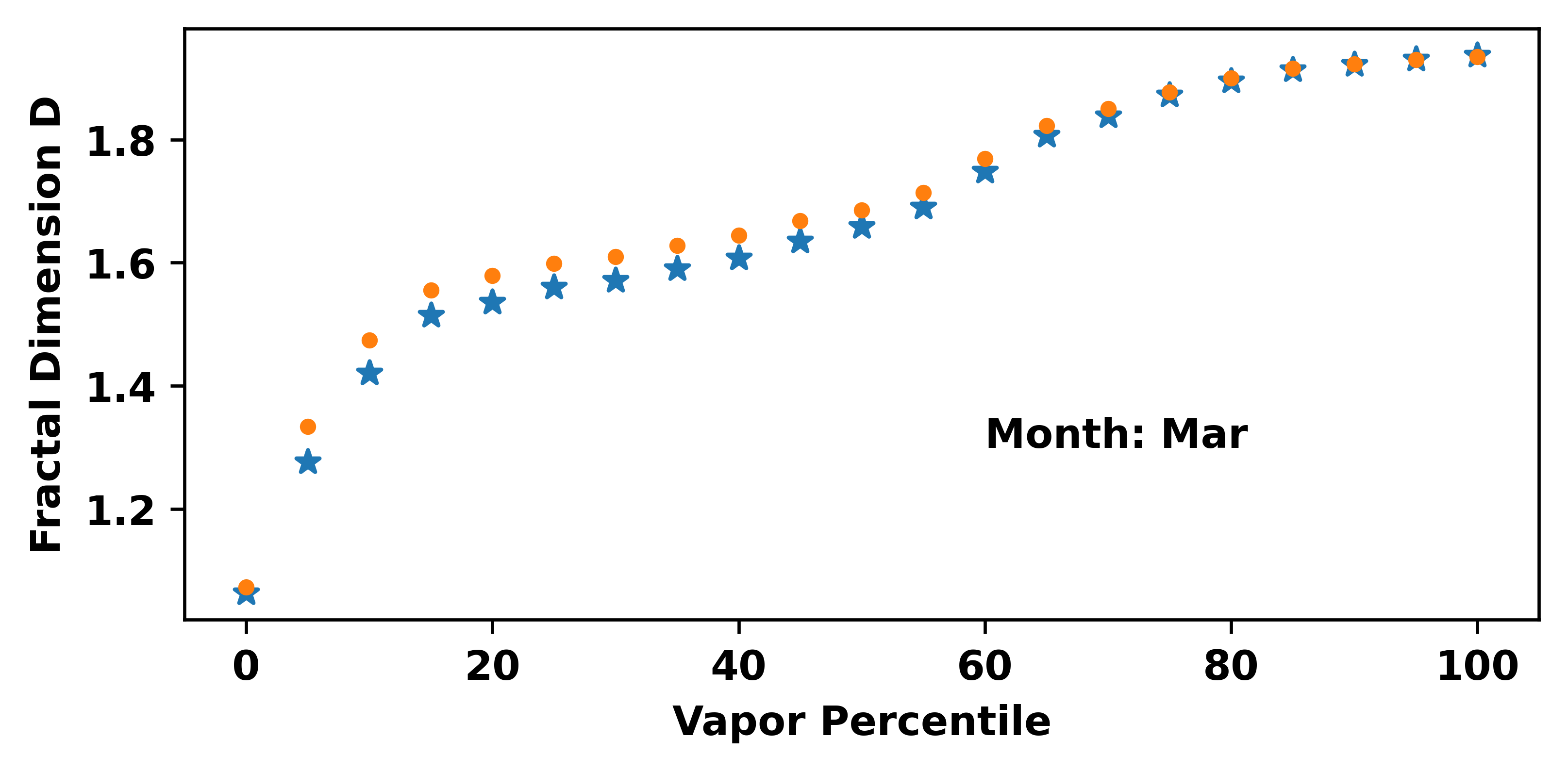} 
	\end{subfigure}		
	\begin{subfigure}[t]{0.3\textwidth}
		\centering
		\includegraphics[width=\linewidth]{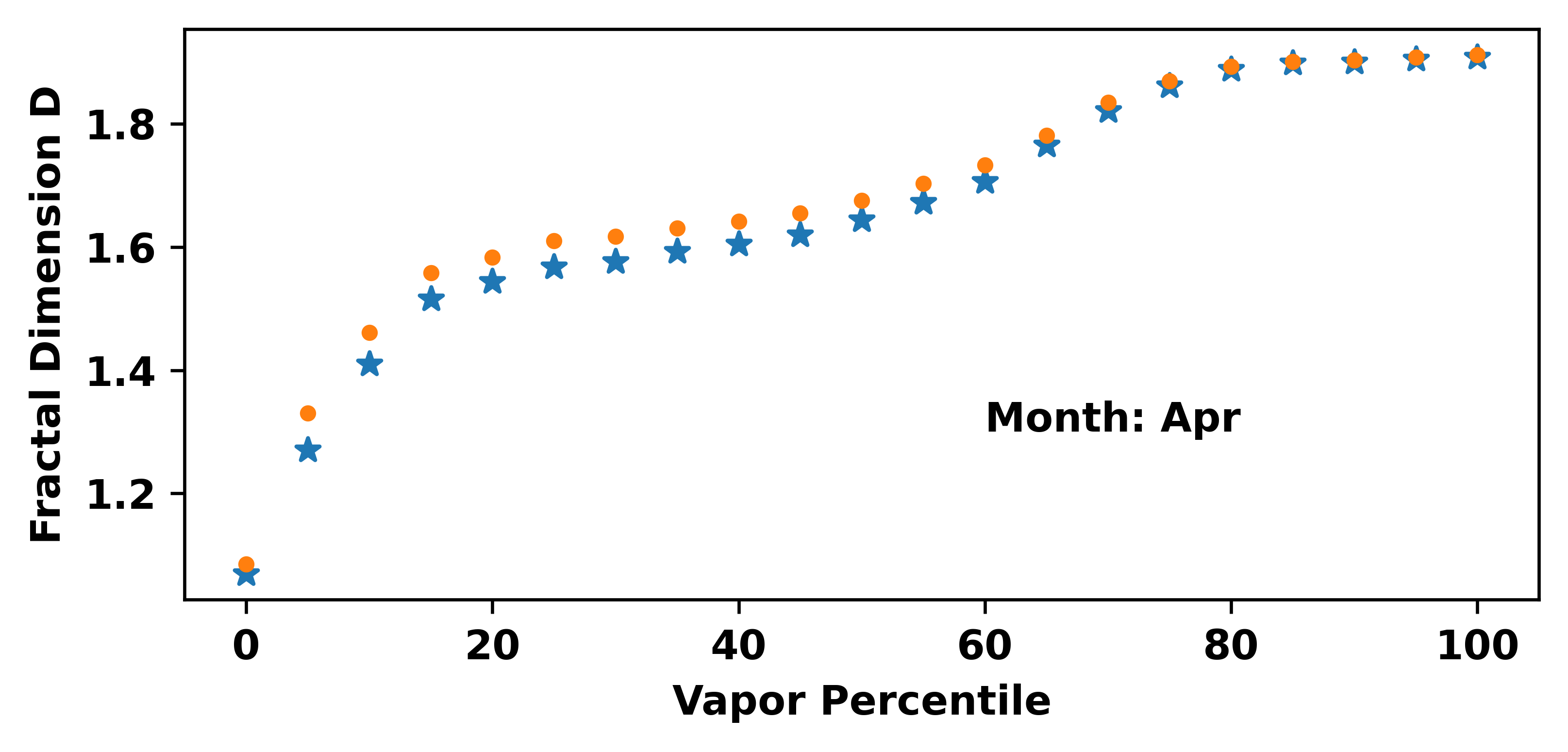} 
	\end{subfigure}
	\begin{subfigure}[t]{0.3\textwidth}
		\centering
		\includegraphics[width=\linewidth]{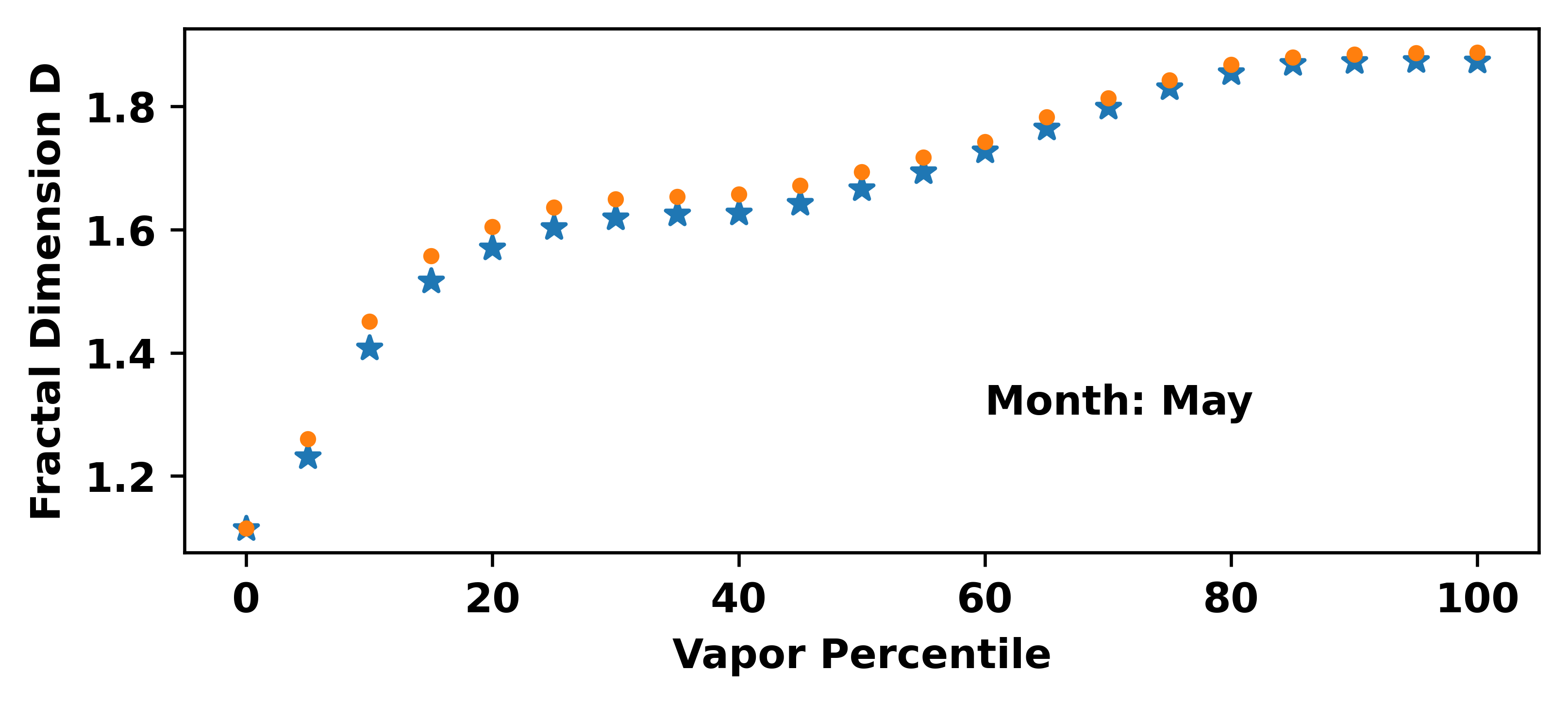} 
	\end{subfigure}
	\begin{subfigure}[t]{0.3\textwidth}
		\centering
		\includegraphics[width=\linewidth]{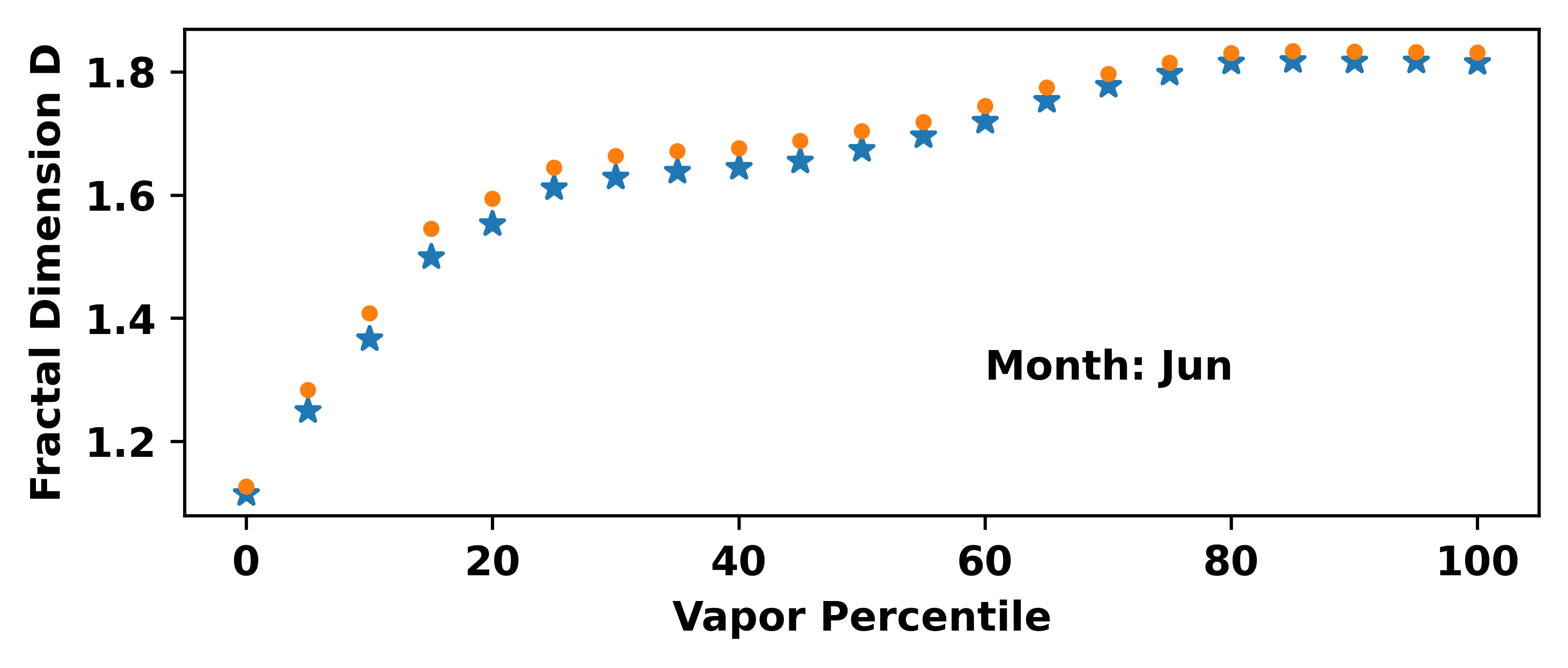} 
	\end{subfigure}		
	\begin{subfigure}[t]{0.3\textwidth}
		\centering
		\includegraphics[width=\linewidth]{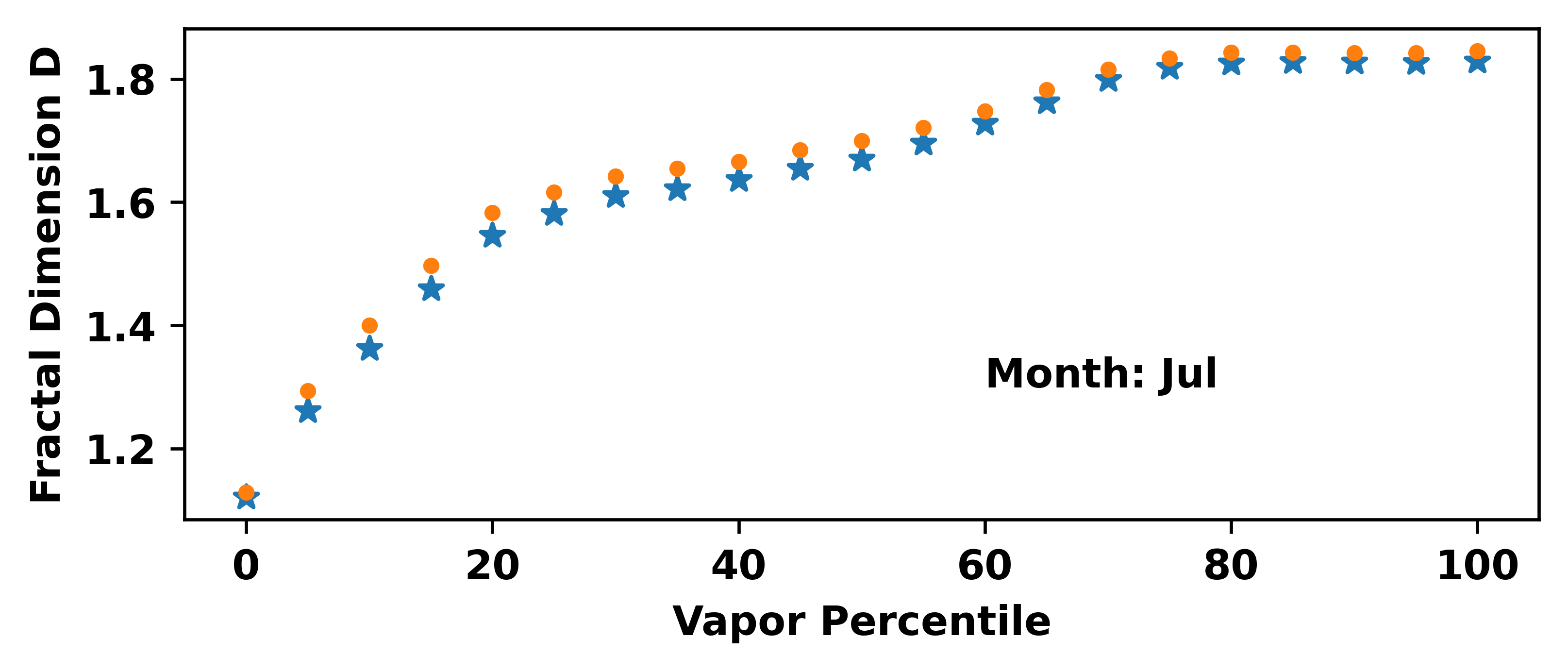} 
	\end{subfigure}
	\begin{subfigure}[t]{0.3\textwidth}
		\centering
		\includegraphics[width=\linewidth]{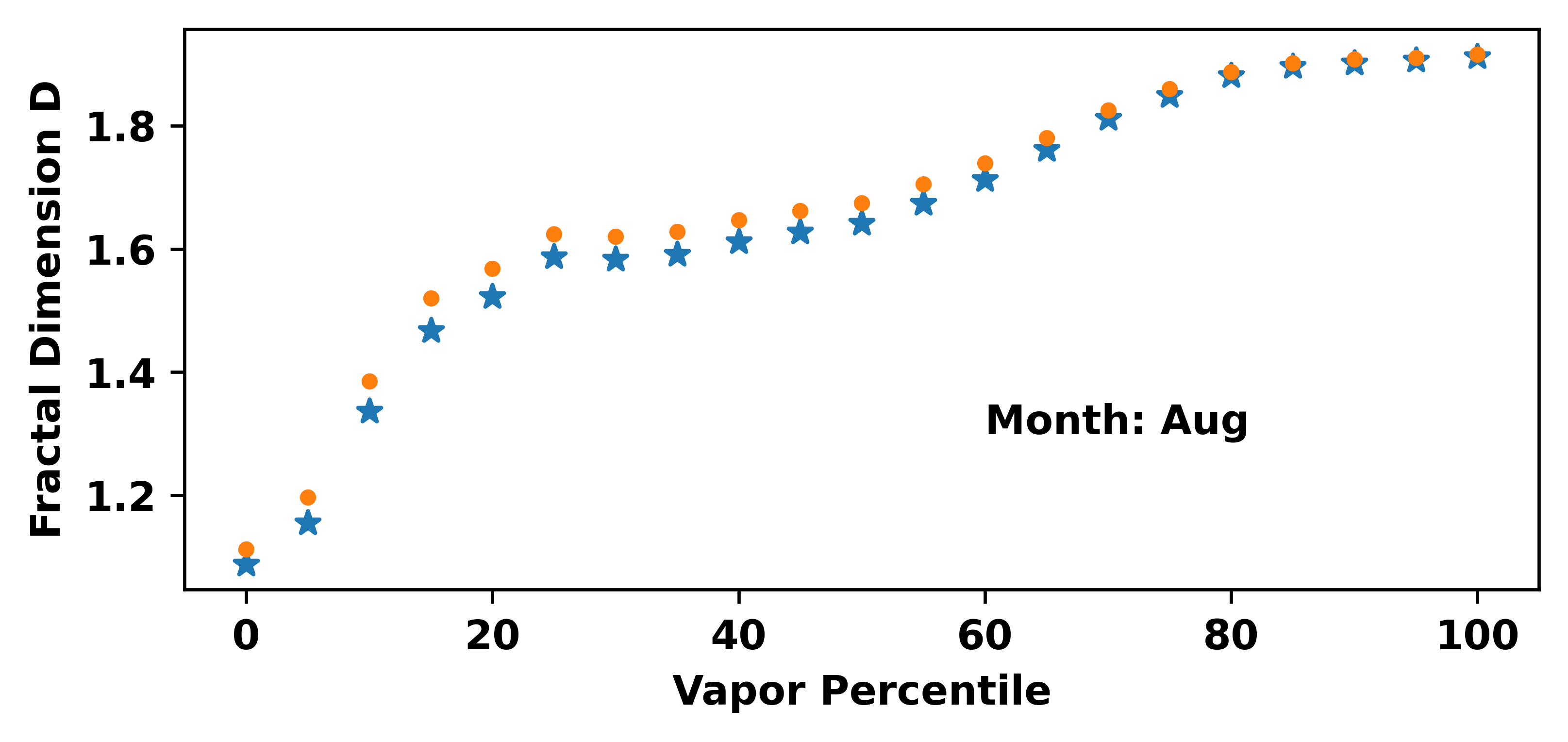} 
	\end{subfigure}
	\begin{subfigure}[t]{0.3\textwidth}
		\centering
		\includegraphics[width=\linewidth]{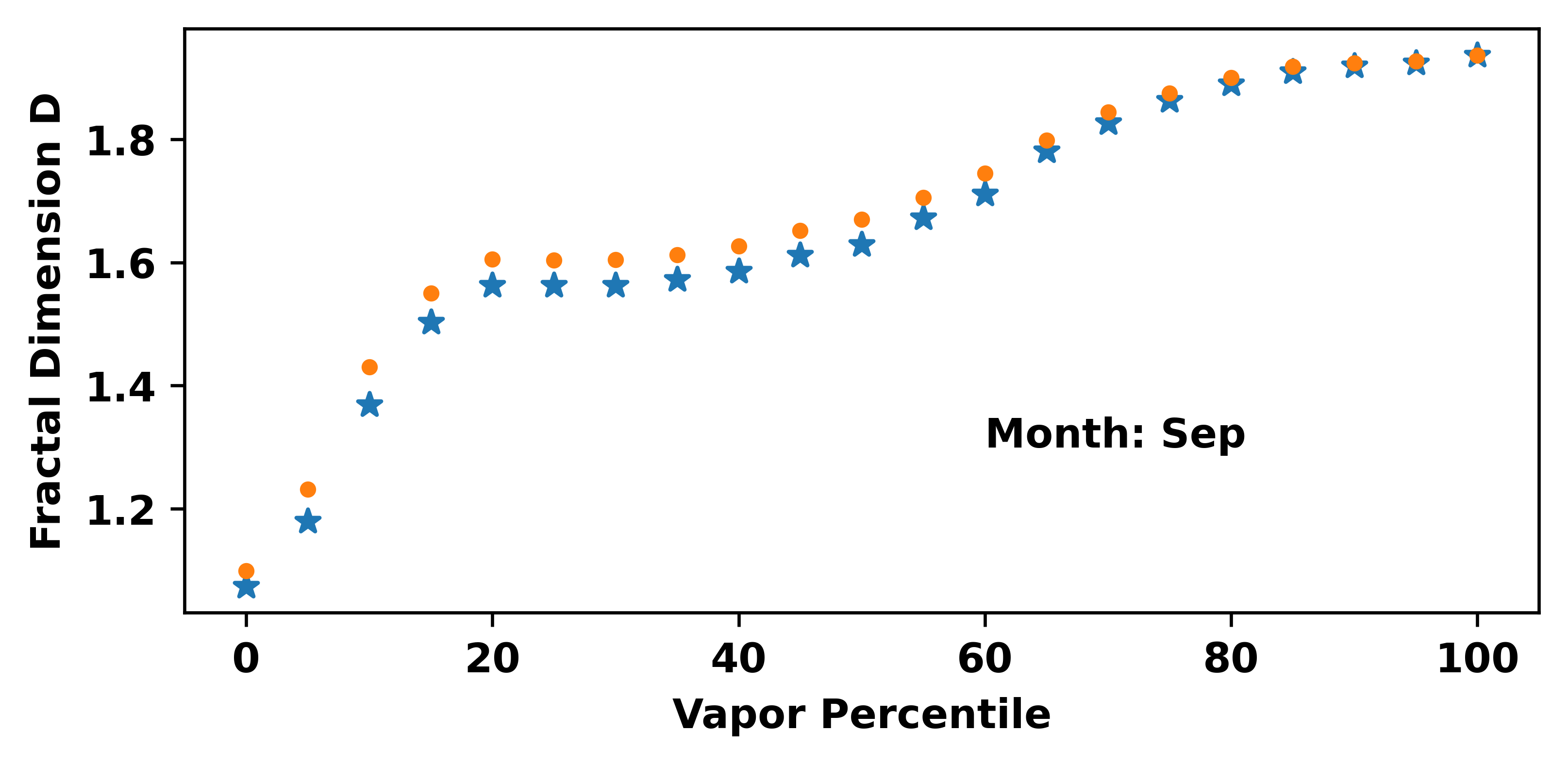} 
	\end{subfigure}		
	\begin{subfigure}[t]{0.3\textwidth}
		\centering
		\includegraphics[width=\linewidth]{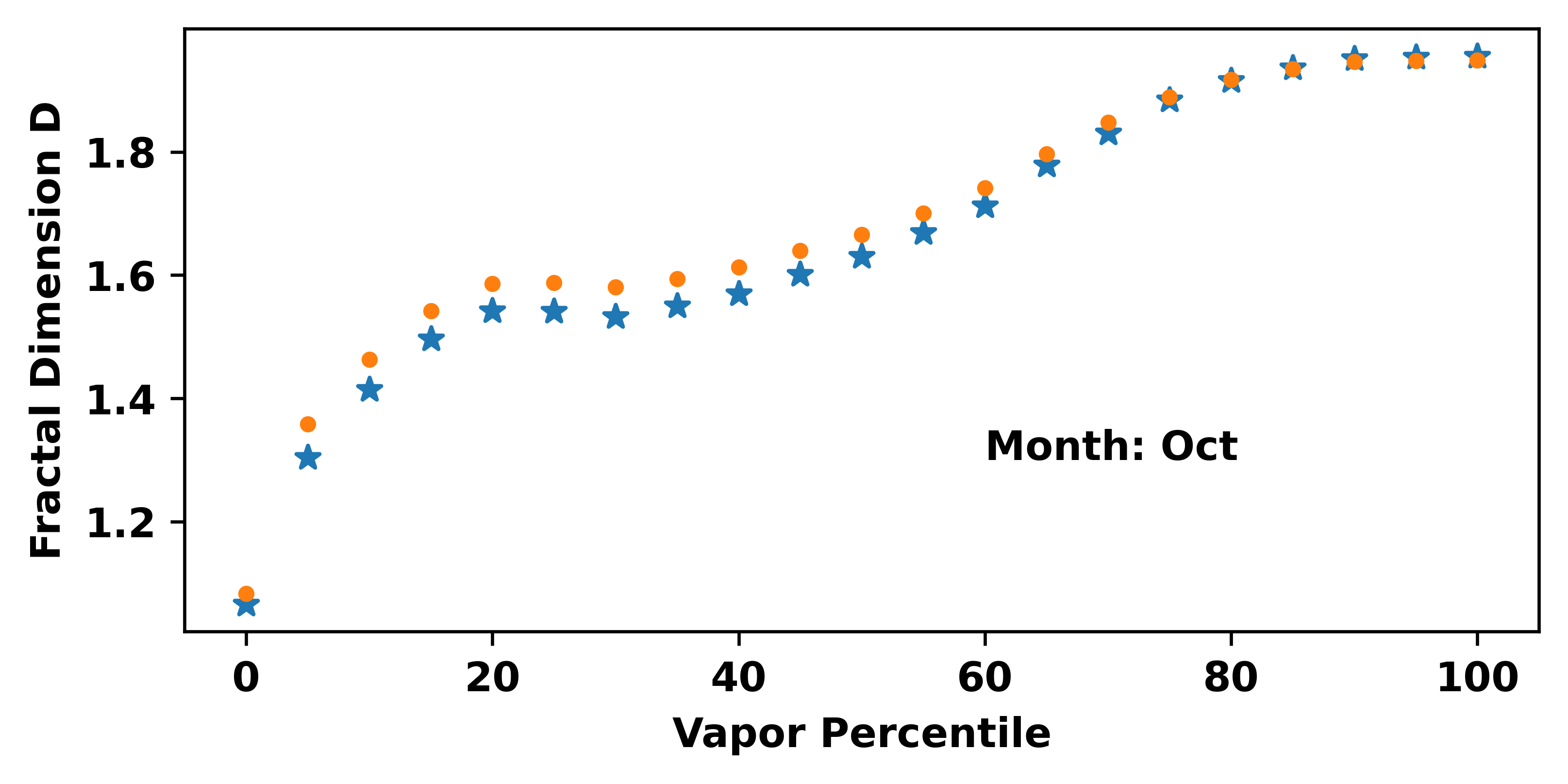} 
	\end{subfigure}
	\begin{subfigure}[t]{0.3\textwidth}
		\centering
		\includegraphics[width=\linewidth]{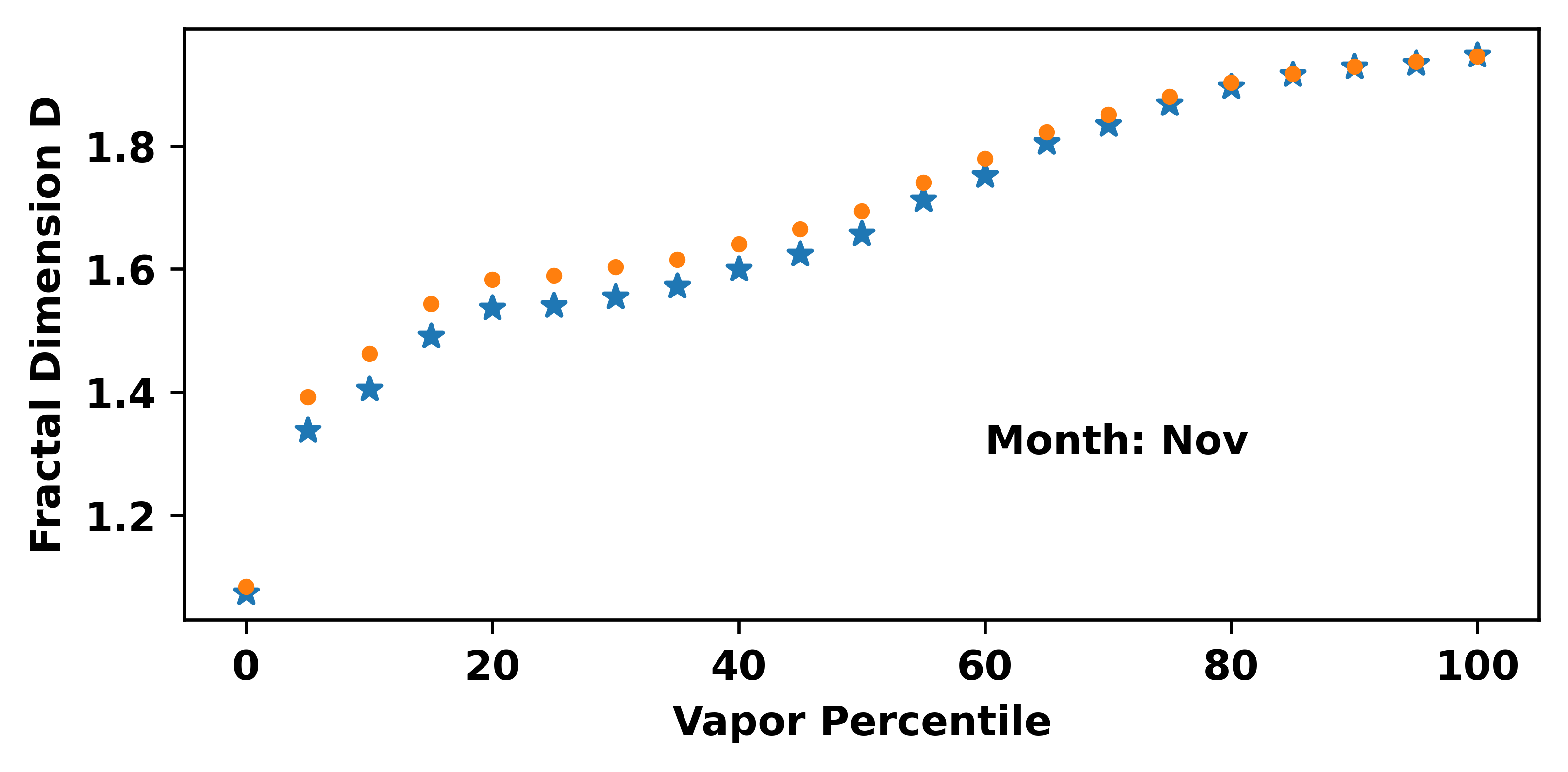} 
	\end{subfigure}
	\begin{subfigure}[t]{0.3\textwidth}
		\centering
		\includegraphics[width=\linewidth]{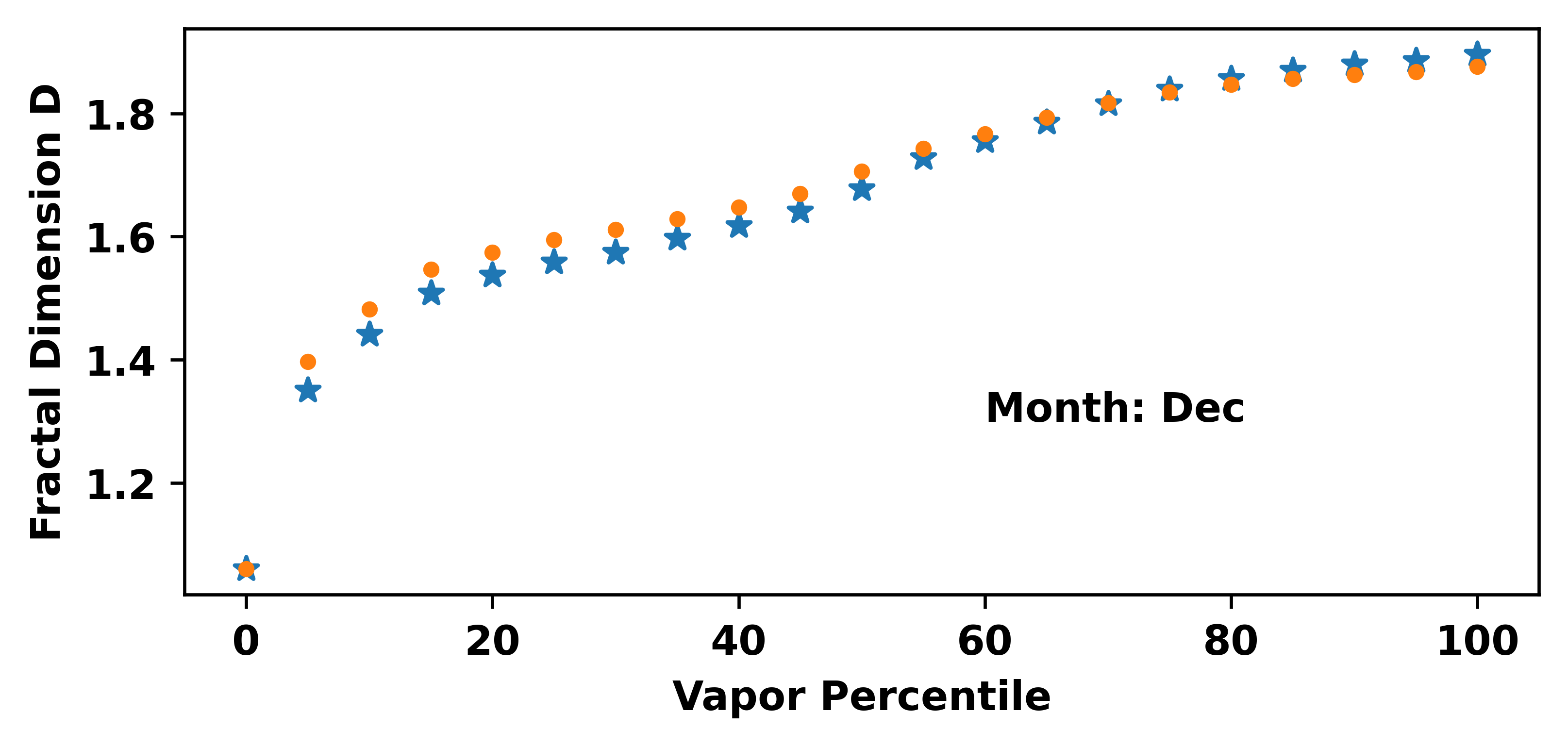} 
	\end{subfigure}		
	\caption{Fractal Dimension vs Vapor Percentile for 12 months in 2021: The blue markers represent the $180\times 360$ pixel data and the orange markers represent the $360\times 720$ pixel data.}
	\label{2}
\end{figure}

We plot the fractal dimensions as a function of water vapor percentiles for the $12$ months in 2021 at both resolutions. The results are shown in Fig. \ref{2}. We observe that the fractal dimensions are in general identical for the cluster distribtuions at both resolutions for almost all the months. We find that the January data presents a disagreement to the previous observation and we allude this discrepancy to an anomaly in the data, since it is not a prevalent phenomenon. The fact that the fractal dimension is independent of resolution is a key result of this work, as the fundamental construction of fractals states that fractal geometries are scale invariant. While we would like to have explored this further at other resolutions, current computational capacity limits this. 

\subsection{Multifractal Analysis}
Multifractality can be viewed as a genaralization of fractals and first appeared in studies on multiplicative cascade models by Mandelbrot. Since then, it has found a wide variety of applications to natural systems, including quantum mechanical systems. While there are a few methods one can employ to calculate the multi-fractal dimension, we focus on the box-counting method. This technique tiles the geo-spatial data file into smaller and smaller boxes and studies the immplications at each scale. On way of understanding this method is that it examines how zooming in on the data alters the observed features. 

We consider the meshing the dataset with boxes of size $\epsilon$, and define the probability of occupation of the $i$th box as $P_i$. This probability is calculated by dividing the number of pixels occupied in each box by the total number of pixels occupied. The generalized dimension $D_q$ then corresponds to the scaling of the $q$th moments of the measure as

\begin{equation}\label{1000}
	D_q = \frac{1}{q-1} \lim_{\epsilon \rightarrow 0} \frac{\sum_i P_i(\epsilon)^q}{\log \epsilon}
\end{equation}

\begin{figure}[H]
	\centering
	\begin{subfigure}[t]{0.3\textwidth}
		\centering
		\includegraphics[width=\linewidth]{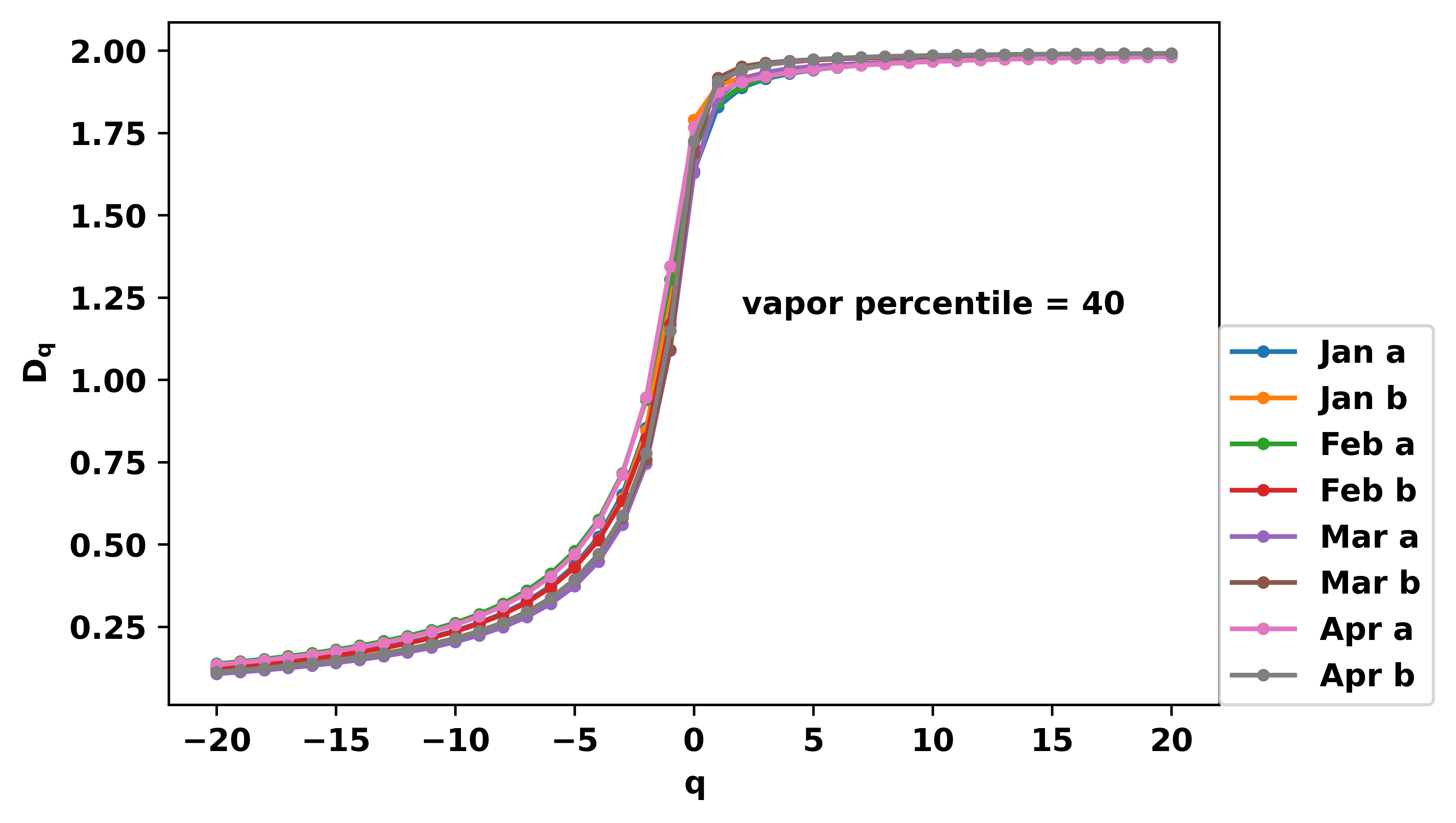} 
	\end{subfigure}
	\begin{subfigure}[t]{0.3\textwidth}
		\centering
		\includegraphics[width=\linewidth]{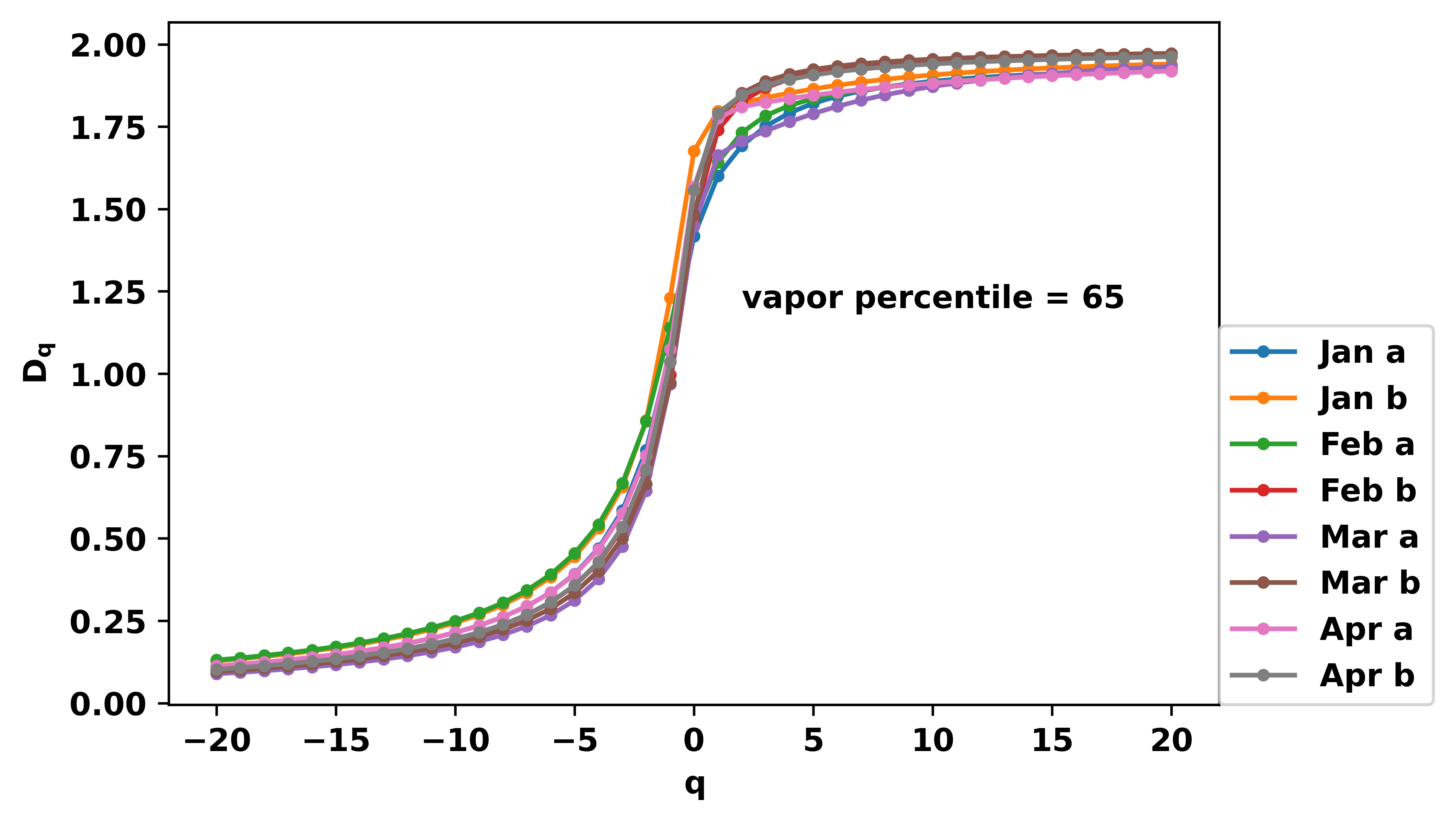} 
	\end{subfigure}		
	\begin{subfigure}[t]{0.3\textwidth}
		\centering
		\includegraphics[width=\linewidth]{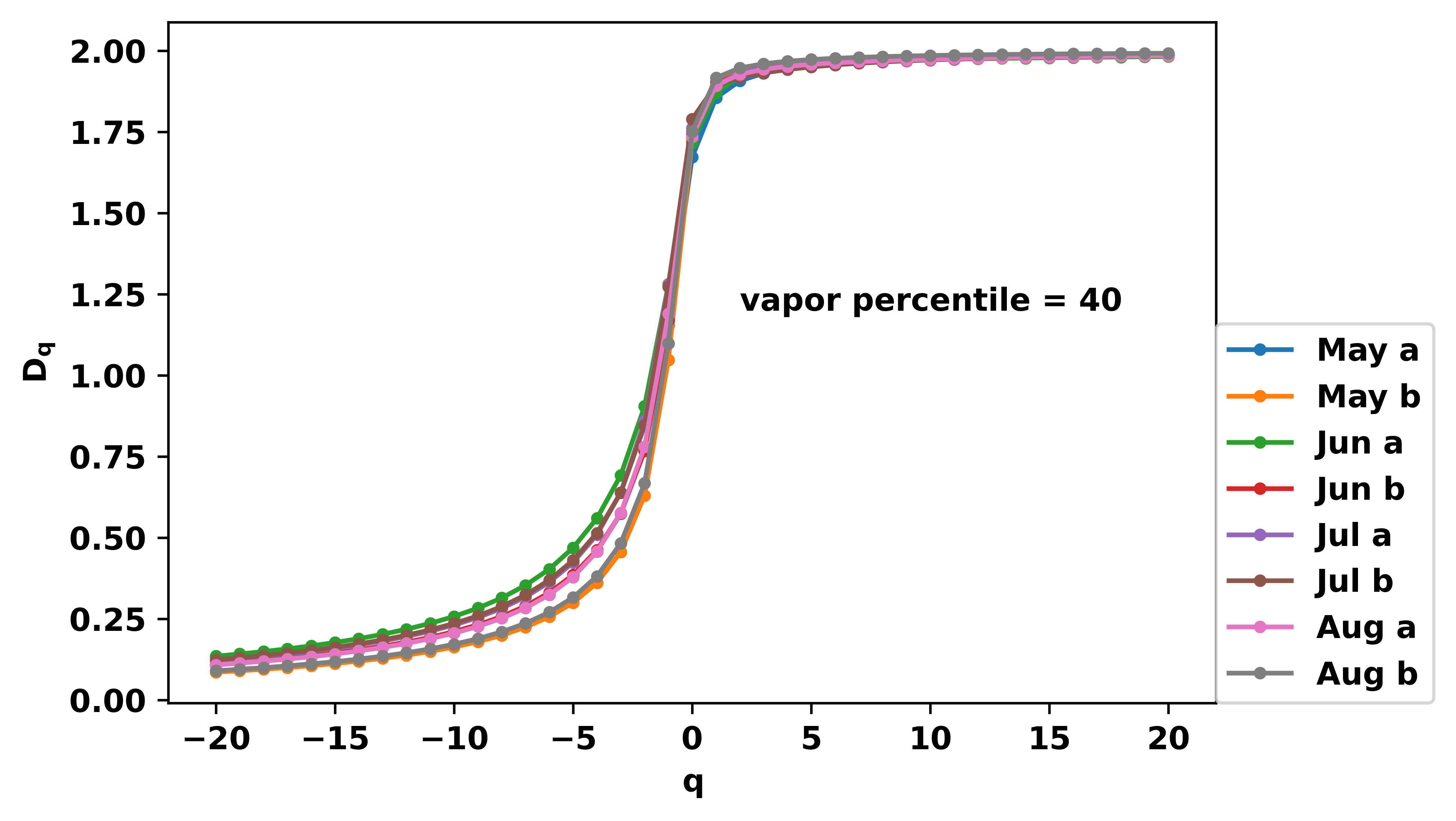} 
	\end{subfigure}
	\begin{subfigure}[t]{0.3\textwidth}
		\centering
		\includegraphics[width=\linewidth]{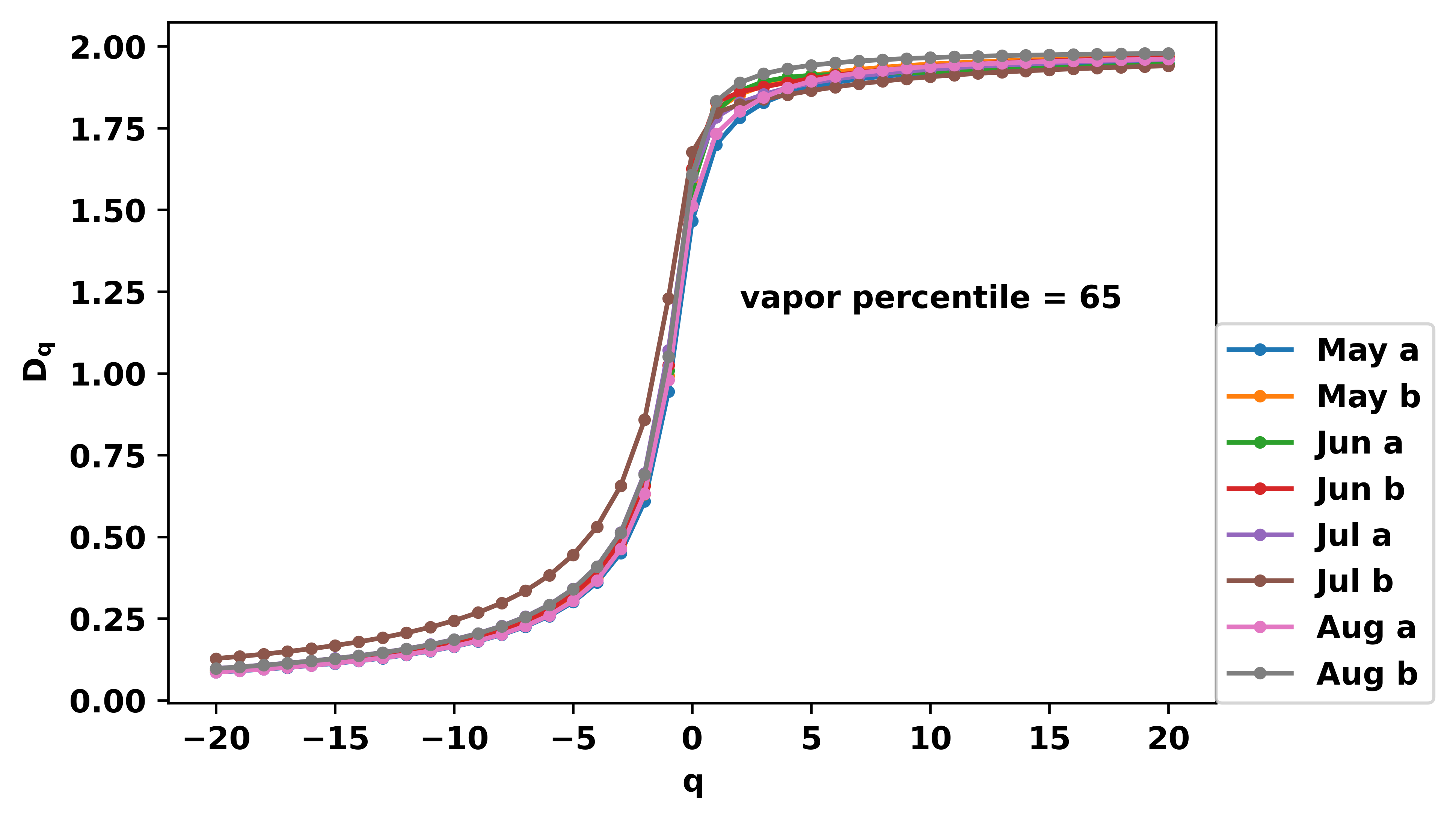} 
	\end{subfigure}		
	\begin{subfigure}[t]{0.3\textwidth}
		\centering
		\includegraphics[width=\linewidth]{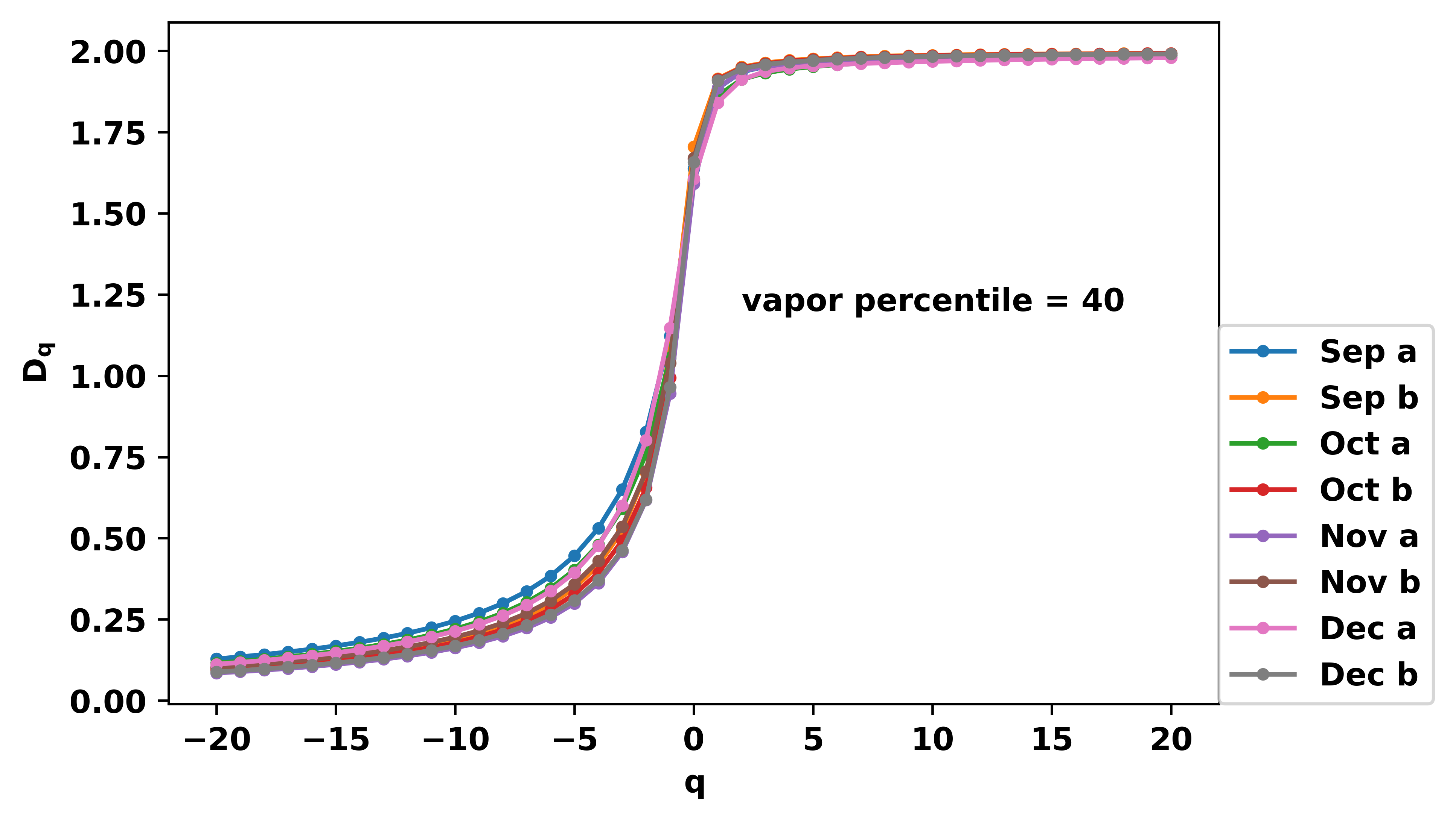} 
	\end{subfigure}
	\begin{subfigure}[t]{0.3\textwidth}
		\centering
		\includegraphics[width=\linewidth]{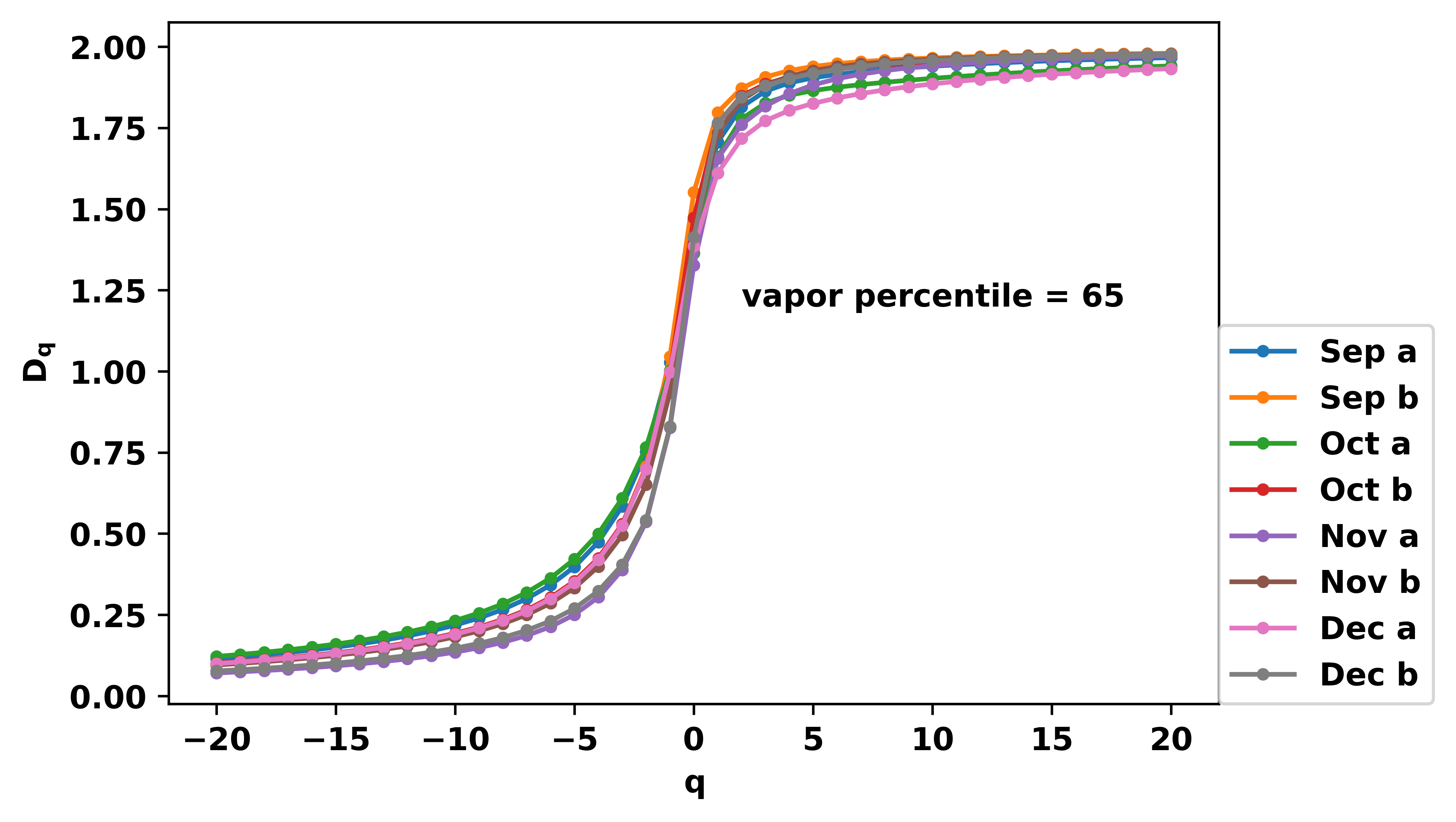} 
	\end{subfigure}		
	\caption{The figures above show the multi-fractal dimension $D_q$ as a function of $q$ at both resolutions for all months of the year 2021. The label 'a' corresponds to the $360 \times 180$ pixel data and the label 'b' corresponds to the $720 \times 360$ pixel data.}
	\label{200}
\end{figure}

$D_0$ is simply the box-counting dimension associated with mono-fractals. $D_1$ is the so called information dimension, and is calculated by applying L'Hospitals rule as $D_1 = \lim_{\epsilon \rightarrow 0} \frac{\sum_i P_i \log P_i}{\log \epsilon}$. $D_2$ is the correlation dimension. In general, multifractal anaylses involve including a distorting factor, defined by the power of $q$, to place greater importance to regions of higher occupancy probability. In practice, Eqn.(\ref{1000}) is evaluated by calculating the slope of the plot of $\log \sum_i P_i^q$ versus $\log \epsilon$. In our case, we have used box sizes of $6 \times 6$ and $10 \times 10$ pixels respectively, and we find that this gives the right box-counting dimension of $2$ at $V=100 \% $, i.e., when all the boxes in the dataset are occupied. We plot $D_q$ as a function of $q$ in the range $q \in [-20,20]$ for all months of the year 2021 at the two resolutions in Fig.\ref{200}. This visualization is at two different vapor percentiles of $V=40 \: \& \: 65 \%$. The curves show an agreement for all months at the two-resolutions. The mutli-fractality of the distribution is apparent from the variation of $D_q$ with $q$ - for monofractals, the relationship would have been linear. The generalized dimension varies between $D_q \sim 2$ for large $q>0$ to $D_q \sim 0$ for large $q<0$.

\subsection{Percolation Phenomena}

The existence of percolation phenomena associated with the vapor distribution clusters may be studied by examining the number of clusters at different vapor percentiles. From general considerations, we expect a percolation phenomenon to occur in our system, since at low $V$ a large subset of pixels are occupied leading a large connected cluster spanning the system size, while at large $V$ very few pixels are occupied leading to small disjoint clusters. Our task here is to determine if the transition lies in the 2D Bernoulli class of transitions and also the nature of the criticality, be it a second order phase transition or otherwise. For the indepth study of  the percolation phenomena and to find the possible percolation threshold , we examine the  number of clusters as well as the size of the largest cluster size formed, as a function of vapor percentile for both the resolutions. 
\subsubsection{Cluster Number}

We examine the {\it scaled} number of clusters as a function of vapor percentile. This is crucial, since we are working with datasets of varying pixel dimesions, i.e., $360 \times 180$ and $720 \times 360$ pixels, and in order to compare their behavior, we need to normalize our data. The scaled number of clusters is defined as follows

\begin{equation}
	\textrm{scaled \# of clusters} (V) = \frac{\textrm{\# of clusters (V)}}{\textrm{maximum \# of clusters (V)}} 
\end{equation}

\begin{figure}[H]
	\centering
	\begin{subfigure}[t]{0.3\textwidth}
		\centering
		\includegraphics[width=\linewidth]{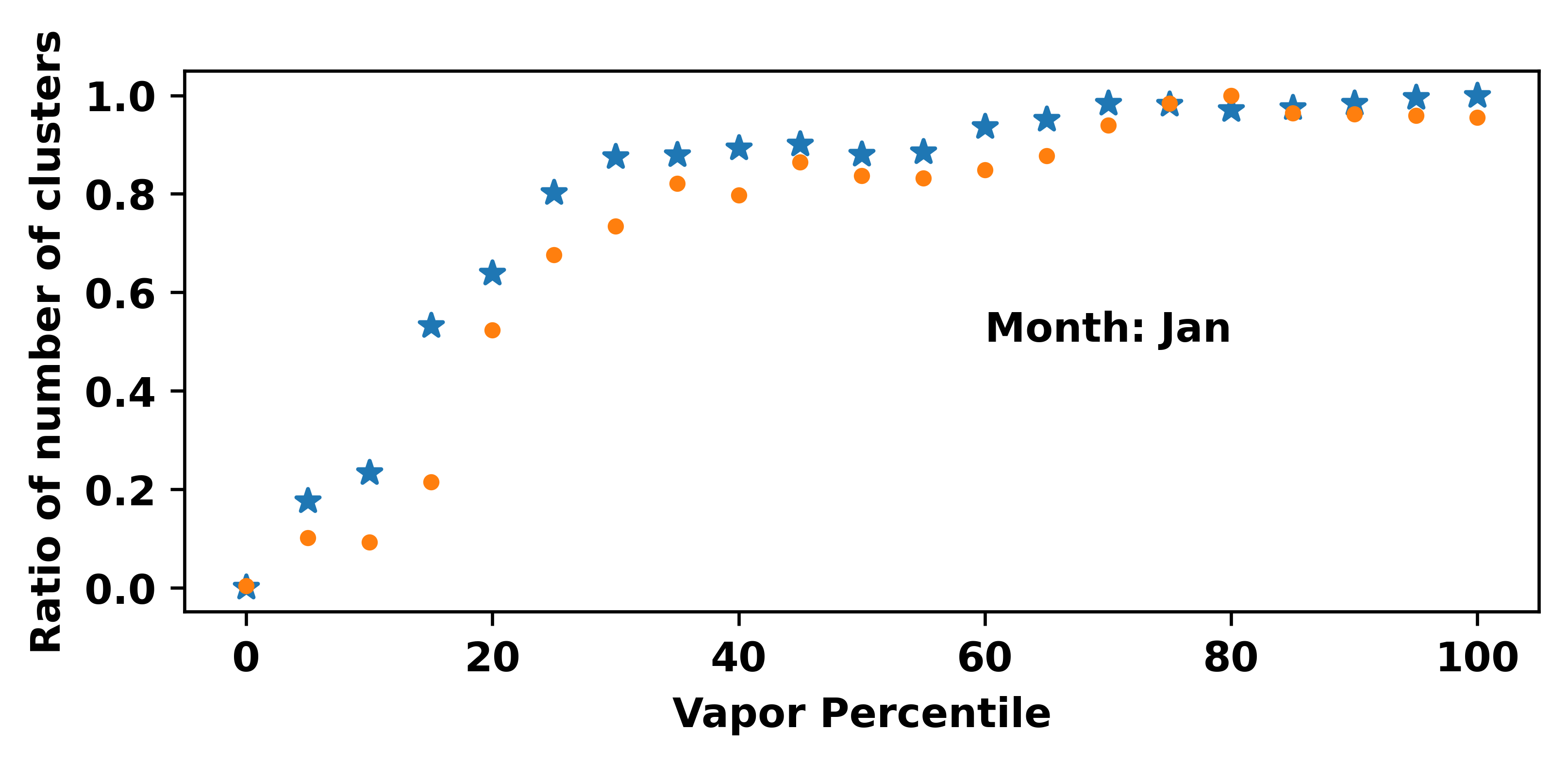} 
	\end{subfigure}
	\begin{subfigure}[t]{0.3\textwidth}
		\centering
		\includegraphics[width=\linewidth]{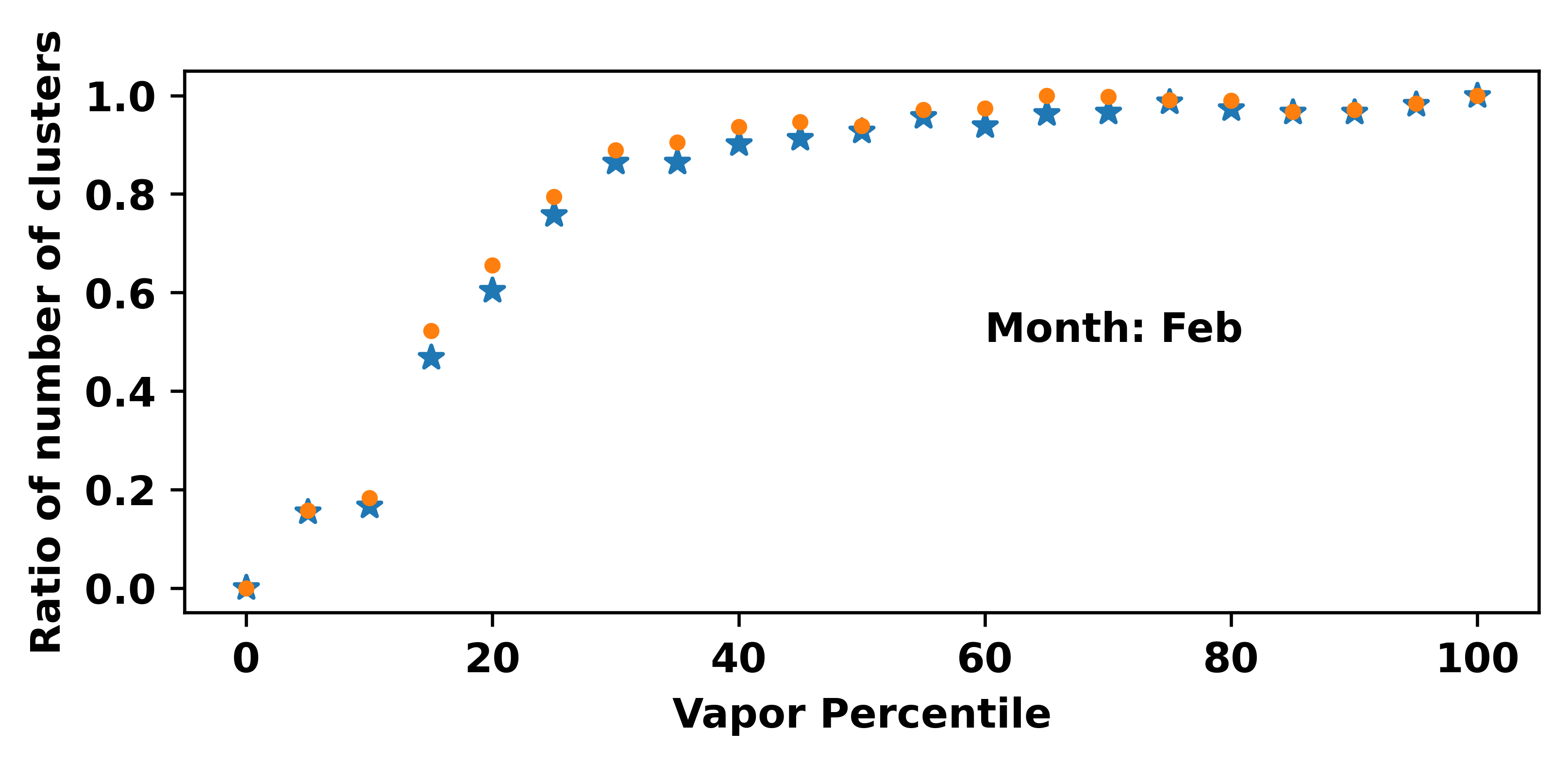} 
	\end{subfigure}
	\begin{subfigure}[t]{0.3\textwidth}
		\centering
		\includegraphics[width=\linewidth]{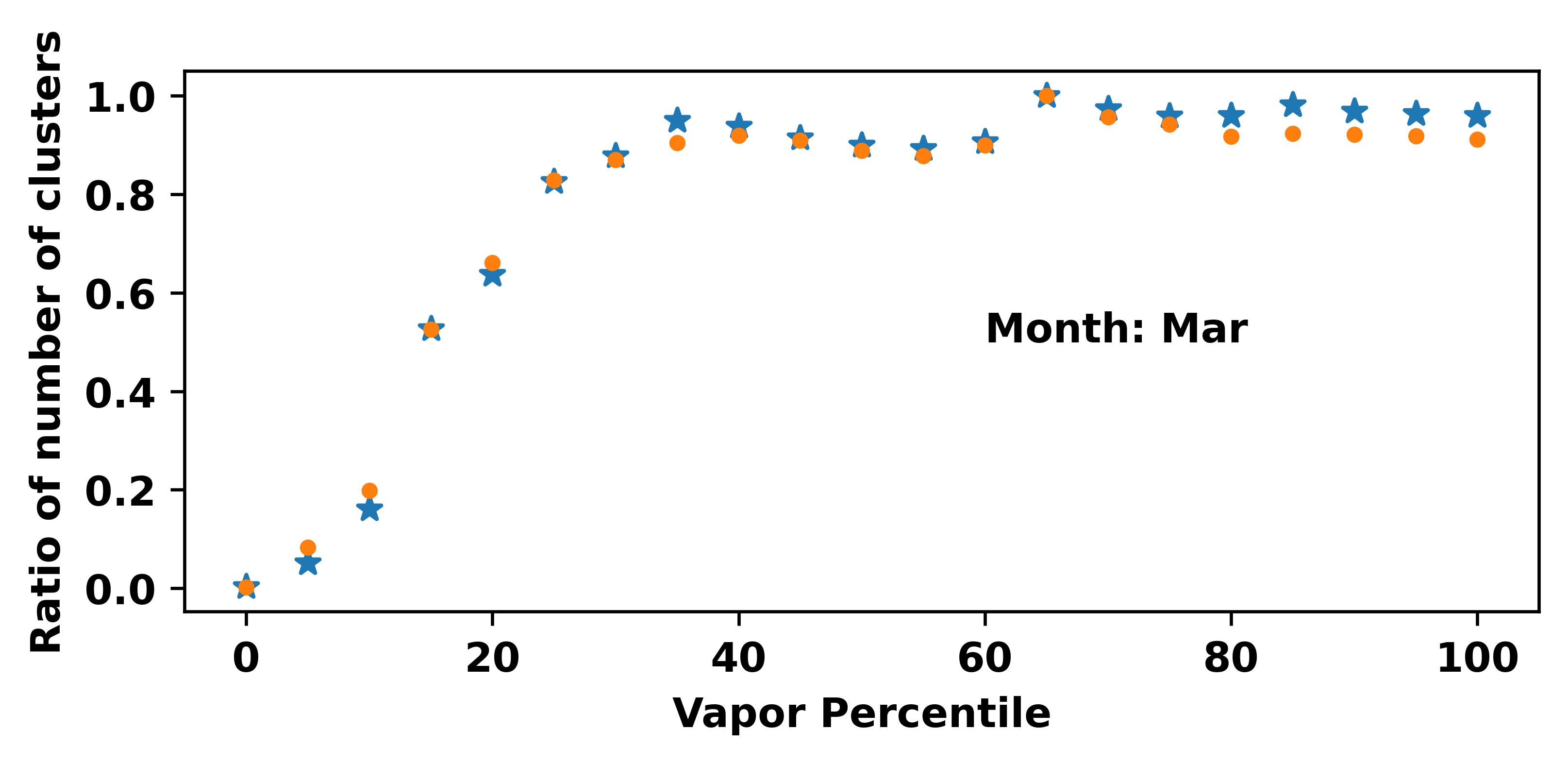} 
	\end{subfigure}		
	\begin{subfigure}[t]{0.3\textwidth}
		\centering
		\includegraphics[width=\linewidth]{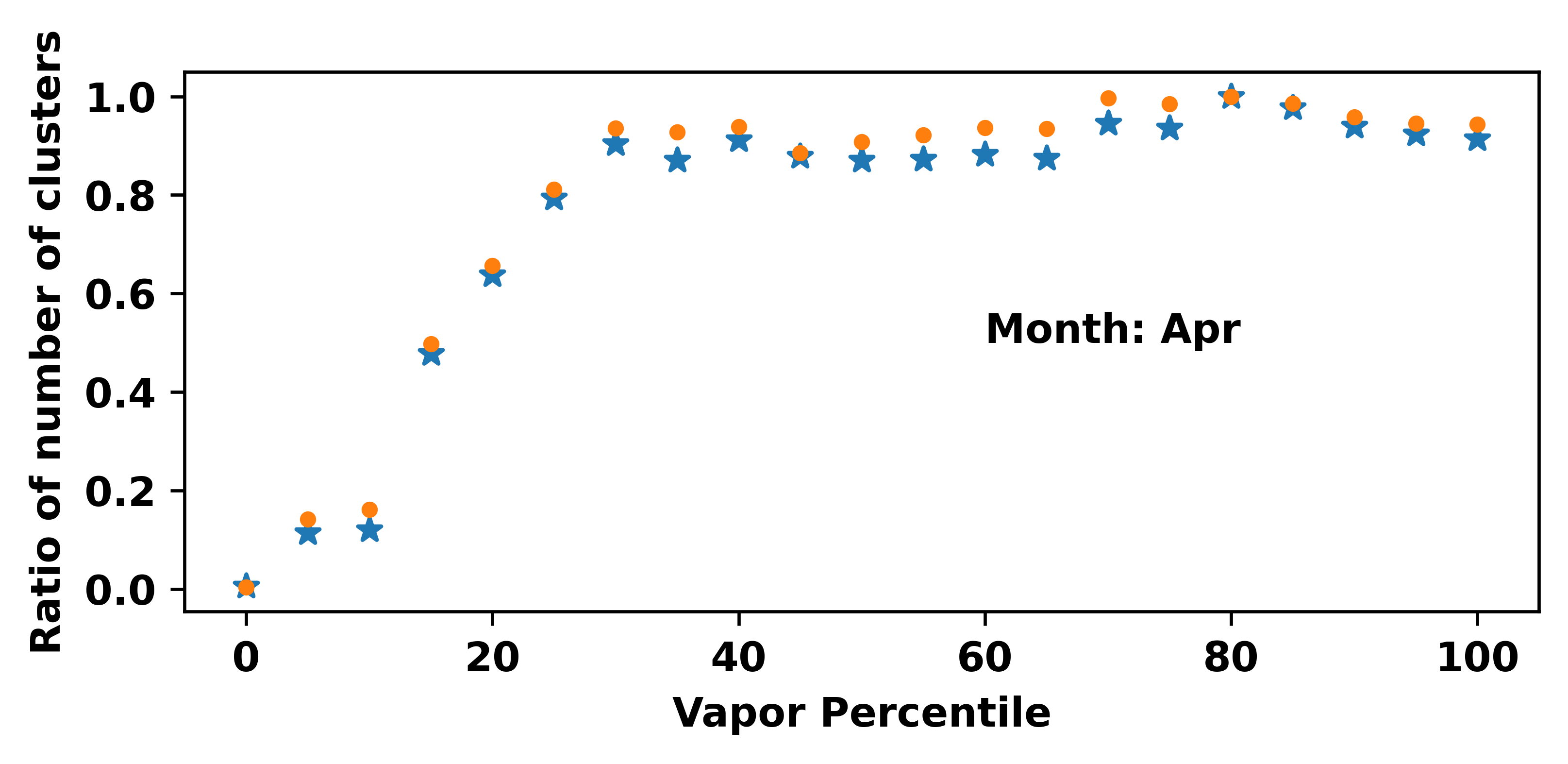} 
	\end{subfigure}
	\begin{subfigure}[t]{0.3\textwidth}
		\centering
		\includegraphics[width=\linewidth]{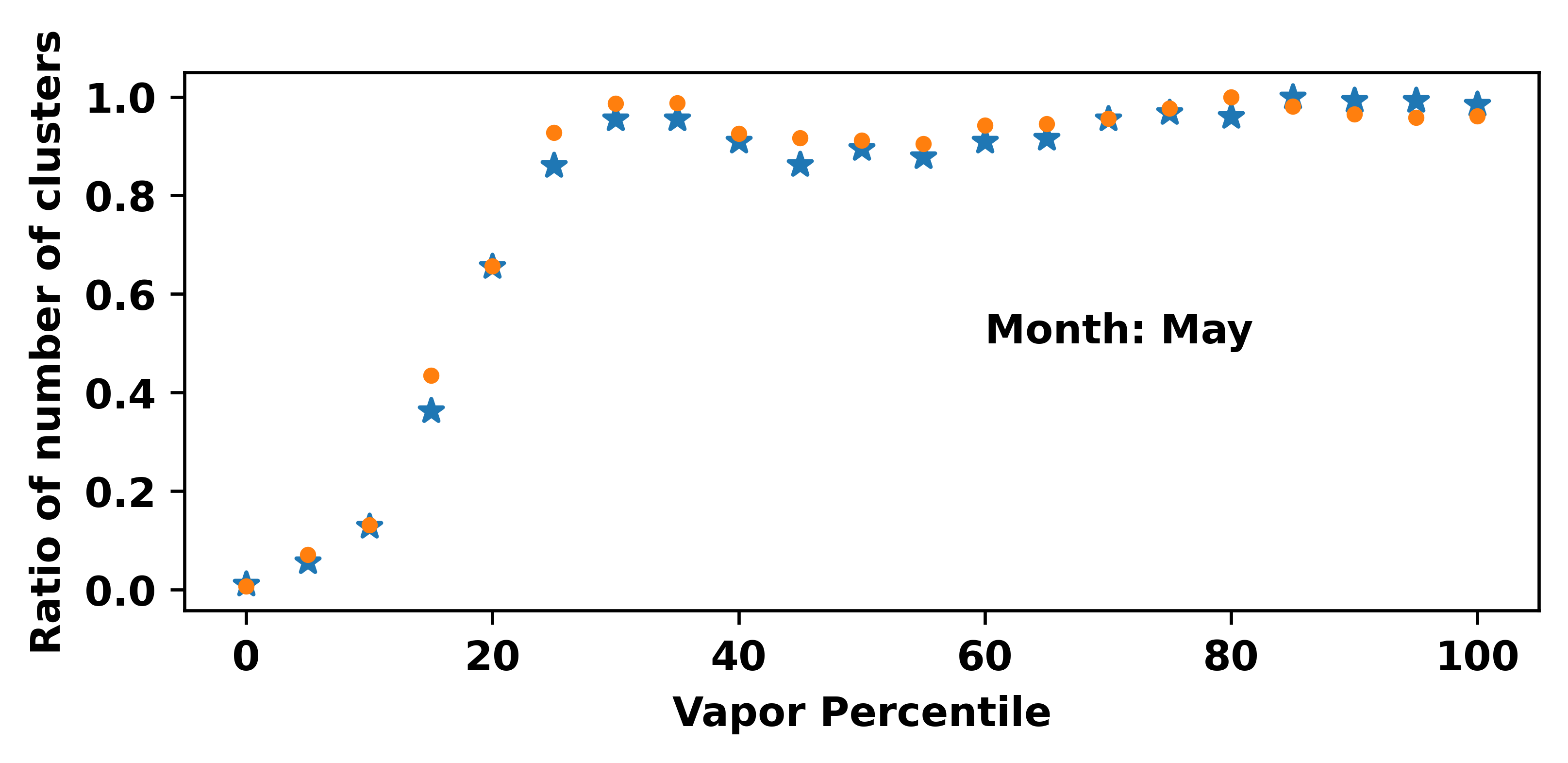} 
	\end{subfigure}
	\begin{subfigure}[t]{0.3\textwidth}
		\centering
		\includegraphics[width=\linewidth]{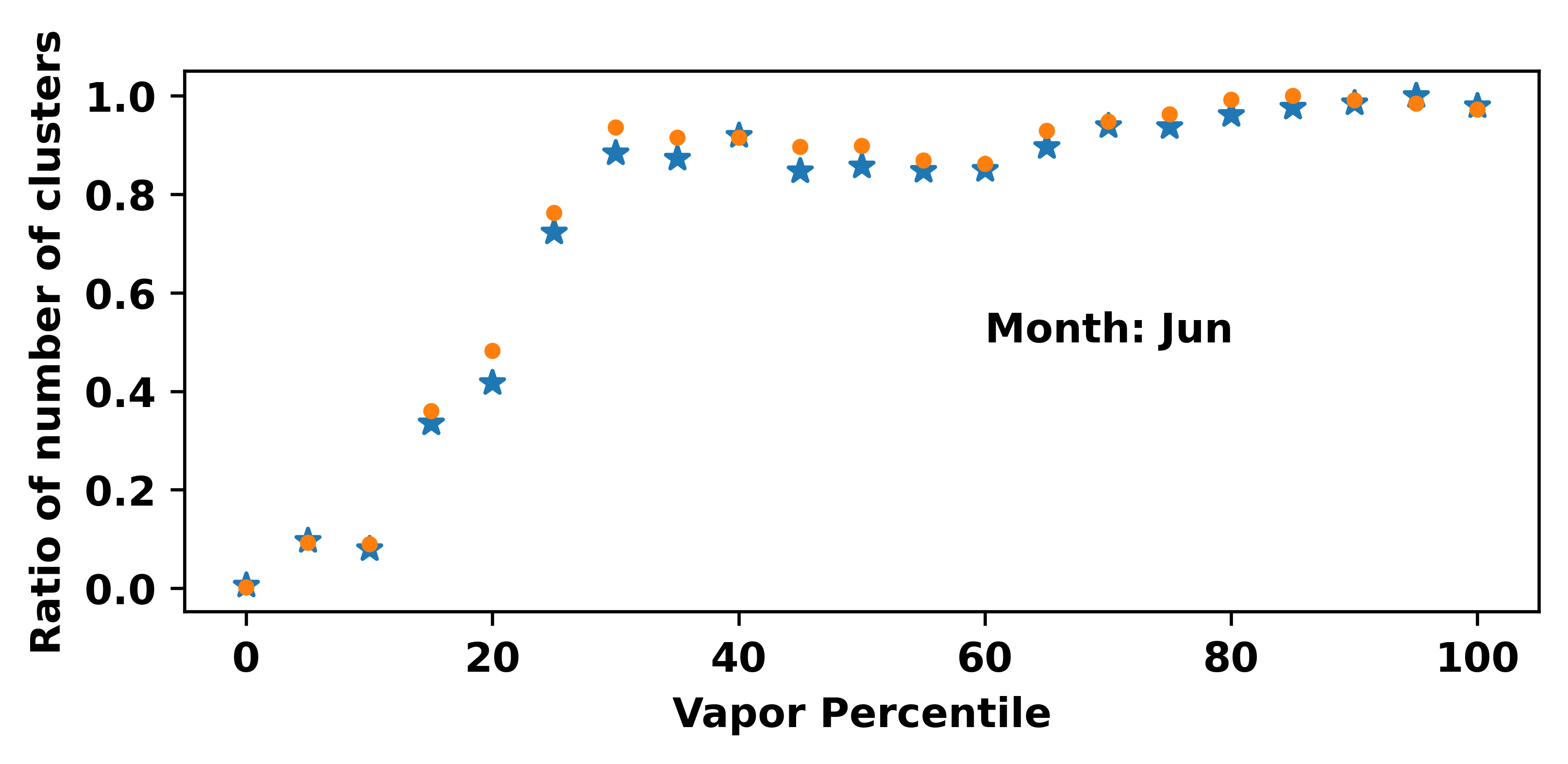} 
	\end{subfigure}		
	\begin{subfigure}[t]{0.3\textwidth}
		\centering
		\includegraphics[width=\linewidth]{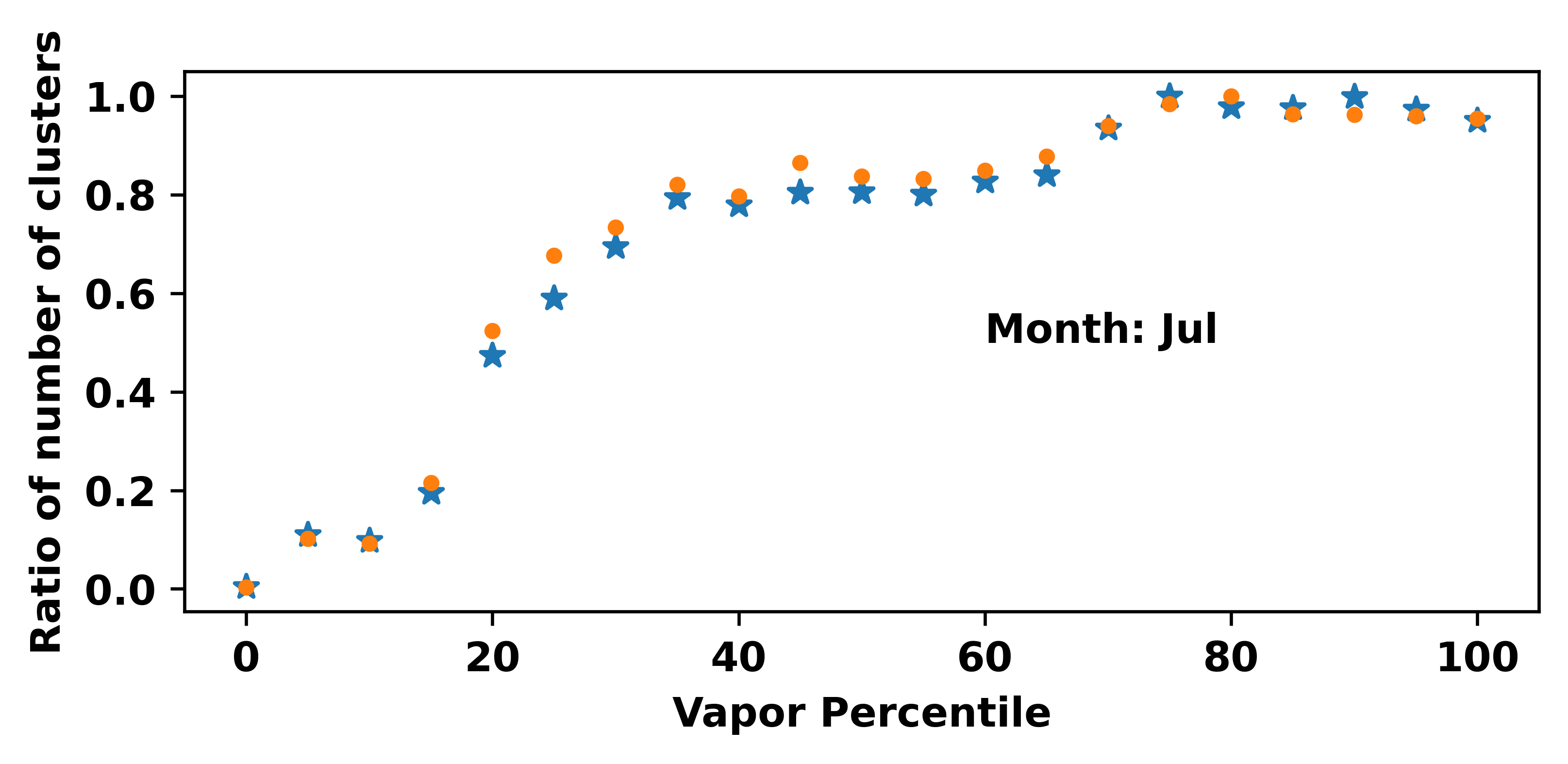} 
	\end{subfigure}
	\begin{subfigure}[t]{0.3\textwidth}
		\centering
		\includegraphics[width=\linewidth]{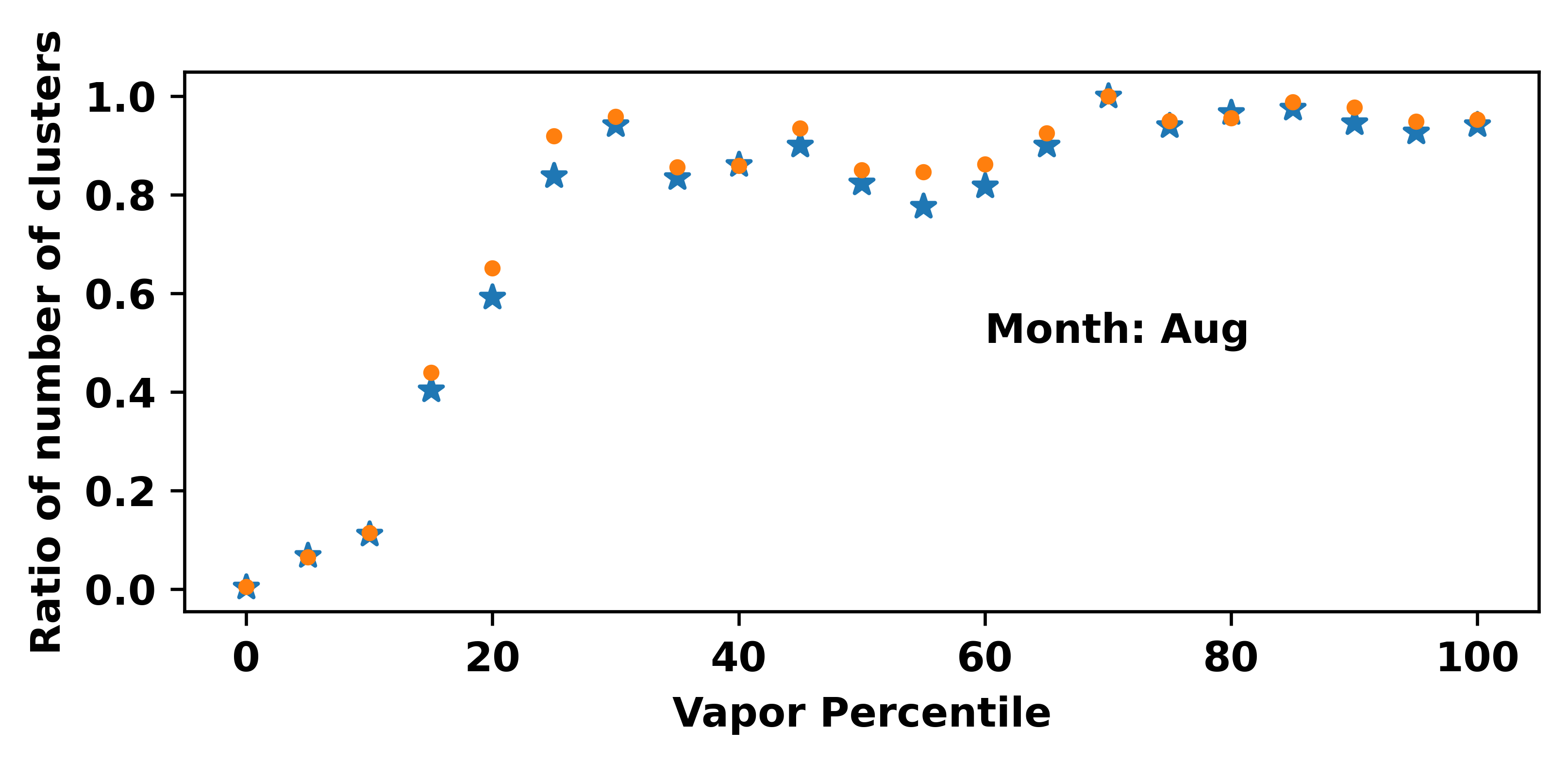} 
	\end{subfigure}
	\begin{subfigure}[t]{0.3\textwidth}
		\centering
		\includegraphics[width=\linewidth]{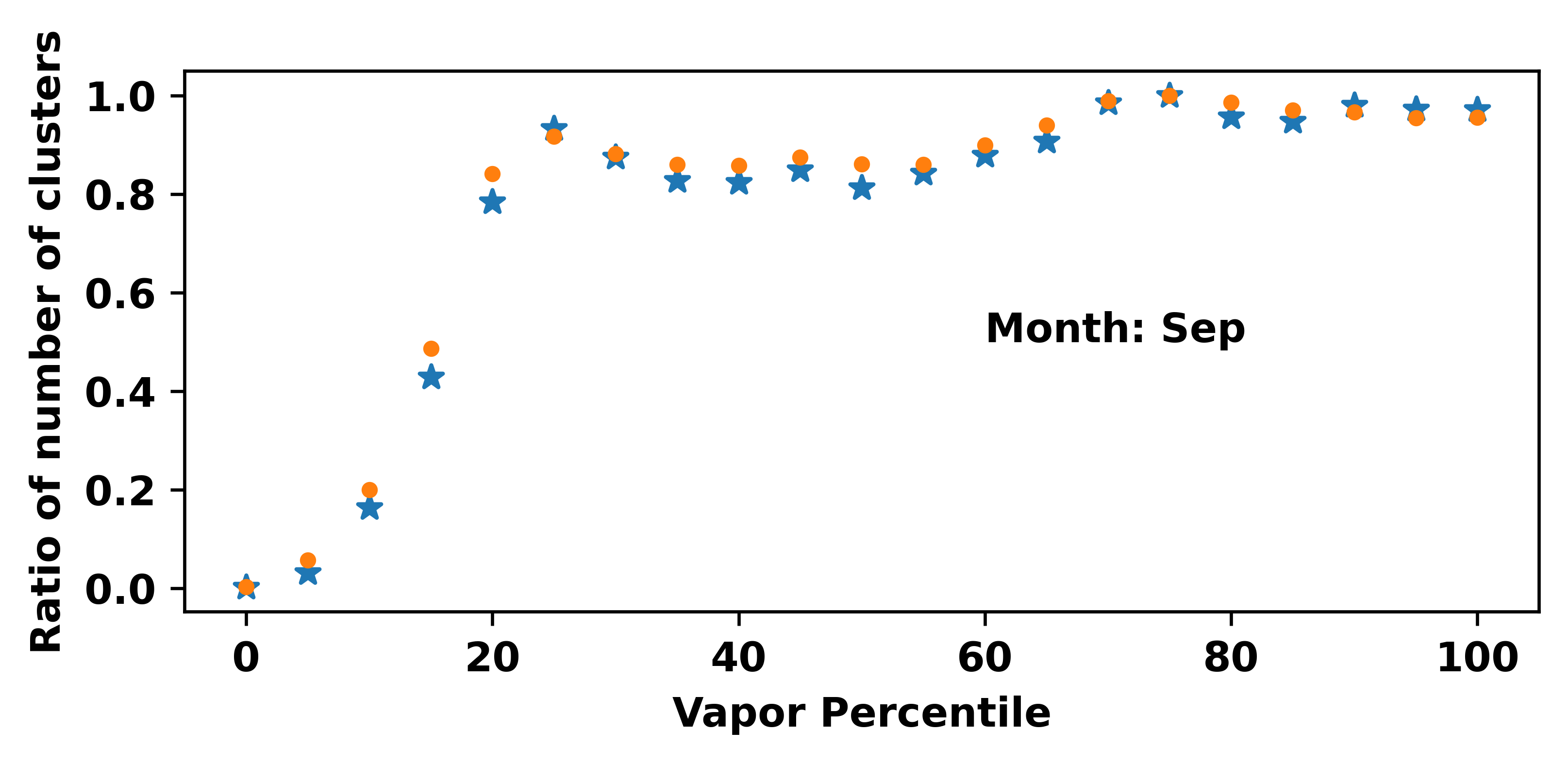} 
	\end{subfigure}		
	\begin{subfigure}[t]{0.3\textwidth}
		\centering
		\includegraphics[width=\linewidth]{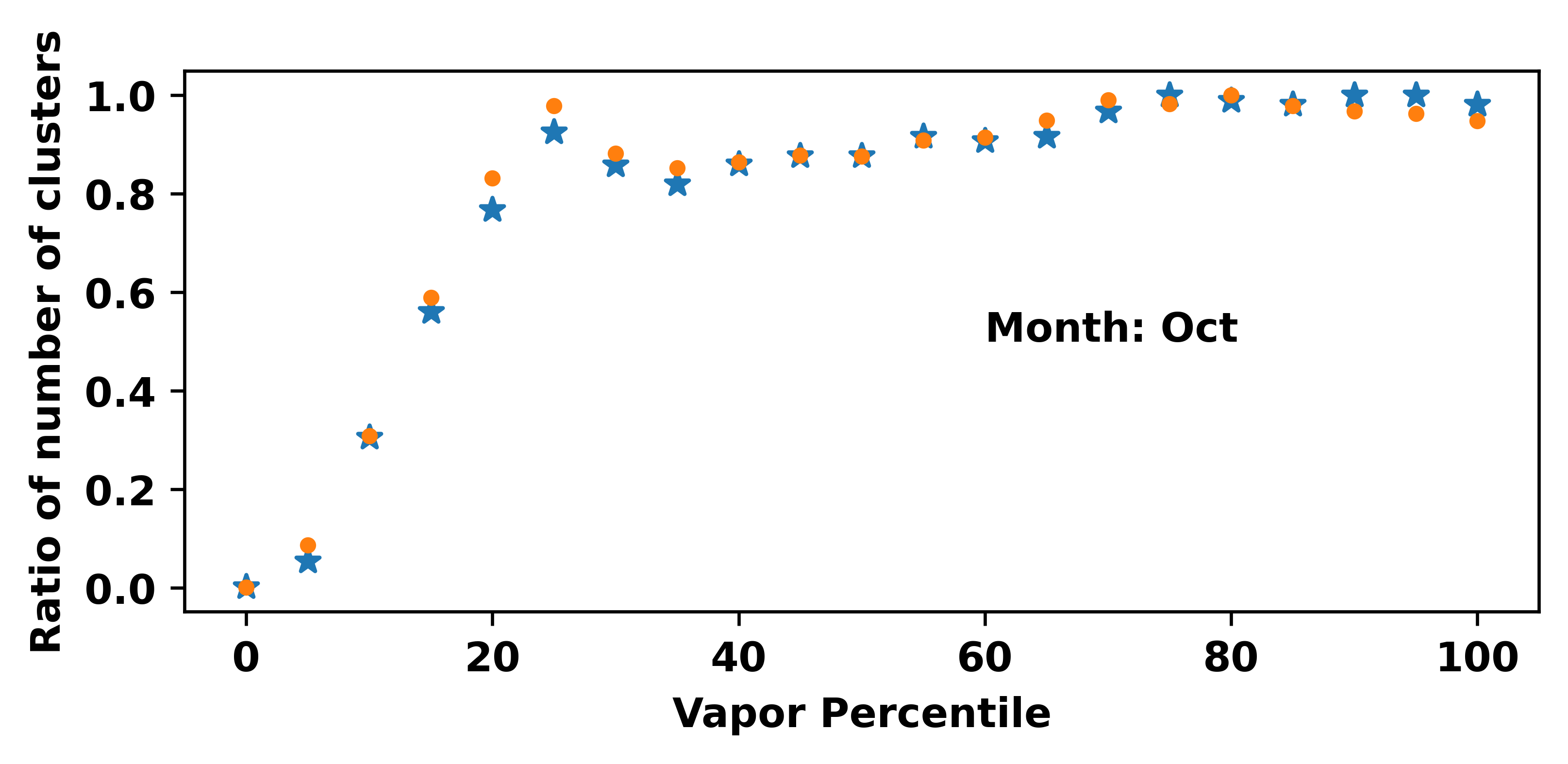} 
	\end{subfigure}
	\begin{subfigure}[t]{0.3\textwidth}
		\centering
		\includegraphics[width=\linewidth]{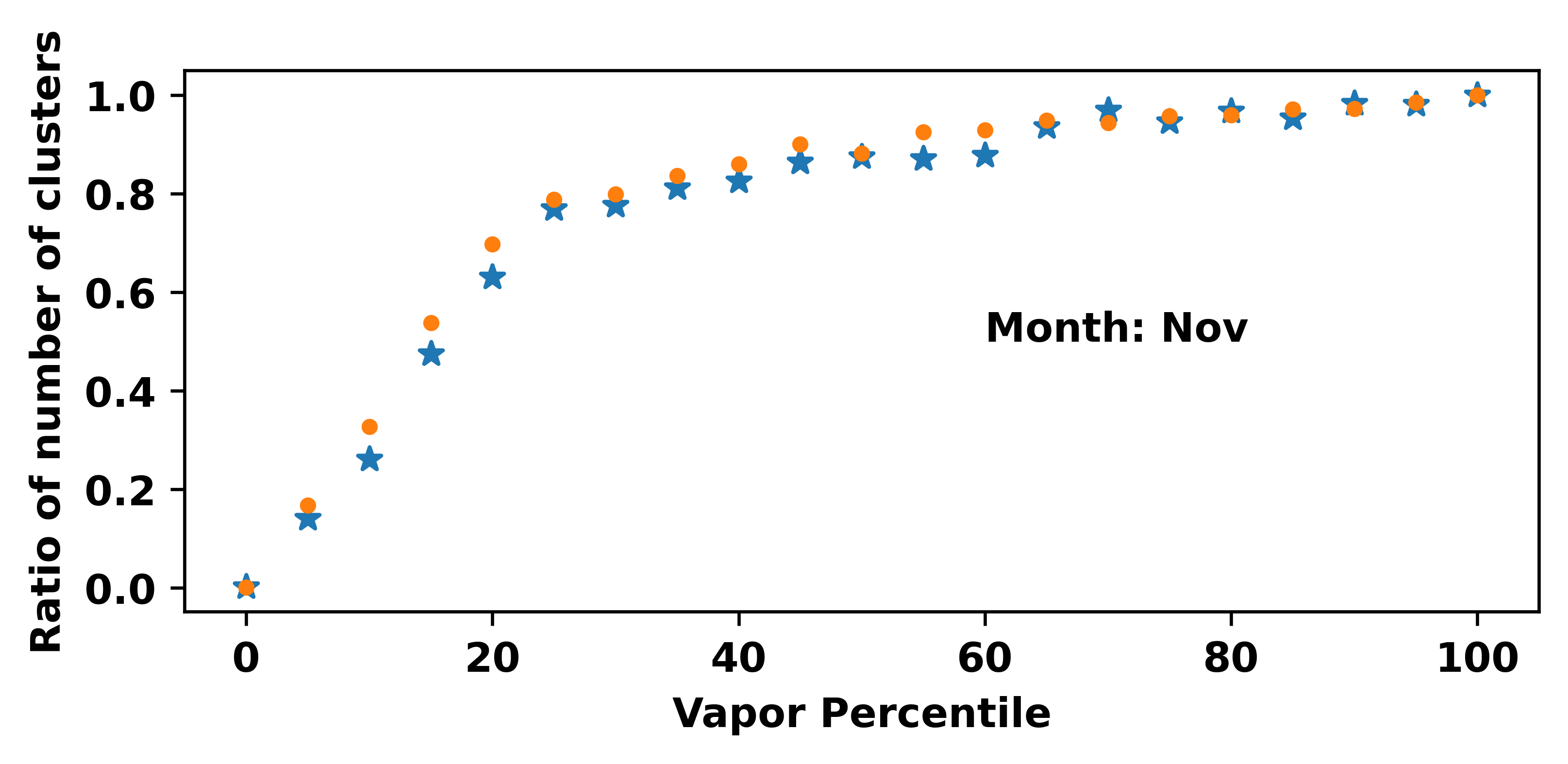} 
	\end{subfigure}
	\begin{subfigure}[t]{0.3\textwidth}
		\centering
		\includegraphics[width=\linewidth]{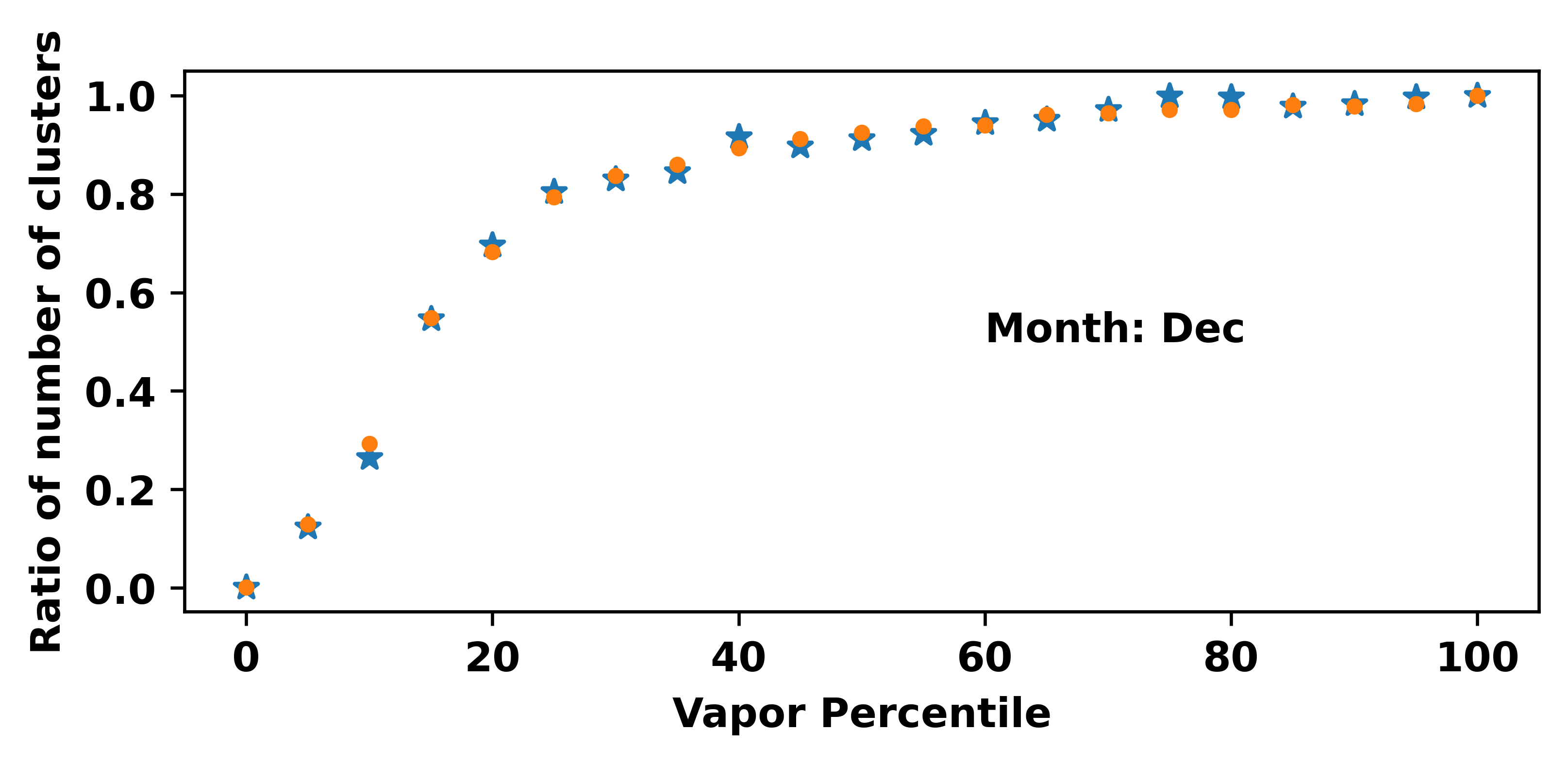} 
	\end{subfigure}		
	\caption{Scaled Number of Clusters vs Vapor Percentile for all 12 months in 2021. The blue markers represent the $180\times 360$ pixel data and the orange markers represent the $360\times 720$ pixel data.}
	\label{3}
\end{figure}

The results are plotted in Fig. \ref{3} for all 12 months in 2021 at both resolutions. Strikingly, we find once again, an almost identical distribution at both resolutions, further strengthening our claim of scale-invariance. We observe a general trend that above $V = 30\%$ the number of clusters remain approximately constant, while below this percentile the number of clusters drop linearly to $1$ at $V=0$. This indicates that a percolation phenomenon exists, such that below the $30$th percentile the clusters aggregate to form larger and larger clusters. The percolation threshold is defined as the vapor percentile at which no cluster exists which connects the top to the bottom of the dataset and the left to the right. It is estimated at the disconnection point visually from the raw data (not shown here) to be at $V_{thres} = 55 \% \pm 5 \%$ for all months. From hereon out, we decide to neglect the $\pm 5 \%$ and refer to the percolation threshold as $V_{thres} = 55\%$, while noting that for our data the threshold is not a point in parameter space but a small range. The occupation probability ($p$) at the percolation threshold percentile is approximately $0.1$ which is very distinct from $p_c = 0.59$ for the Bernoulli class of transitions. Thus we find that this transition is not in the same universality class as the Bernoulli transition. We make no attempt in this work to determine the critical exponents associated with this transition, but relegate it to a future work. The anomaly presented in the previous subsection is also present in this case for the month of January.

\subsubsection{Maximum Cluster Size}

\begin{figure}[H]
	\centering
	\begin{subfigure}[t]{0.3\textwidth}
		\centering
		\includegraphics[width=\linewidth]{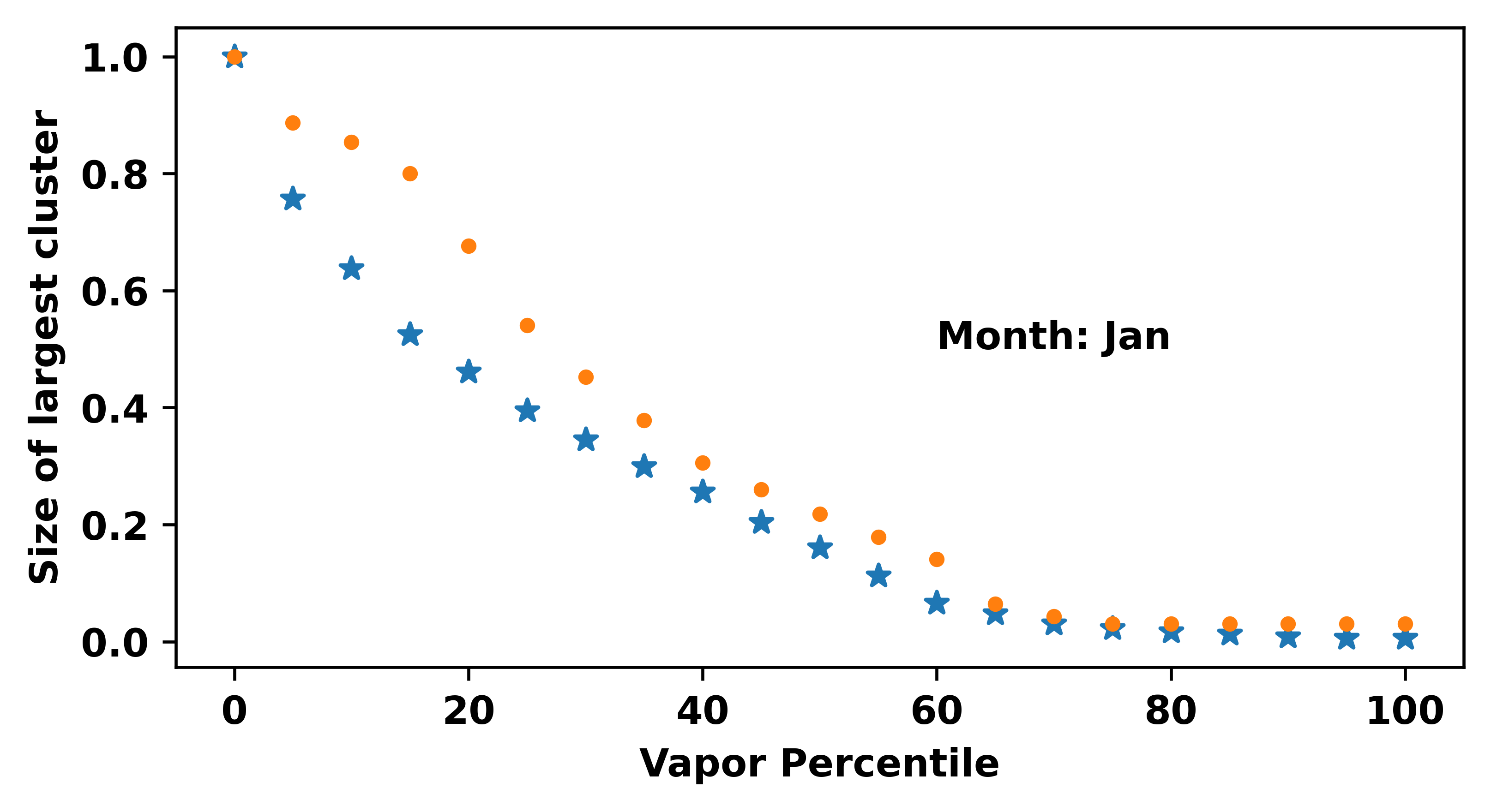} 
	\end{subfigure}
	\begin{subfigure}[t]{0.3\textwidth}
		\centering
		\includegraphics[width=\linewidth]{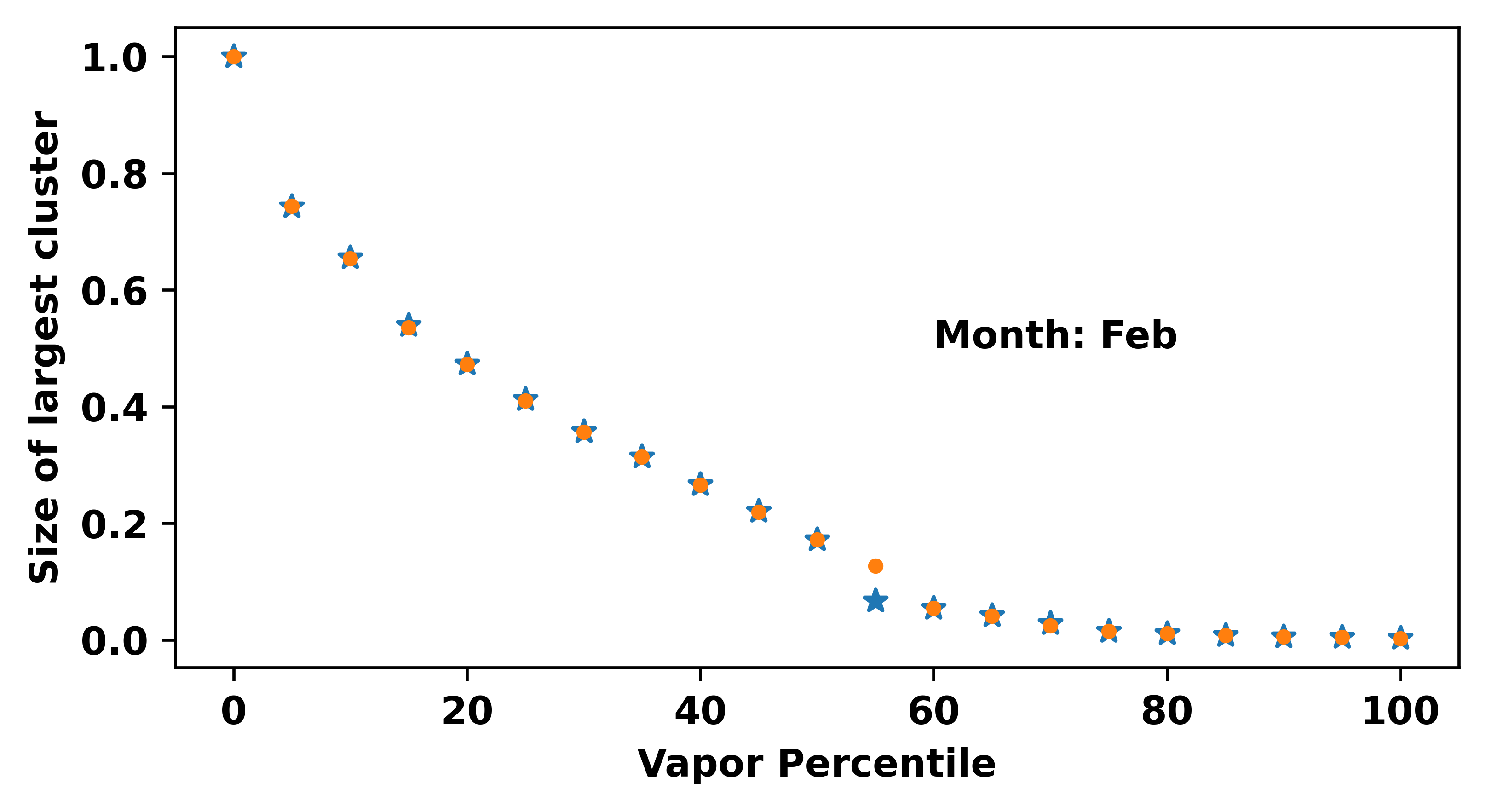} 
	\end{subfigure}
	\begin{subfigure}[t]{0.3\textwidth}
		\centering
		\includegraphics[width=\linewidth]{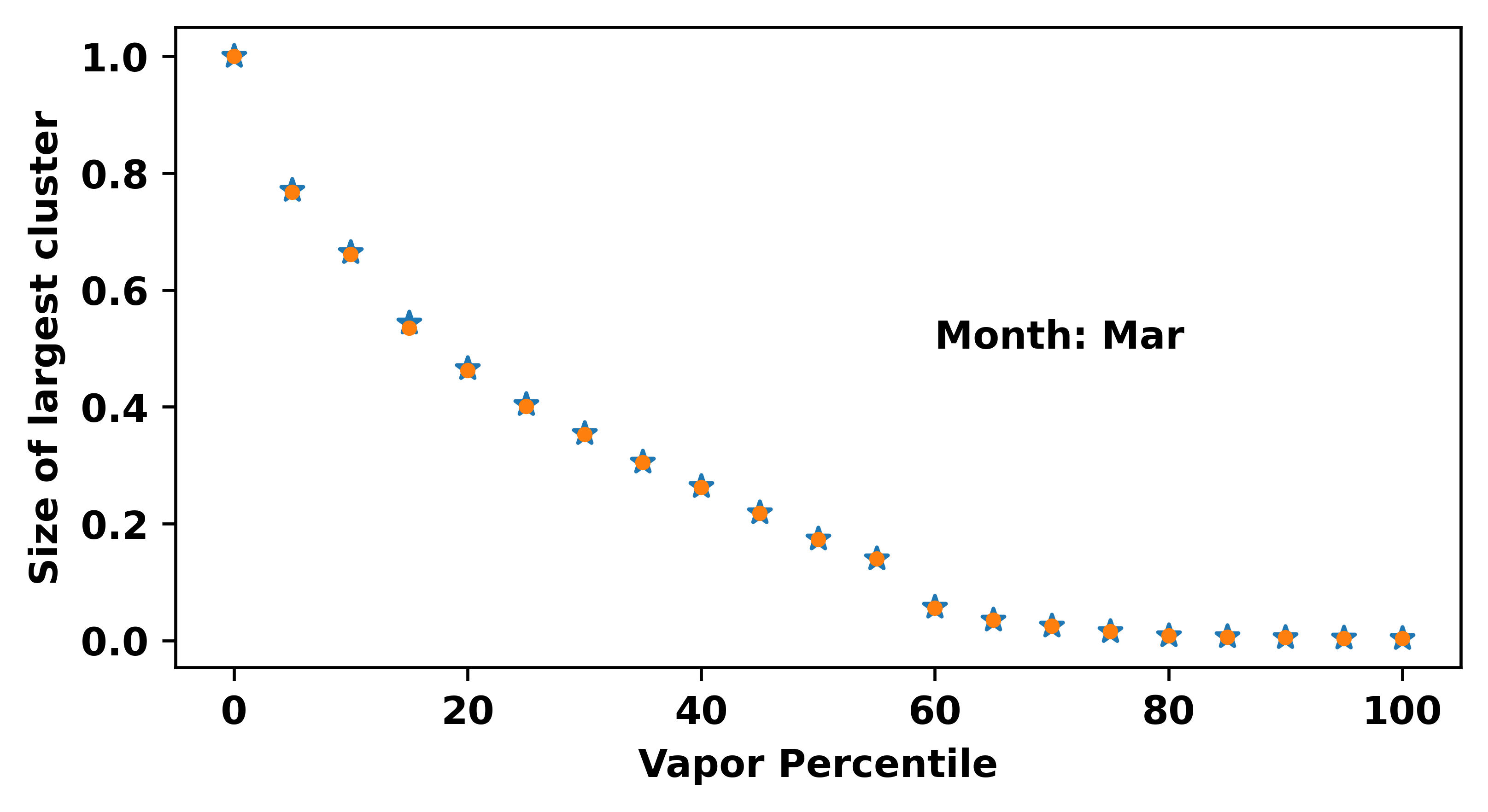} 
	\end{subfigure}		
	\begin{subfigure}[t]{0.3\textwidth}
		\centering
		\includegraphics[width=\linewidth]{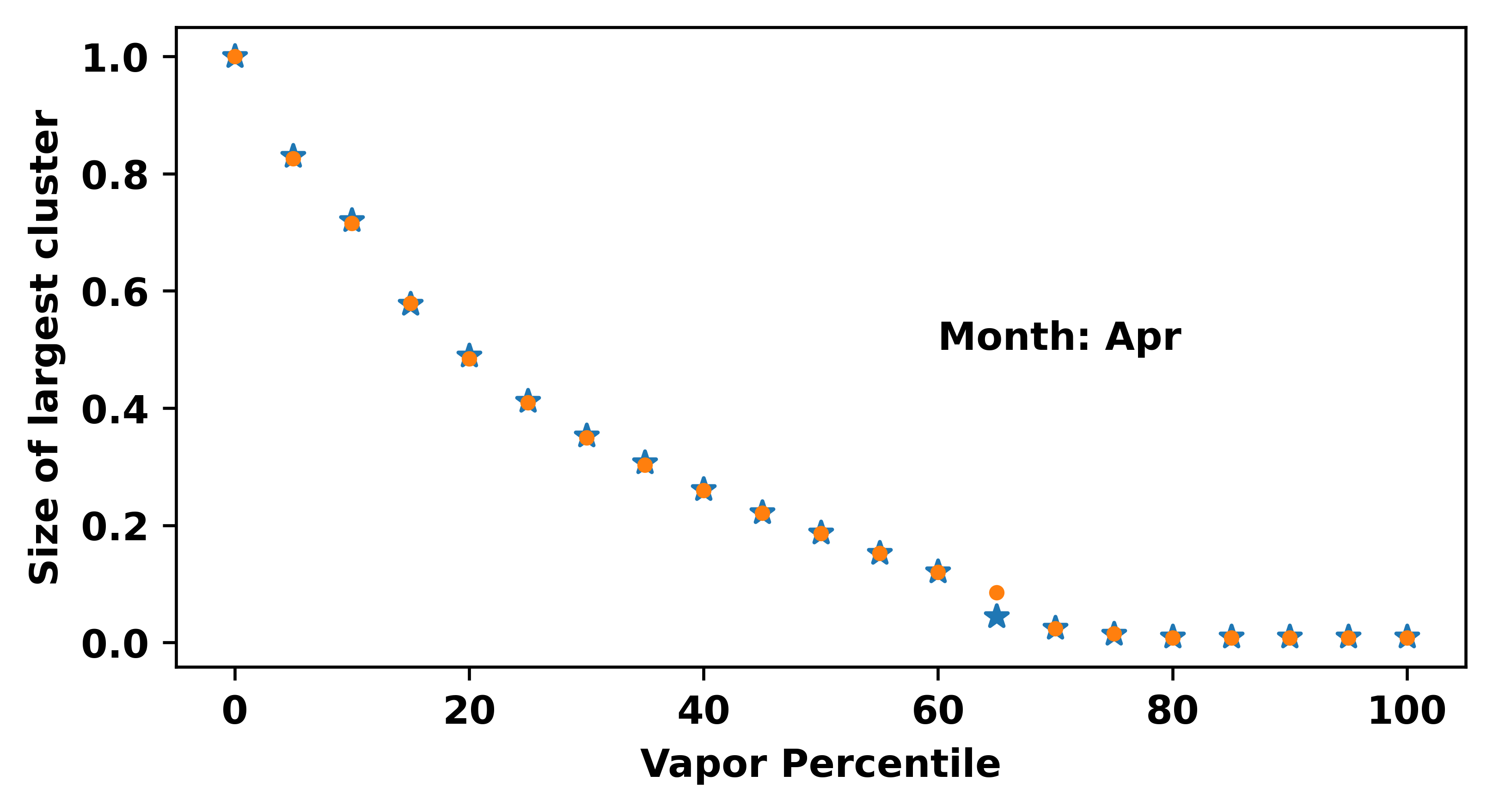} 
	\end{subfigure}
	\begin{subfigure}[t]{0.3\textwidth}
		\centering
		\includegraphics[width=\linewidth]{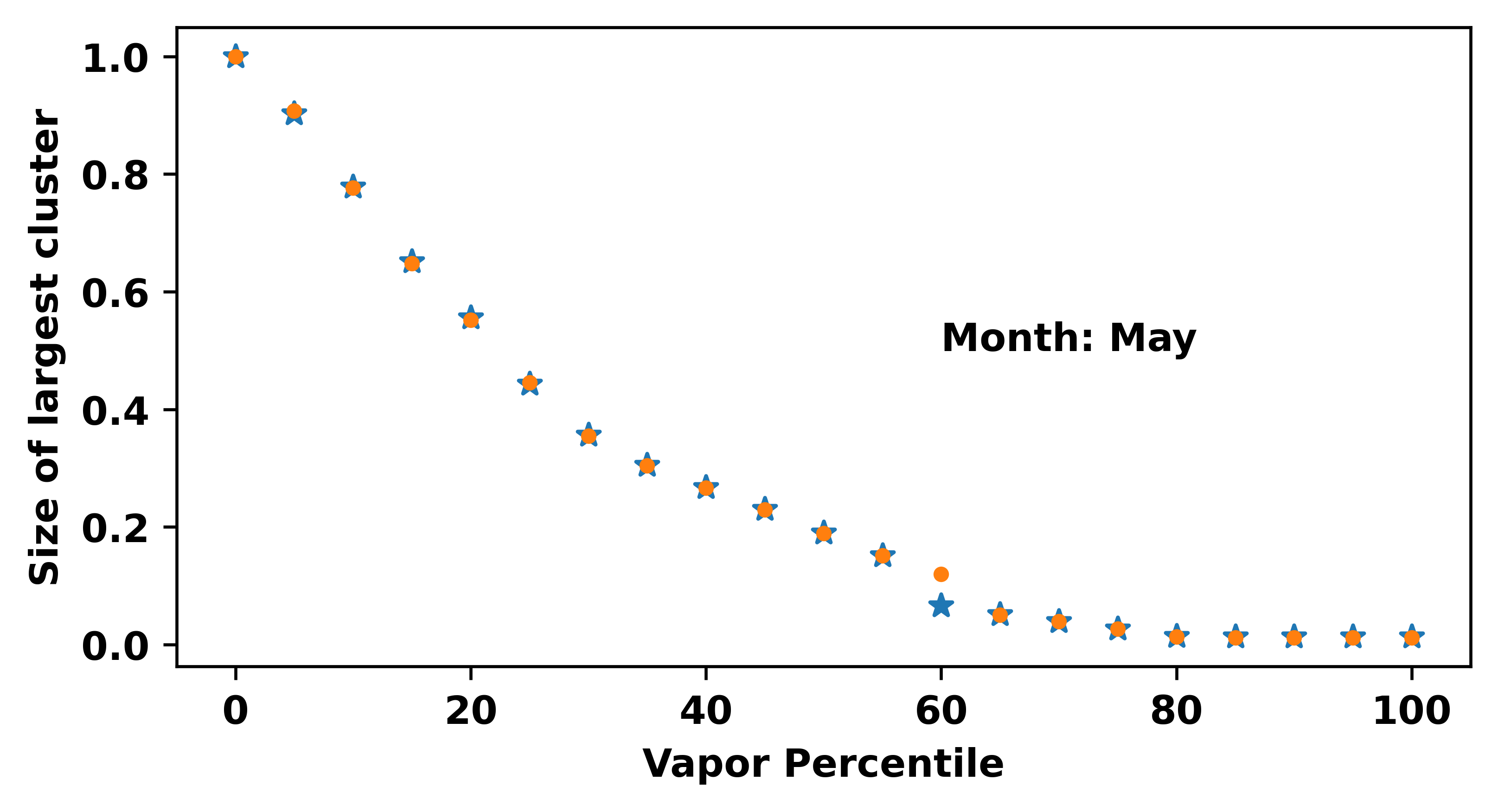} 
	\end{subfigure}
	\begin{subfigure}[t]{0.3\textwidth}
		\centering
		\includegraphics[width=\linewidth]{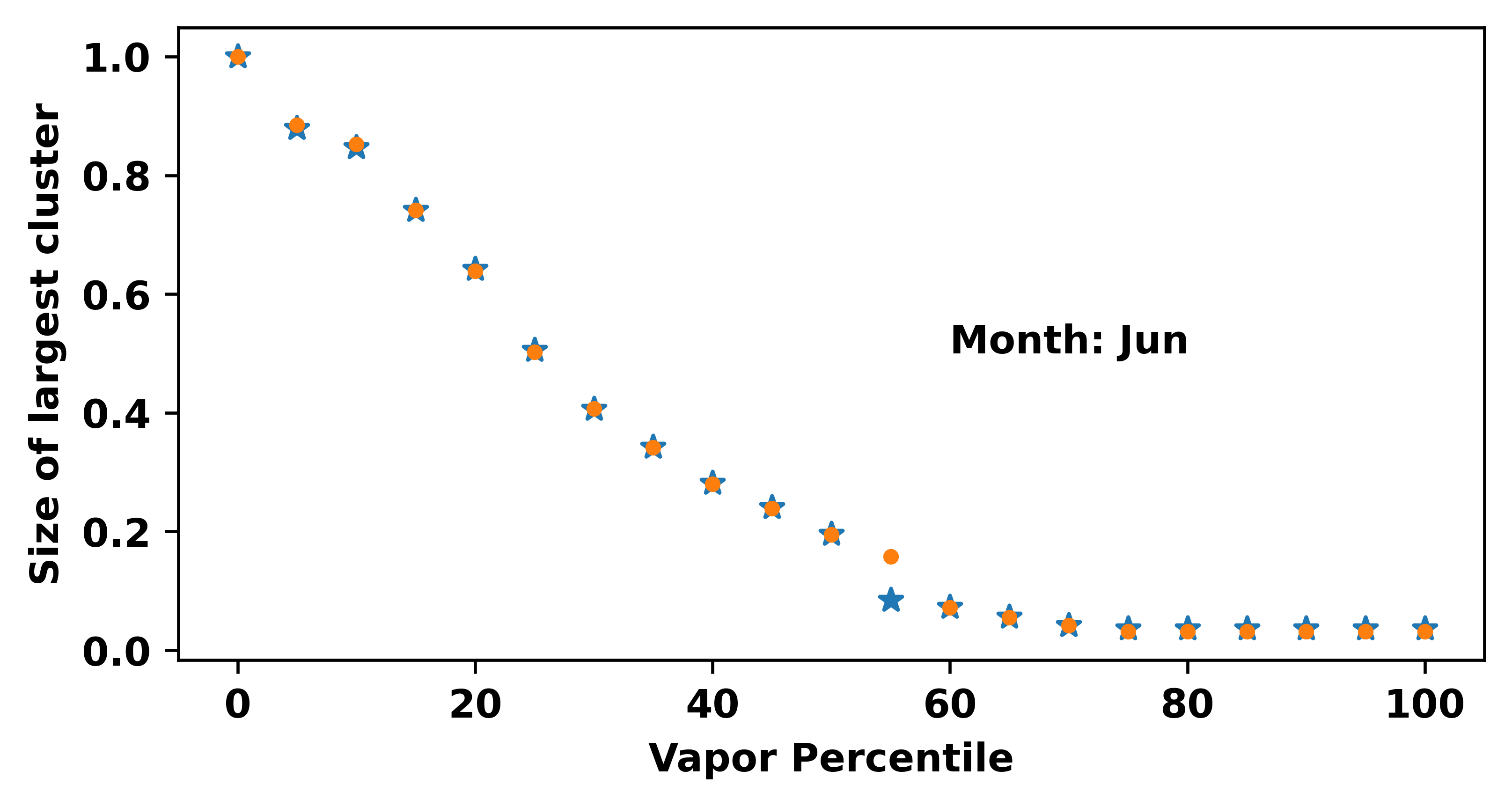} 
	\end{subfigure}		
	\begin{subfigure}[t]{0.3\textwidth}
		\centering
		\includegraphics[width=\linewidth]{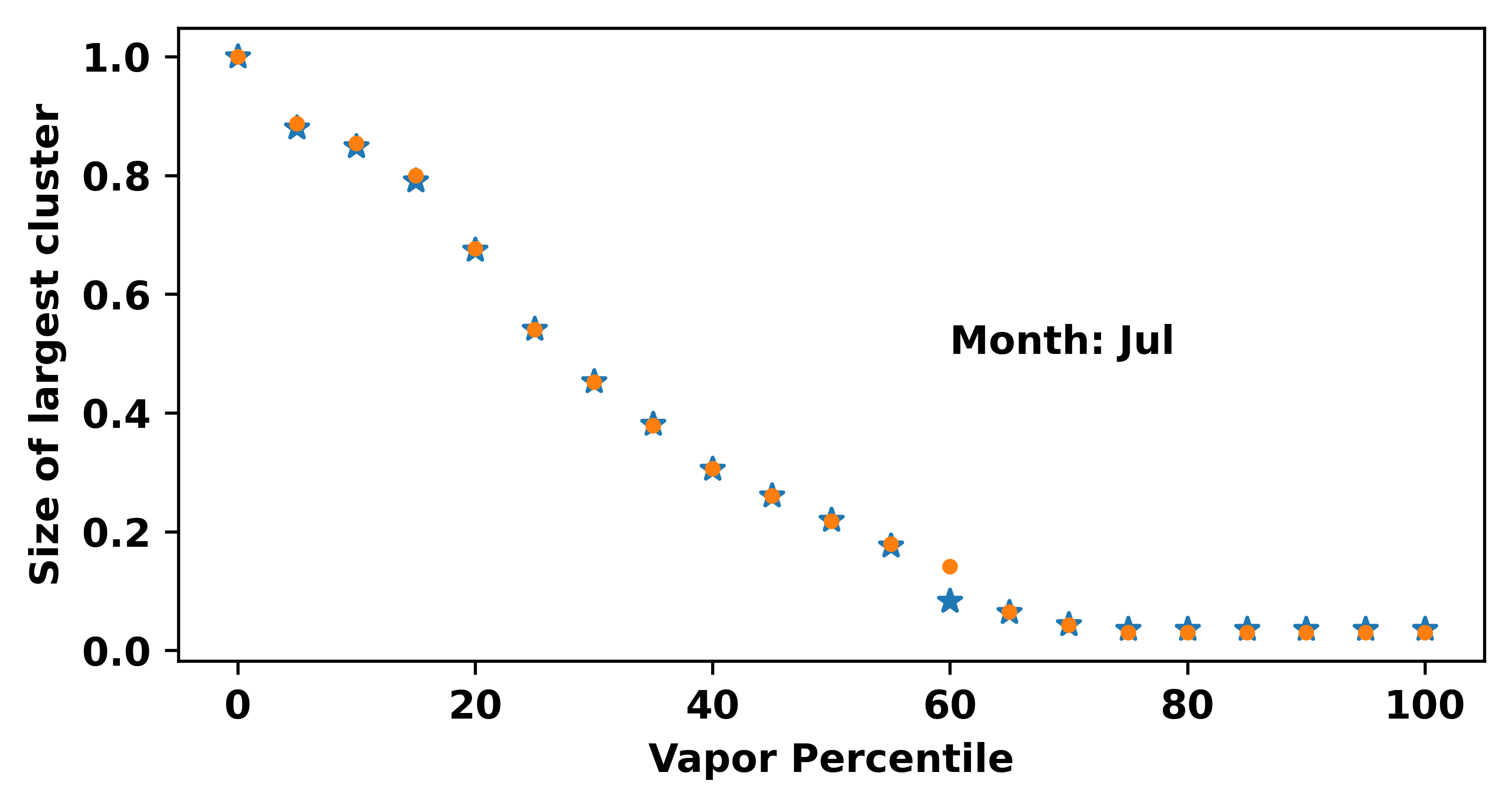} 
	\end{subfigure}
	\begin{subfigure}[t]{0.3\textwidth}
		\centering
		\includegraphics[width=\linewidth]{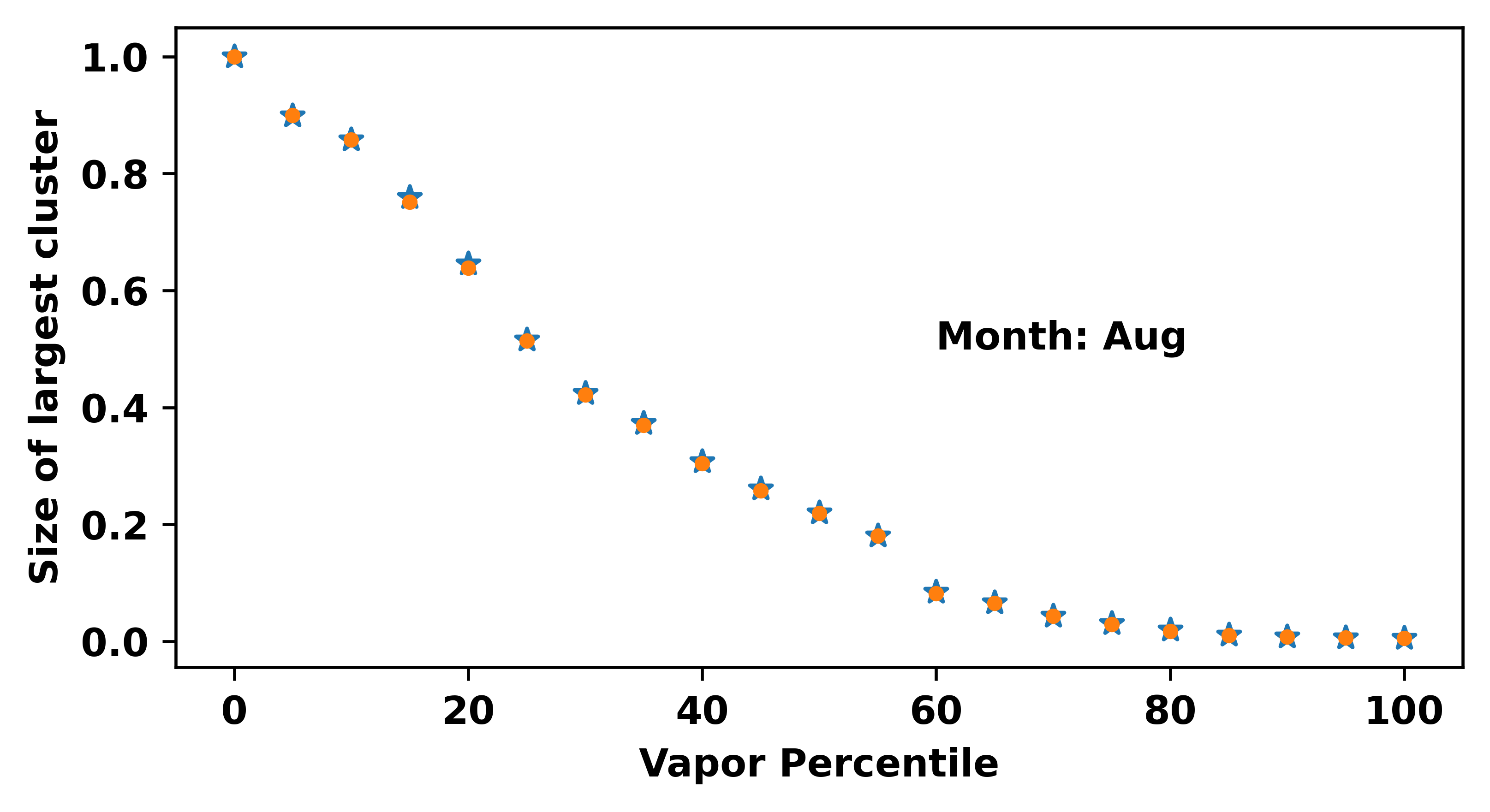} 
	\end{subfigure}
	\begin{subfigure}[t]{0.3\textwidth}
		\centering
		\includegraphics[width=\linewidth]{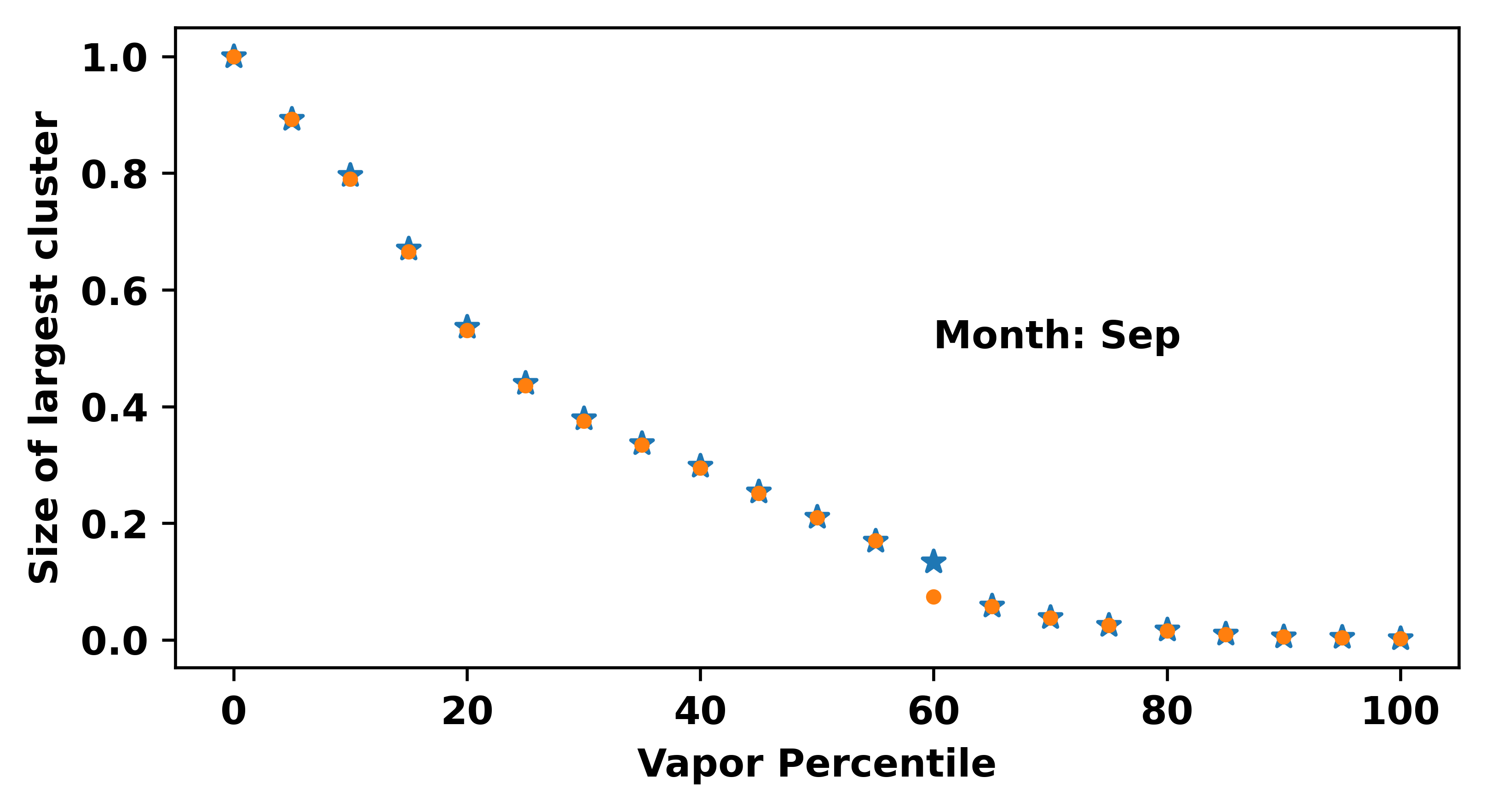} 
	\end{subfigure}		
	\begin{subfigure}[t]{0.3\textwidth}
		\centering
		\includegraphics[width=\linewidth]{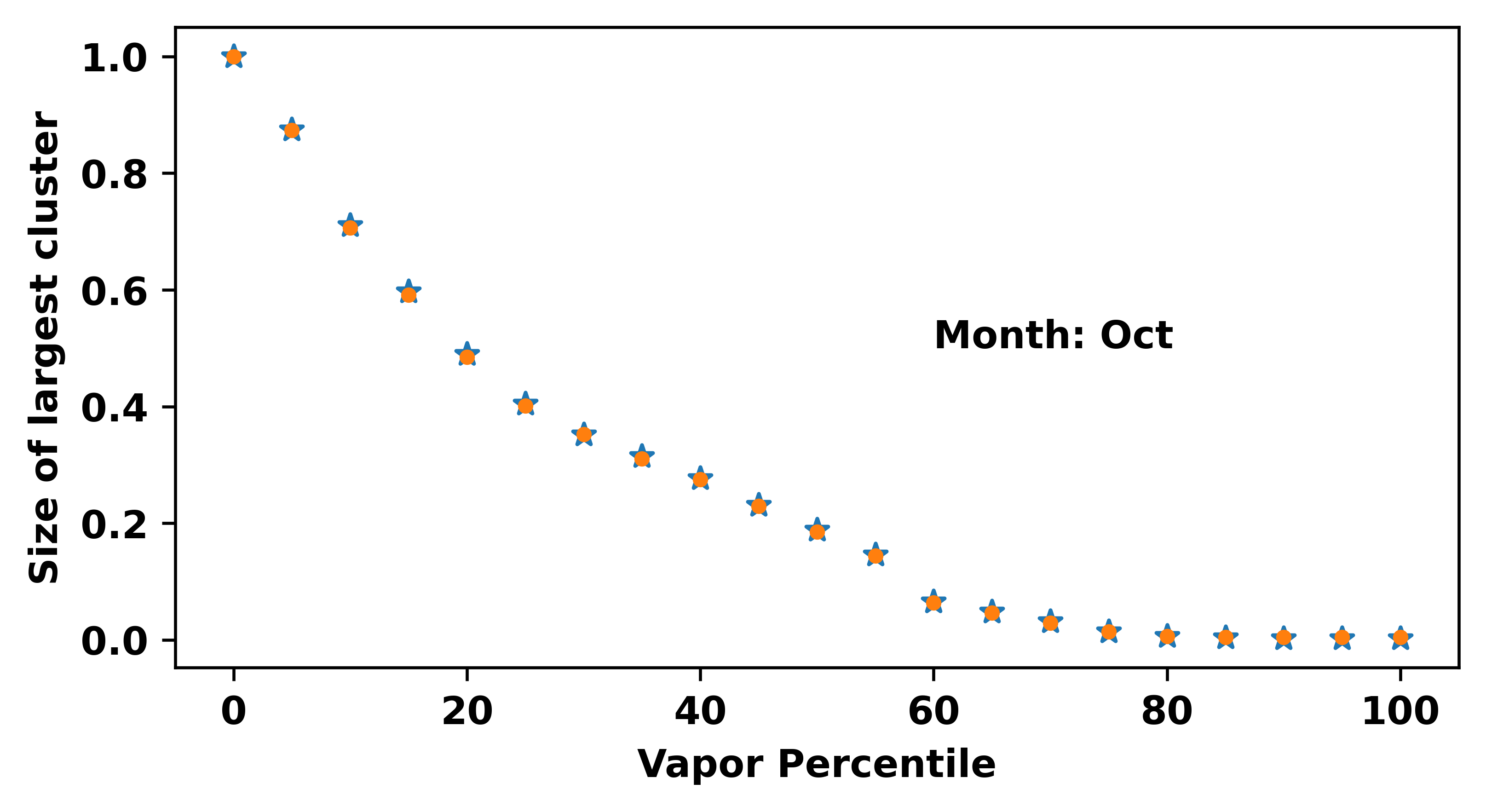} 
	\end{subfigure}
	\begin{subfigure}[t]{0.3\textwidth}
		\centering
		\includegraphics[width=\linewidth]{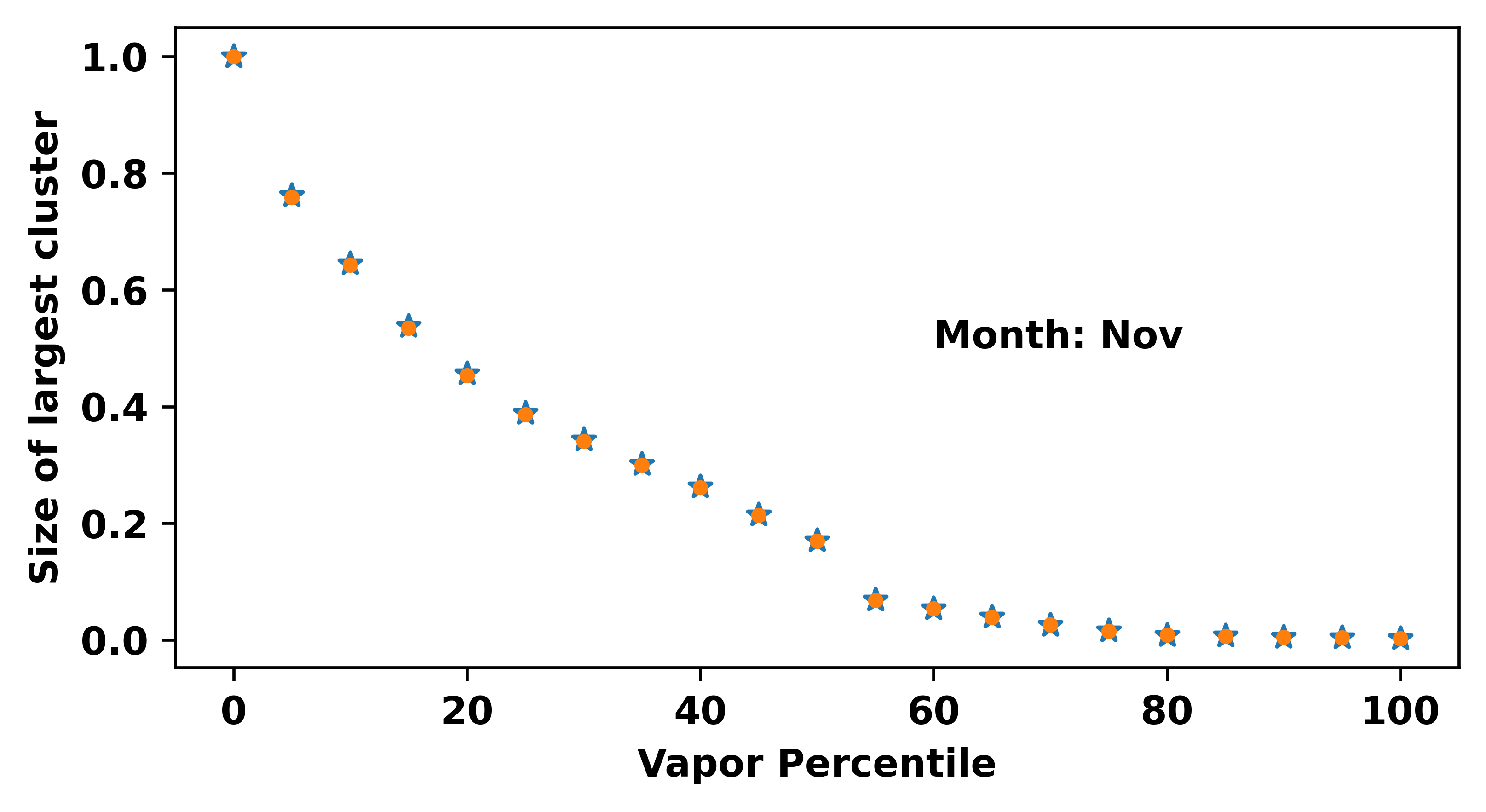} 
	\end{subfigure}
	\begin{subfigure}[t]{0.3\textwidth}
		\centering
		\includegraphics[width=\linewidth]{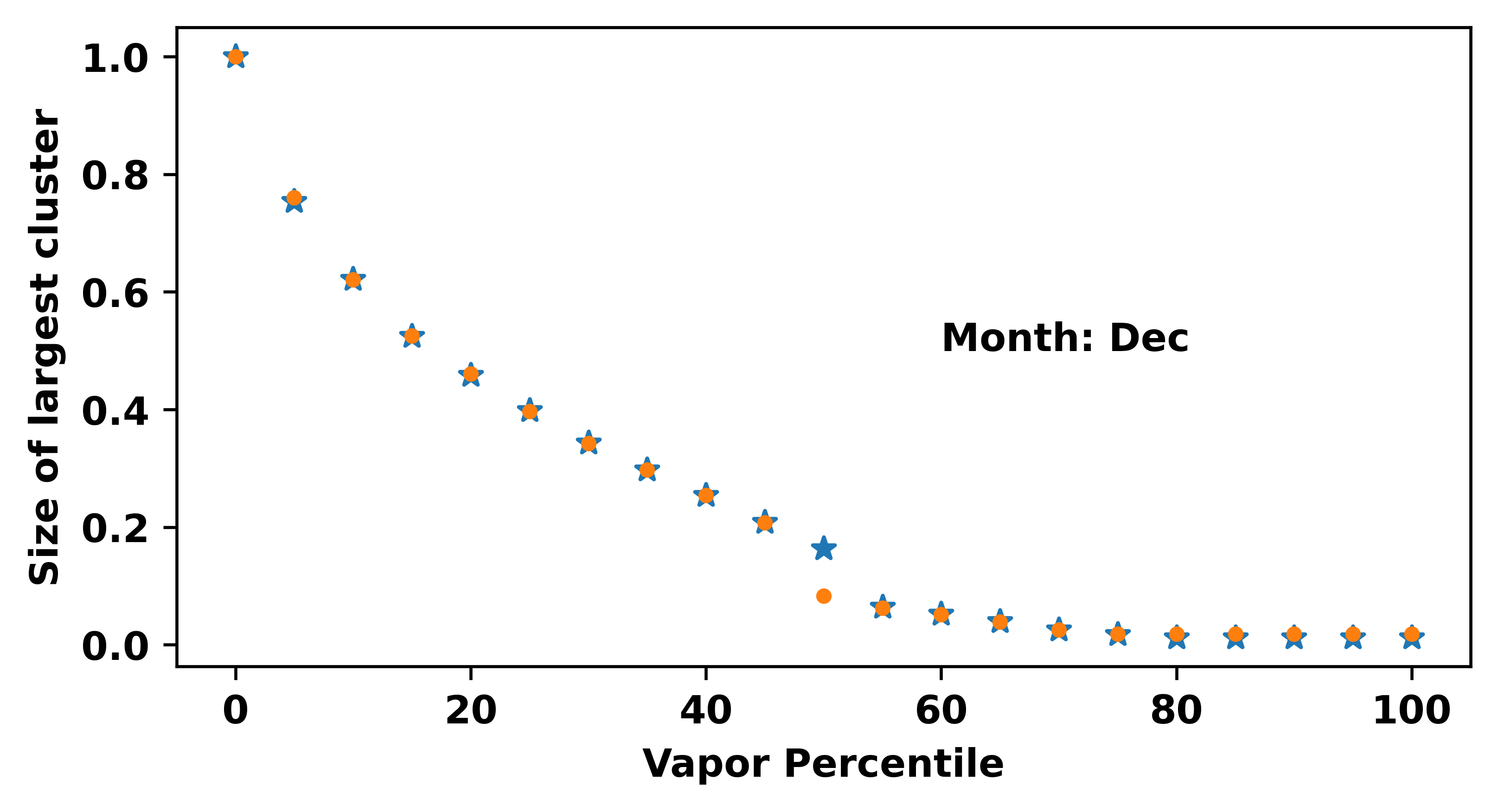} 
	\end{subfigure}		
	\caption{Scaled Maximum Cluster Size vs Vapor Percentile for all 12 months in 2021. The blue markers represent the $180\times 360$ pixel data and the orange markers represent the $360\times 720$ pixel data.}
	\label{4}
\end{figure}

We also examine the variation of the {\it scaled} maximum cluster size as a function of the vapor percentile and the results for both resolutions is presented in Fig. \ref{4}. This analysis allows us to see the size of largest cluster as a function of vapor percentile. Specifically, as $V \rightarrow 0$, there will be a single cluster spanning all pixels of the data, and as $V \rightarrow 100$, the clusters will be rather small, mostly single pixel objects. The scaled maximum cluster size is defined as 

\begin{equation}
	\textrm{scaled maximum size of clusters} = \frac{\textrm{local maximum cluster size}}{\textrm{global maximum cluster size}} 
\end{equation}

In the definition above, the cluster sizes have been scaled by the largest cluster size (global maximum cluster size) in the entire dataset considering all percentiles. This is important to obtain a meaningful comparison between the two resolutions, and serves as a normalization. The data in the plots below corroborate well with the story presented in the previous subsection, with the scaled maximum cluster size decreasing from $1$ to $0$ as the vapor percentile increases from $0$ to $100$. As such, this shows that the clusters are largest at $0\%$ and get smaller as the vapor percentile increases. The curves presented in Fig. \ref{4} appear to be decaying exponentials though we have not attempted to fit the data to an exponentially decaying fit function. We note that there appears to be a dip in the scaled maximum cluster size for all months at the $60$th percentile, in correspondence with the percolation transition. Also, while comparing to Figs. \ref{22} \& \ref{2}, we find that for large $V$ the fractal dimension tends to $D = 2$ which is indicative of a large number of elongated clusters in the distribution. The exact reason behind this phenomenon is not known to the authors, but it maybe possible to allude it to climactic considerations. 

\subsection{Korcak's Law and Scaling Exponents}

In fractal landscape structures, clusters are statistically self-similar at the percolation threshold over certain size-ranges. The cluster areas follow a probability distribution with a power-law tail \cite{19} at the percolation threshold. The physcist Korcak made this prediction, and the law is now known as Korcak's law \cite{20}. We wish to examine the applicability of Korcak's law to our distribution, wherein the statement of the law reads

\begin{equation}
	P(A \geq a)\propto a^{1-\beta}
\end{equation}

where $a$ represents the area of a cluster, $P(a)$ represents the probability of randomly selecting a cluster of size $A \geq a$, and $\beta$ is the scaling exponent. Tempering may occur as a result of deviations from the powerlaw above the percolation threshold \cite{21}. 

\begin{figure}[H]
	\centering
	\begin{subfigure}[t]{0.45\textwidth}
		\centering
		\includegraphics[width=\linewidth]{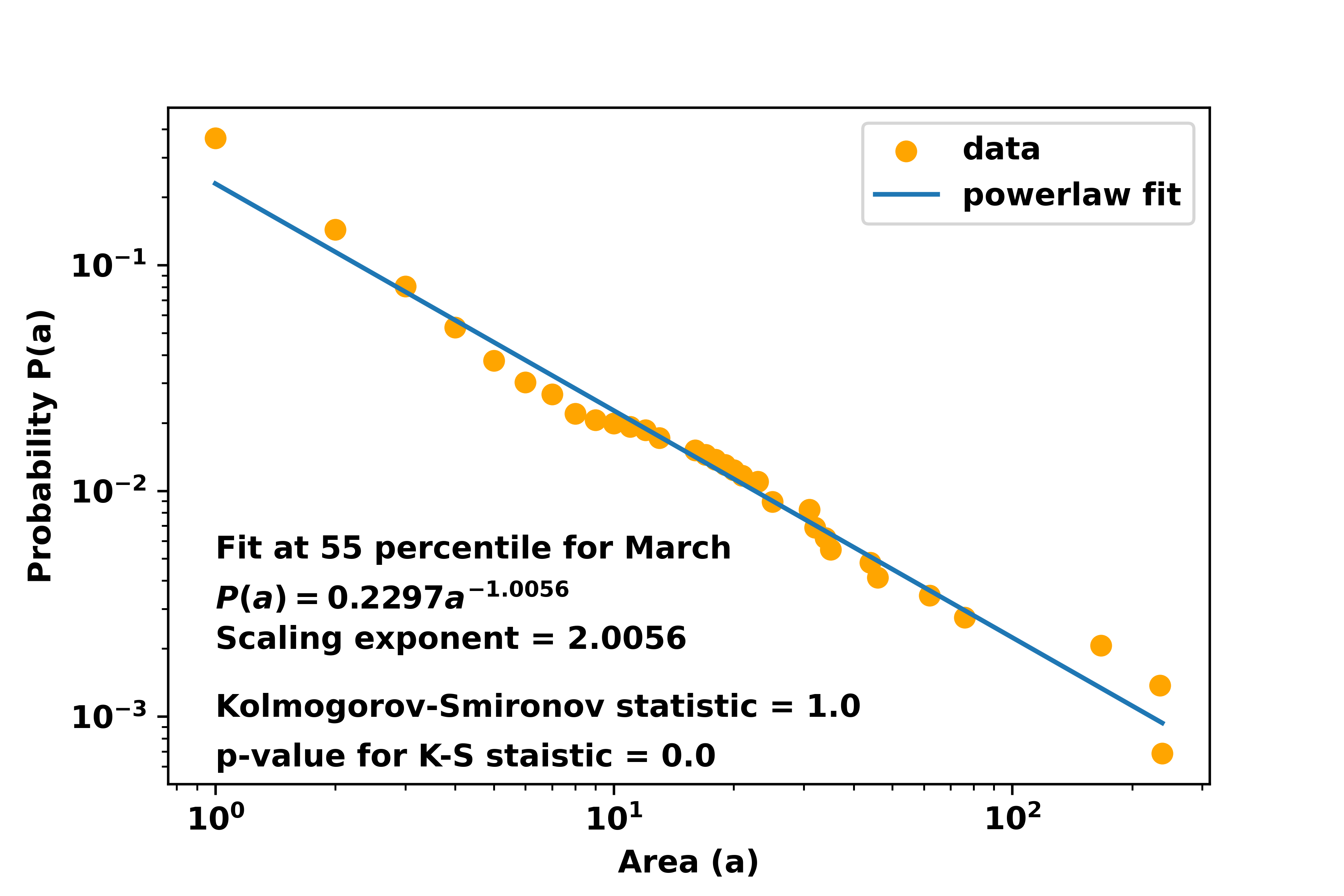} 
	\end{subfigure}
	\begin{subfigure}[t]{0.45\textwidth}
		\centering
		\includegraphics[width=\linewidth]{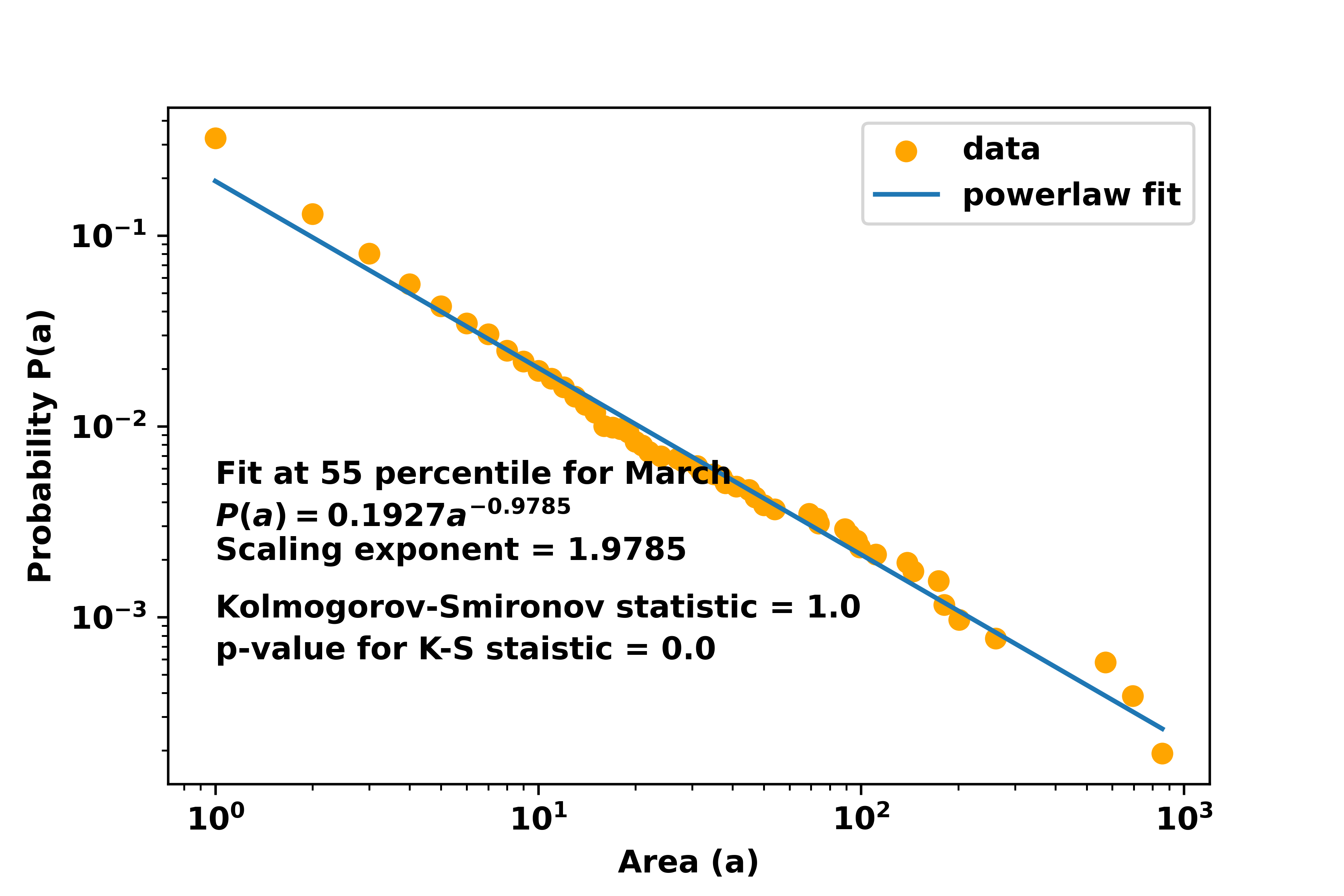} 
	\end{subfigure}
	\caption{Power law fit: The plots show the power-law fit at two resolutions of the data for March 2021, $180\times 360$ pixels (left) and $360 \times 720$ pixels (right). }
	\label{15}
\end{figure}

We plot the probabilities for March 2021 as a function of the cluster areas in Fig. \ref{15} at the percolation threshold for both resolutions, $180\times 360$ pixels on the left and $360\times 720$ pixels on the right. The data is plotted as the orange scatter points, and the power-law fit is represented by the blue line. For this we employ a least-squared fitting procedure for the logarithms of the data. We test for the quality of the fit by using the Kolmogorov-Smirnoff (KS) statistic. These values are stated in the plots below. As we can see, the fit to the data is reasonable, with the KS statistic being approximately $1$ (up to machine precision) and the p-value being $0$. This verifies that our data is in agreement with Korcak's law.

\begin{figure}[H]
		\centering
		\includegraphics[scale = 0.7]{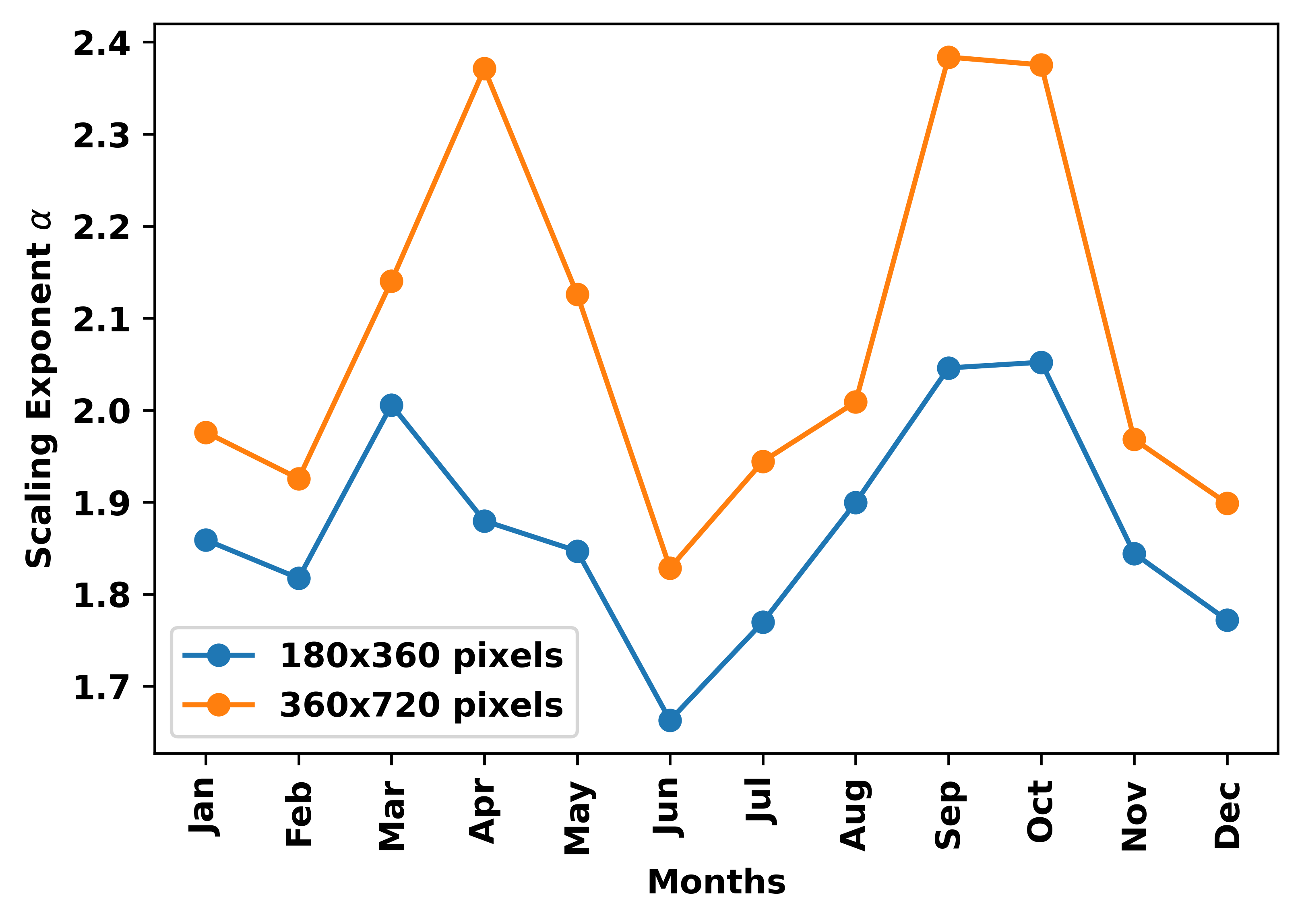} 
	\caption{Scaling Exponents: The plot shows the variation of the scaling exponents as a function of months in the year 2021 at both resolutions. The blue markers show the scaling exponents for the $180\times 360$ pixel data and the orange dots show the same for the $360\times 720$ data.}
	\label{14}
\end{figure}

We also plot the scaling exponents for the data at the percolation threshold for all months of the year 2021 in Fig. \ref{14}. The blue markers show the scaling exponents for the $180\times 360$ pixel data and the orange dots show the same for the $360\times 720$ pixel data. The scaling exponents are not identical for each month at the two resolutions, however, notably they seem to follow the same trend. The range exhibited by the scaling exponent is $\beta \in (1.6, 2.4)$. We have verified that in each case the KS statistic is very close to $1$ and the p-value is roughly $0$.

\section{Vapor Percentile based Numerical Analysis}

In this section, we examine the fractal nature of the water vapor distribution as a function of vapor percentile. We conduct the analysis at a data resolution of $180 \times 360$, and $10$ years worth of monthly data is considered in the period $2012-2021$. We proceed with the analysis as in the previous section and the results are presented below.

\subsection{Area-Perimeter Relation and Fractal Dimension}
\begin{figure}[H]
	\centering
	\begin{subfigure}[t]{0.45\textwidth}
		\centering
		\includegraphics[width=\linewidth]{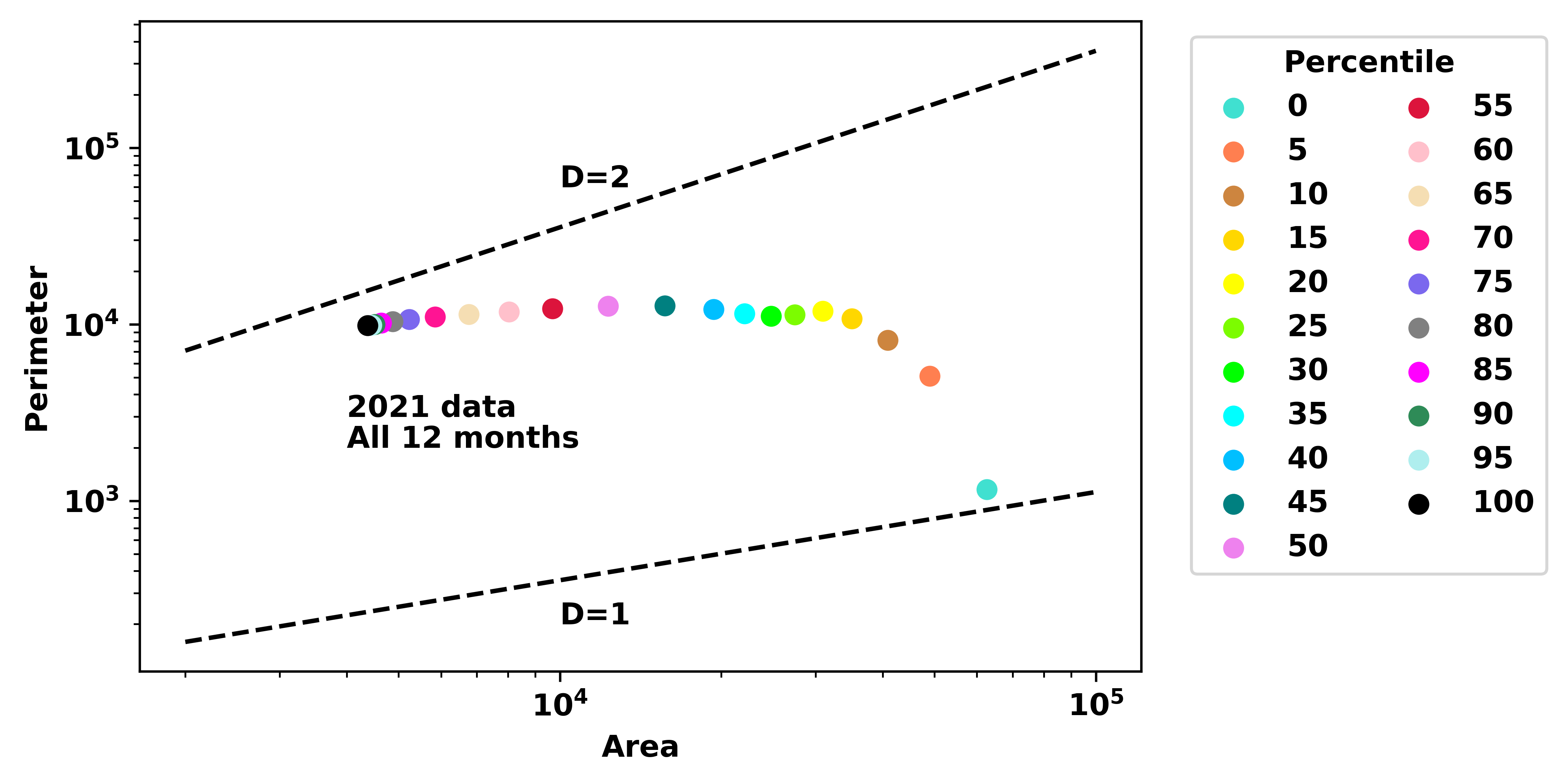} 
	\end{subfigure}
\begin{subfigure}[t]{0.45\textwidth}
	\centering
	\includegraphics[width=\linewidth]{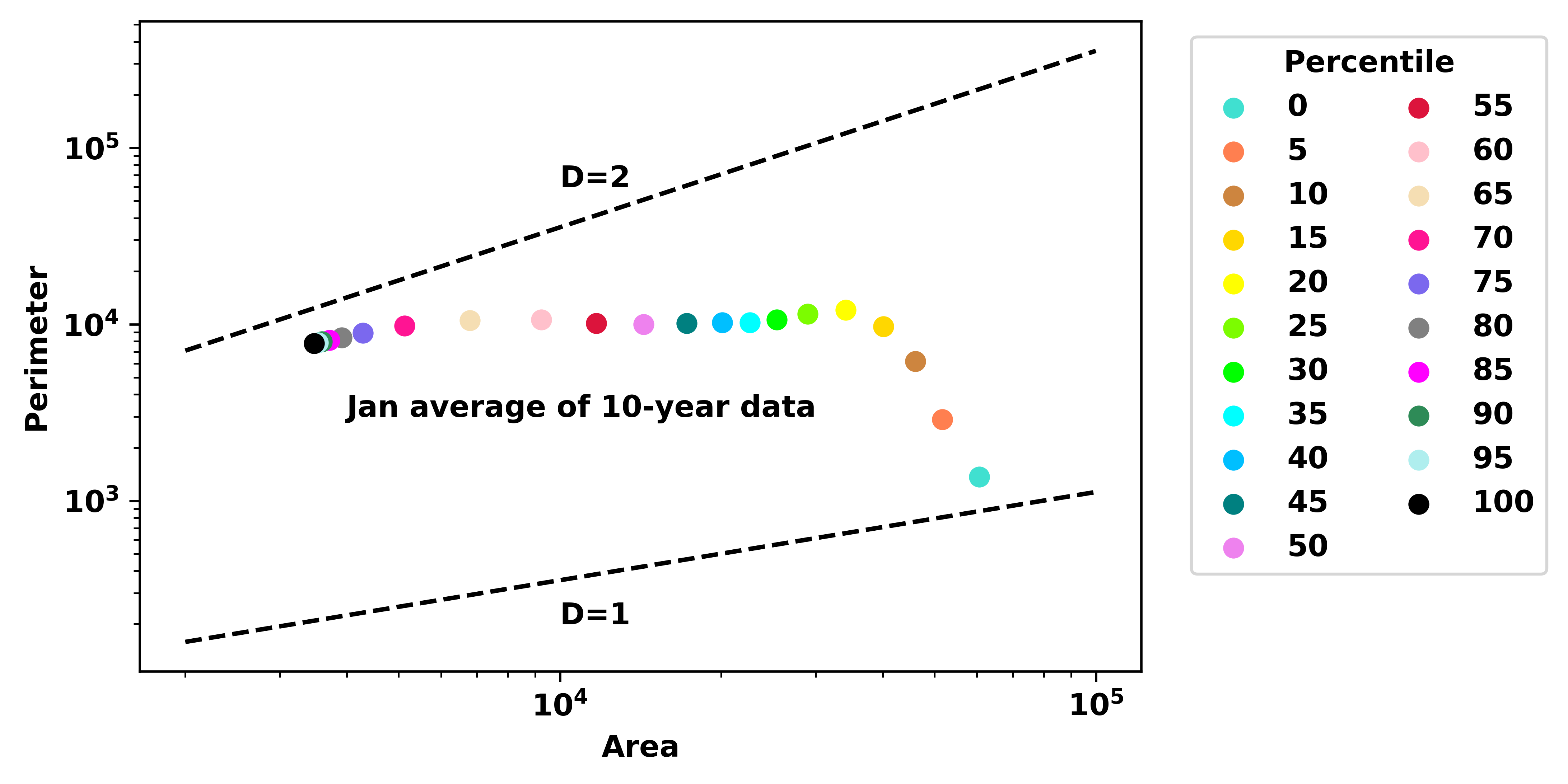} 
\end{subfigure}
\caption{Aggregated Area-Perimeter log-log plot for 10 years of data from 2012 to 2021 at a resolution of $180 \times 360$ pixels.}
\label{5}
\end{figure}

We start by plotting the cumulative perimeter versus the cumulative area on a log-log plot. The plots are presented in Fig. \ref{5} where we average over each month on the left plot and over each year for the plot on the right. The distributions are very similar with the fractal dimension lying between $1$ and $2$ which are the limiting cases of a circle and a straight line. This is the first indication that there is some fractal character to the distribution, since the aggregated area-perimeter fractal dimension is not an integer.

\begin{figure}[H]
	\centering
	\begin{subfigure}[t]{0.3\textwidth}
		\centering
		\includegraphics[width=\linewidth]{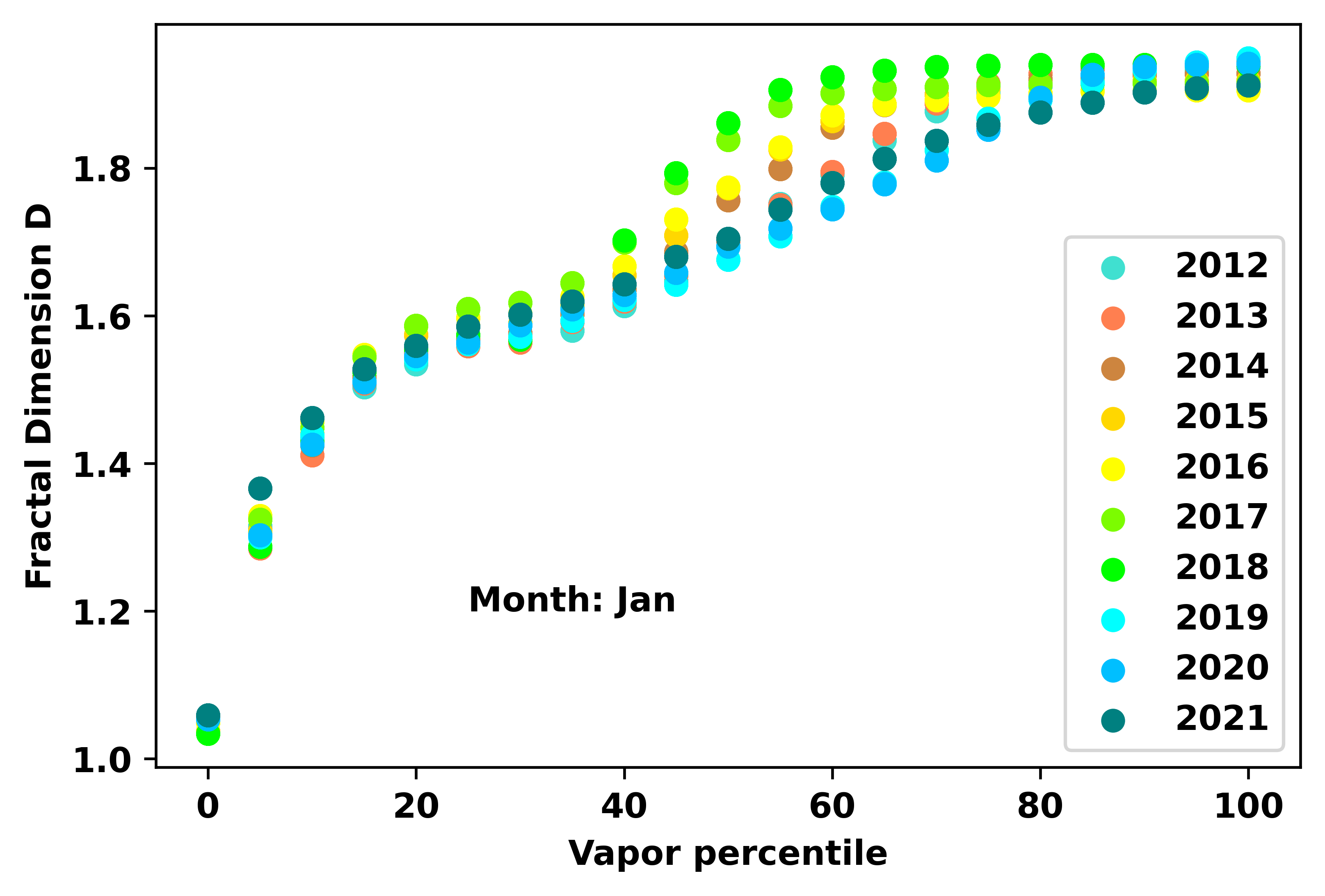} 
	\end{subfigure}
	\begin{subfigure}[t]{0.3\textwidth}
		\centering
		\includegraphics[width=\linewidth]{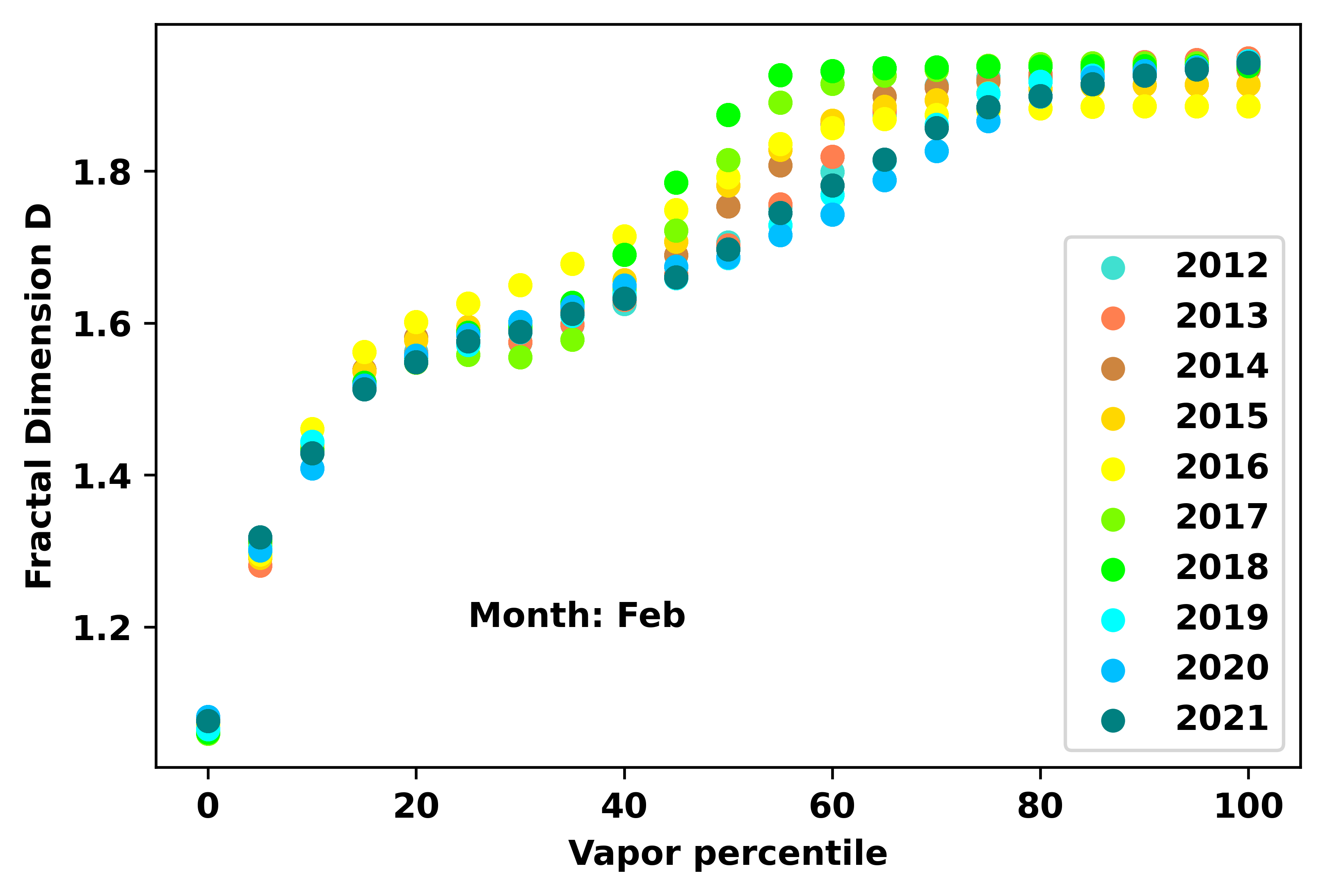} 
	\end{subfigure}
	\begin{subfigure}[t]{0.3\textwidth}
		\centering
		\includegraphics[width=\linewidth]{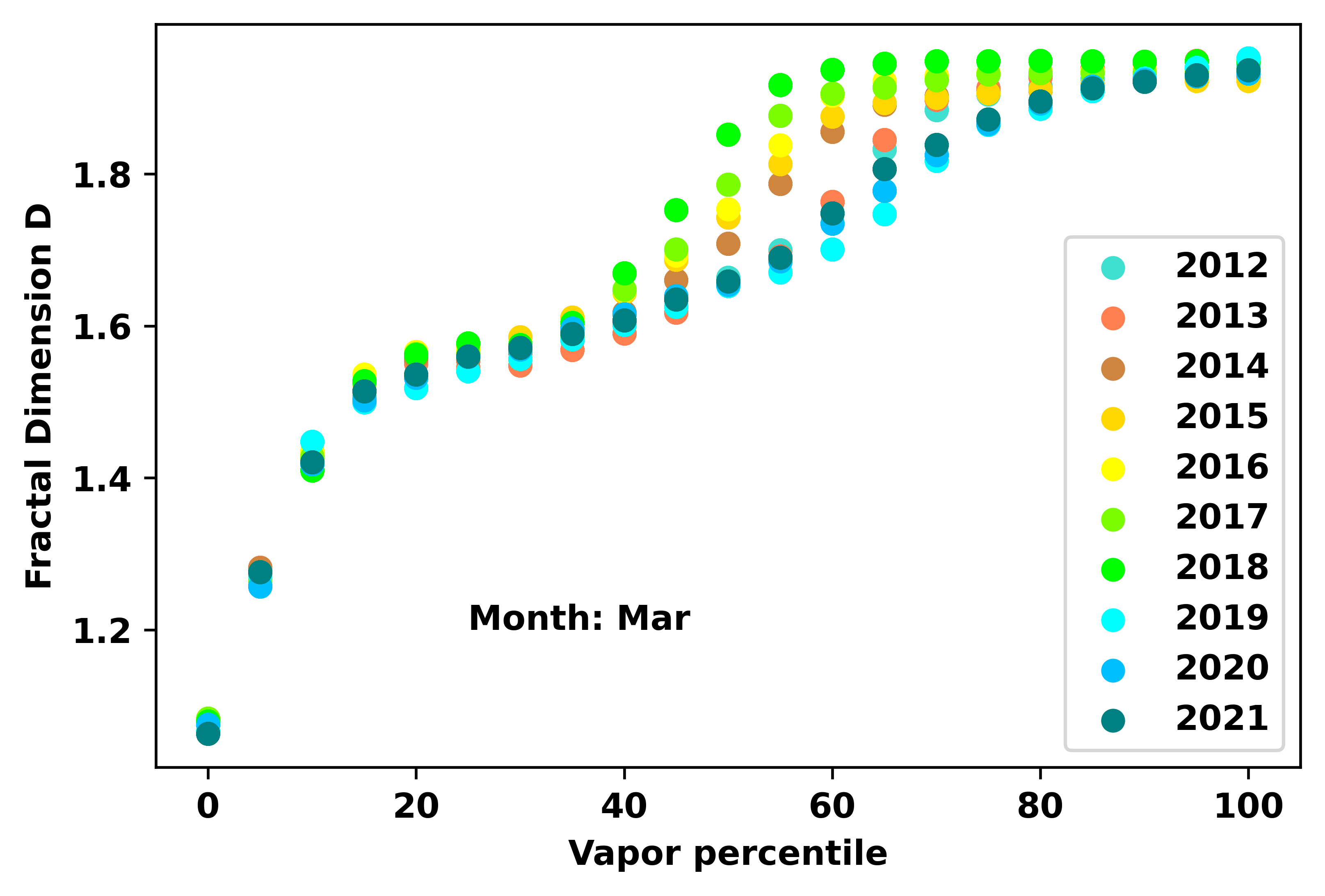} 
	\end{subfigure}		
	\begin{subfigure}[t]{0.3\textwidth}
		\centering
		\includegraphics[width=\linewidth]{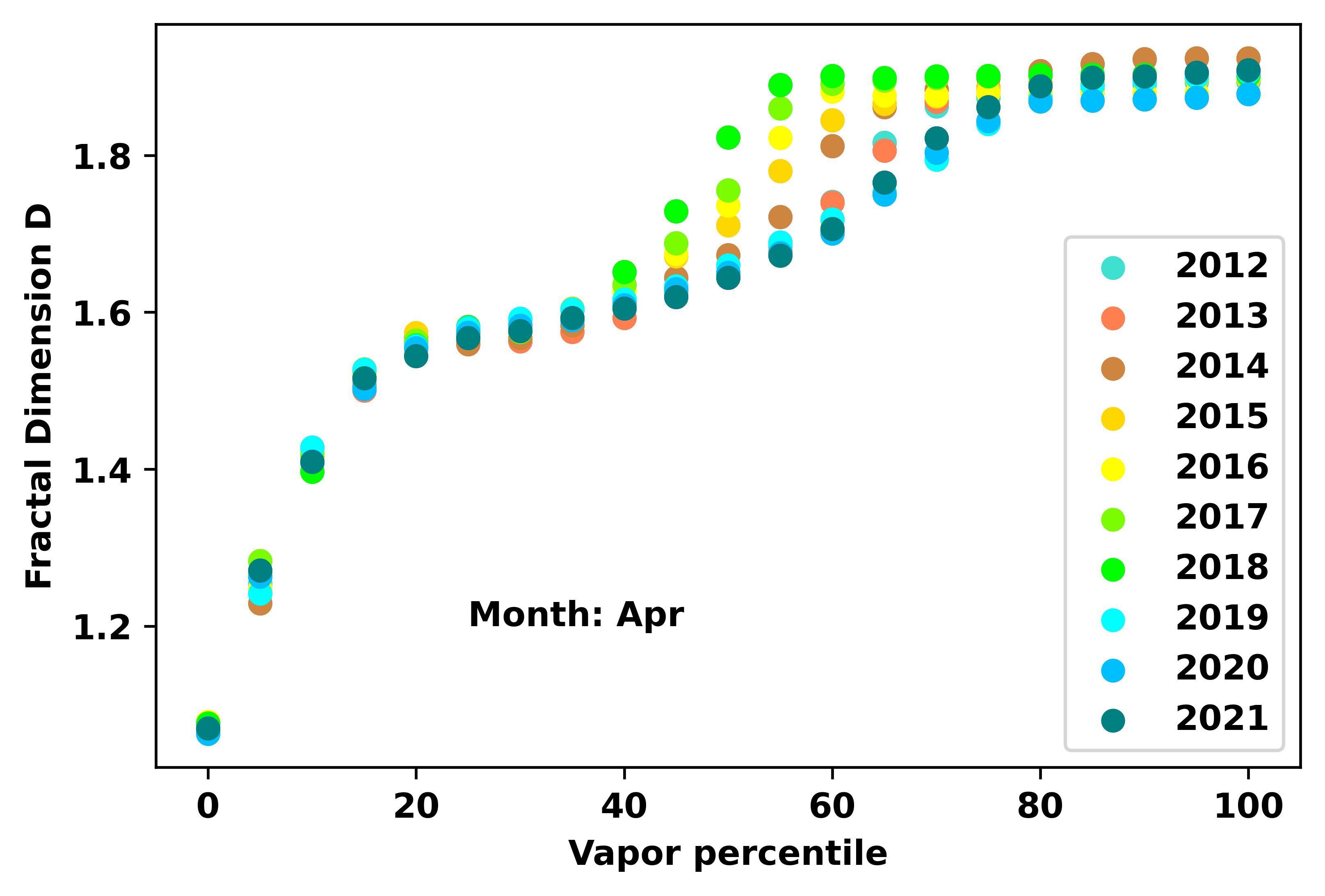} 
	\end{subfigure}
	\begin{subfigure}[t]{0.3\textwidth}
		\centering
		\includegraphics[width=\linewidth]{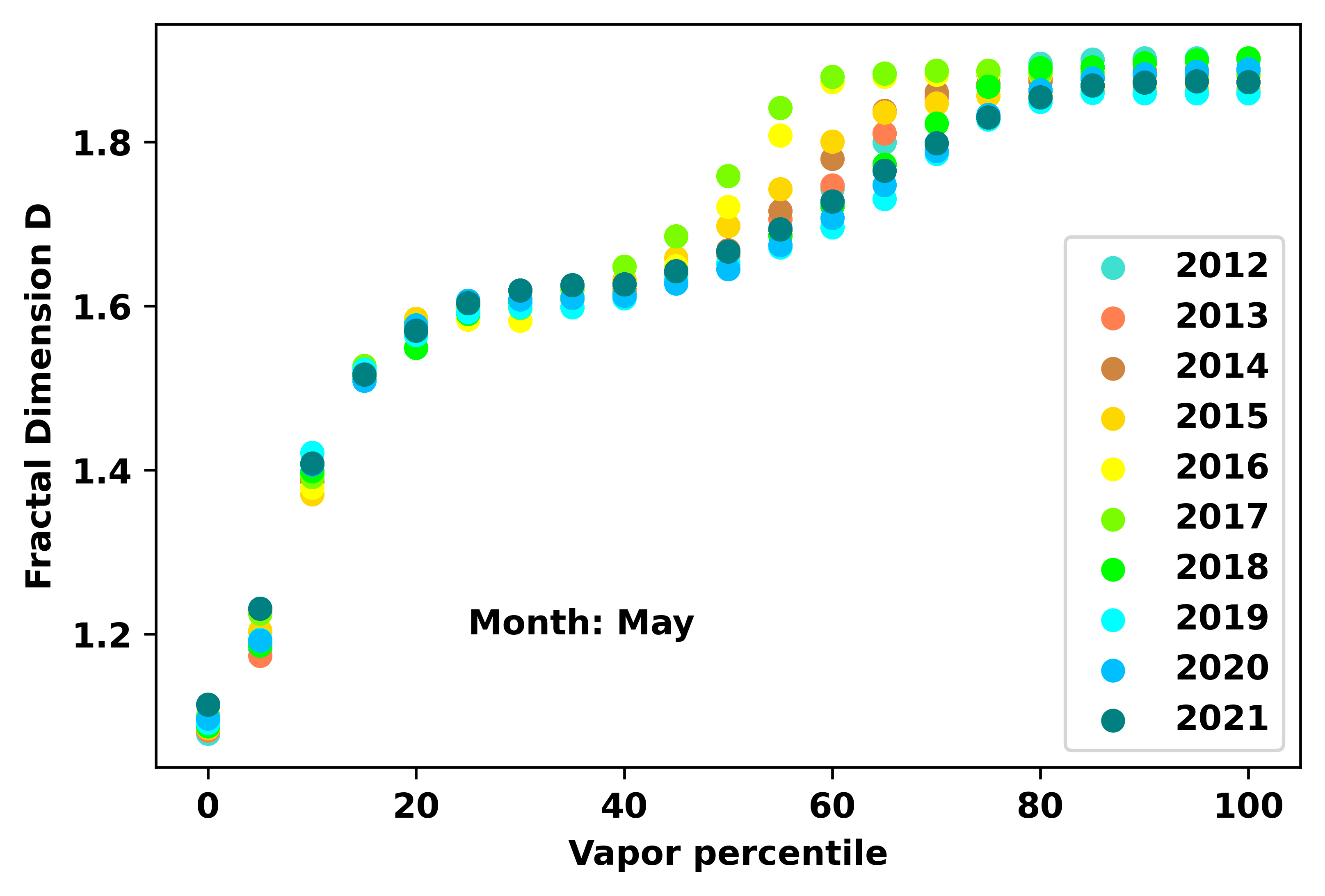} 
	\end{subfigure}
	\begin{subfigure}[t]{0.3\textwidth}
		\centering
		\includegraphics[width=\linewidth]{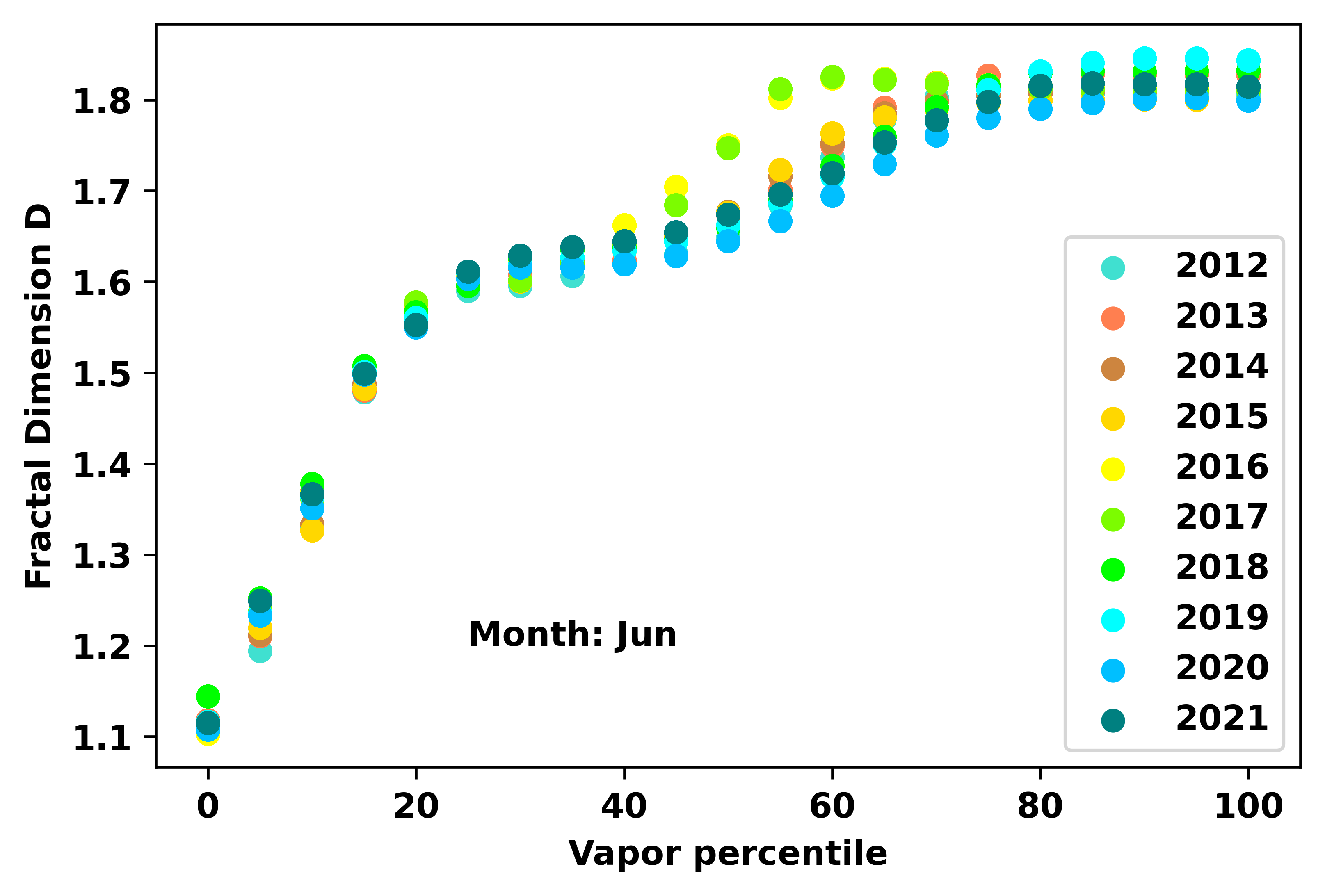} 
	\end{subfigure}		
	\begin{subfigure}[t]{0.3\textwidth}
		\centering
		\includegraphics[width=\linewidth]{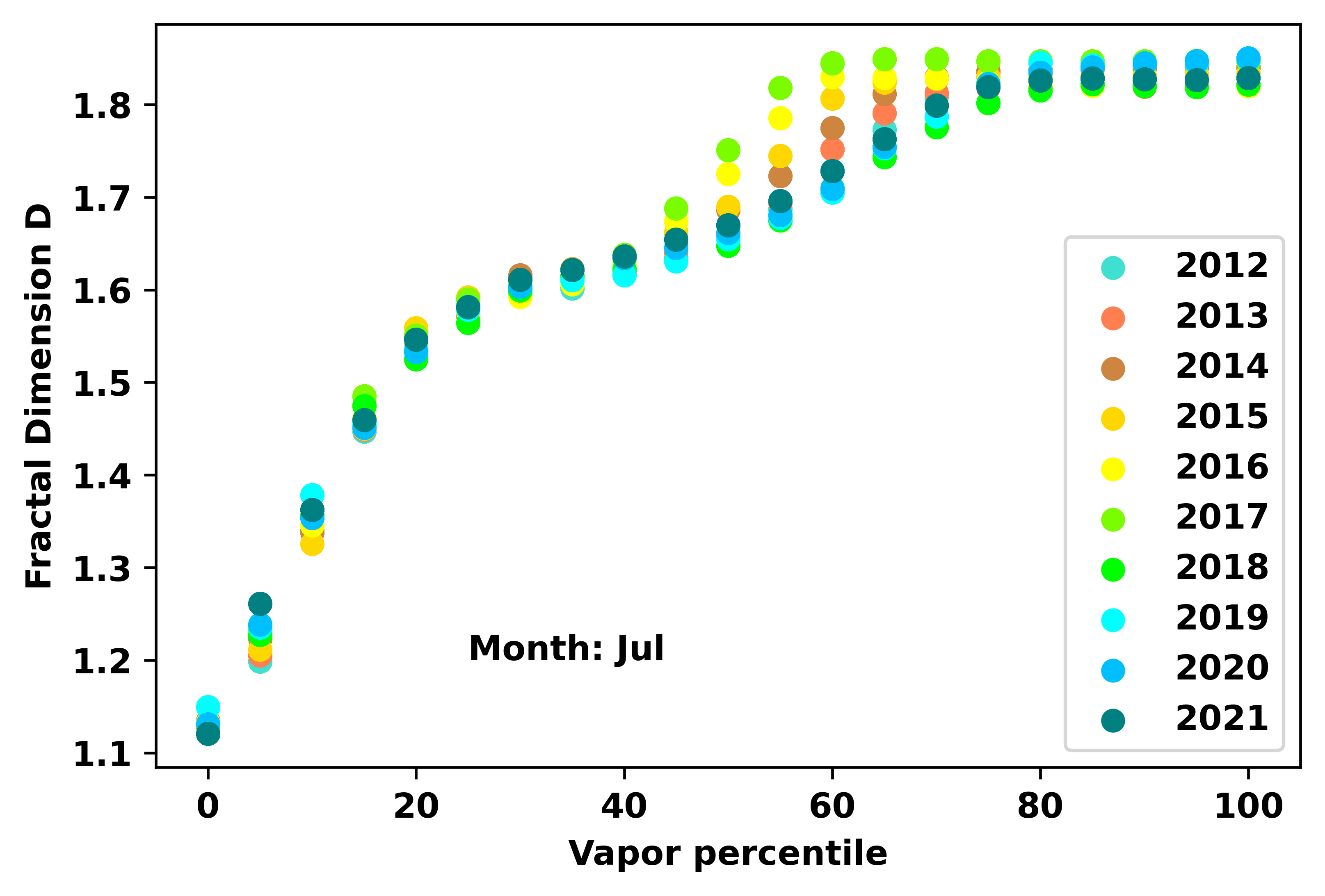} 
	\end{subfigure}
	\begin{subfigure}[t]{0.3\textwidth}
		\centering
		\includegraphics[width=\linewidth]{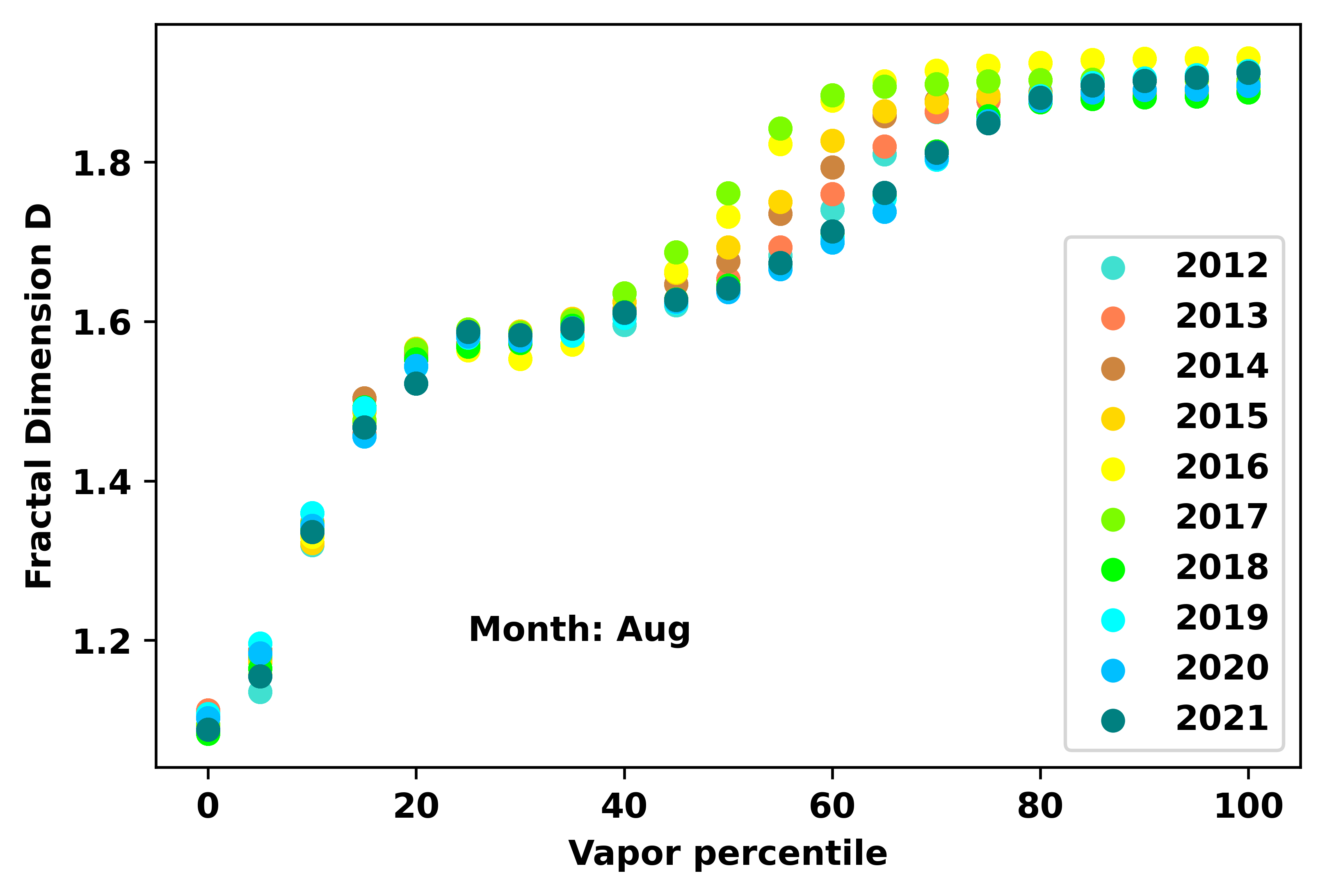} 
	\end{subfigure}
	\begin{subfigure}[t]{0.3\textwidth}
		\centering
		\includegraphics[width=\linewidth]{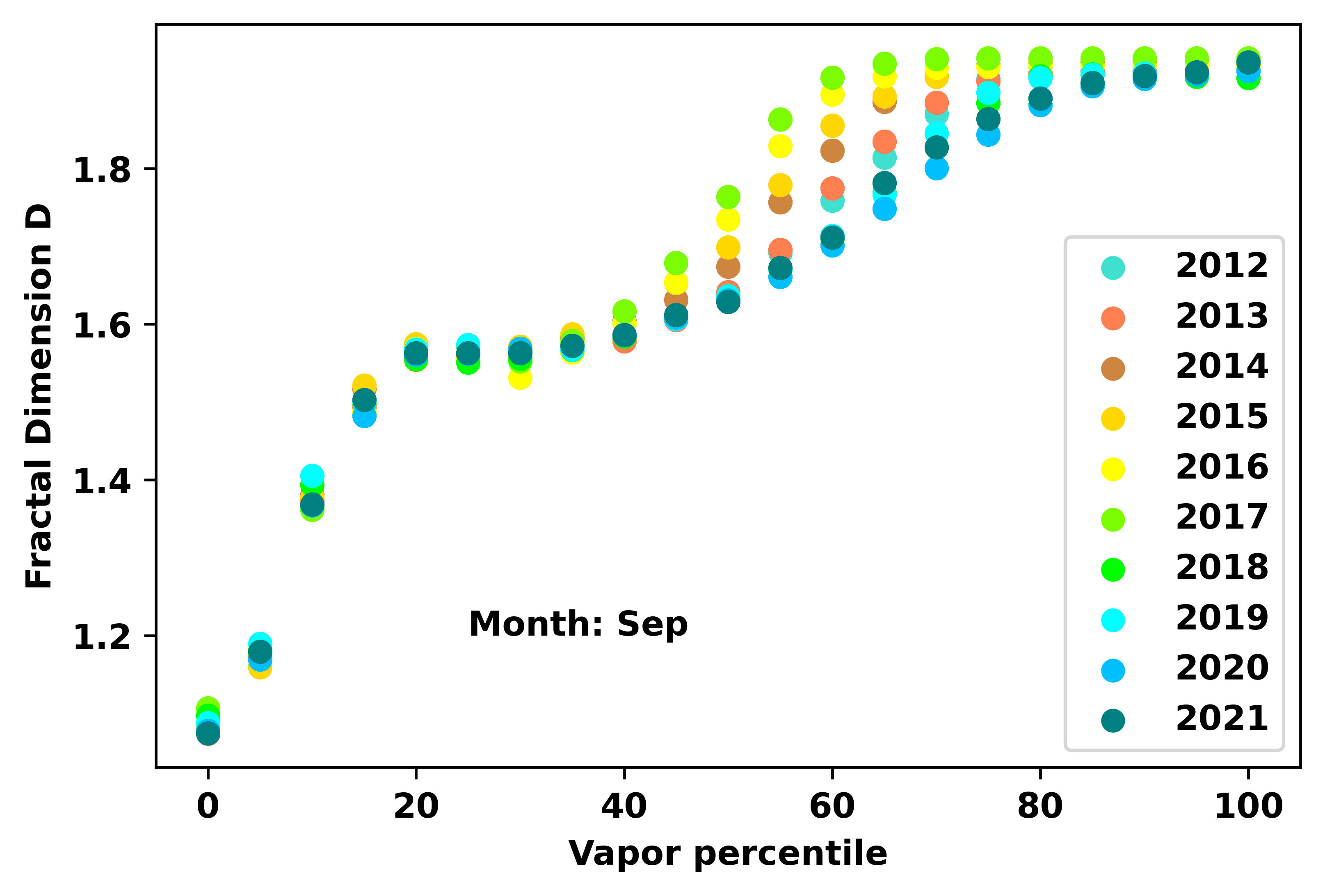} 
	\end{subfigure}		
	\begin{subfigure}[t]{0.3\textwidth}
		\centering
		\includegraphics[width=\linewidth]{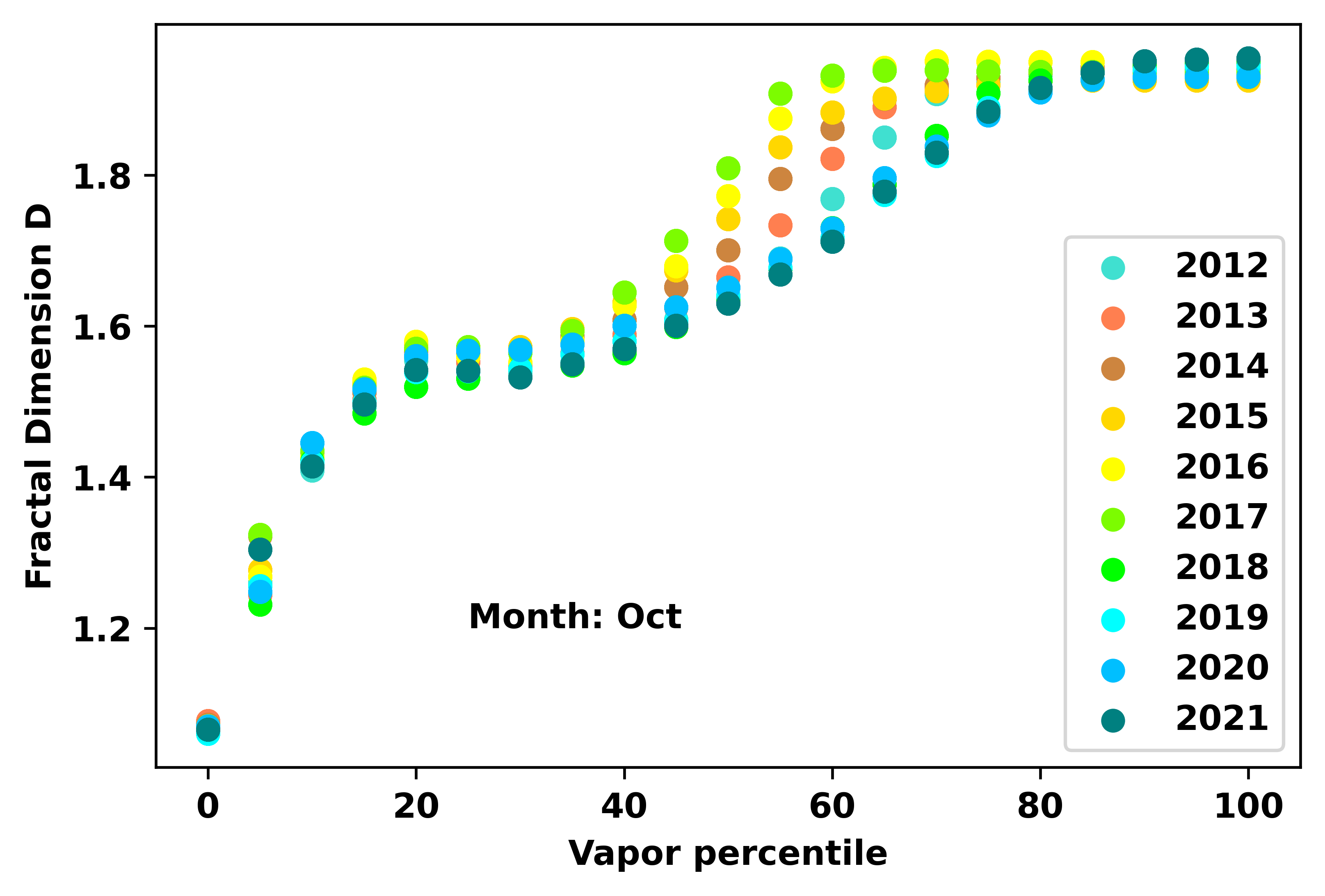} 
	\end{subfigure}
	\begin{subfigure}[t]{0.3\textwidth}
		\centering
		\includegraphics[width=\linewidth]{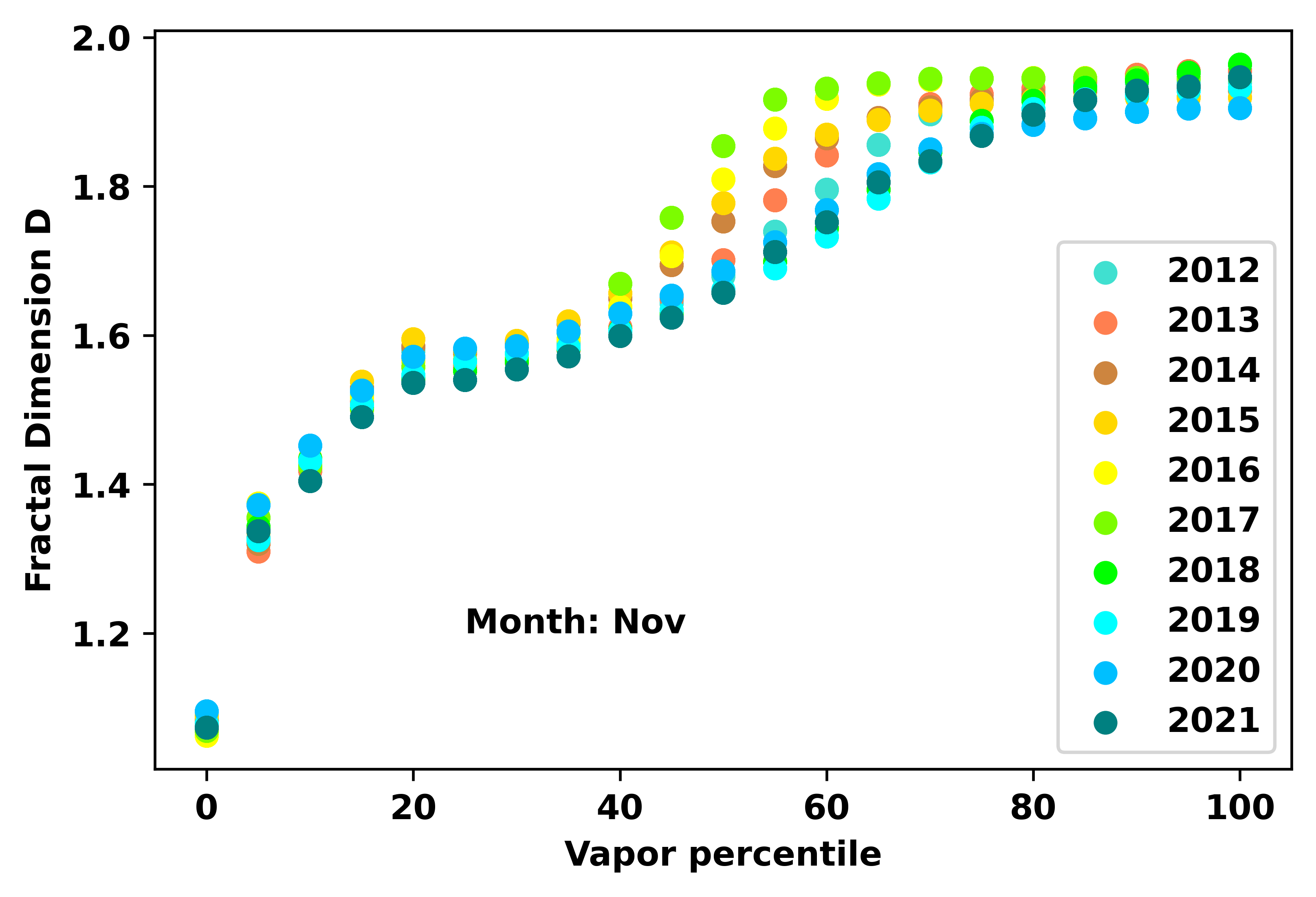} 
	\end{subfigure}
	\begin{subfigure}[t]{0.3\textwidth}
		\centering
		\includegraphics[width=\linewidth]{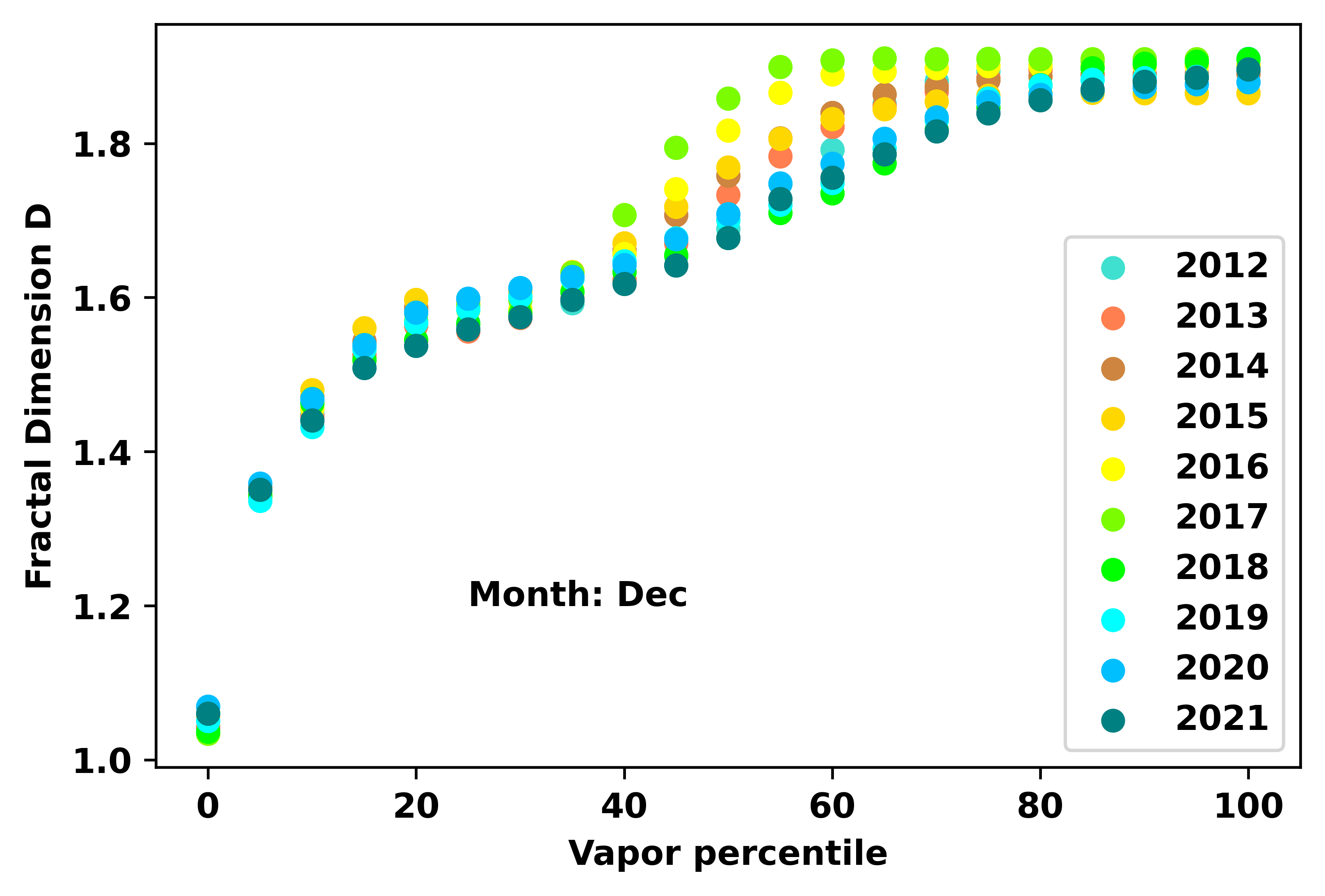} 
	\end{subfigure}		
	\caption{Fractal Dimension vs Vapor Percentile for each month and all 10 years, $2012-2021$, at a resolution of $180\times 360$ pixels.}
	\label{6}
\end{figure}

We next examine the fractal dimensions for all $12$ months of the years $2012-2021$ as a function of the vapor percentile. Each subplot in Fig. \ref{6} contains the data for all $10$ years for a given month. We notice an overall trend in the behavior of the fratal dimension while the exact numerical values are case dependent, but continue to lie in the interval $1 < D < 2$. For each of the subplots in Fig. \ref{6} we observe that the fractal dimension increases rapidly for vapor percentiles $V\leq 30\%$, and then stabilizes to an approximately fixed value in the range $30 < V < 50$. Then the fractal dimensions increase again in the interval $50 \leq V < 80$, but more gradually than before. Finally, the fractal dimensions remain steady for percentiles $80 \leq V \leq 100$.

\begin{figure}[H]
	\centering
	\begin{subfigure}[t]{0.45\textwidth}
		\centering
		\includegraphics[width=\linewidth]{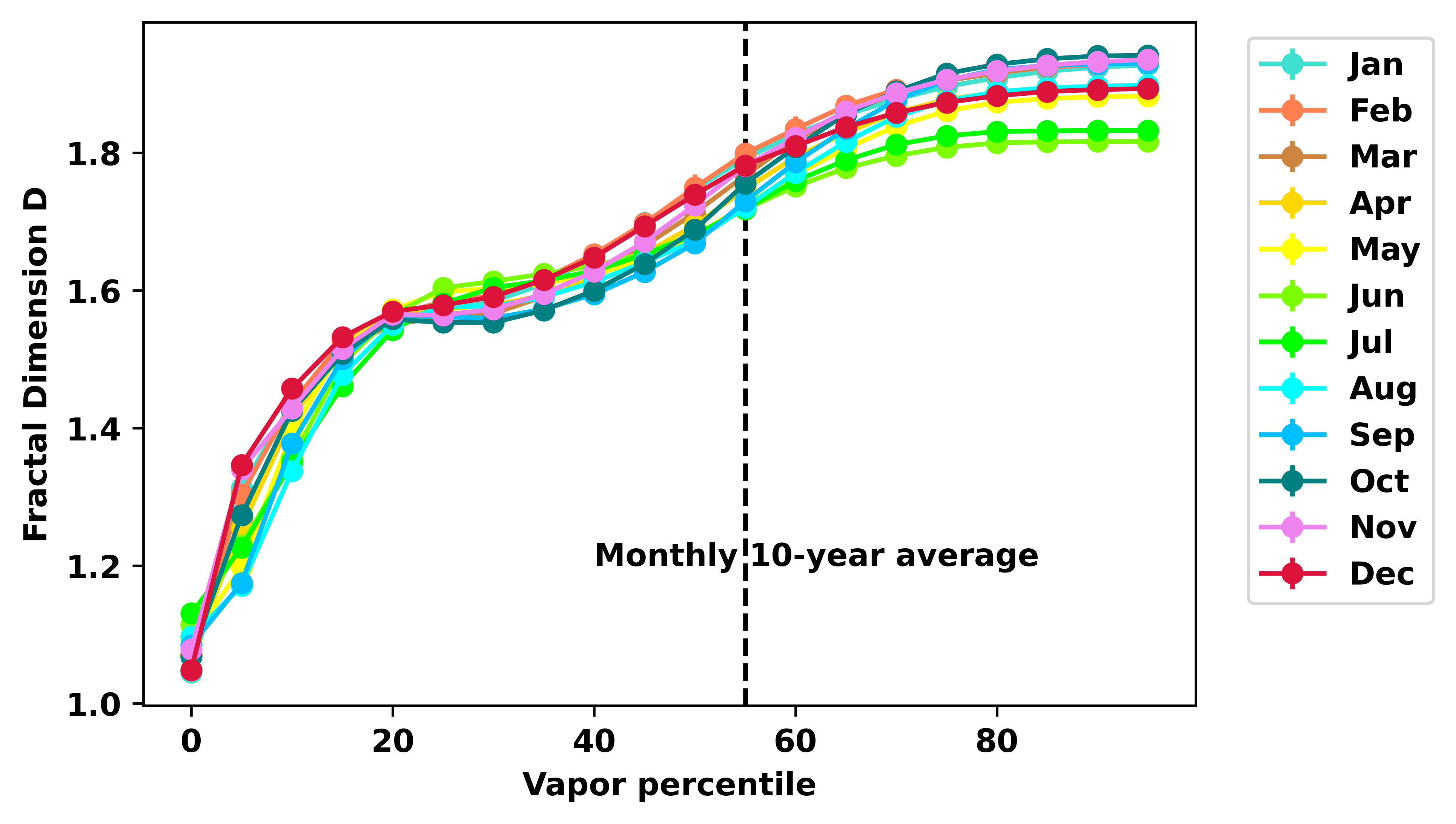} 
	\end{subfigure}
	\begin{subfigure}[t]{0.45\textwidth}
		\centering
		\includegraphics[width=\linewidth]{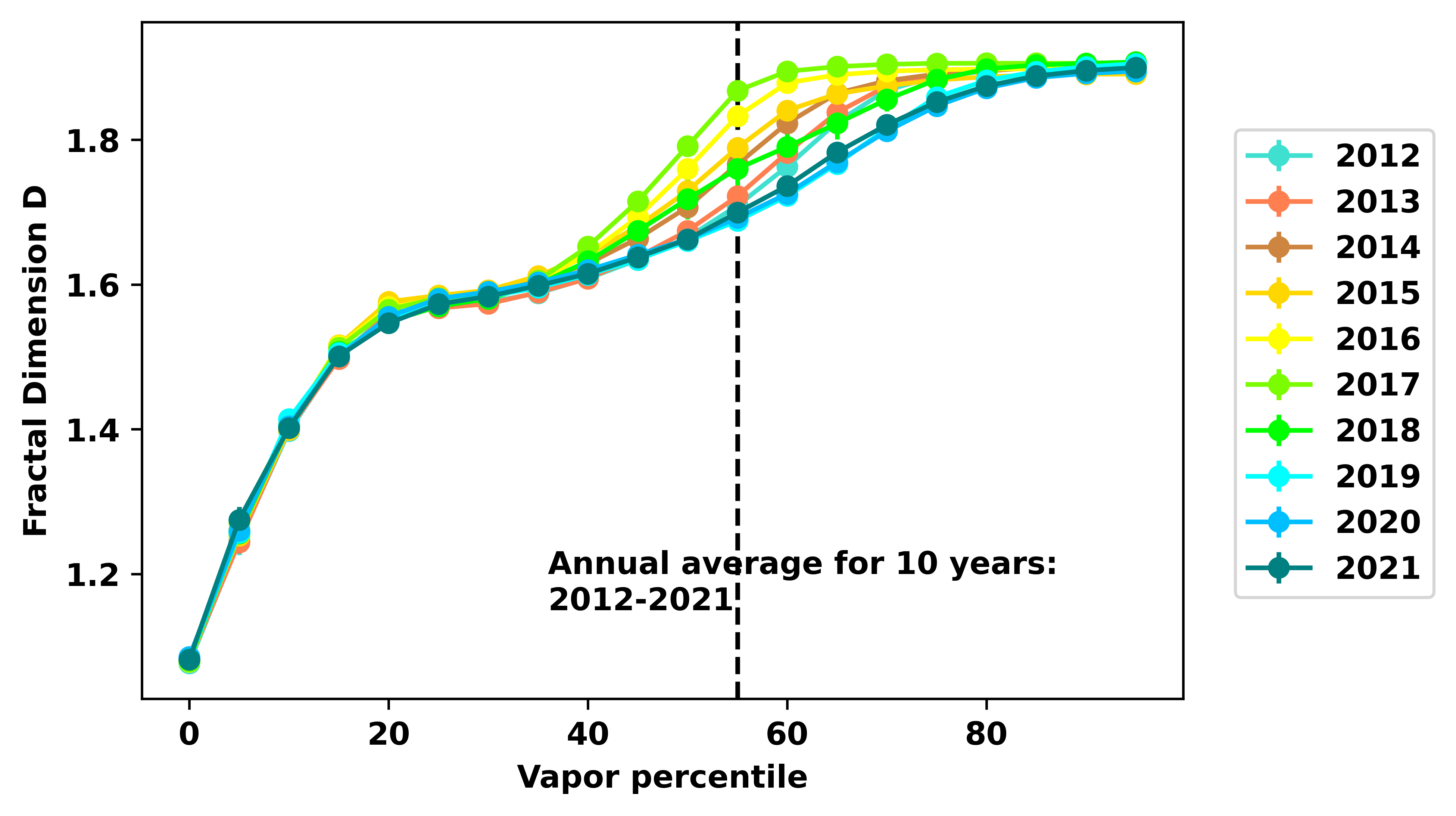} 
	\end{subfigure}
	\caption{Fractal Dimension vs Vapor Percentile ($V$): the data is averaged monthly and yearly for all 10 years, $2012-2021$, at a resolution of $180\times 360$ pixels.}
	\label{7}
\end{figure}

We can thus conclude that the fractal dimension remains constant in two regimes of vapor percentiles, and so the clusters exhibit an approximate fractal behavior as a function of threshold in these two regimes. Outside these regimes, since the fractal dimensions vary with vapor percentile, we cannot conclude an approximate fractal behavior.

We further plot the monthly and yearly averaged distributions of fractal dimensions as a function of vapor percentile in Fig. \ref{7}. Some interesting trends arise as a result of this data representation. Firstly, we see that on the left plot of Fig. \ref{7} that fractal dimensions show a broad distribution at $80\% \leq V \leq 100\%$ with the hottest and driest months of the year in the Northern hemisphere, May and June showing the lowest values. A comparatively drier month, December, shows the next highest value of fractal dimension in this regime, while the rest of the months are clumped together. The yearly or annual averages also show unobvious trends, with the fractal dimension distributions being almost identical in the ranges $0\% \leq V \leq 40\%$ and $80\% \leq V \leq 100\%$. In the intermediate range $40\% < V < 80\%$, the distributions grow in magnitude at different rates for different years, with the peak growth rate being observed for the year 2018. The plot on the right shows that the peak fractal dimension occurs for the year 2017 and this value is reached at the $60$th percentile.

\subsection{Multifractal Analysis}
\begin{figure}[H]
	\centering
	\begin{subfigure}[t]{0.3\textwidth}
		\centering
		\includegraphics[width=\linewidth]{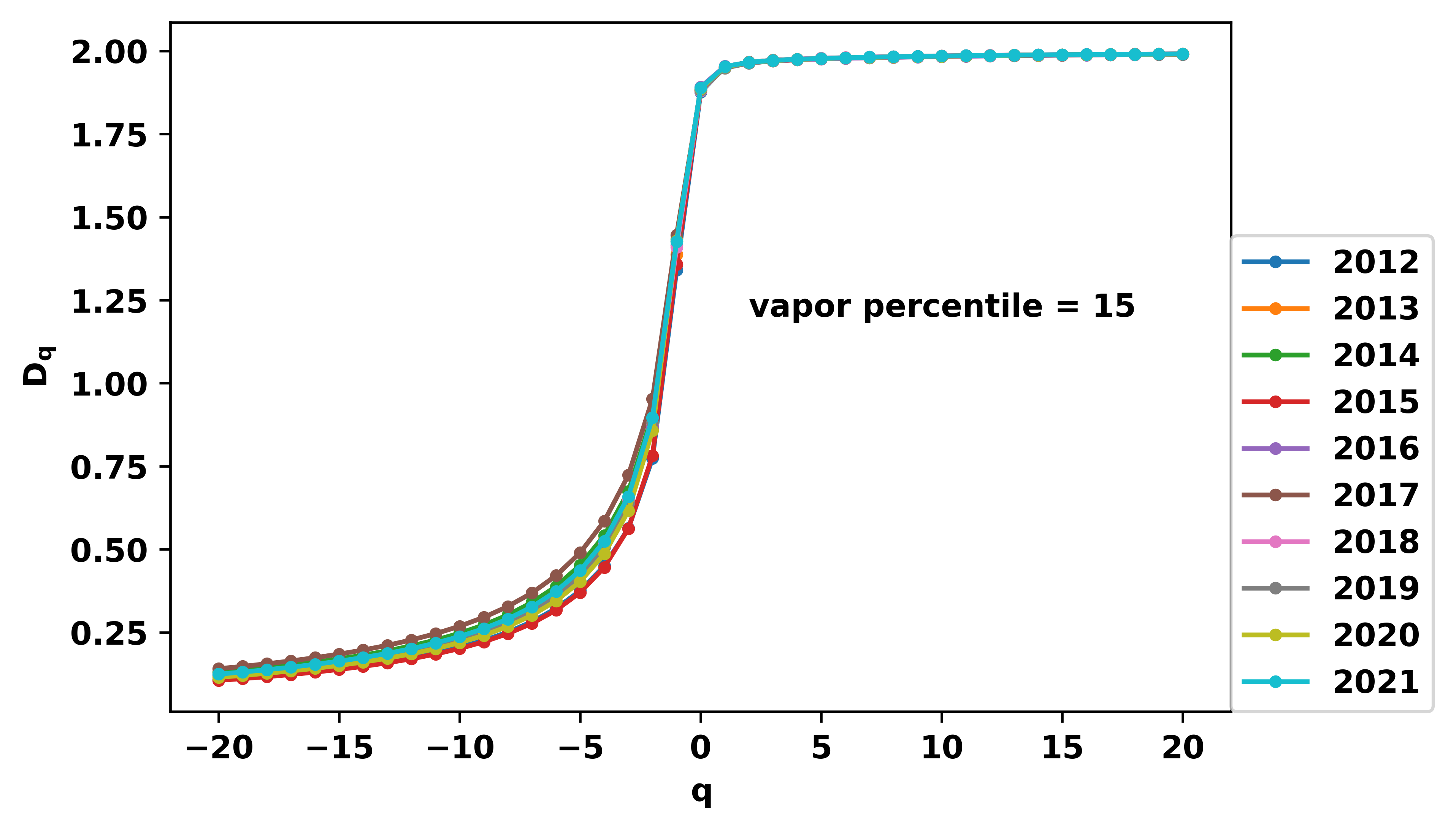} 
	\end{subfigure}
	\begin{subfigure}[t]{0.3\textwidth}
		\centering
		\includegraphics[width=\linewidth]{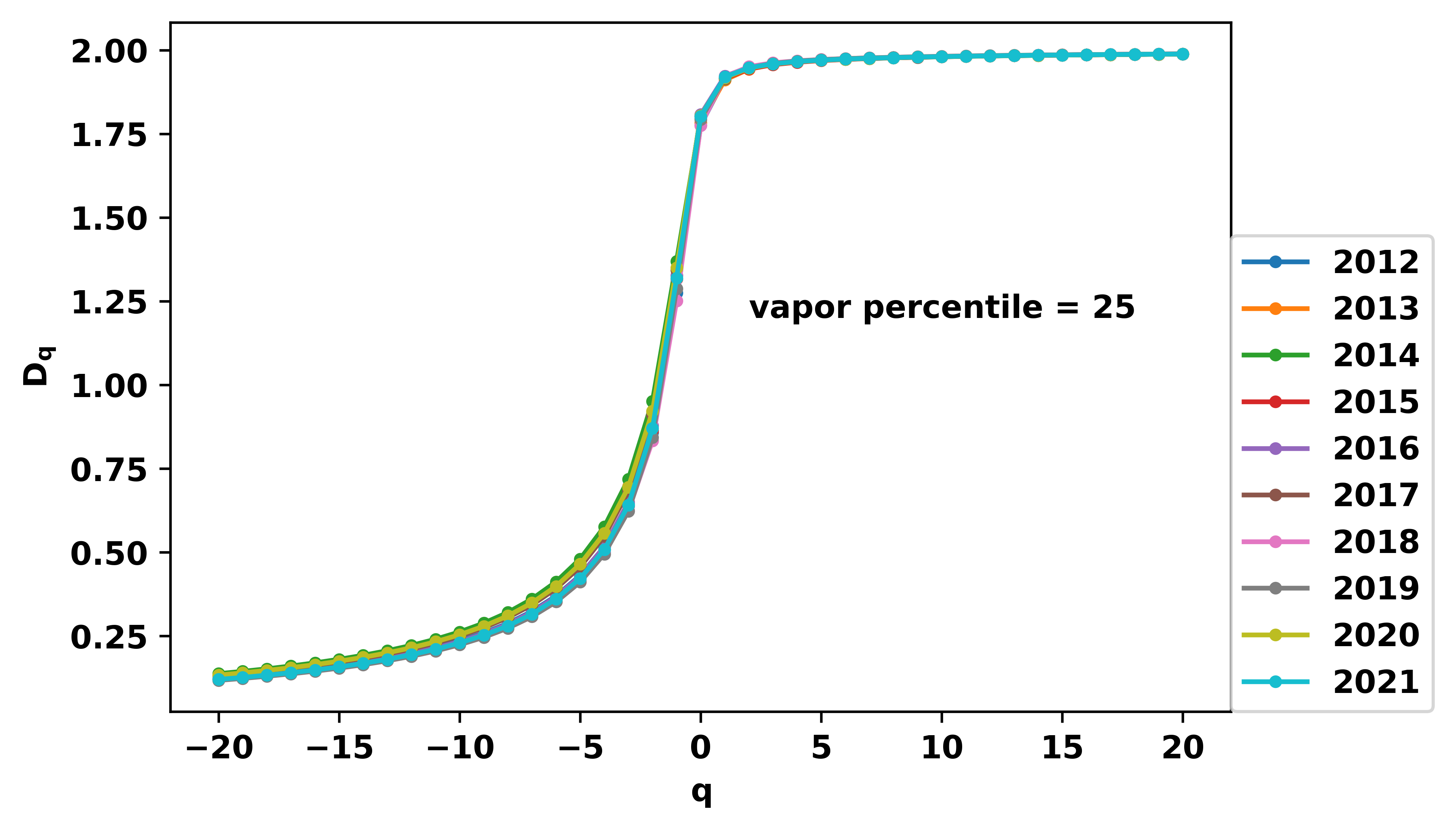} 
	\end{subfigure}
	\begin{subfigure}[t]{0.3\textwidth}
		\centering
		\includegraphics[width=\linewidth]{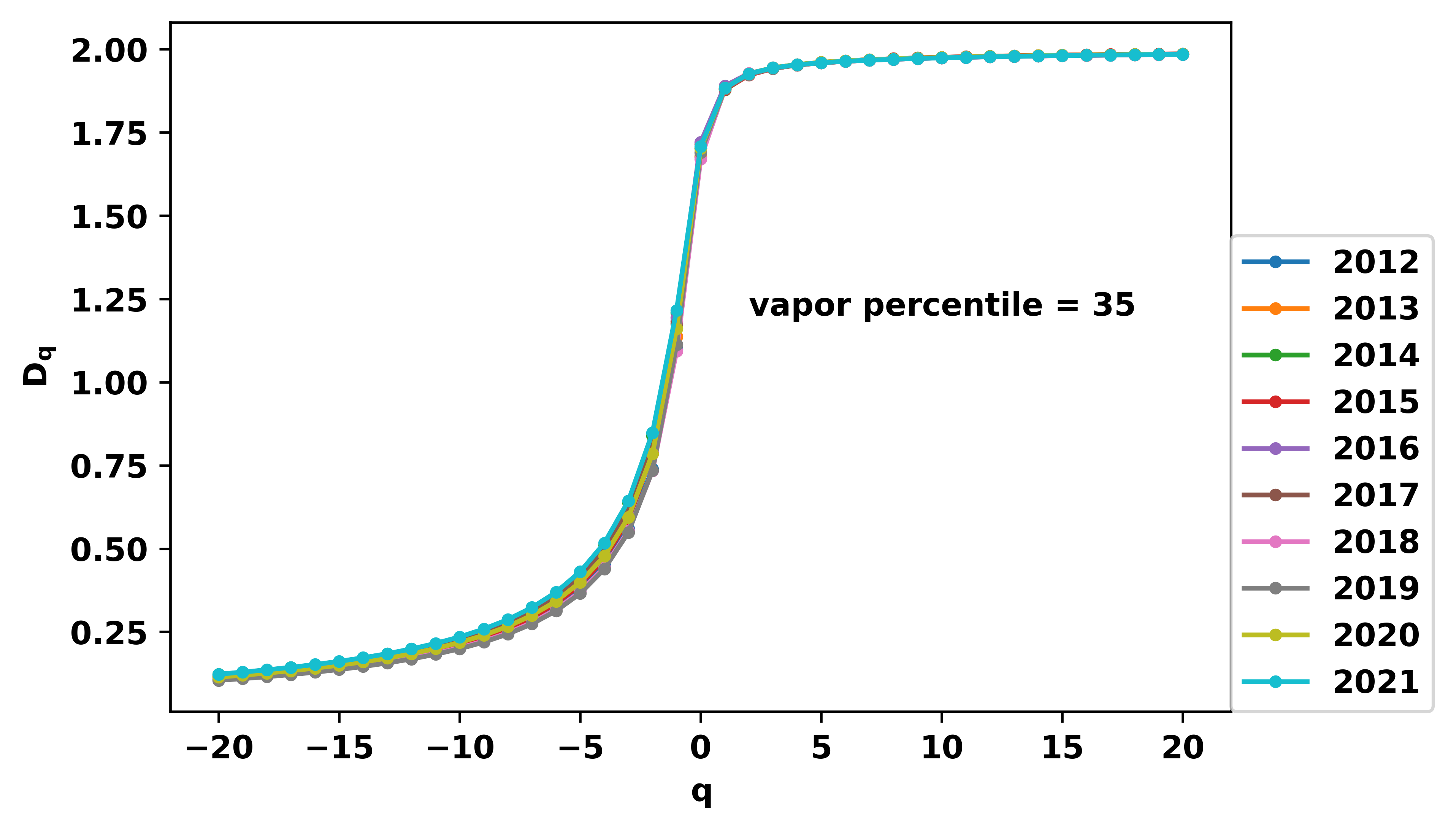} 
	\end{subfigure}		
	\begin{subfigure}[t]{0.3\textwidth}
		\centering
		\includegraphics[width=\linewidth]{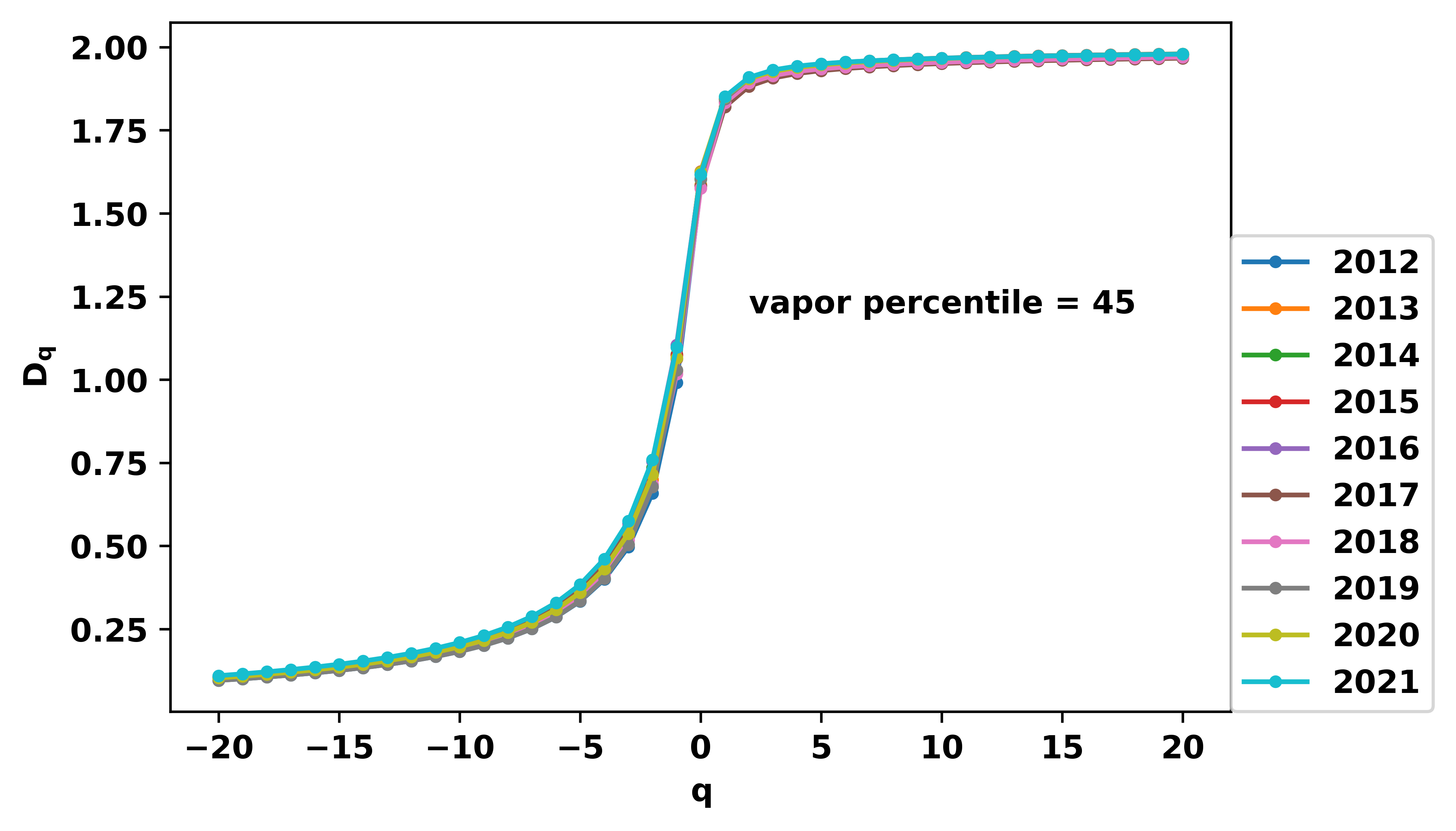} 
	\end{subfigure}
	\begin{subfigure}[t]{0.3\textwidth}
		\centering
		\includegraphics[width=\linewidth]{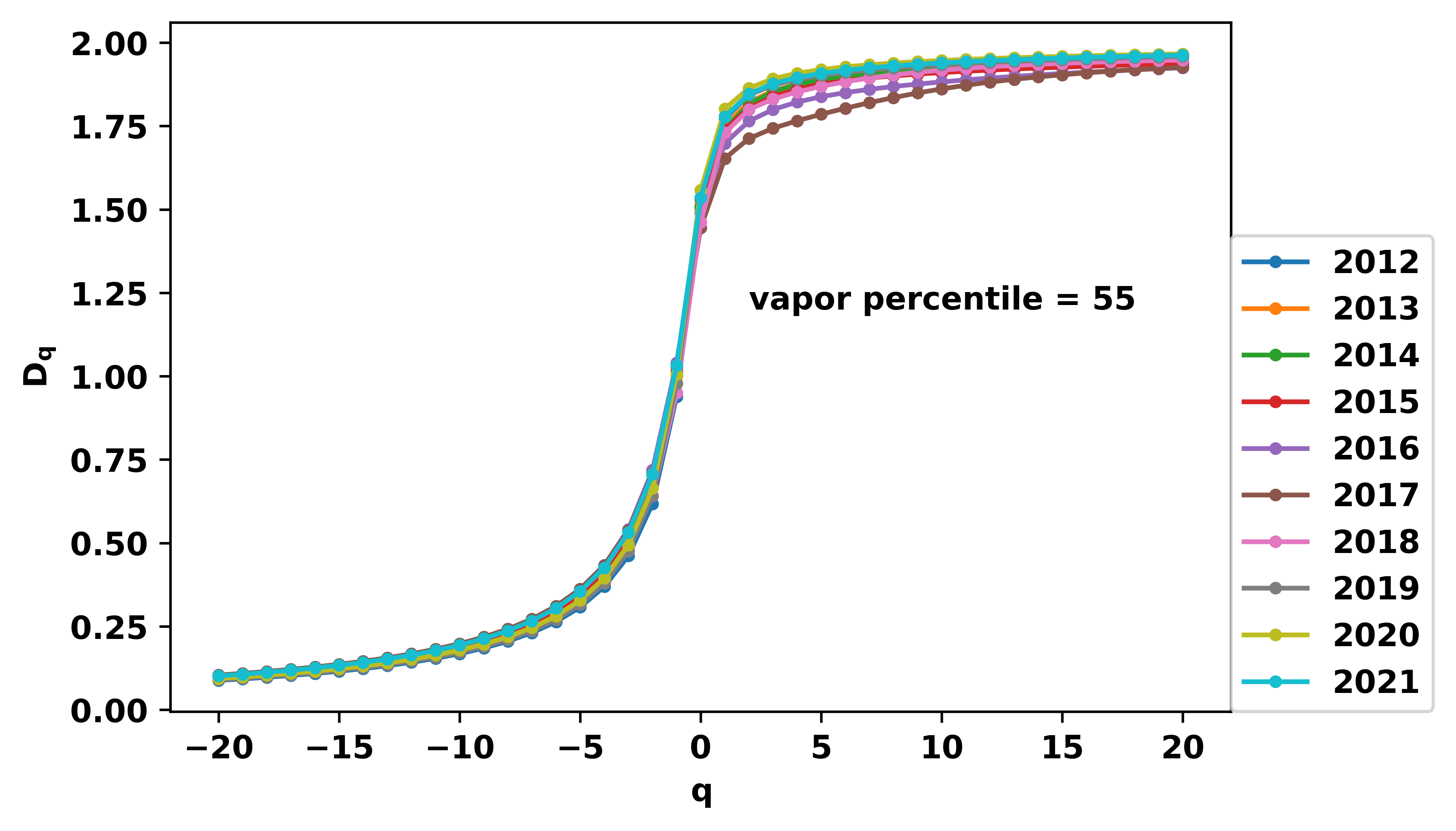} 
	\end{subfigure}
	\begin{subfigure}[t]{0.3\textwidth}
		\centering
		\includegraphics[width=\linewidth]{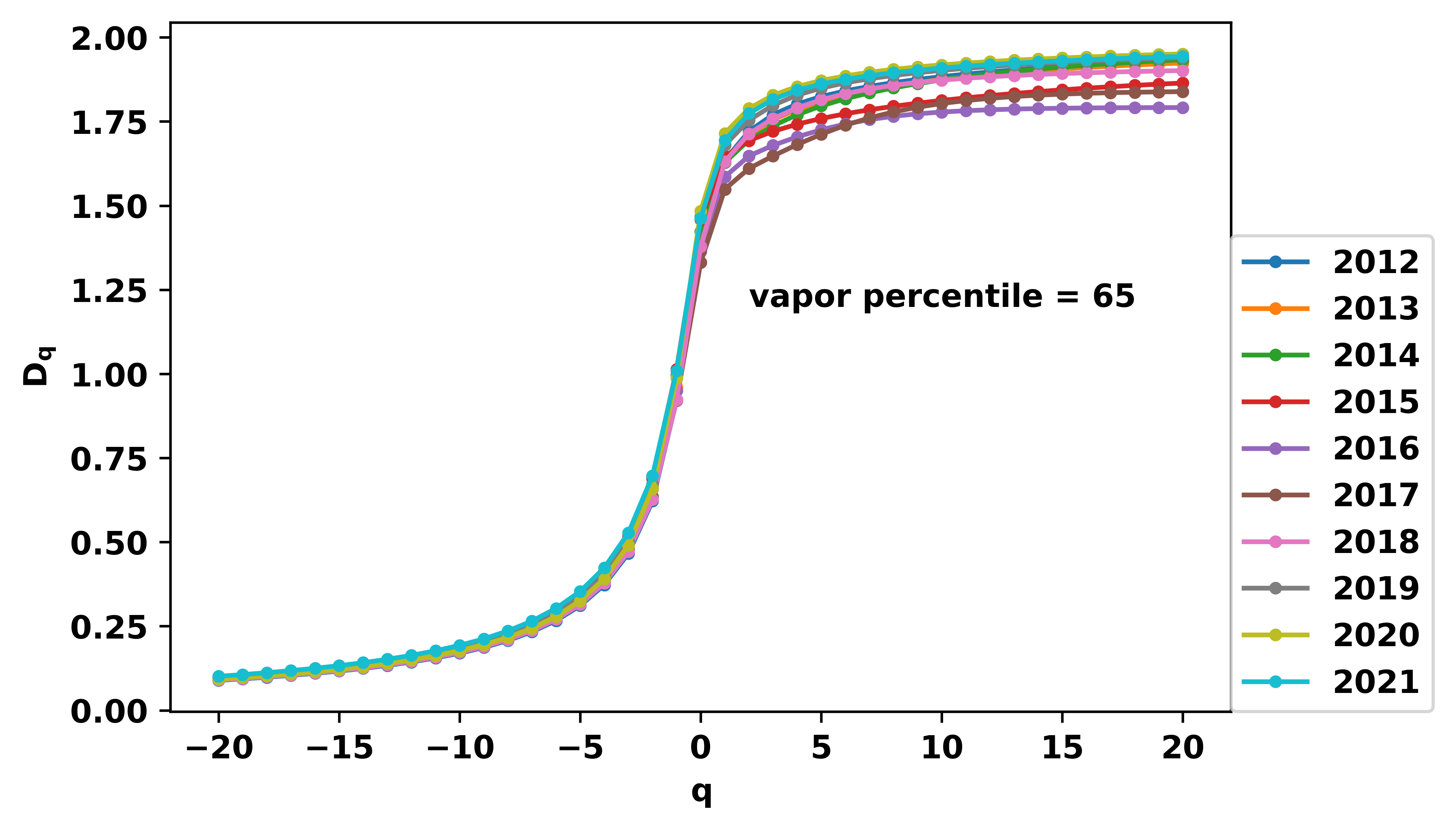} 
	\end{subfigure}		
	\begin{subfigure}[t]{0.3\textwidth}
		\centering
		\includegraphics[width=\linewidth]{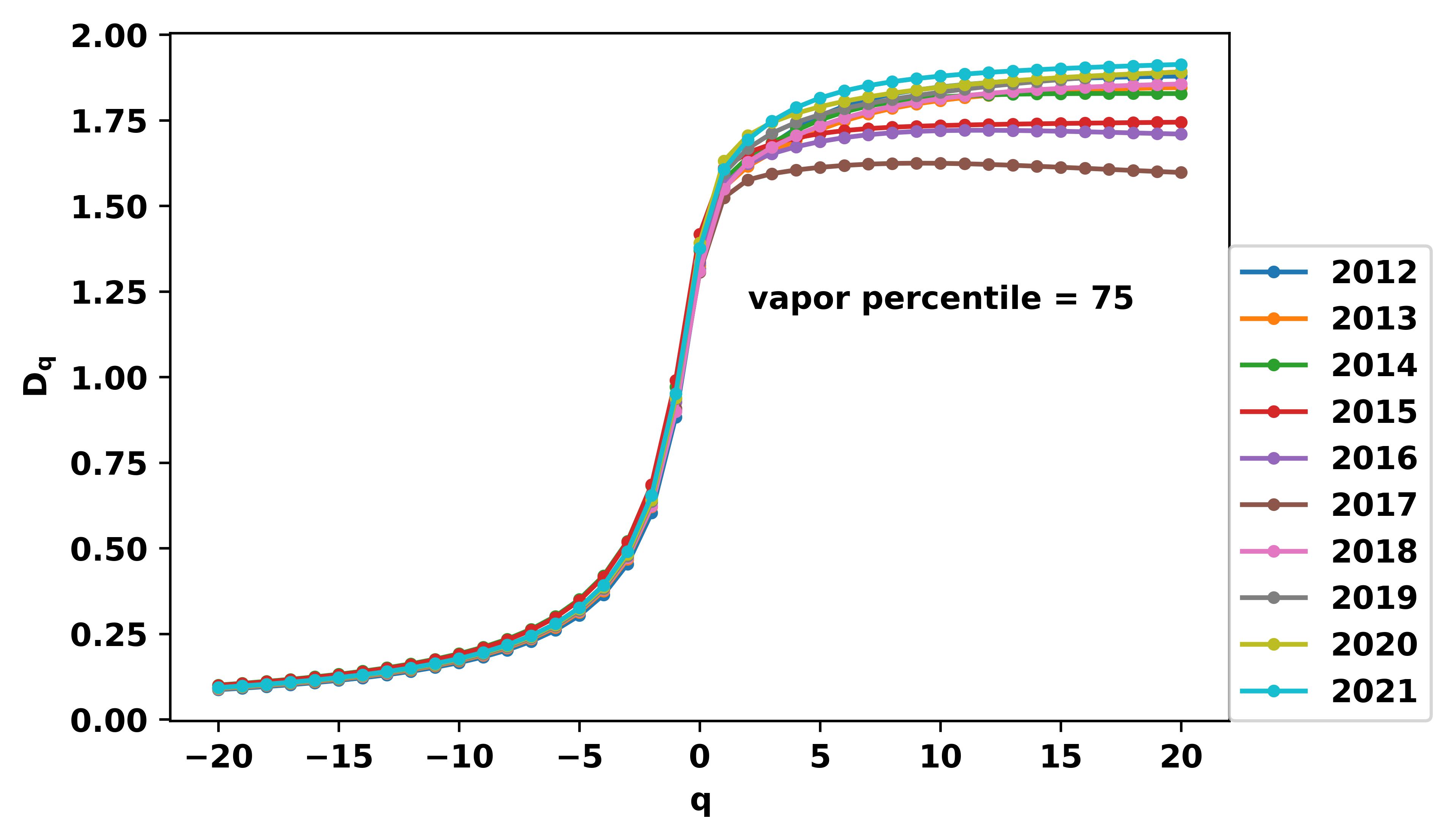} 
	\end{subfigure}
	\begin{subfigure}[t]{0.3\textwidth}
		\centering
		\includegraphics[width=\linewidth]{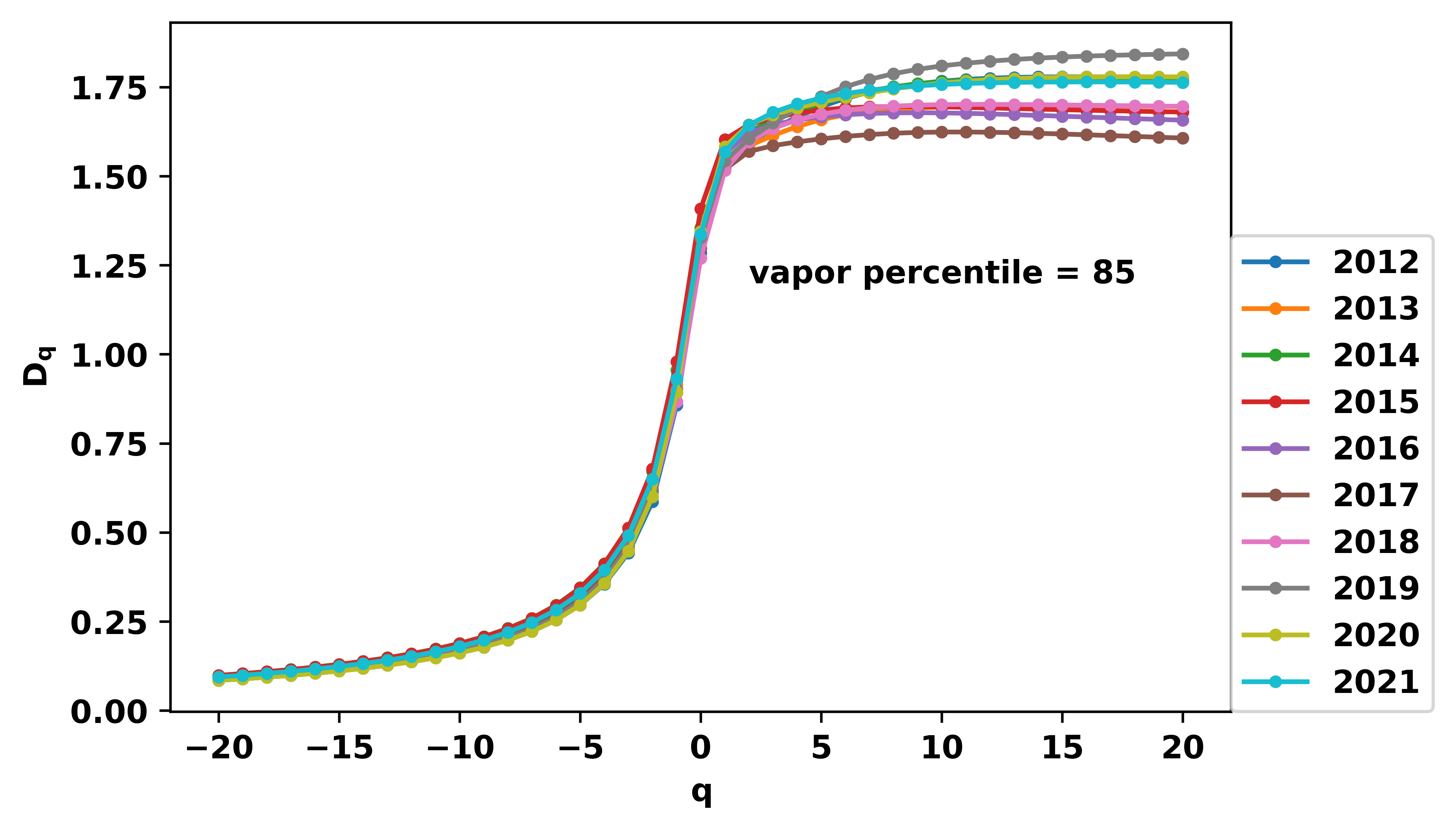} 
	\end{subfigure}
	\caption{The multi-fractal dimension $D_q$ for all 10 years (averaged over all months of each year) are plotted as a function of $q$ at different vapor percentiles. We see that a branching in the distribution occurs at the percolation threshold.}
	\label{300}
\end{figure}

\begin{figure}[H]
	\centering
	\begin{subfigure}[t]{0.3\textwidth}
		\centering
		\includegraphics[width=\linewidth]{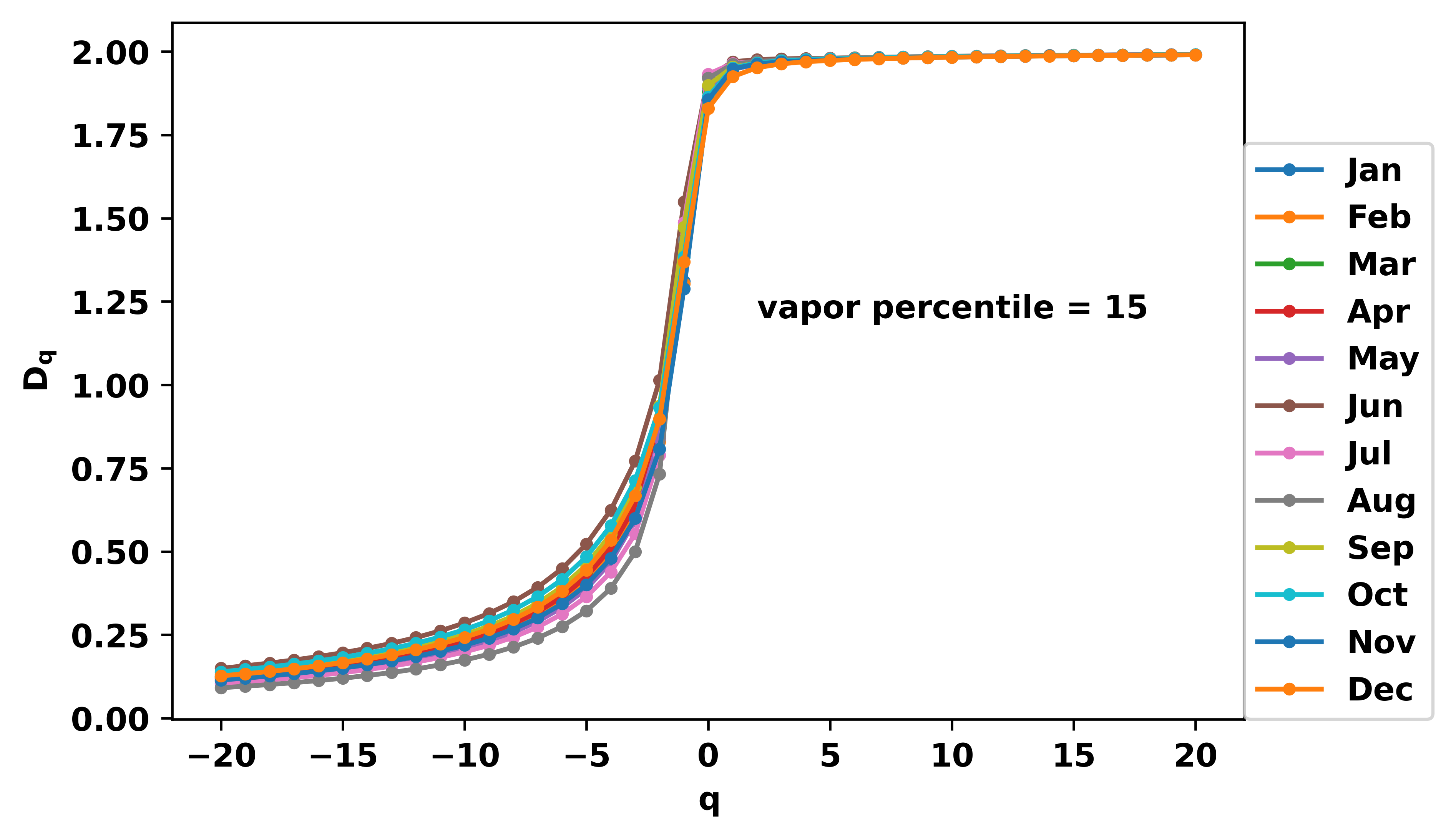} 
	\end{subfigure}
	\begin{subfigure}[t]{0.3\textwidth}
		\centering
		\includegraphics[width=\linewidth]{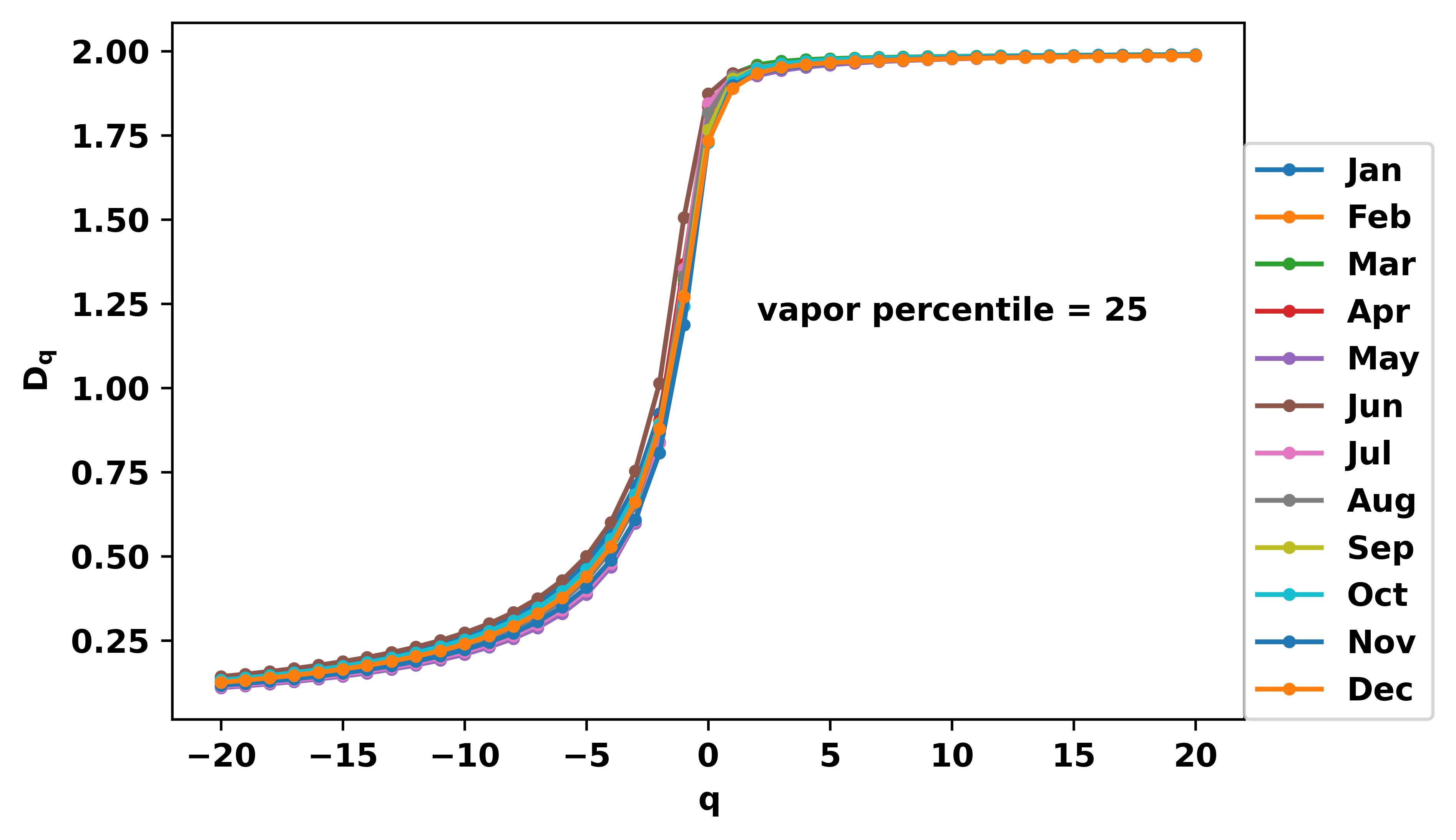} 
	\end{subfigure}
	\begin{subfigure}[t]{0.3\textwidth}
		\centering
		\includegraphics[width=\linewidth]{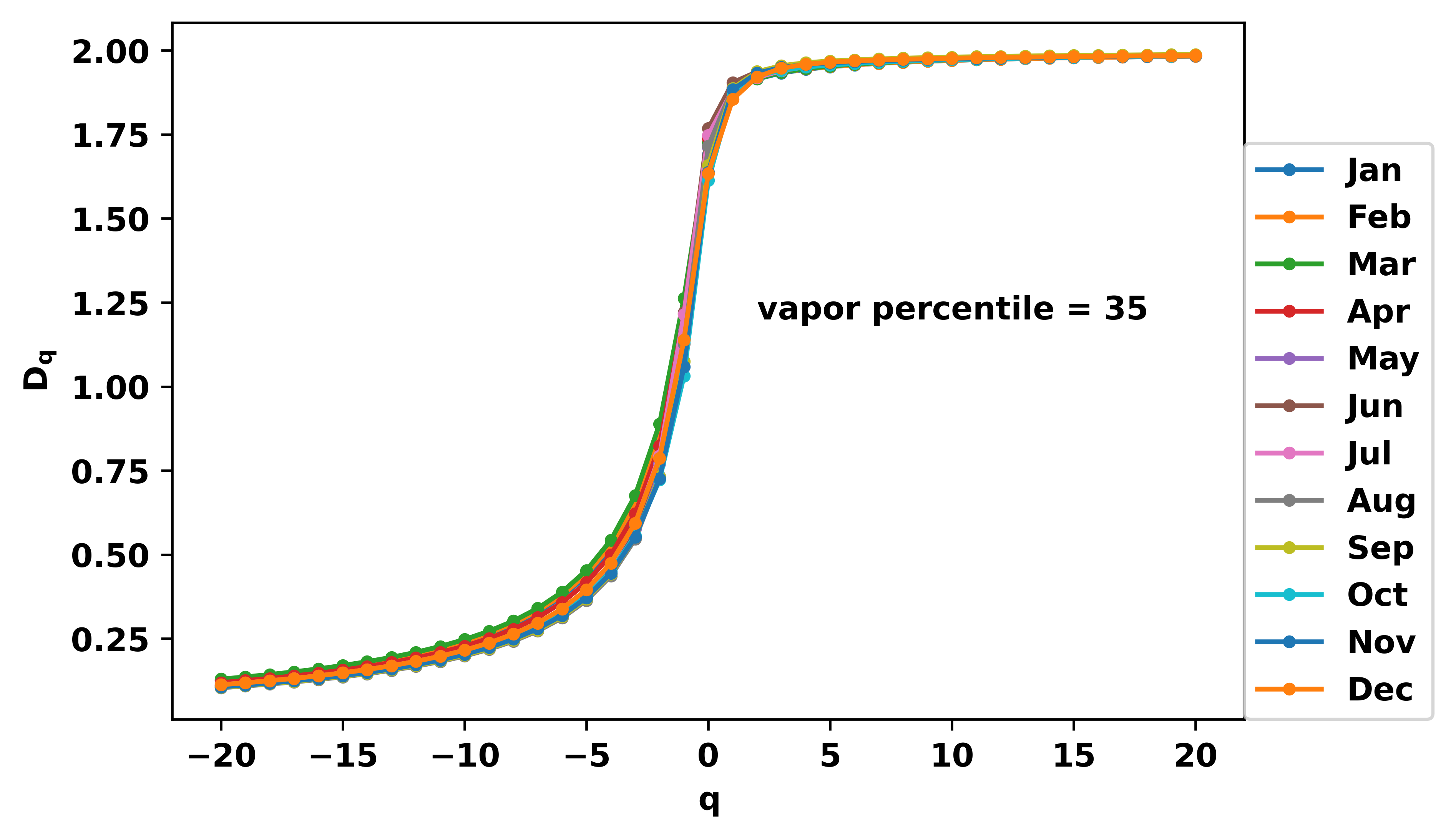} 
	\end{subfigure}		
	\begin{subfigure}[t]{0.3\textwidth}
		\centering
		\includegraphics[width=\linewidth]{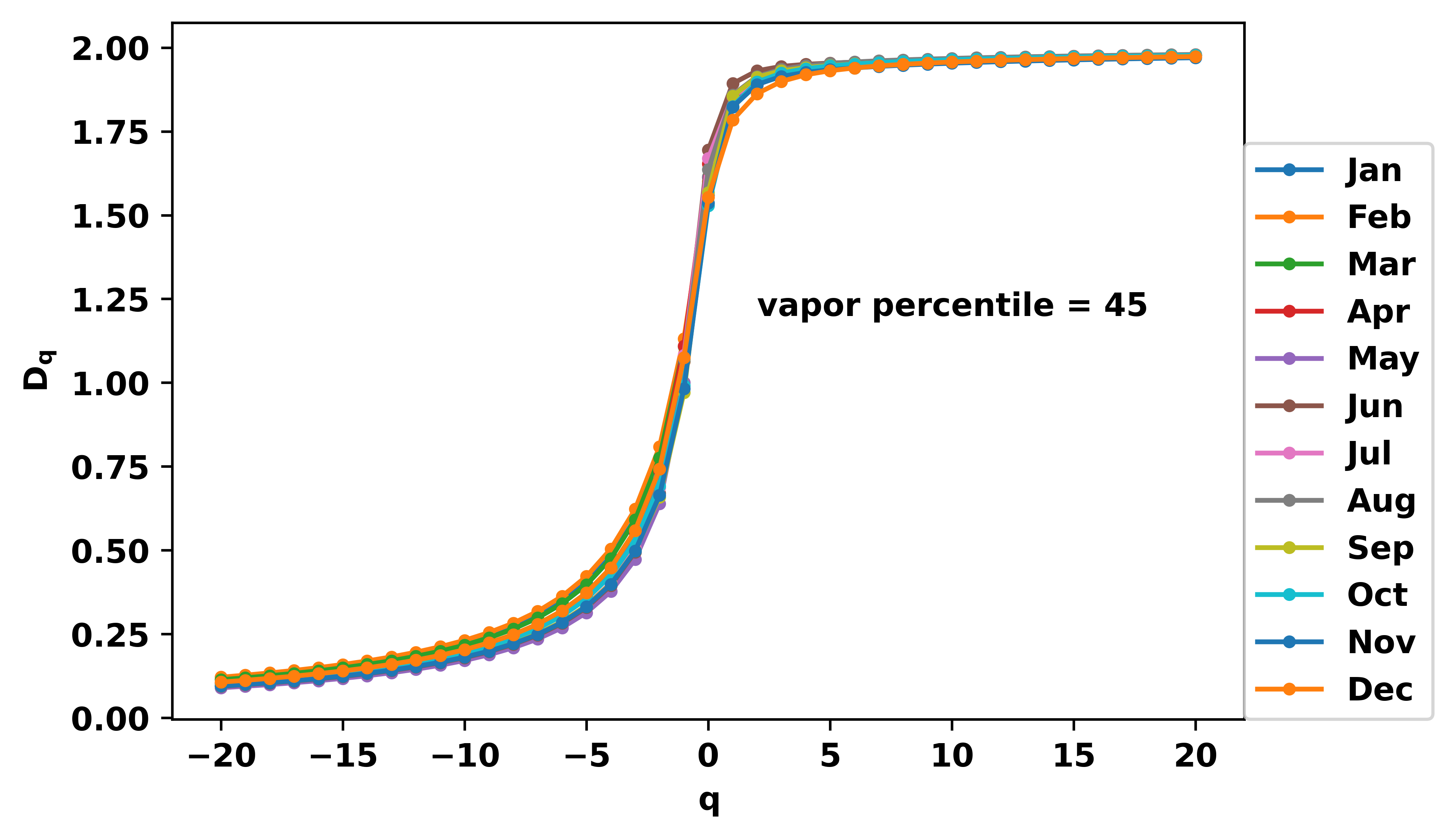} 
	\end{subfigure}
	\begin{subfigure}[t]{0.3\textwidth}
		\centering
		\includegraphics[width=\linewidth]{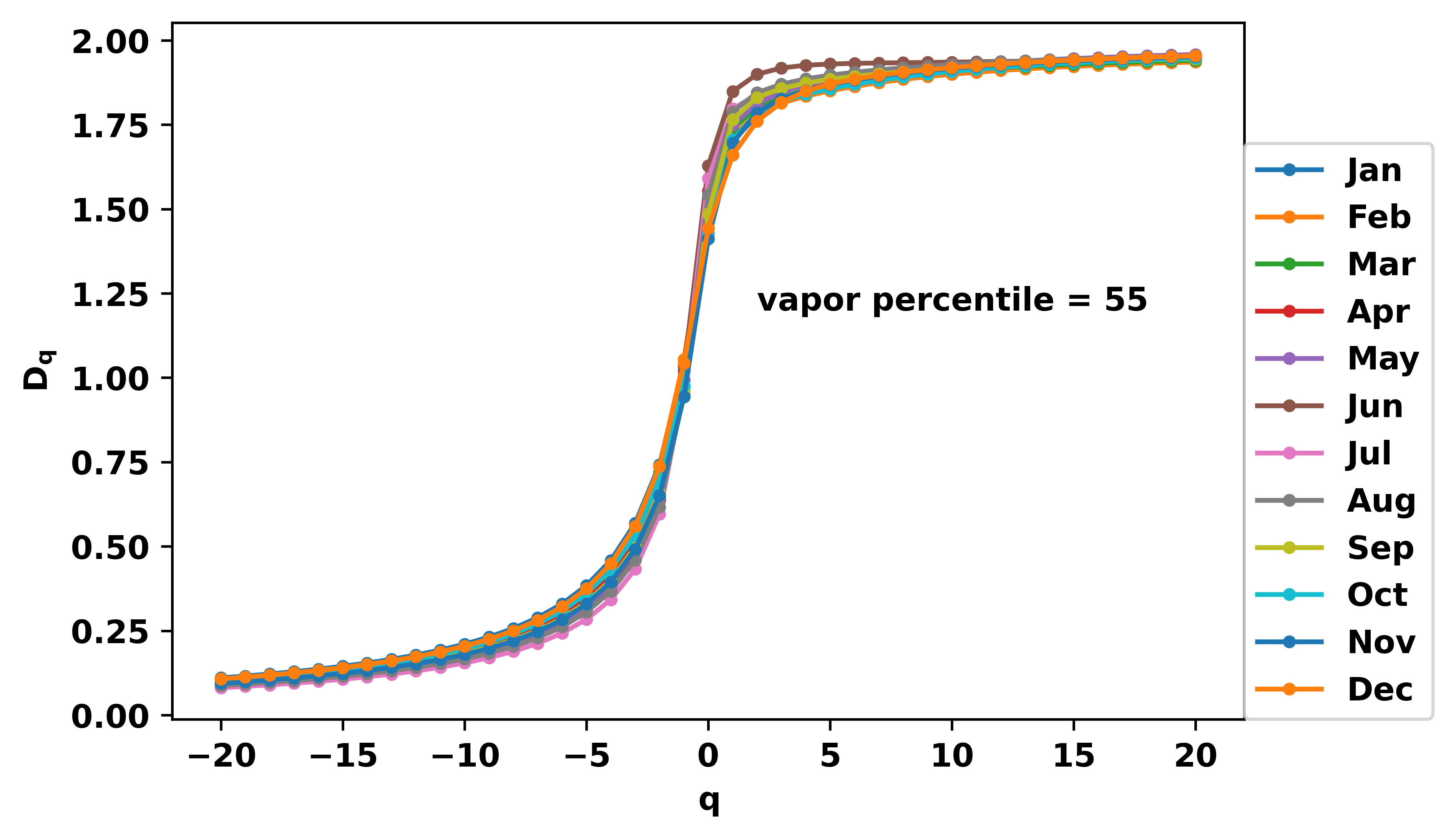} 
	\end{subfigure}
	\begin{subfigure}[t]{0.3\textwidth}
		\centering
		\includegraphics[width=\linewidth]{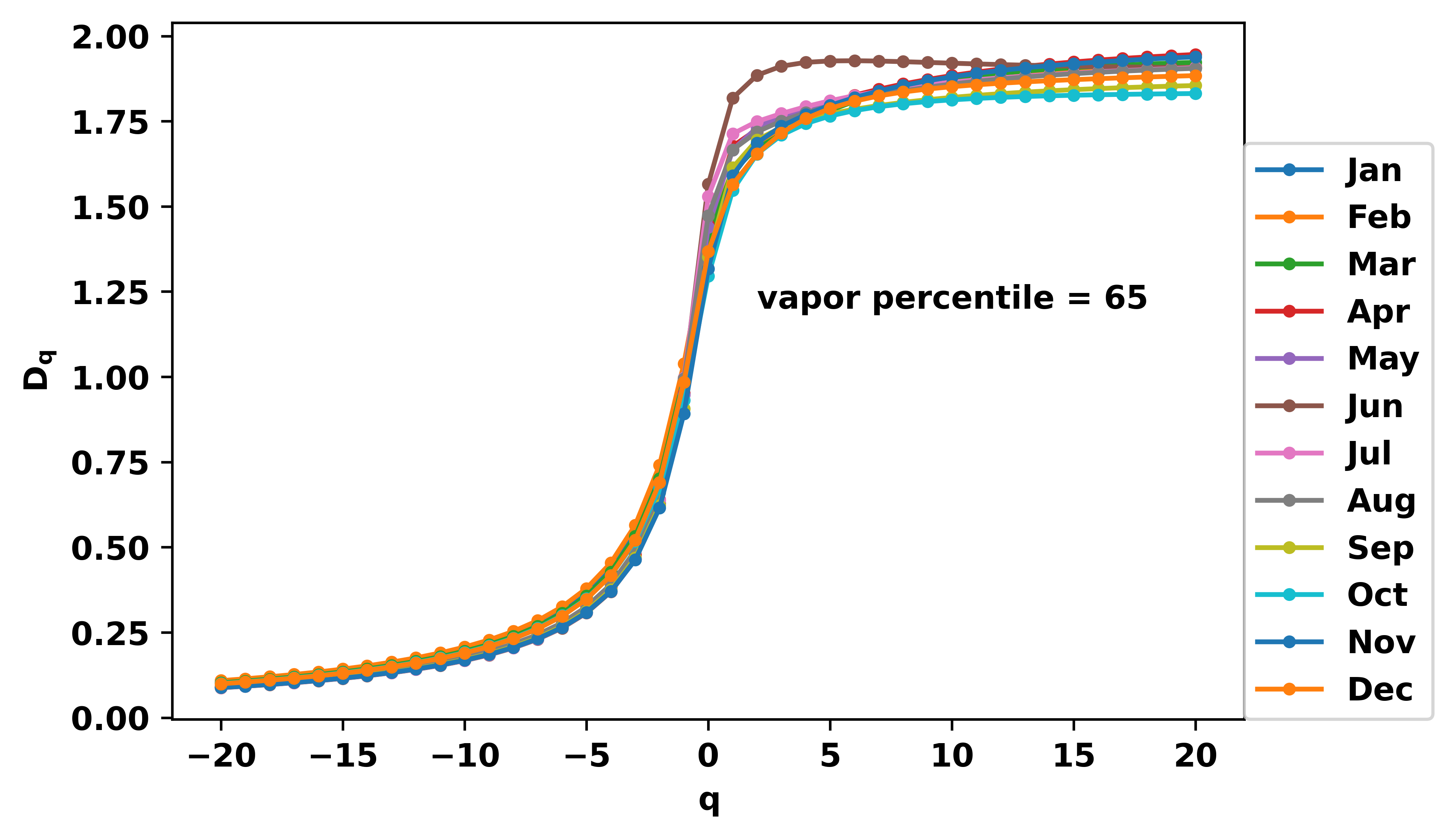} 
	\end{subfigure}		
	\begin{subfigure}[t]{0.3\textwidth}
		\centering
		\includegraphics[width=\linewidth]{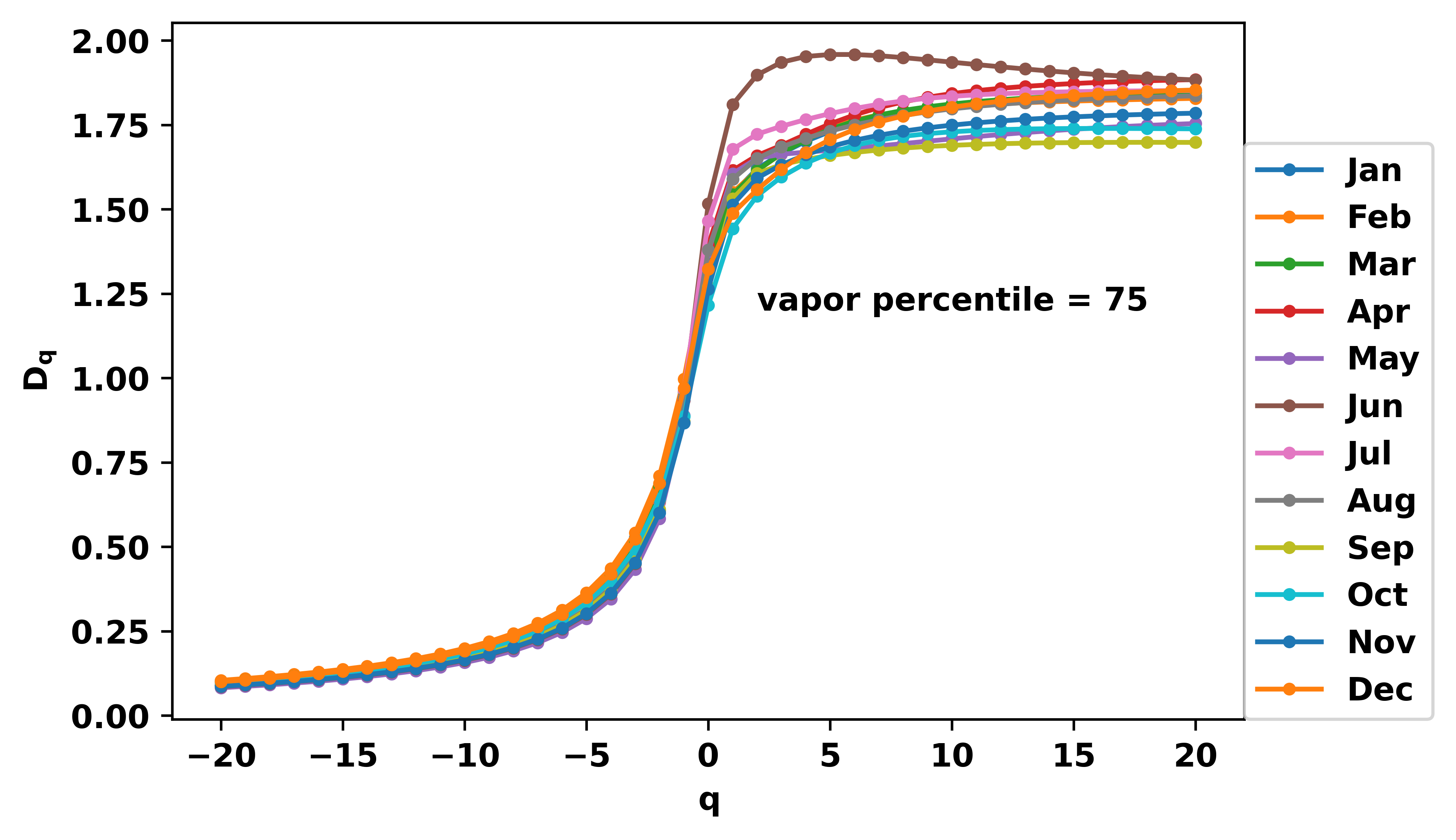} 
	\end{subfigure}
	\begin{subfigure}[t]{0.3\textwidth}
		\centering
		\includegraphics[width=\linewidth]{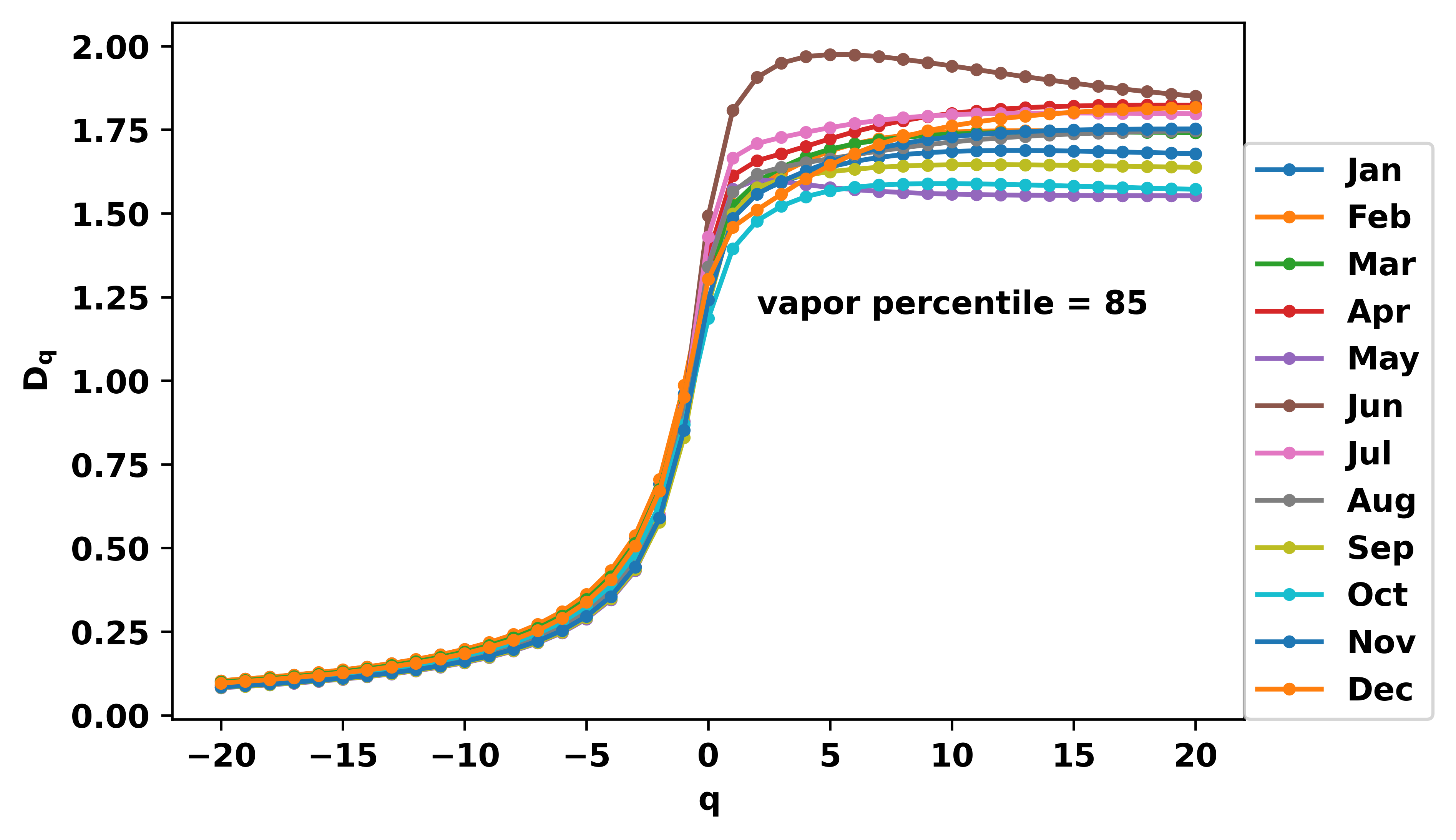} 
	\end{subfigure}
	\caption{The multi-fractal dimension $D_q$ for all 12 months (averaged over all 10 years) are plotted as a function of $q$ at different vapor percentiles. We see that a branching in the distribution occurs at the percolation threshold.}
	\label{400}
\end{figure}

As in the previous section, we compute the multifractal dimension $D_q$ for our data as a function of $q$ and also as a function of the vapor percentile $V$. The technique followed is the box-counting method and we use the formula in Eqn.\ref{1000} to determine $D_q$. In Fig.\ref{300}, we plot $D_q$ as a function of $q$ for different vapor percentiles, averaged over all months of each of the $10$ years shown. We find that for lower vapor percentiles, the plots have a ubiquitous character, while for larger vapor percentiles $V \geq 55$ they exhibit a branching based on the year in question for $q>0$. Incidentally, the onset of this branching occurs near the percolation threshold (as discussed in the next section), making this a possible signature of the phase transition. Similiar behaviour if found for the plots in Fig.\ref{400} where we average over all 10 years of data for each month shown. 

\begin{figure}
	\centering
	\begin{subfigure}[t]{0.3\textwidth}
		\centering
		\includegraphics[width=\linewidth]{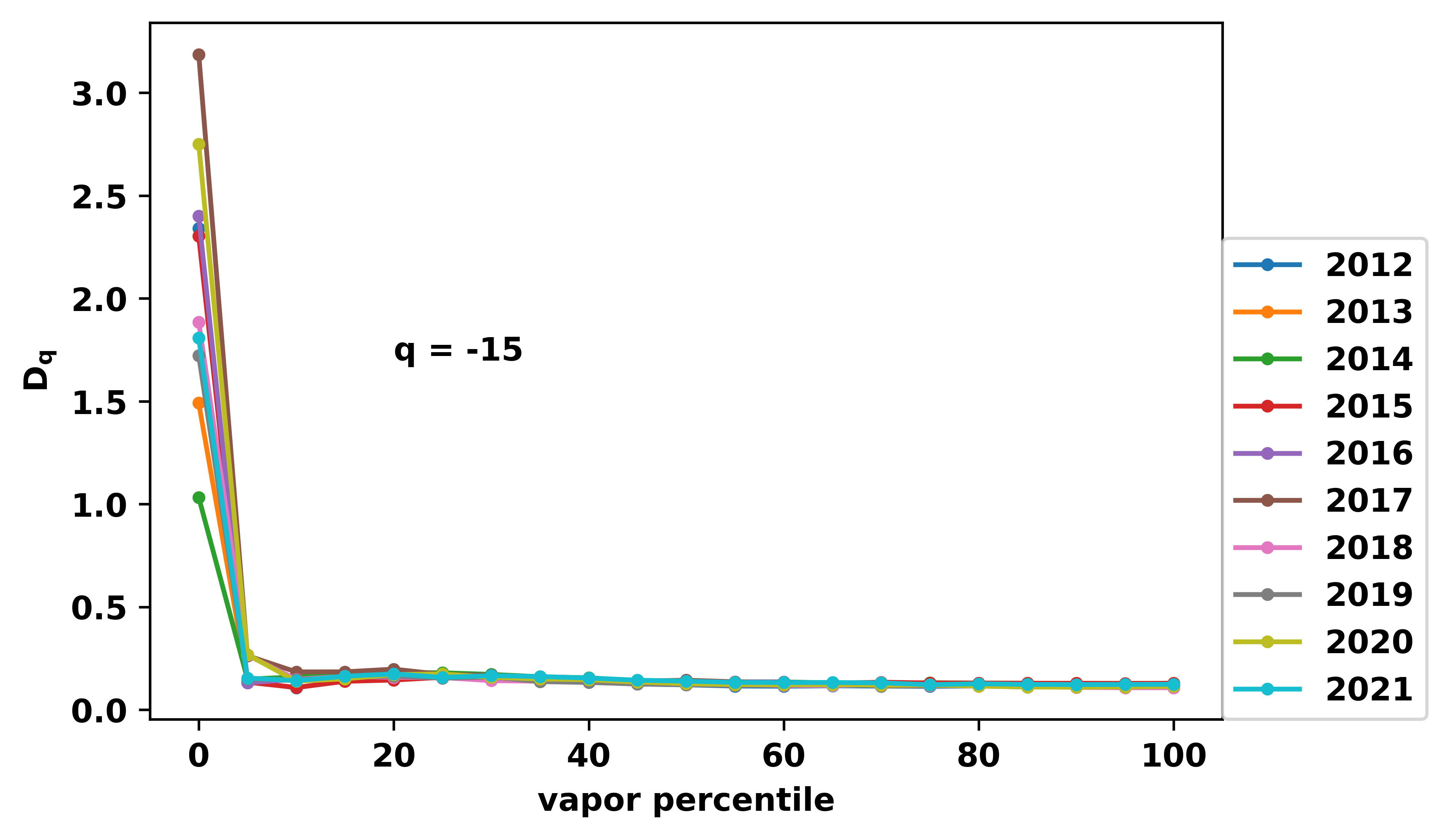} 
	\end{subfigure}
	\begin{subfigure}[t]{0.3\textwidth}
		\centering
		\includegraphics[width=\linewidth]{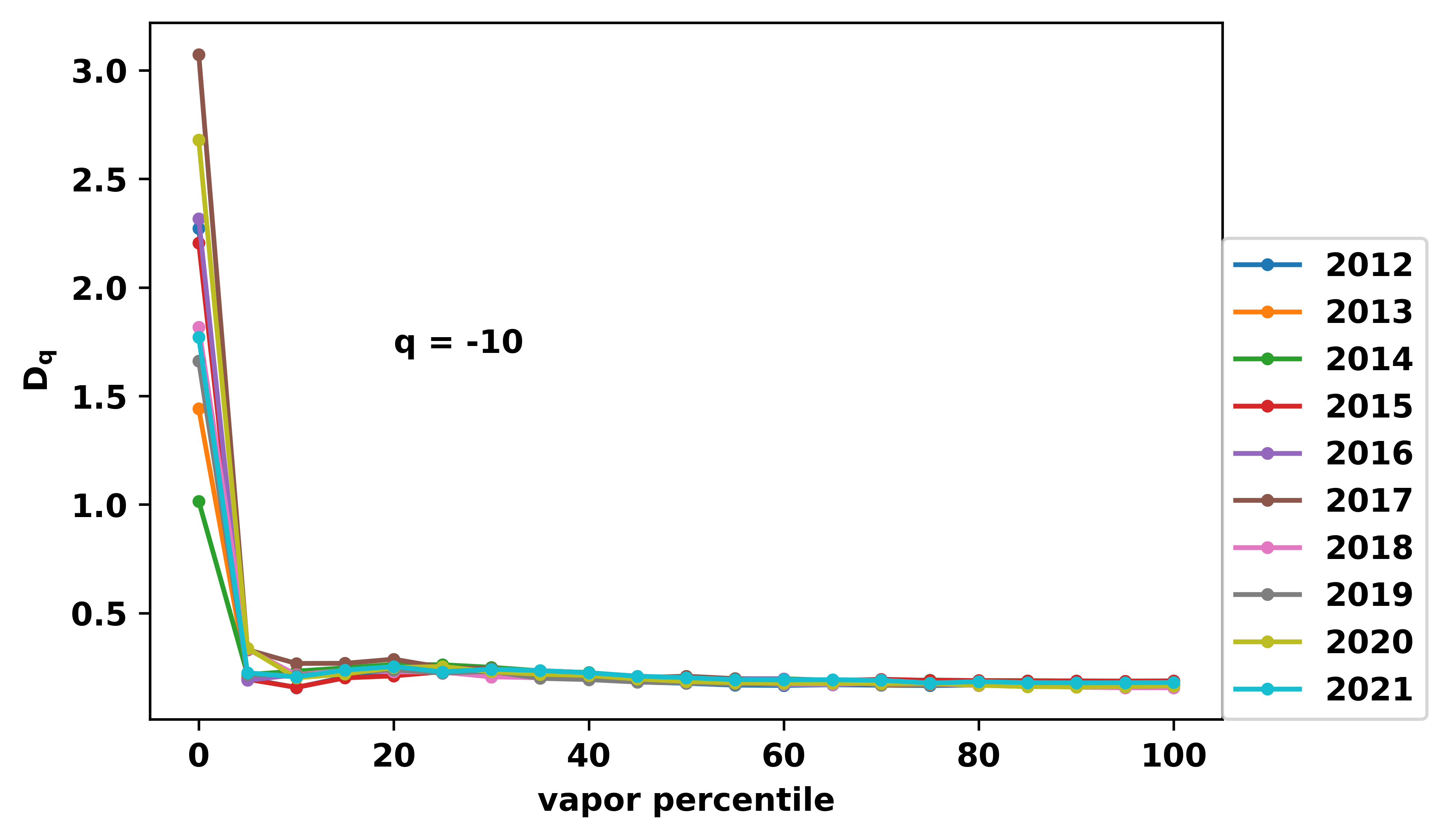} 
	\end{subfigure}
	\begin{subfigure}[t]{0.3\textwidth}
		\centering
		\includegraphics[width=\linewidth]{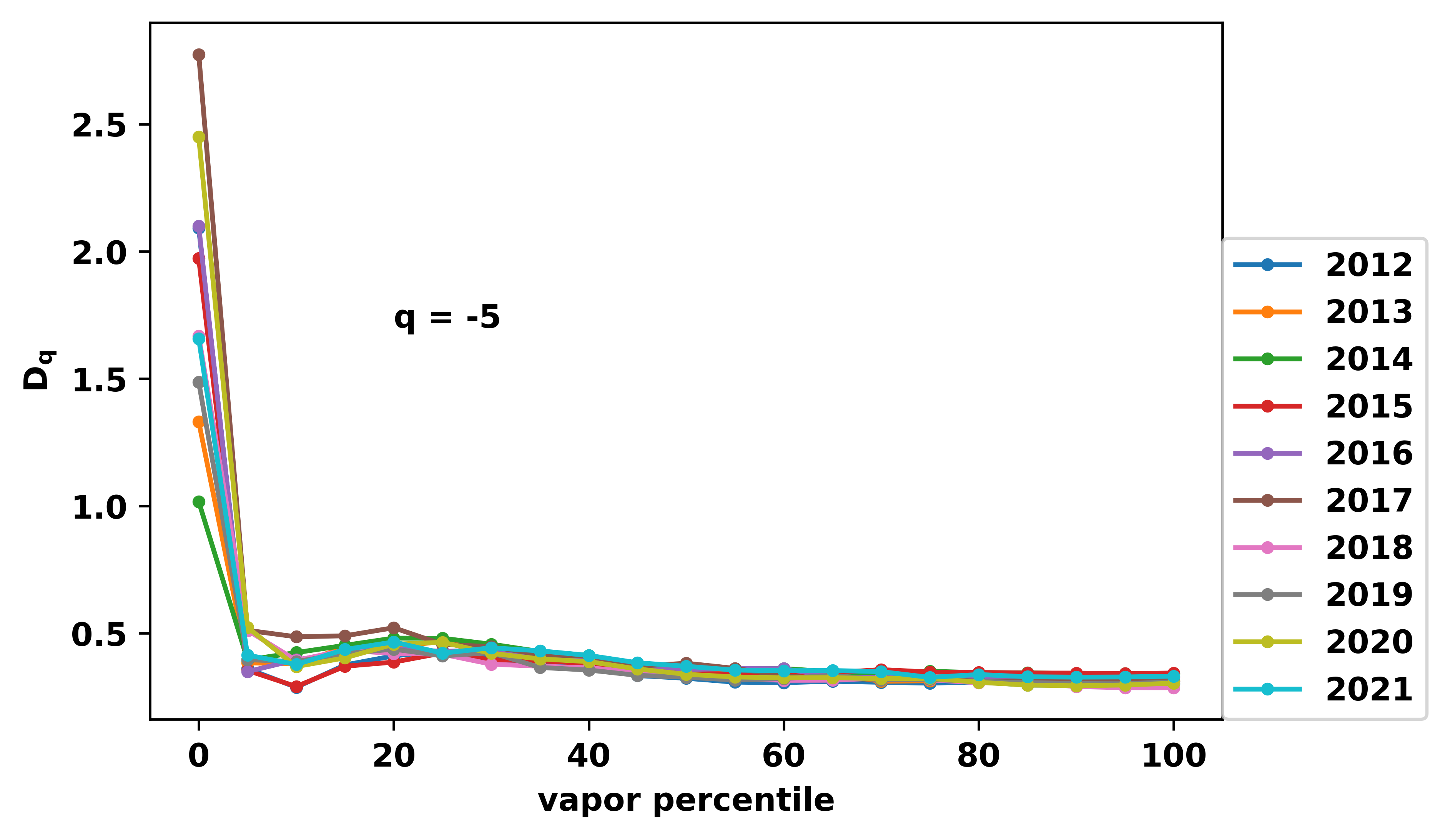} 
	\end{subfigure}		
	\begin{subfigure}[t]{0.3\textwidth}
		\centering
		\includegraphics[width=\linewidth]{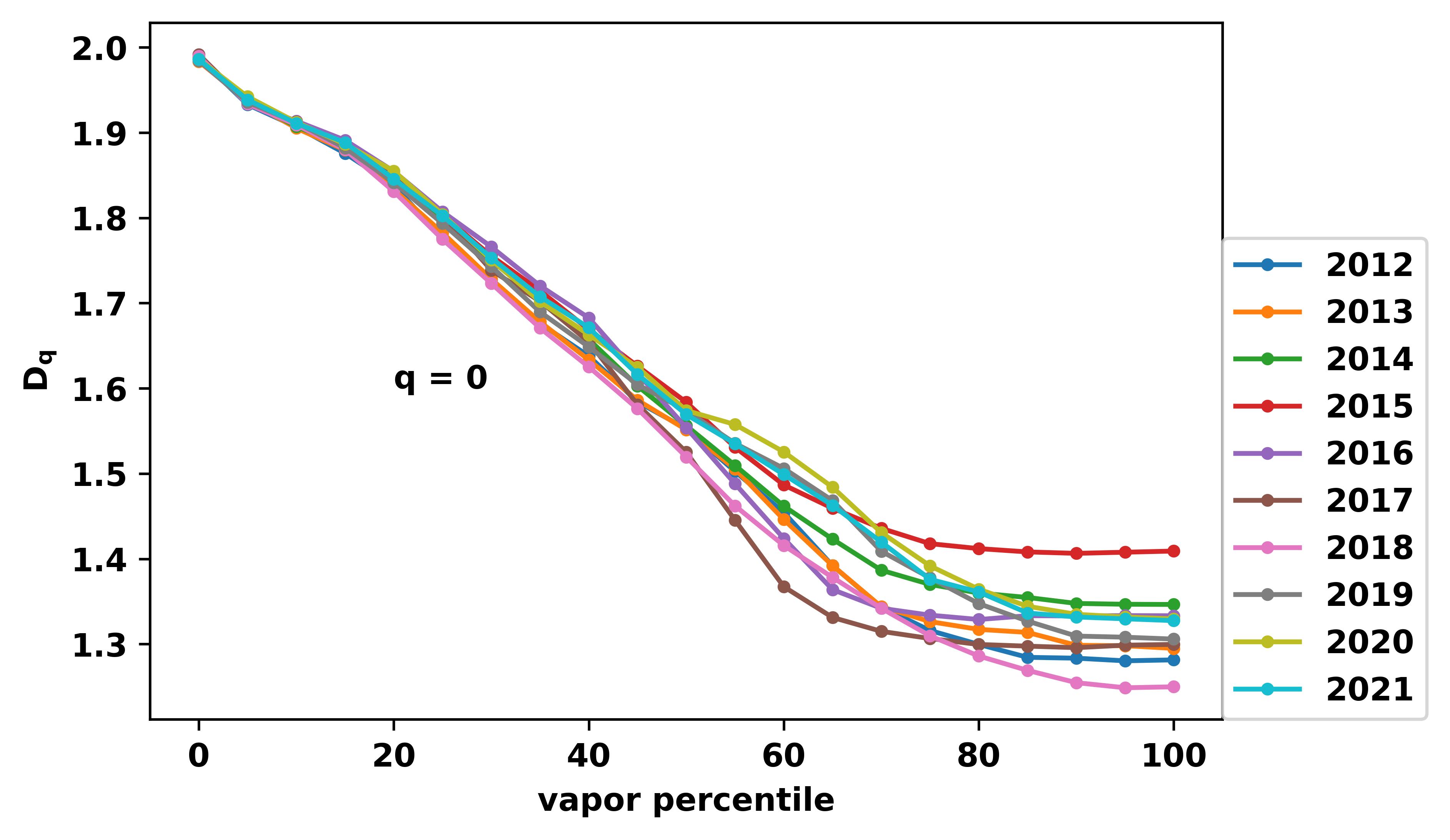} 
	\end{subfigure}
	\begin{subfigure}[t]{0.3\textwidth}
		\centering
		\includegraphics[width=\linewidth]{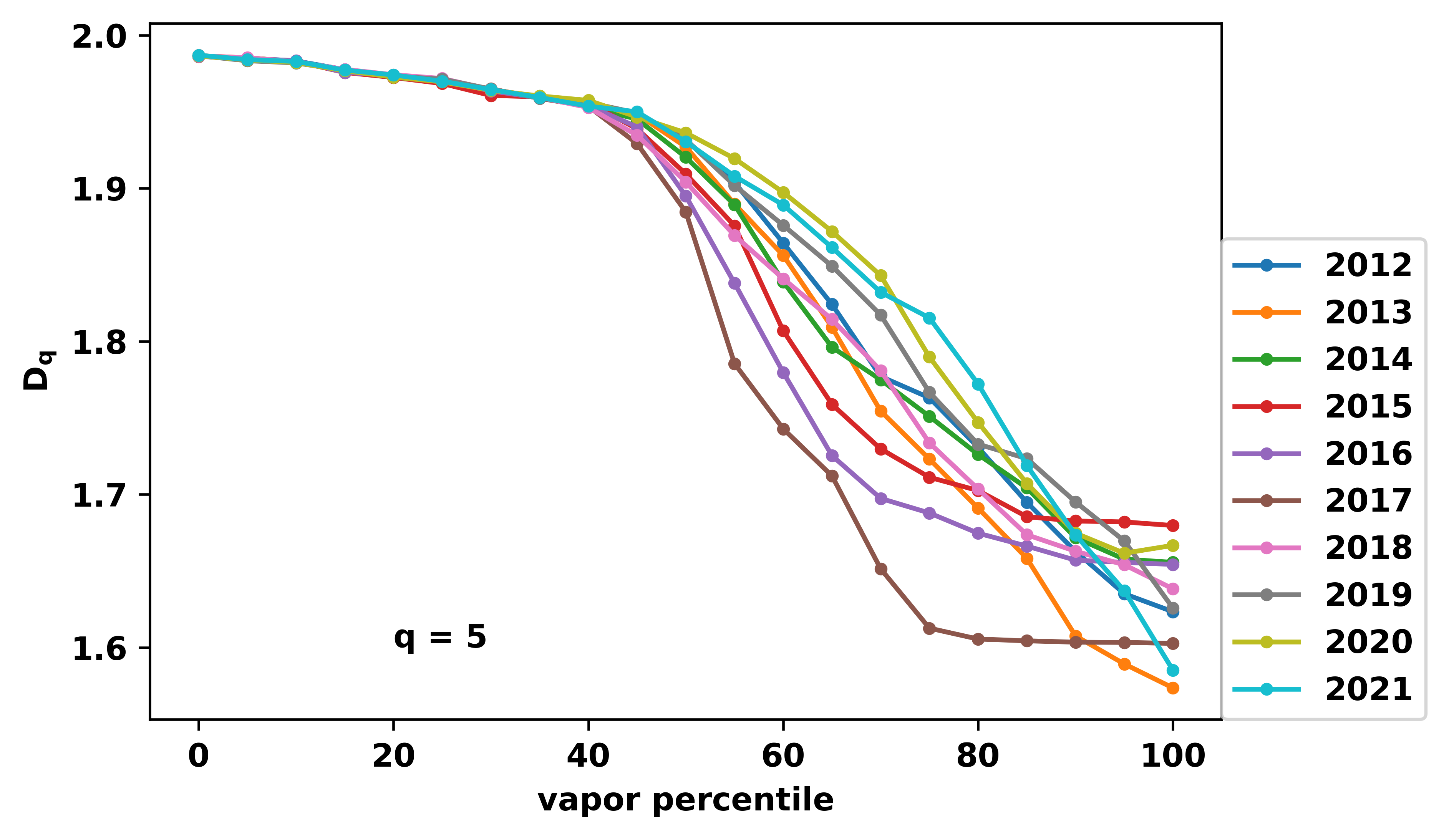} 
	\end{subfigure}
	\begin{subfigure}[t]{0.3\textwidth}
		\centering
		\includegraphics[width=\linewidth]{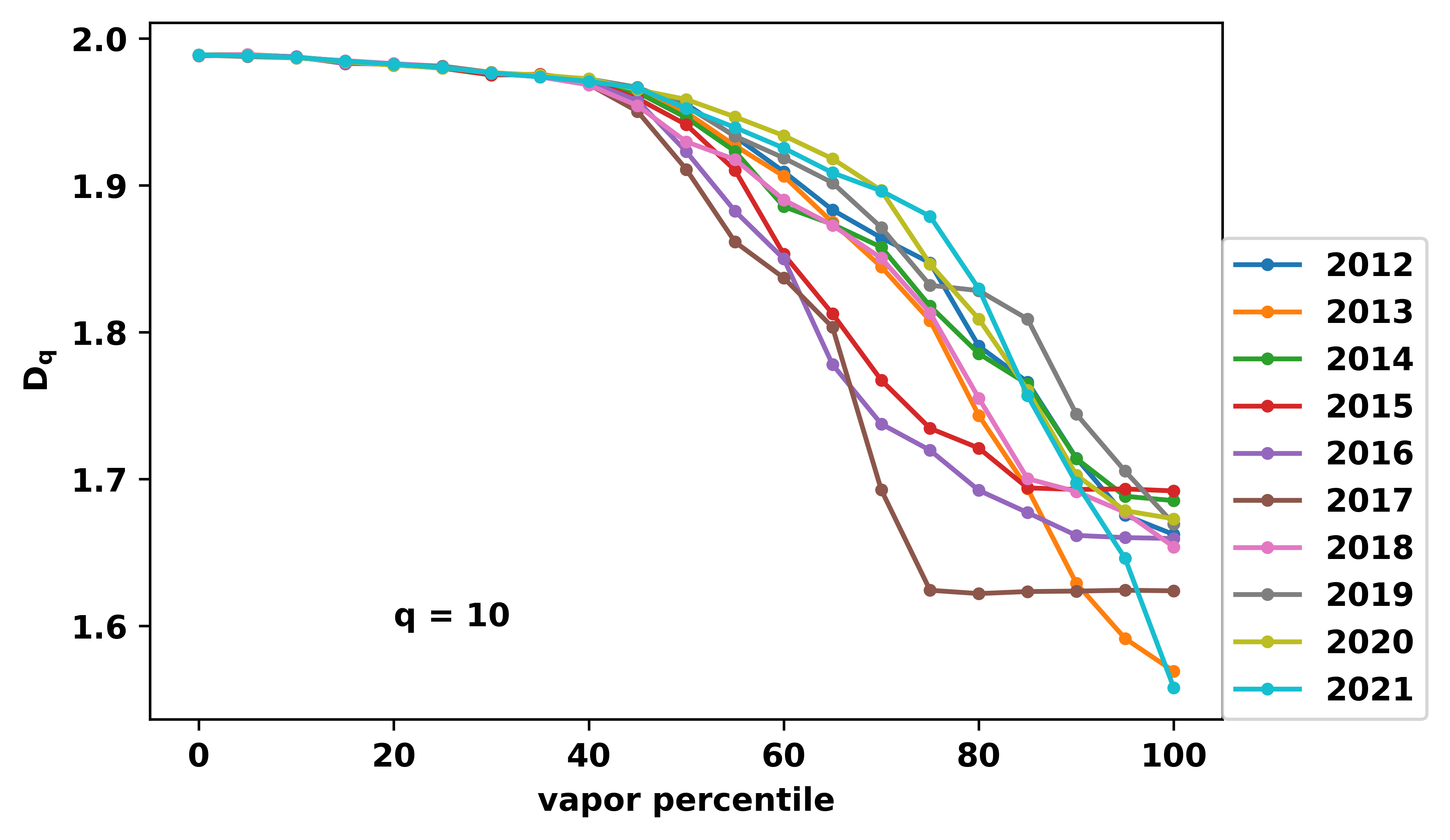} 
	\end{subfigure}		
	\begin{subfigure}[t]{0.3\textwidth}
		\centering
		\includegraphics[width=\linewidth]{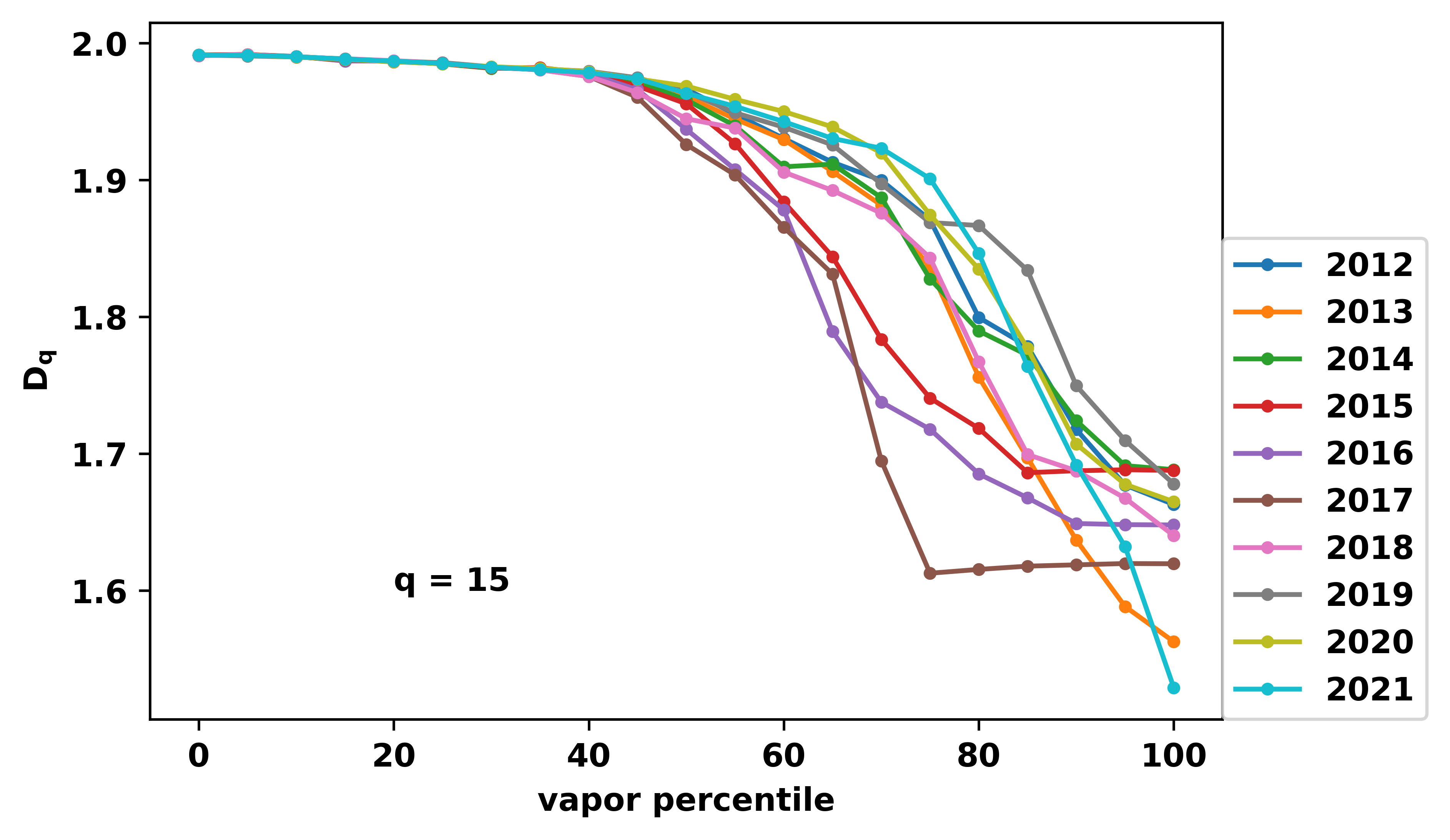} 
	\end{subfigure}
	\caption{The multi-fractal dimension $D_q$ for all 10 years (averaged over all months of each year) are plotted as a function of $V$ at different moments $q$. We see that a branching in the distribution occurs at the percolation threshold.}
	\label{500}
\end{figure}

\begin{figure}
	\centering
	\begin{subfigure}[t]{0.3\textwidth}
		\centering
		\includegraphics[width=\linewidth]{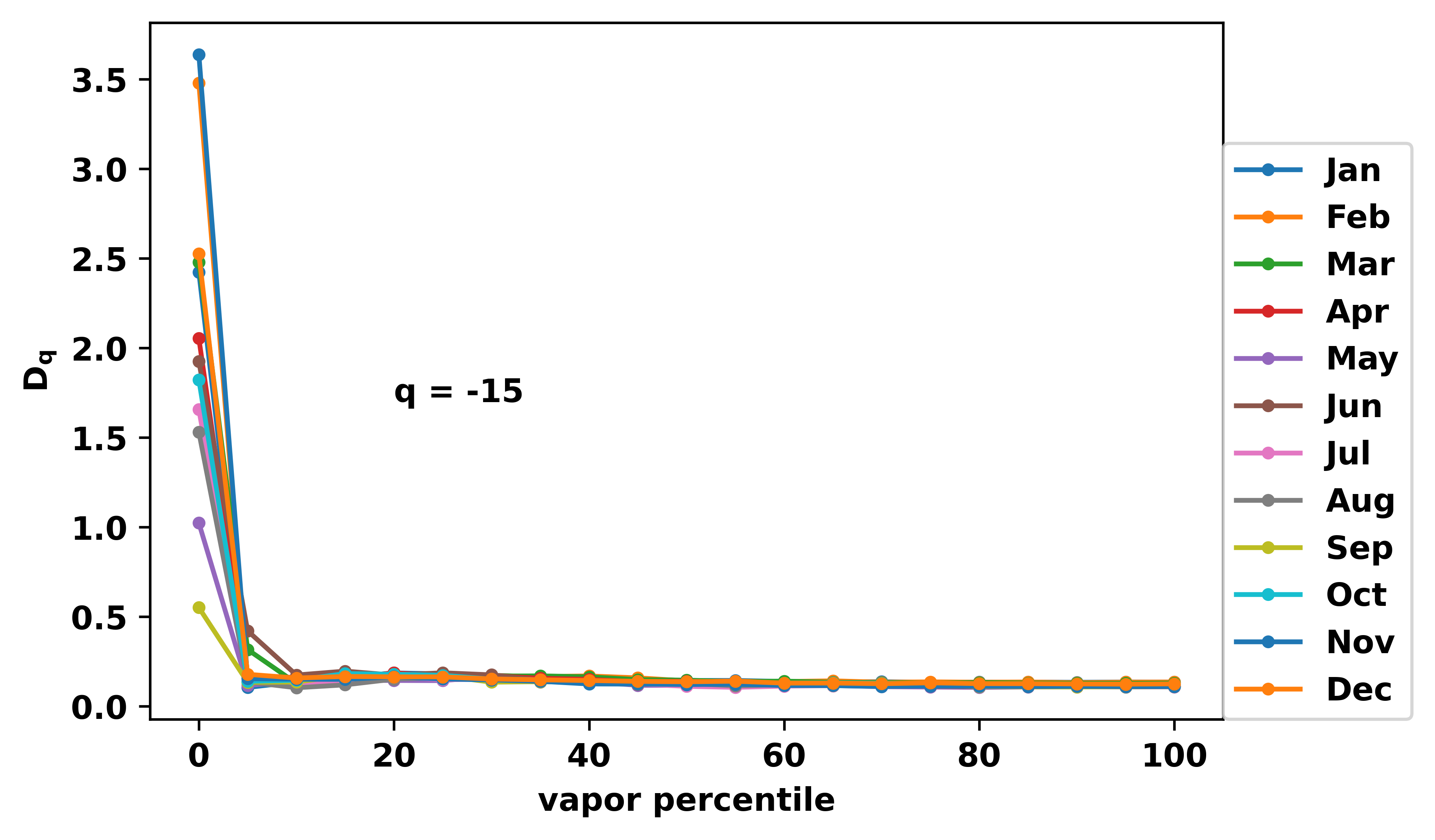} 
	\end{subfigure}
	\begin{subfigure}[t]{0.3\textwidth}
		\centering
		\includegraphics[width=\linewidth]{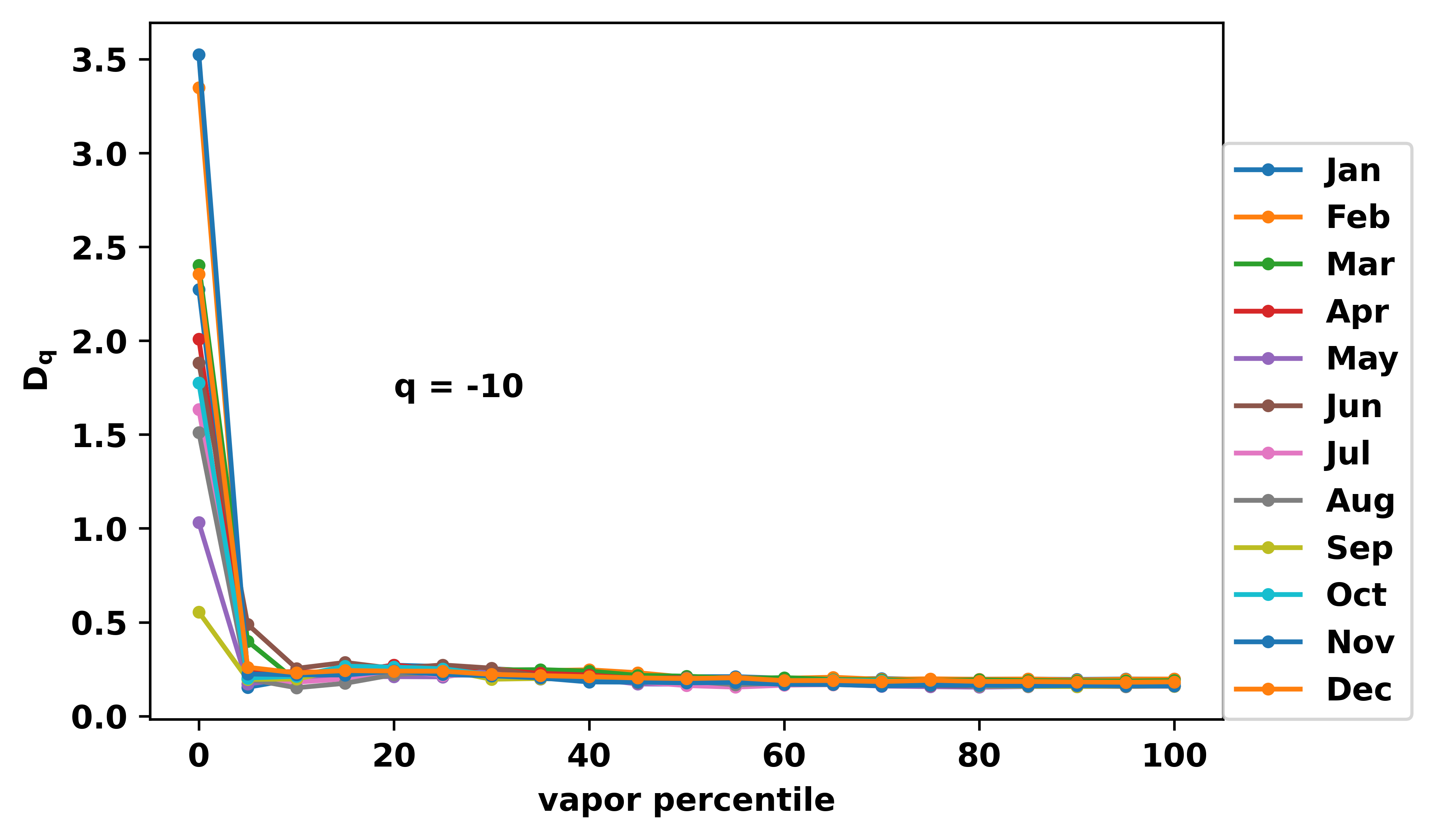} 
	\end{subfigure}
	\begin{subfigure}[t]{0.3\textwidth}
		\centering
		\includegraphics[width=\linewidth]{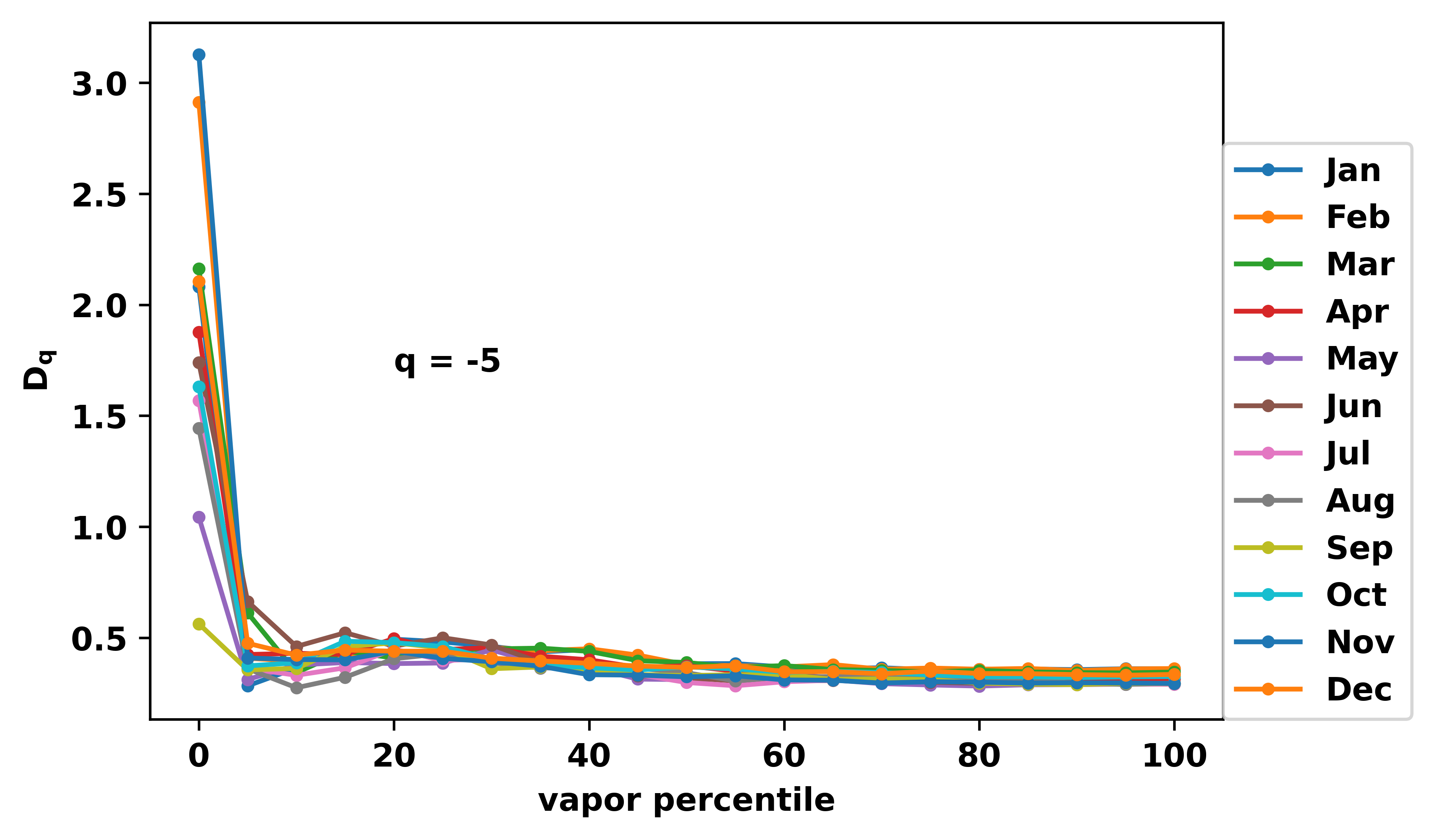} 
	\end{subfigure}		
	\begin{subfigure}[t]{0.3\textwidth}
		\centering
		\includegraphics[width=\linewidth]{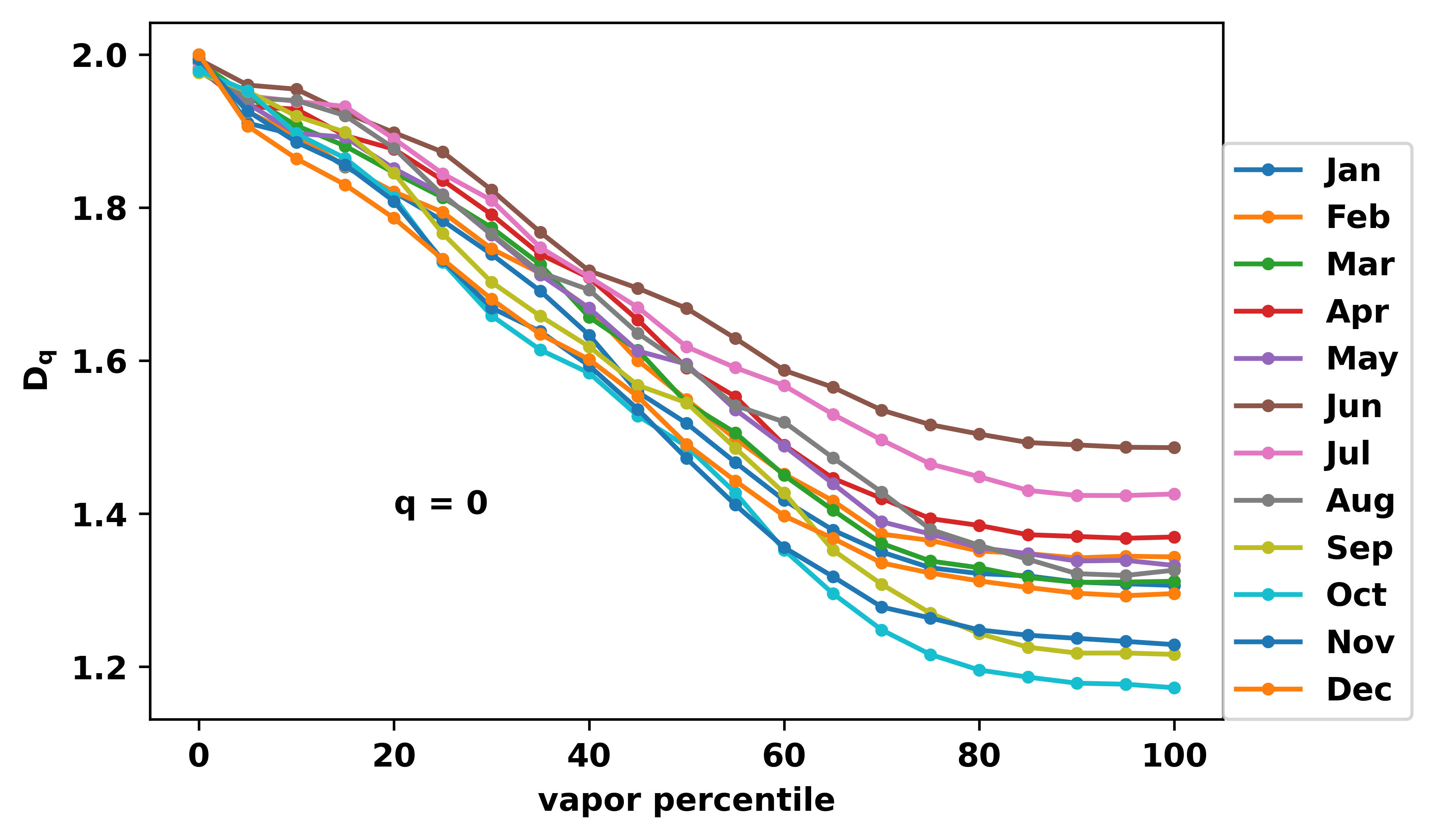} 
	\end{subfigure}
	\begin{subfigure}[t]{0.3\textwidth}
		\centering
		\includegraphics[width=\linewidth]{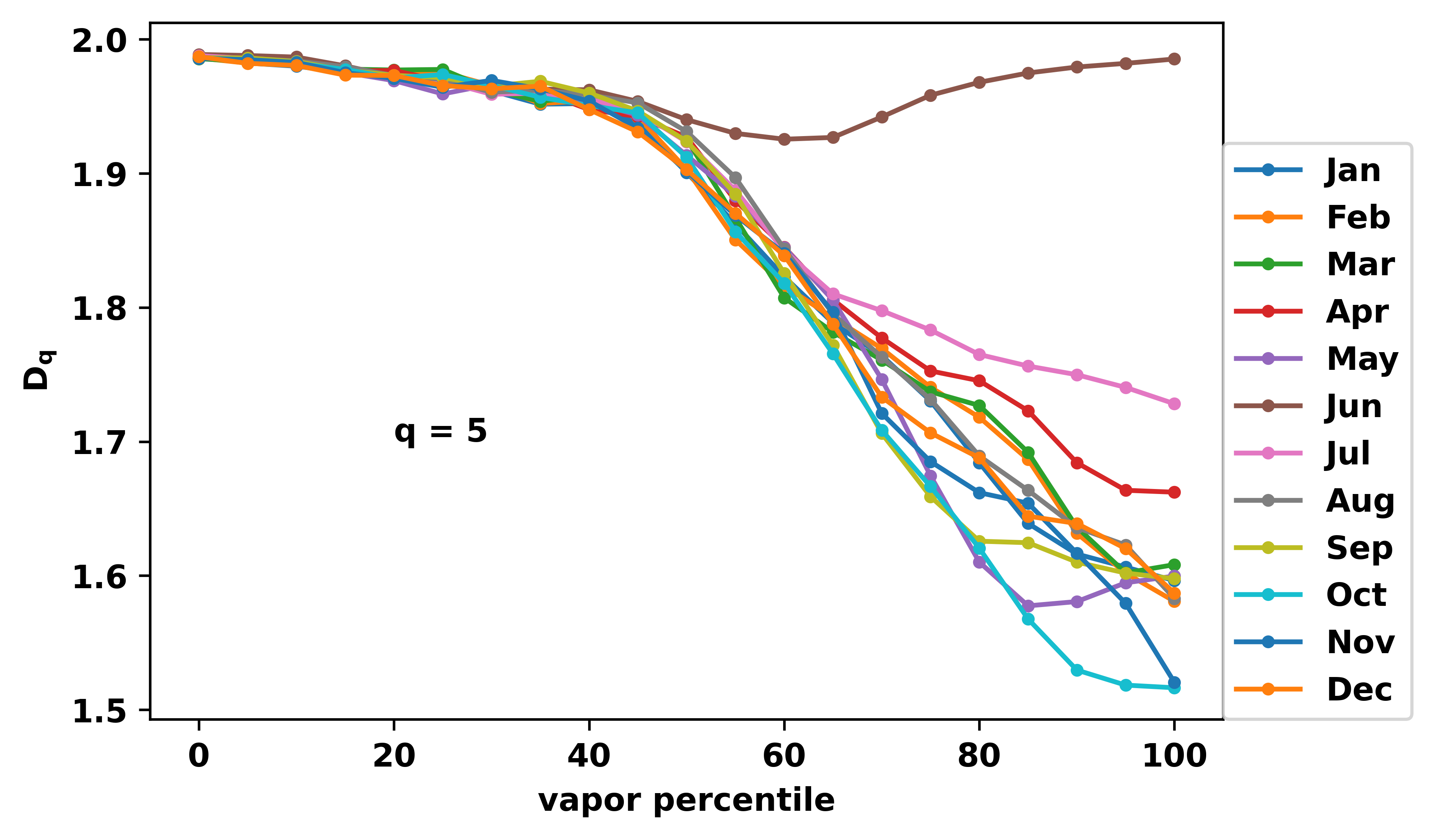} 
	\end{subfigure}
	\begin{subfigure}[t]{0.3\textwidth}
		\centering
		\includegraphics[width=\linewidth]{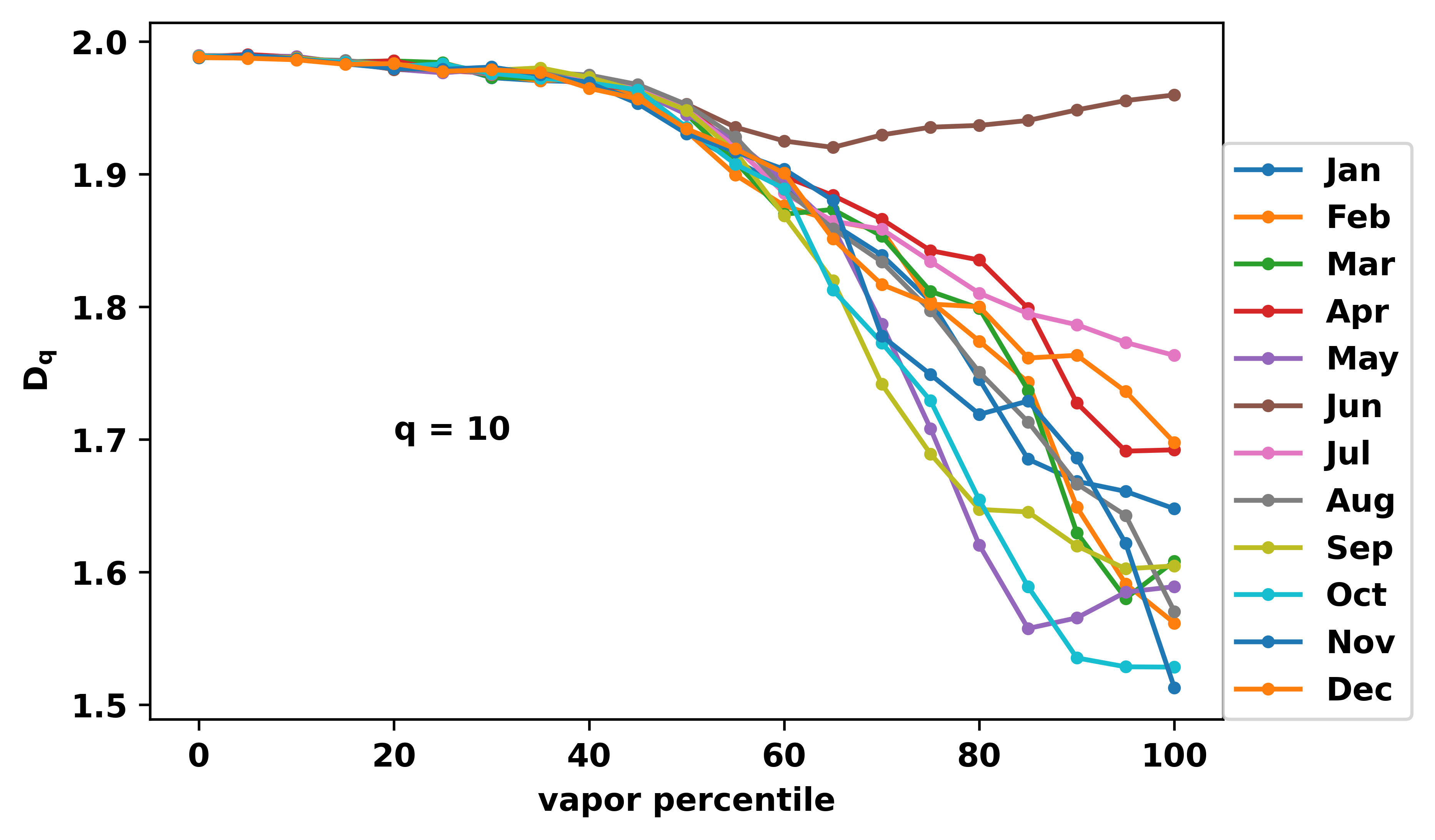} 
	\end{subfigure}		
	\begin{subfigure}[t]{0.3\textwidth}
		\centering
		\includegraphics[width=\linewidth]{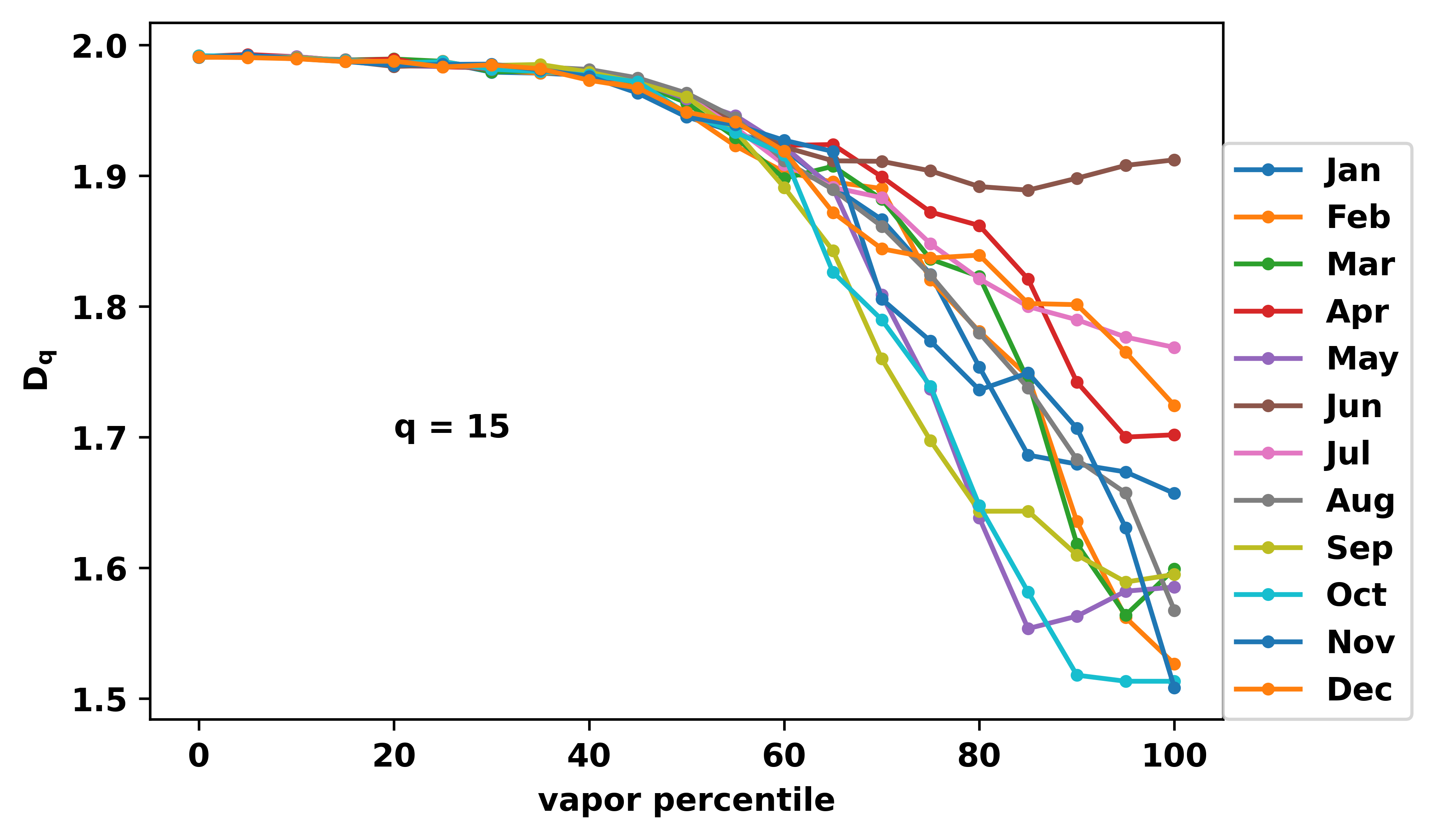} 
	\end{subfigure}
	\caption{The multi-fractal dimension $D_q$ for all 12 months (averaged over all 10 years) are plotted as a function of $V$ at different moments $q$. We see that a branching in the distribution occurs at the percolation threshold.}
	\label{600}
\end{figure}

In Figs.\ref{500} \& \ref{600}, we plot $D_q$ as a function of vapor percentile $V$ for various values of $q$. Fig.\ref{500} shows the yearly averages of $D_q$ for 10 years, and we observe that for $q < 0$, the plots are essentially featureless and drop to zero for very small values of $V$. At $q=0$ the plot shows a sharp change in behavior with continious flow lines from $D_0 = 2$ to $D_0 = 0$ for each of the years in different trajectories. For $q>0$, we see that the $D_q$'s are the same for all years up to $V=55\% $, which corresponds to the percolation threshold. For larger values of $V$, we see a branching in behavior of $D_q$ for all the years. Similar observations are made in Fig.\ref{600}, where we average over all 10 years for each of the months shown in the plots. 

\subsection{Percolation Phenomena}
In this section, we re-examine the percolation phenomenon alluded to previously, this time focusing on 10-years worth of data from 2012 to 2021 at a resolution of $360 \times 180$ pixels. 
\subsubsection{Cluster Number}
\begin{figure}
		\centering
		\begin{subfigure}[t]{0.3\textwidth}
			\centering
			\includegraphics[width=\linewidth]{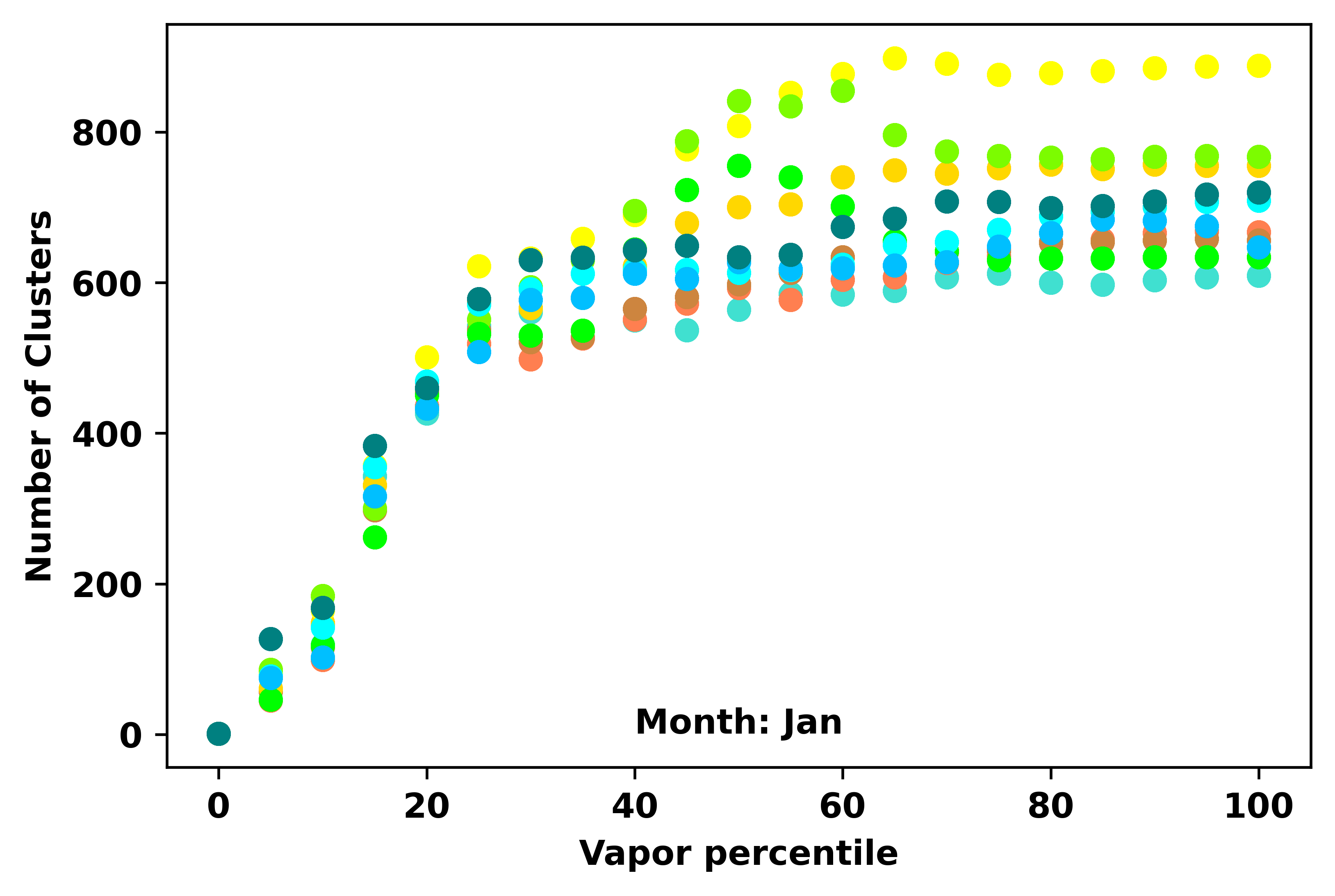} 
		\end{subfigure}
		\begin{subfigure}[t]{0.3\textwidth}
			\centering
			\includegraphics[width=\linewidth]{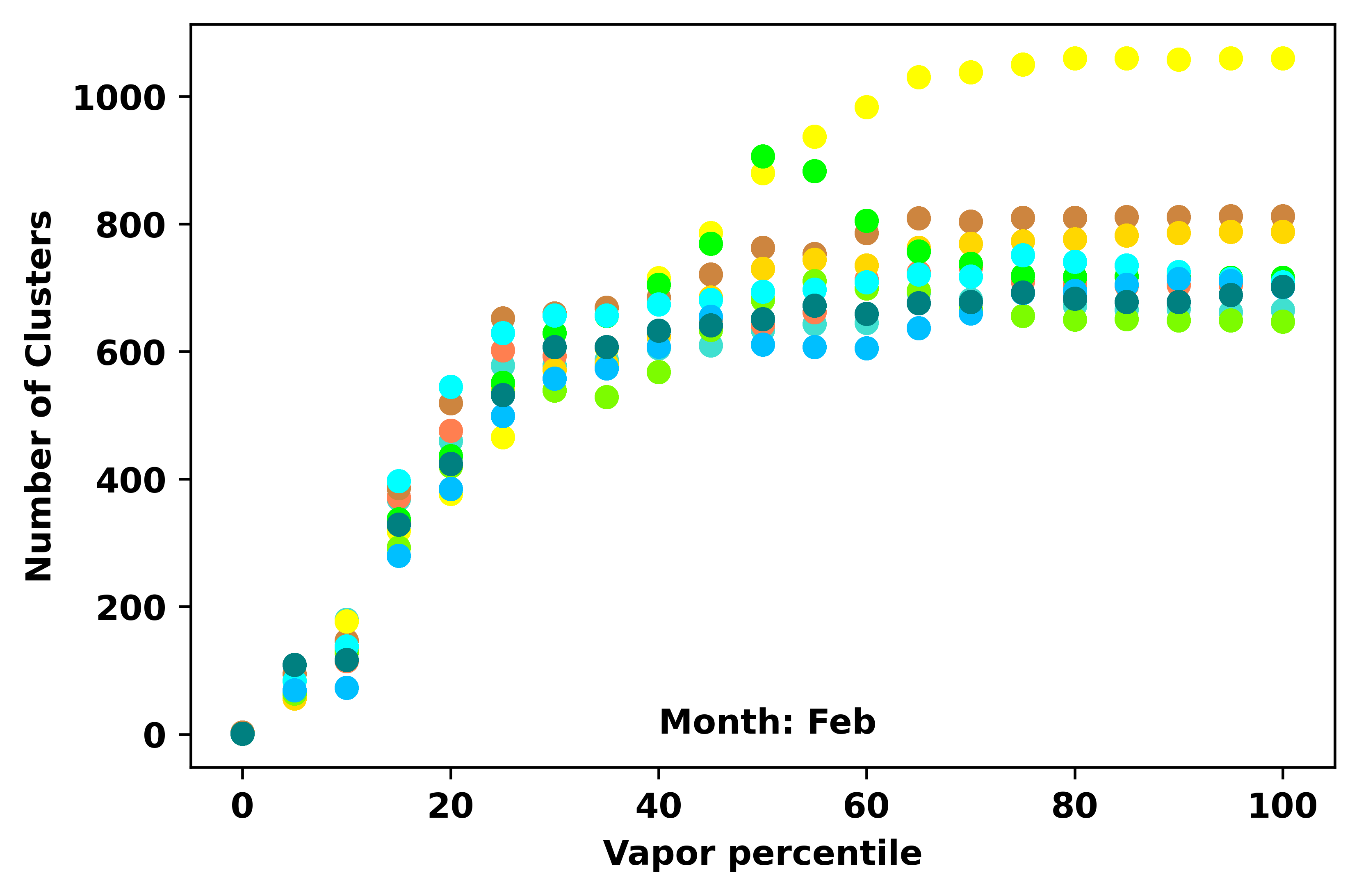} 
		\end{subfigure}
		\begin{subfigure}[t]{0.3\textwidth}
			\centering
			\includegraphics[width=\linewidth]{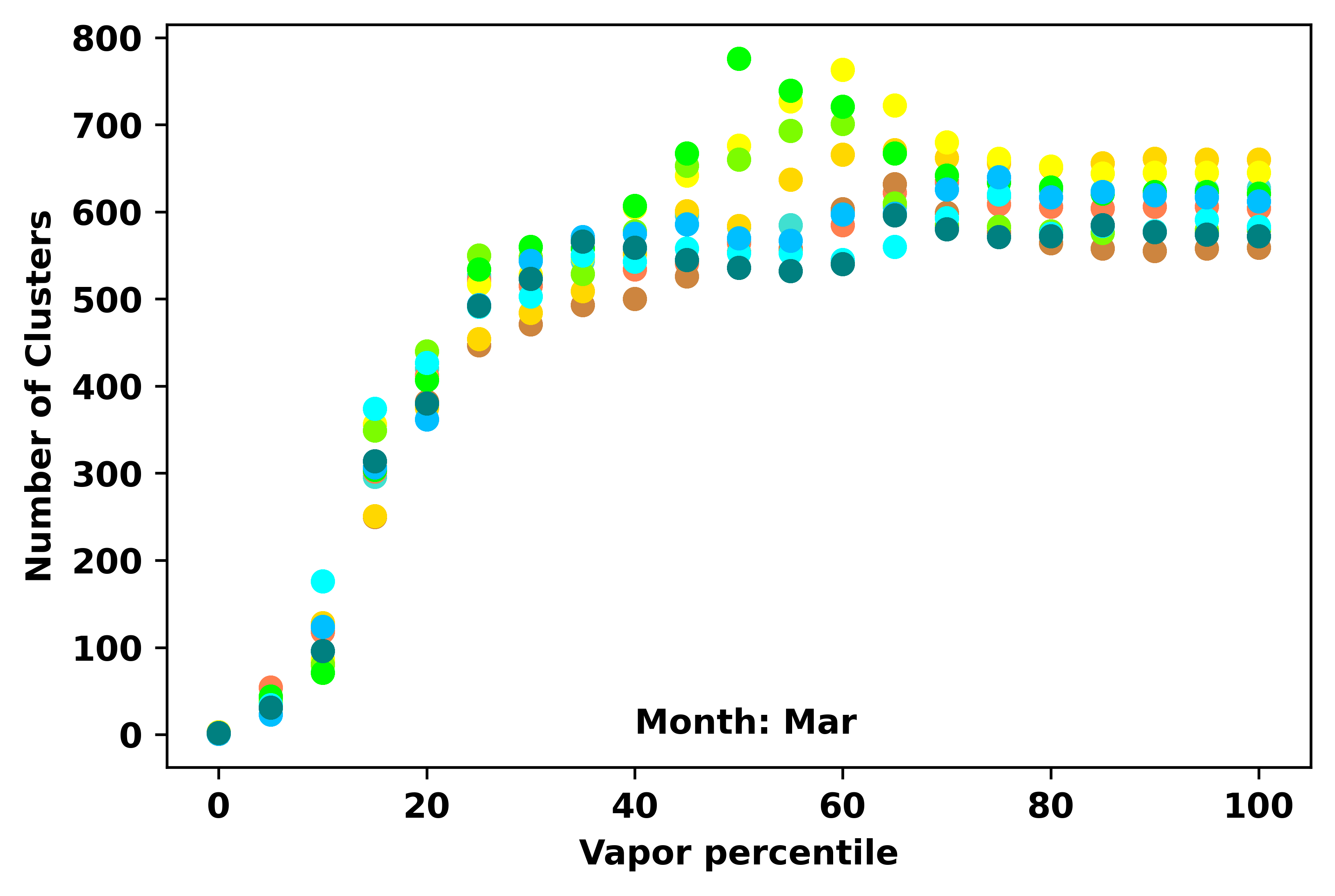} 
		\end{subfigure}		
			\begin{subfigure}[t]{0.3\textwidth}
		\centering
		\includegraphics[width=\linewidth]{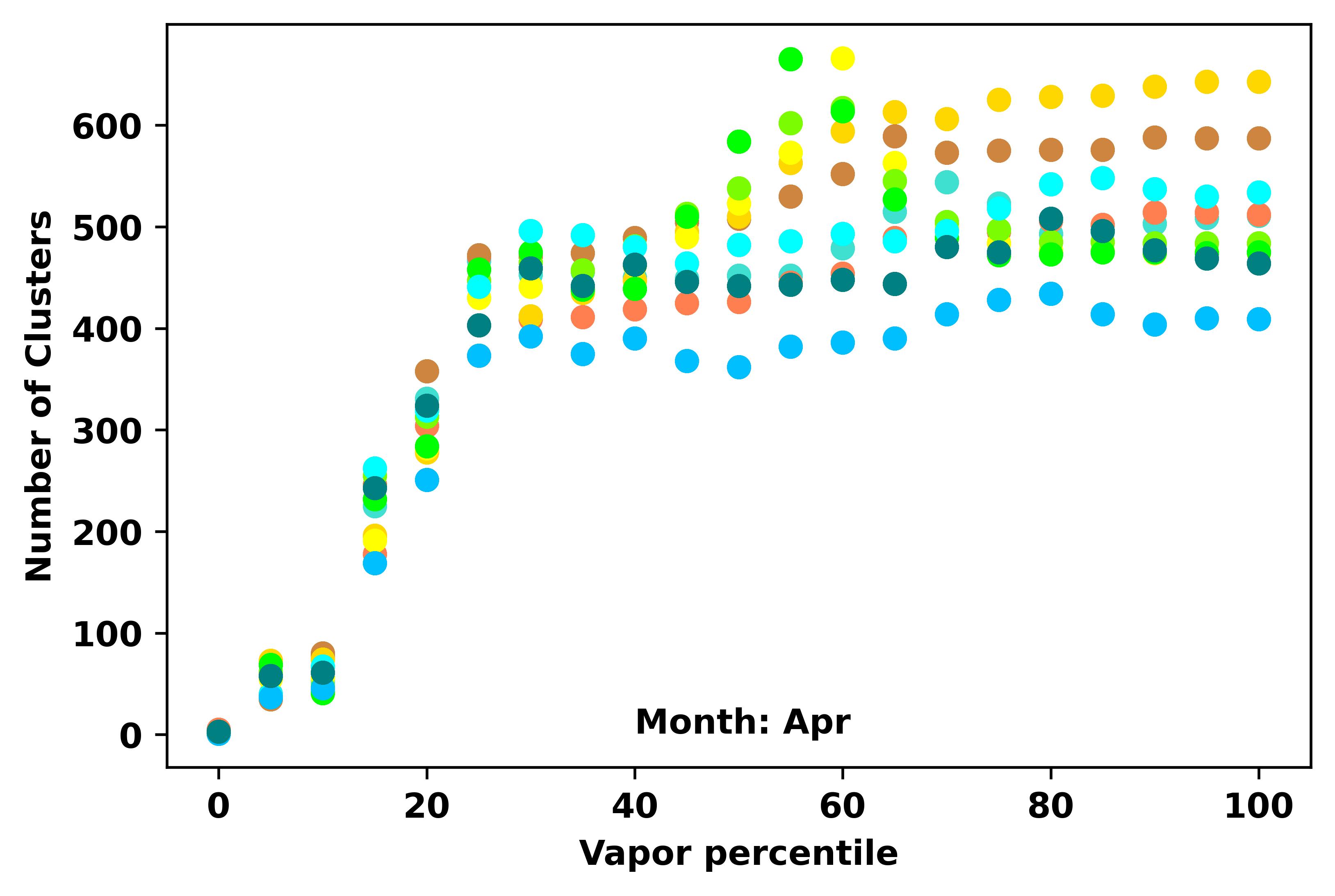} 
	\end{subfigure}
	\begin{subfigure}[t]{0.3\textwidth}
		\centering
		\includegraphics[width=\linewidth]{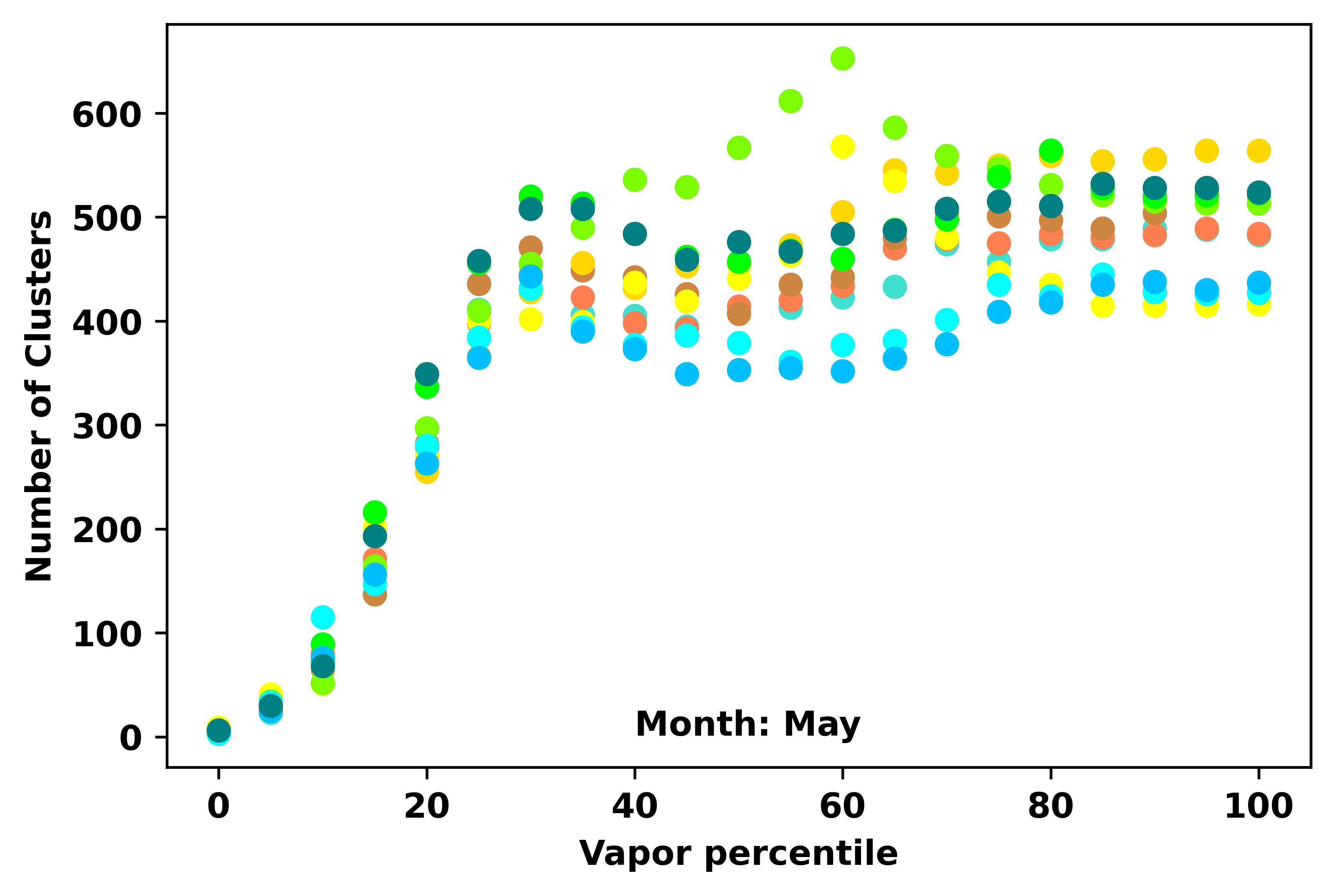} 
	\end{subfigure}
	\begin{subfigure}[t]{0.3\textwidth}
		\centering
		\includegraphics[width=\linewidth]{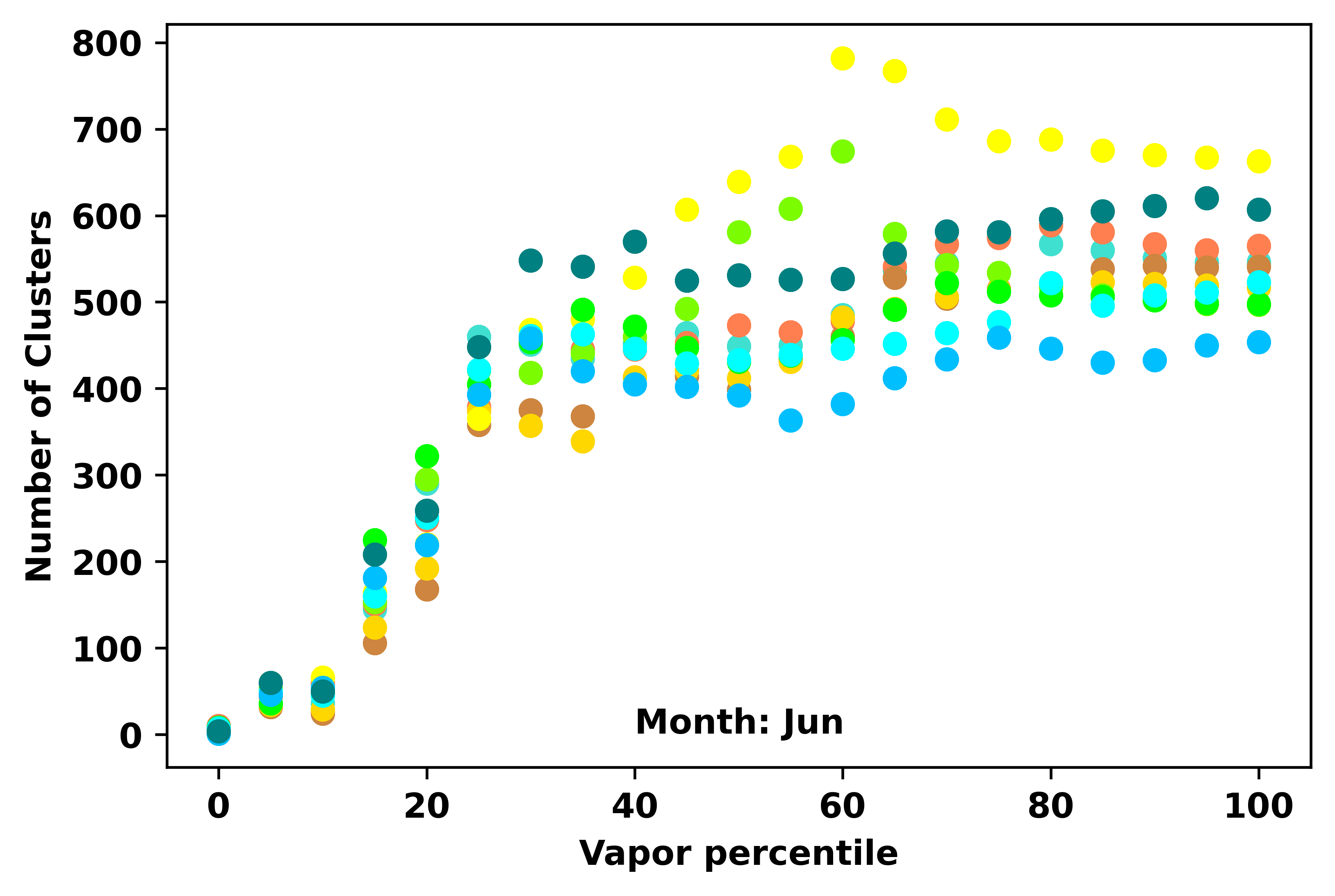} 
	\end{subfigure}		
		\begin{subfigure}[t]{0.3\textwidth}
	\centering
	\includegraphics[width=\linewidth]{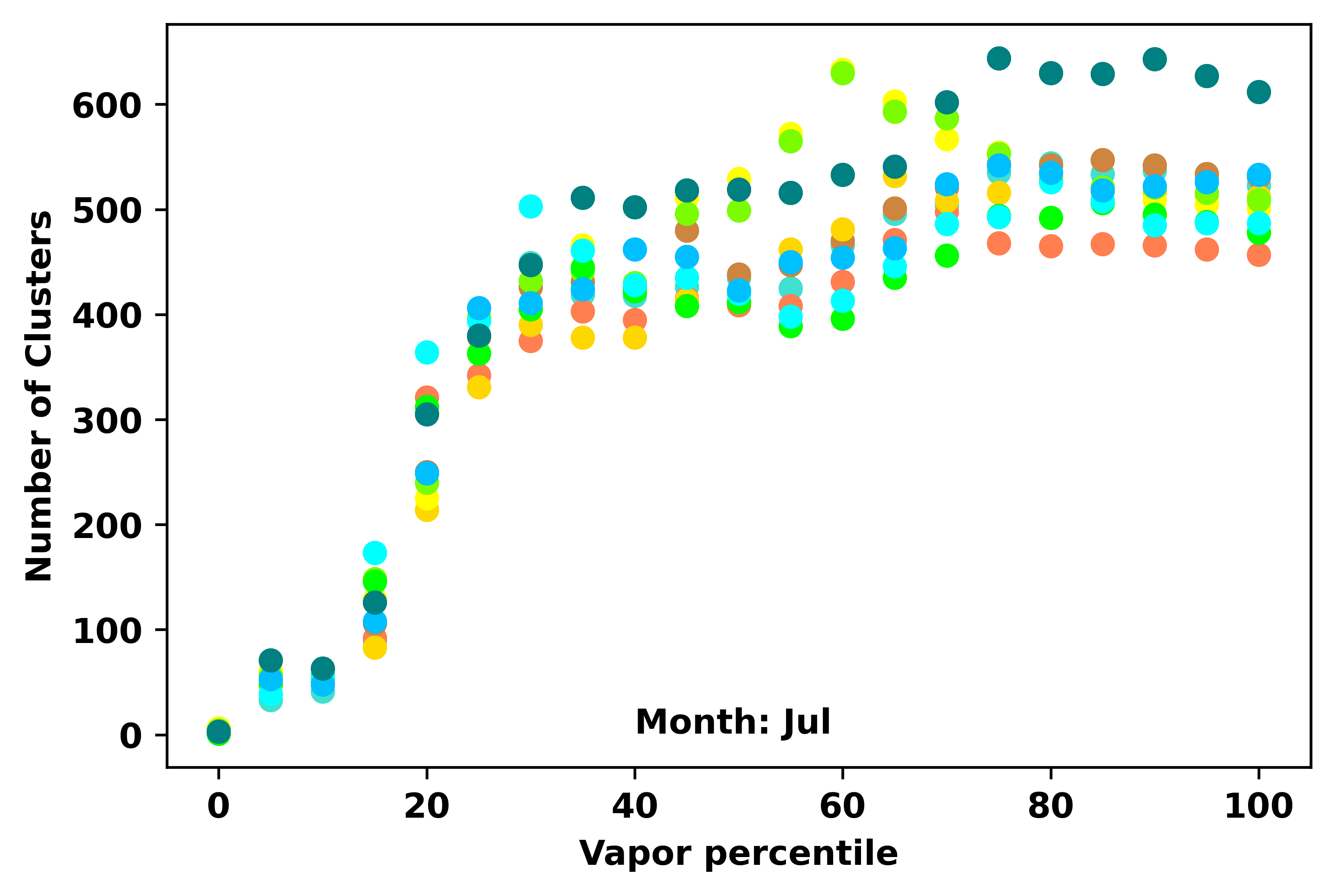} 
\end{subfigure}
\begin{subfigure}[t]{0.3\textwidth}
	\centering
	\includegraphics[width=\linewidth]{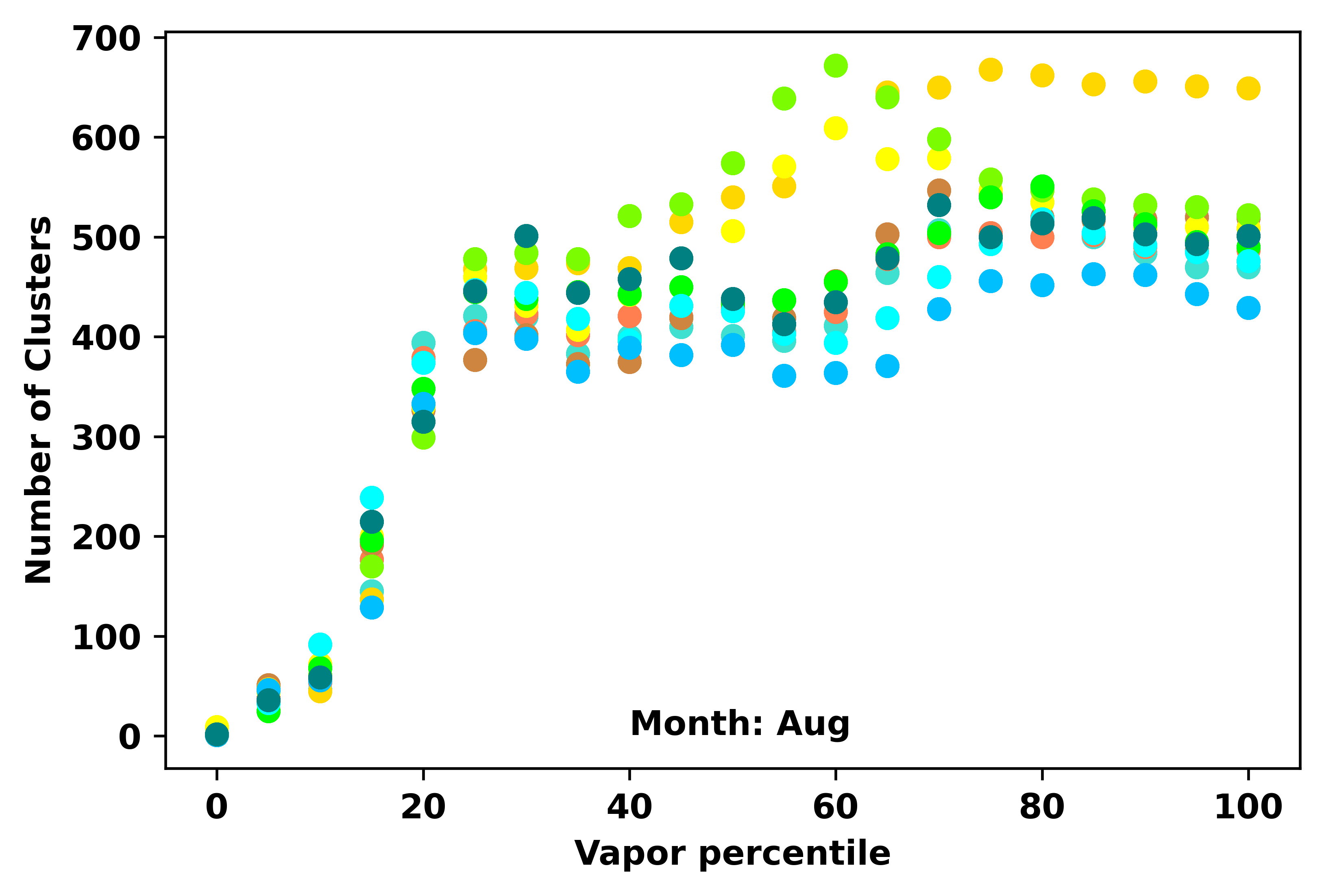} 
\end{subfigure}
\begin{subfigure}[t]{0.3\textwidth}
	\centering
	\includegraphics[width=\linewidth]{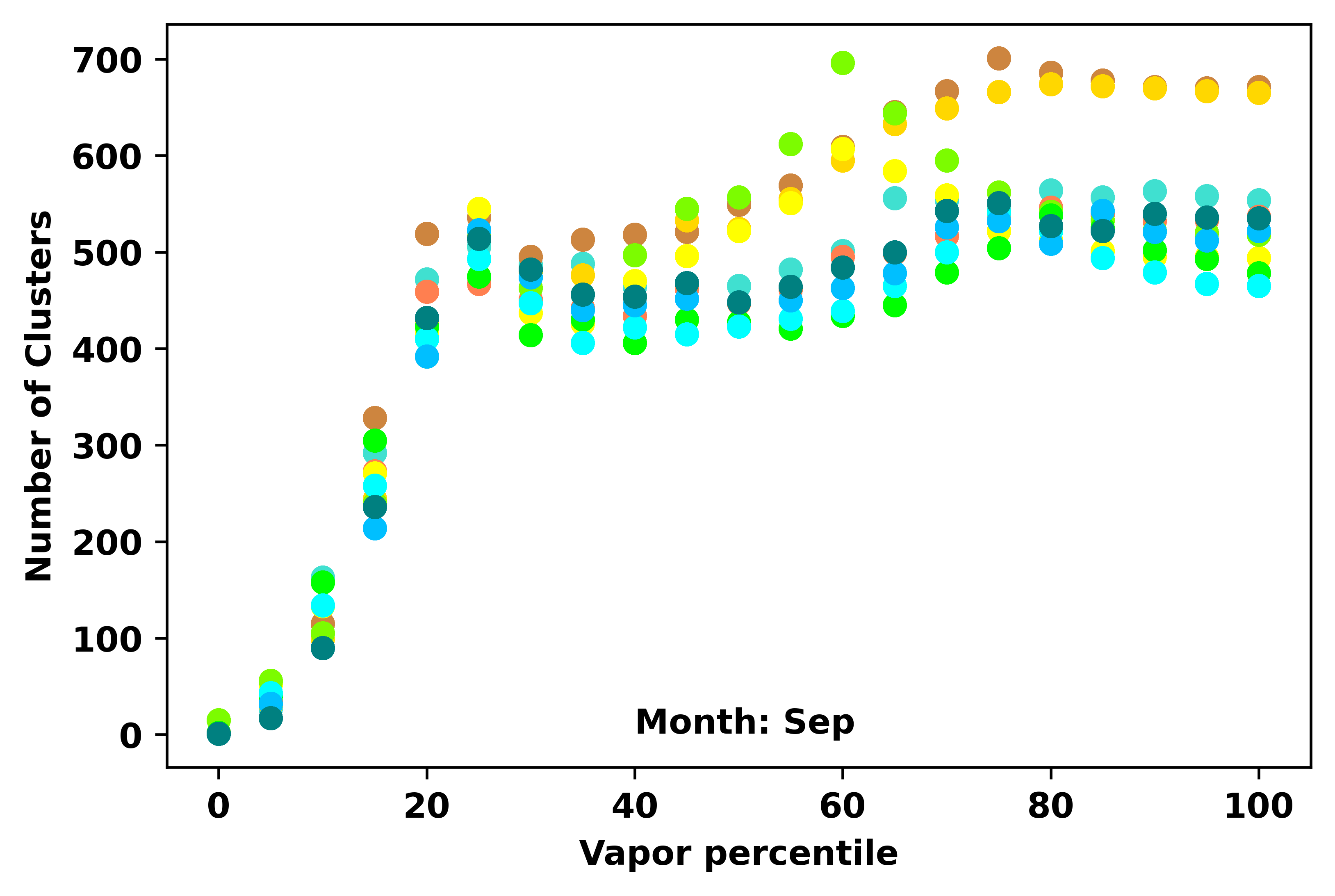} 
\end{subfigure}		
		\begin{subfigure}[t]{0.3\textwidth}
	\centering
	\includegraphics[width=\linewidth]{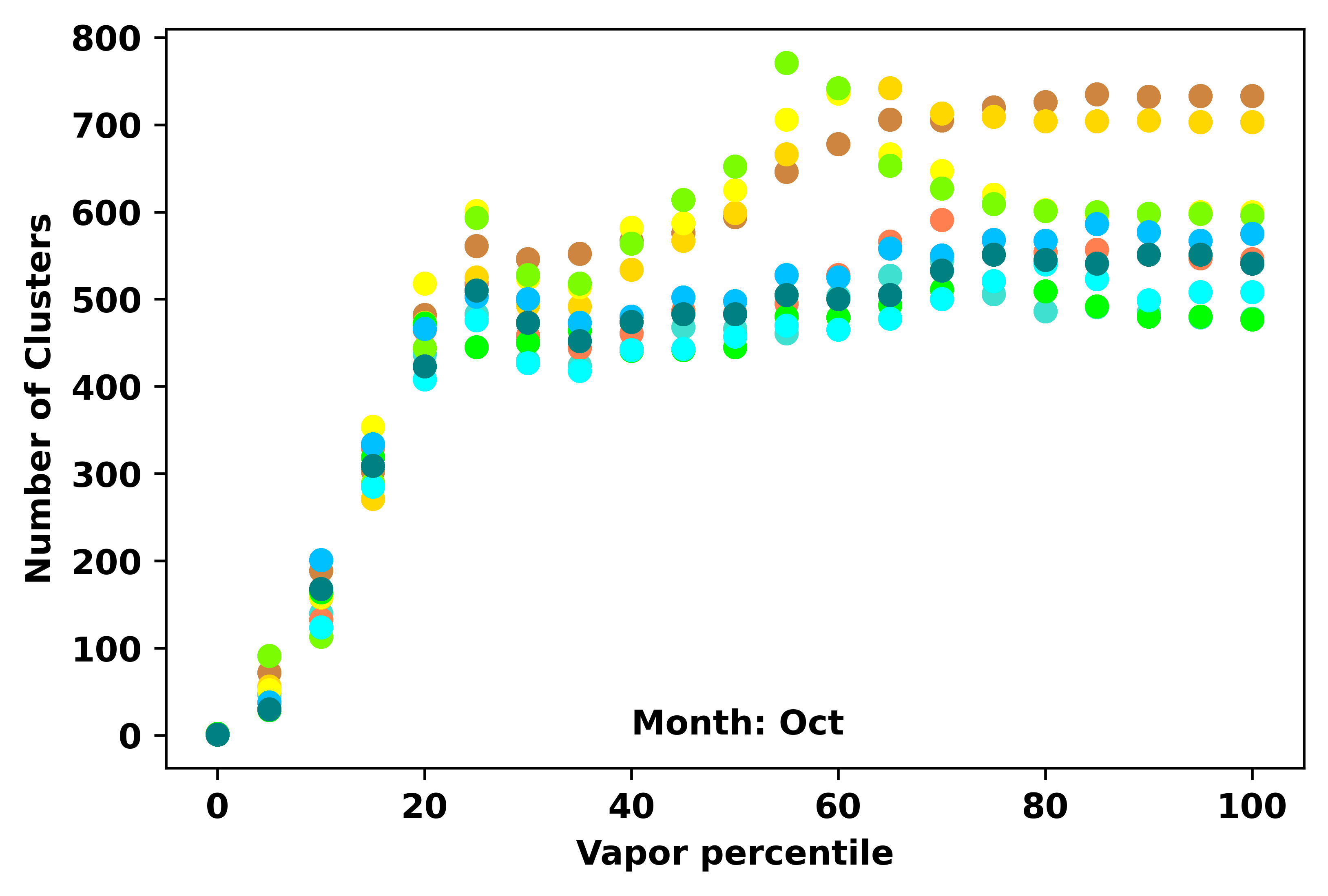} 
\end{subfigure}
\begin{subfigure}[t]{0.3\textwidth}
	\centering
	\includegraphics[width=\linewidth]{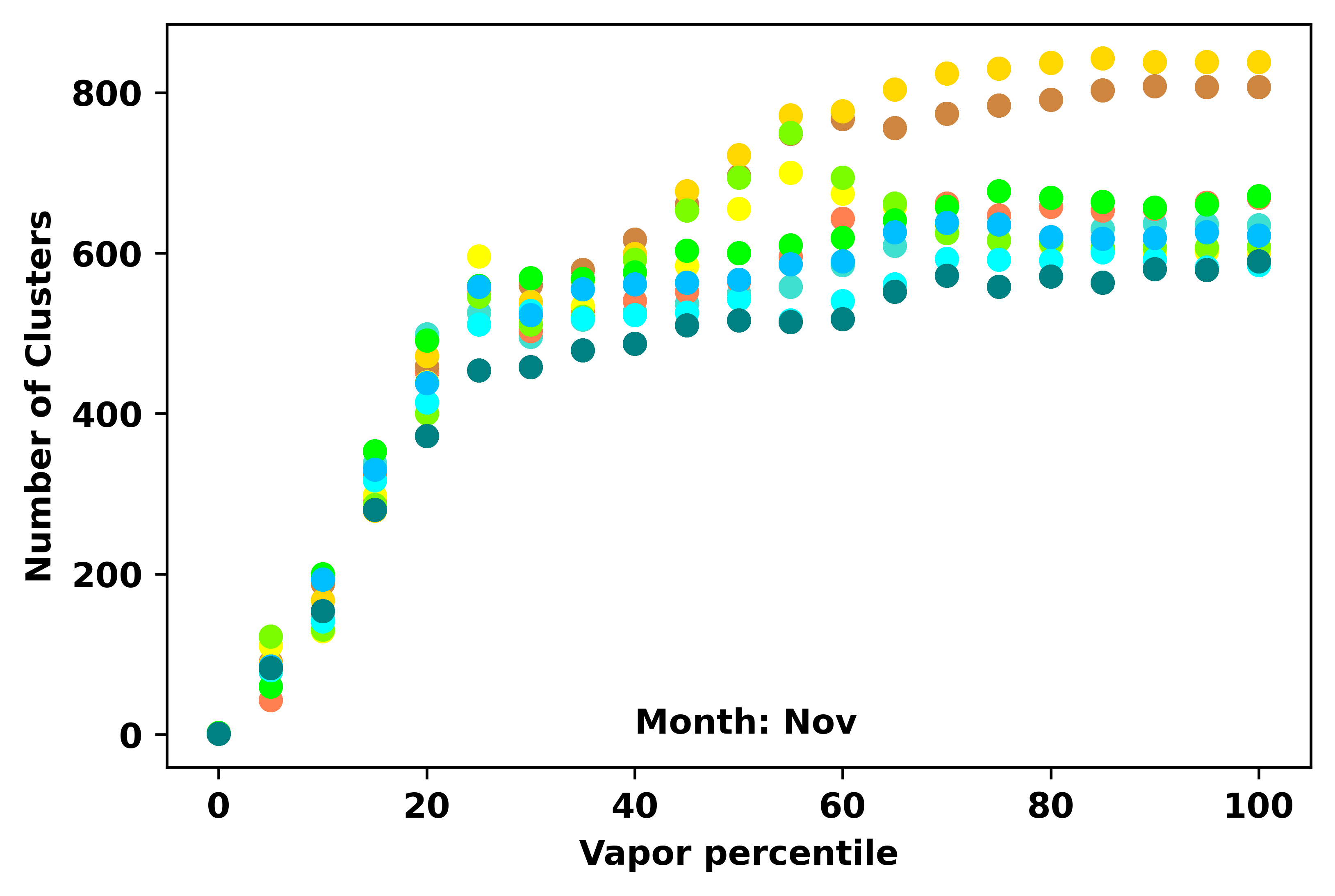} 
\end{subfigure}
\begin{subfigure}[t]{0.3\textwidth}
	\centering
	\includegraphics[width=\linewidth]{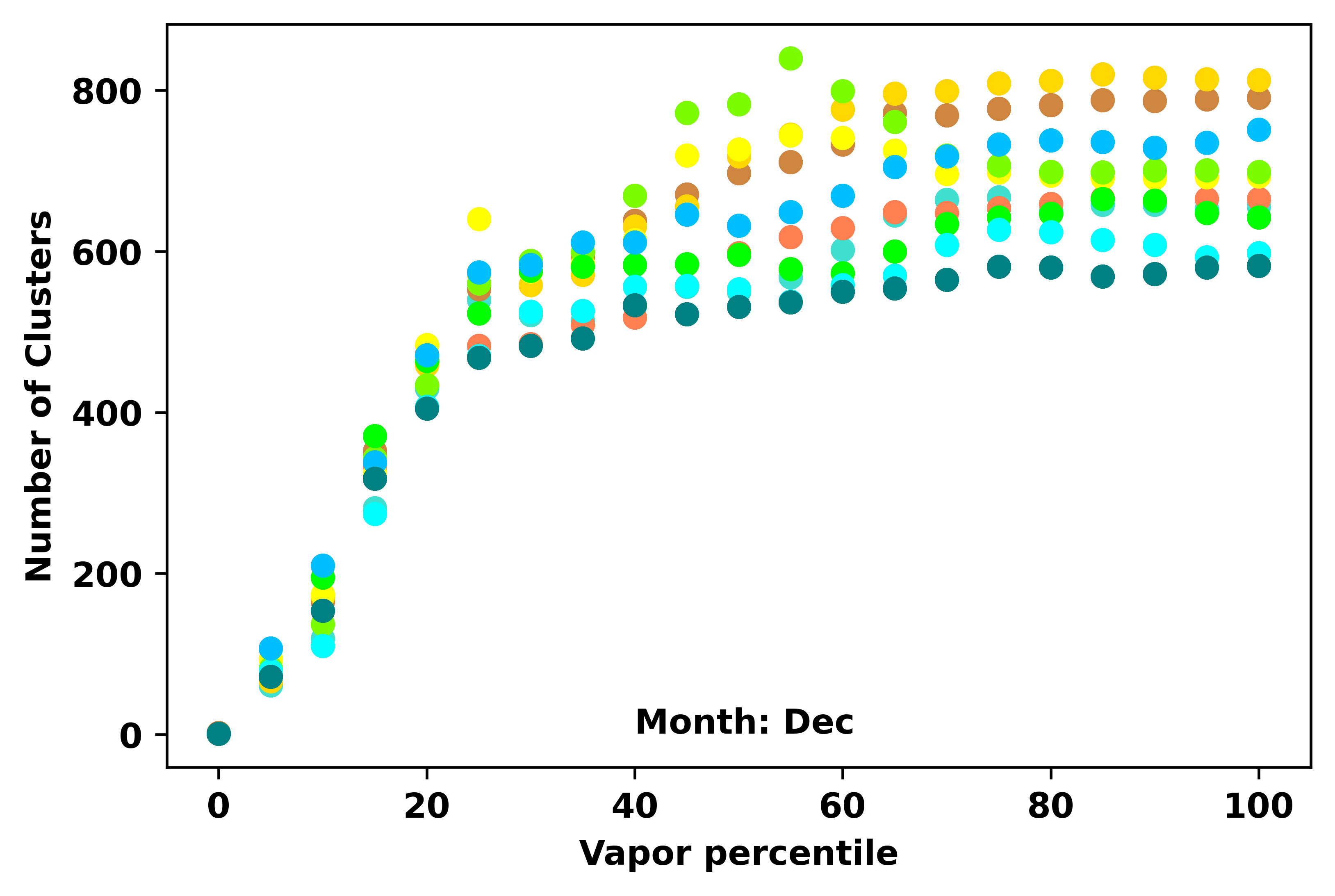} 
\end{subfigure}		
	\caption{Number of Clusters vs Vapor Percentile for each month for all 10 years, $2012-2021$ at $180\times 360$ pixels.}
	\label{8}
	\end{figure}

We plot the number of clusters as a function of vapor percentile for each month for all 10 years, $2012-2021$ in the plots in Fig. \ref{8}. In this case, the data is not scaled, as it was in the previous section, because here we are dealing with the same resolution in every case. While the distributions of cluster numbers are narrow for each month for percentiles $V<30\%$, they form a broad distribution above this percentile. The phenomenon of percolation is once again observed here and we determine the peroclation threshold percentile as $V_{thres}=55\%$. Above the $30$th percentile, the number of clusters is approximately constant for each year, while below it the number of clusters falls in a linear manner to $1$ at $V=0\%$. Other features exist in the plots in Fig. \ref{8}, however, we can make no ubiquitous statements about these features and so we decide not to discuss them in any detail.

\begin{figure}
	\centering
	\begin{subfigure}[t]{0.45\textwidth}
		\centering
		\includegraphics[width=\linewidth]{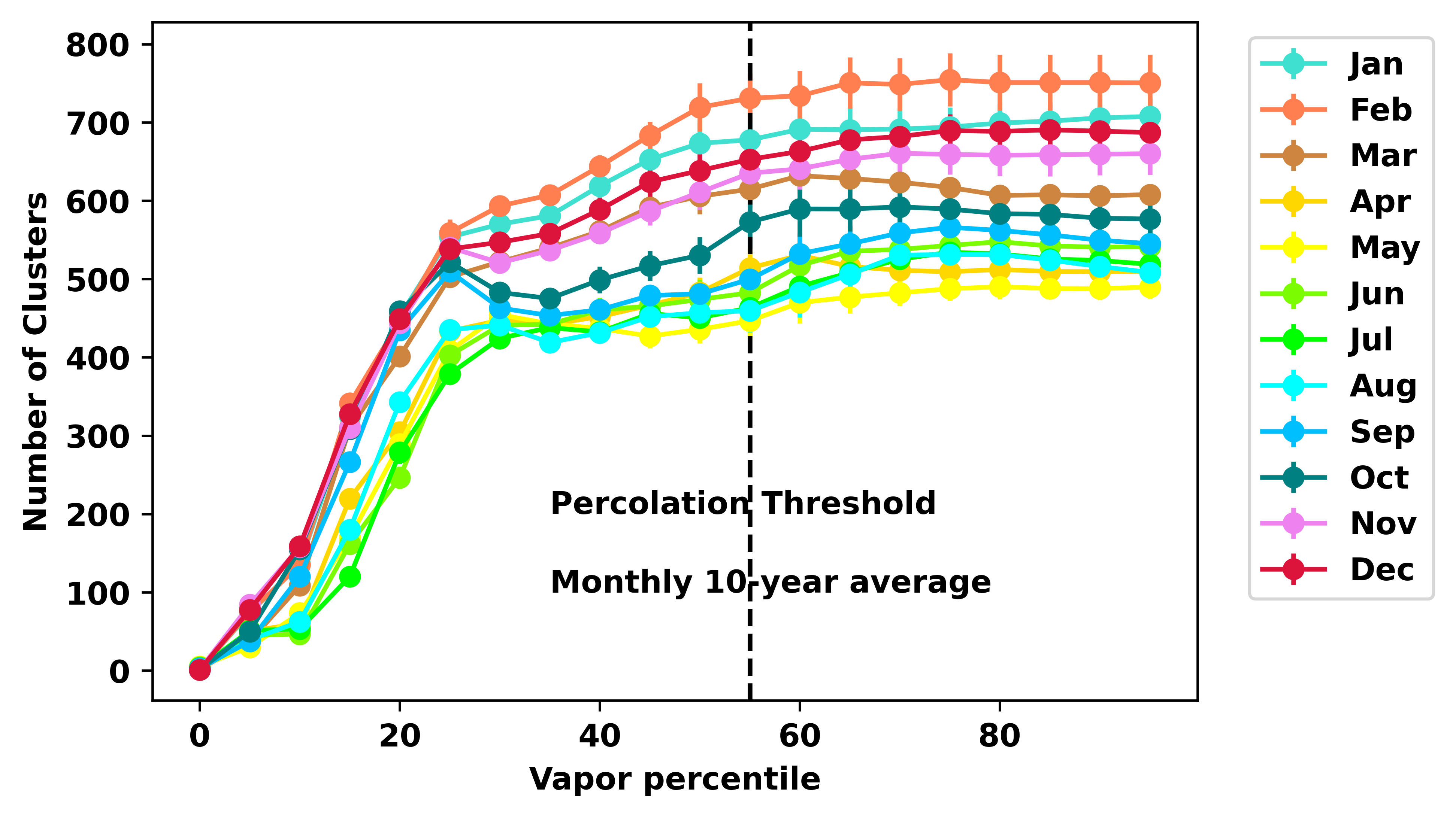} 
	\end{subfigure}
	\begin{subfigure}[t]{0.45\textwidth}
		\centering
		\includegraphics[width=\linewidth]{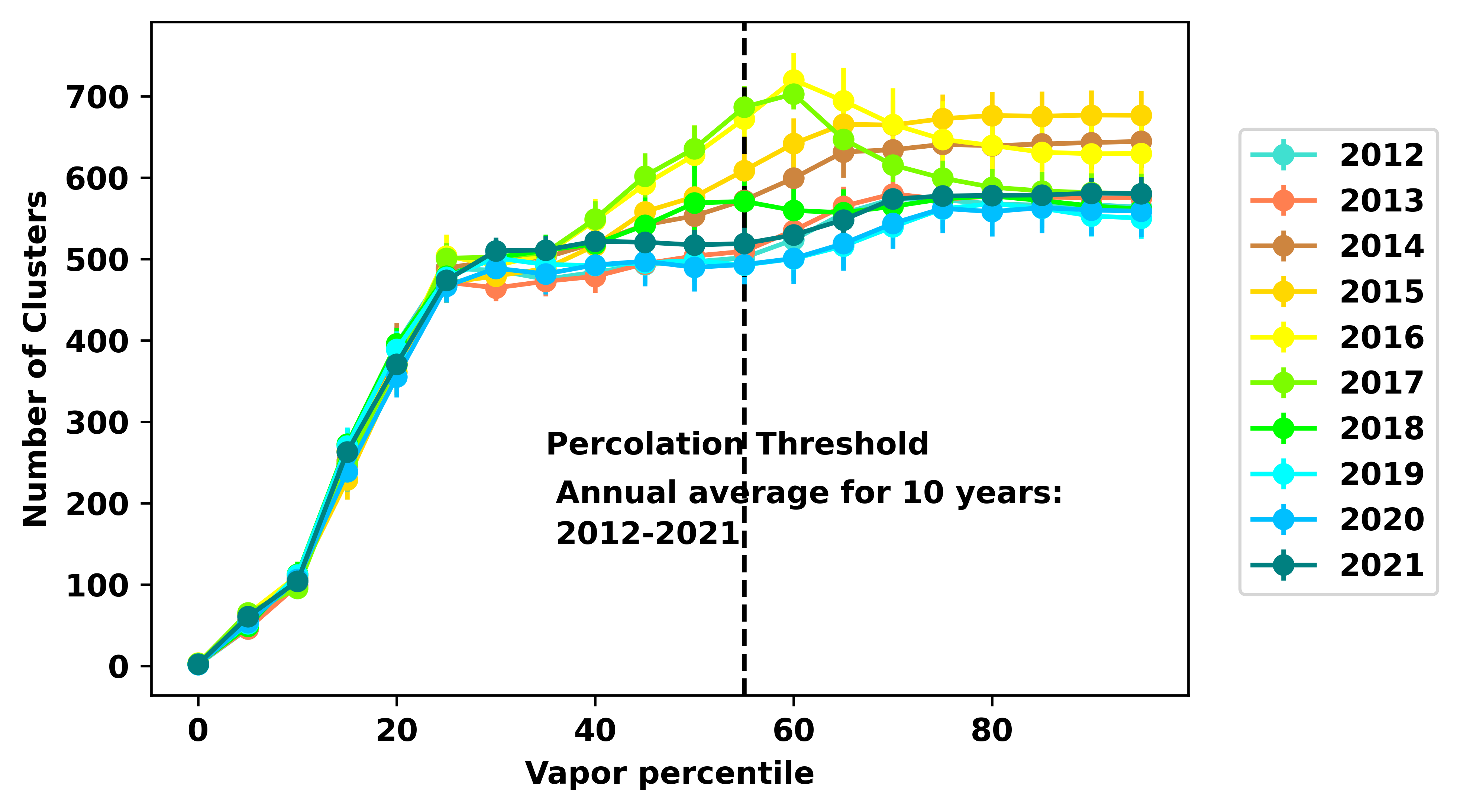} 
	\end{subfigure}
	\caption{Number of Clusters vs Vapor Percentile averaged over months and years for 10 years, $2012-2021$, at a resolution of $180\times 360$ pixels.}
	\label{9}
\end{figure}

In Fig. \ref{9}, we plot the same distributions as in Fig. \ref{8} averaged over years and months on the left and right plots, respectively. For the plot on the left, the cluster counts show a broad distribution at percentiles higher than $V = 50\%$. The general trend is for the number of clusters to increase with the increase in vapor percentile, with the highest value being held by April at the $100$th percentile. The plot on the right shows an overlap for the data at percentiles lower than $V = 30\%$. At percentiles higher than $V_{thres}$, the distribution assumes a broader character, with the highest cluster count being exhibited by 2015 at the $100$th percentile.

\subsubsection{Maximum Cluster Size}
\begin{figure}
	\centering
	\begin{subfigure}[t]{0.3\textwidth}
		\centering
		\includegraphics[width=\linewidth]{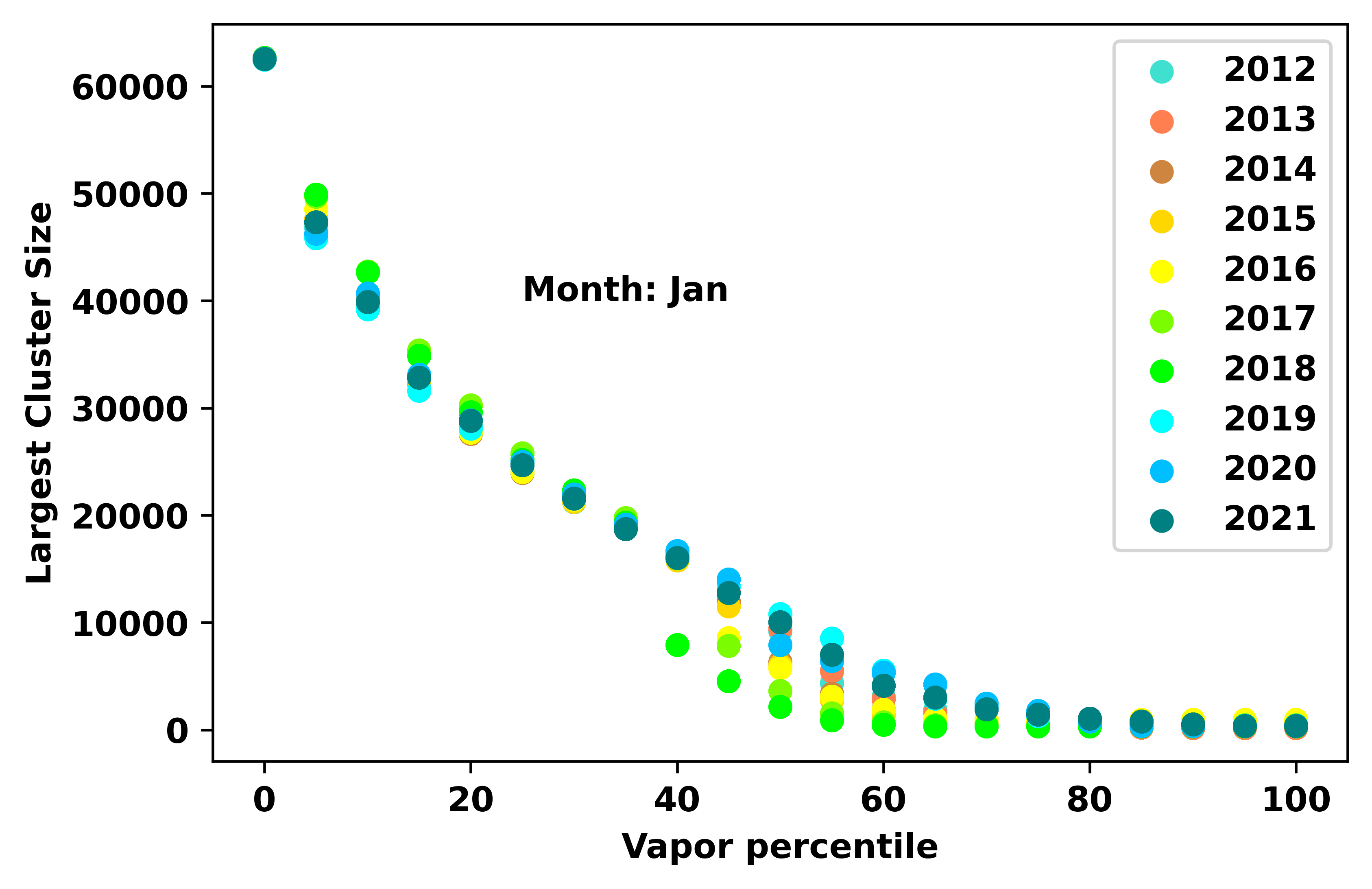} 
	\end{subfigure}
	\begin{subfigure}[t]{0.3\textwidth}
		\centering
		\includegraphics[width=\linewidth]{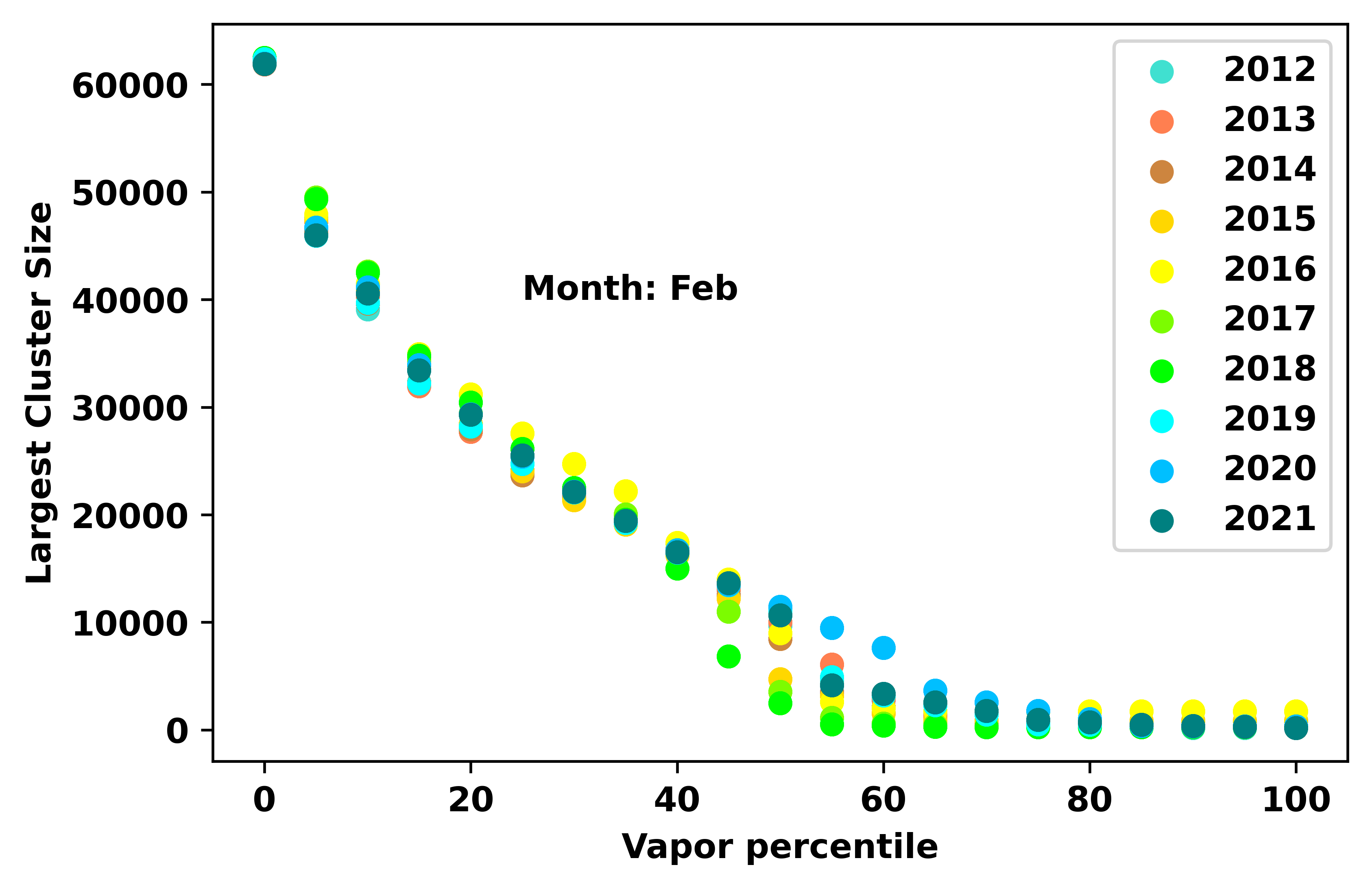} 
	\end{subfigure}
	\begin{subfigure}[t]{0.3\textwidth}
		\centering
		\includegraphics[width=\linewidth]{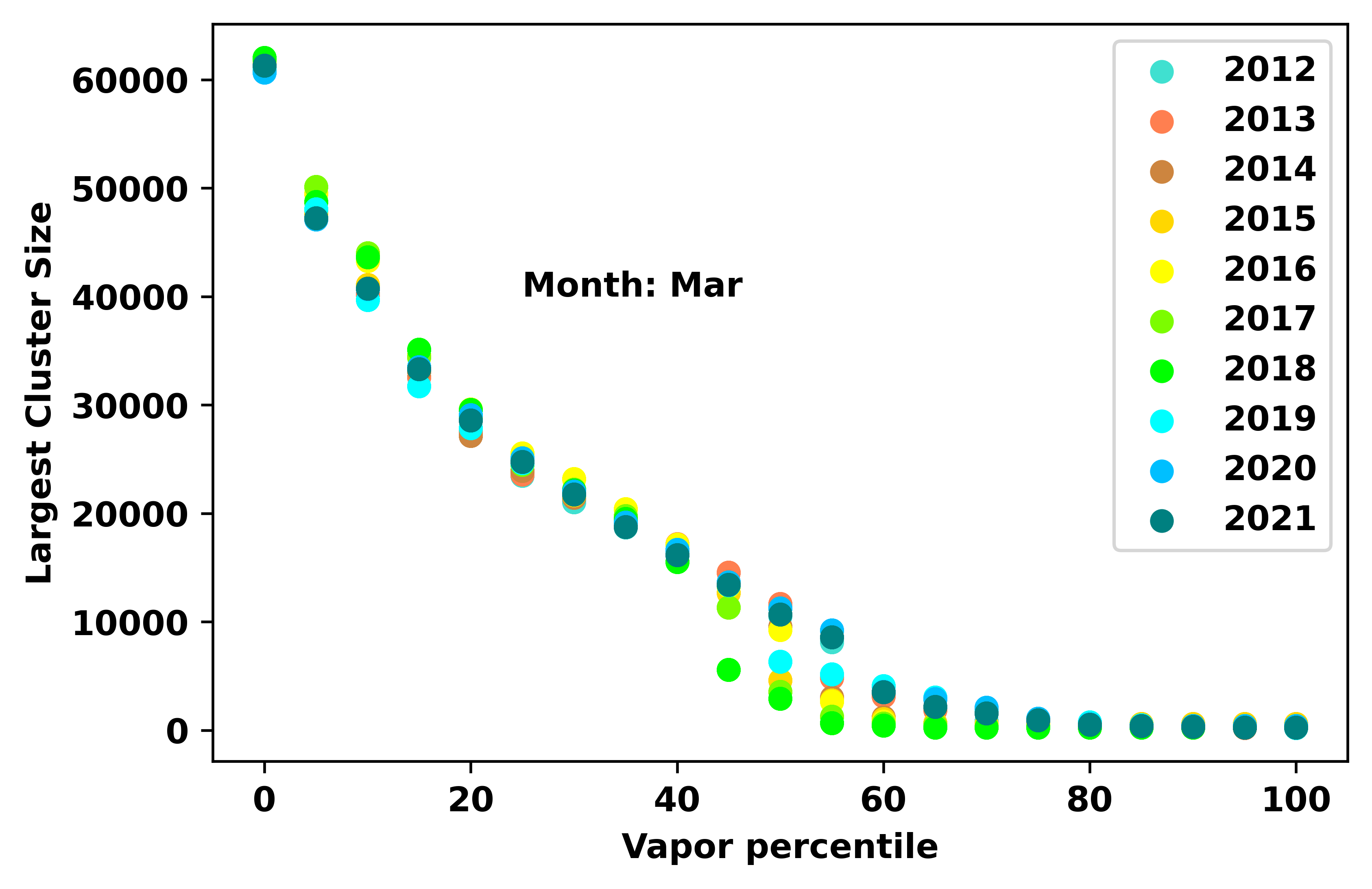} 
	\end{subfigure}		
	\begin{subfigure}[t]{0.3\textwidth}
		\centering
		\includegraphics[width=\linewidth]{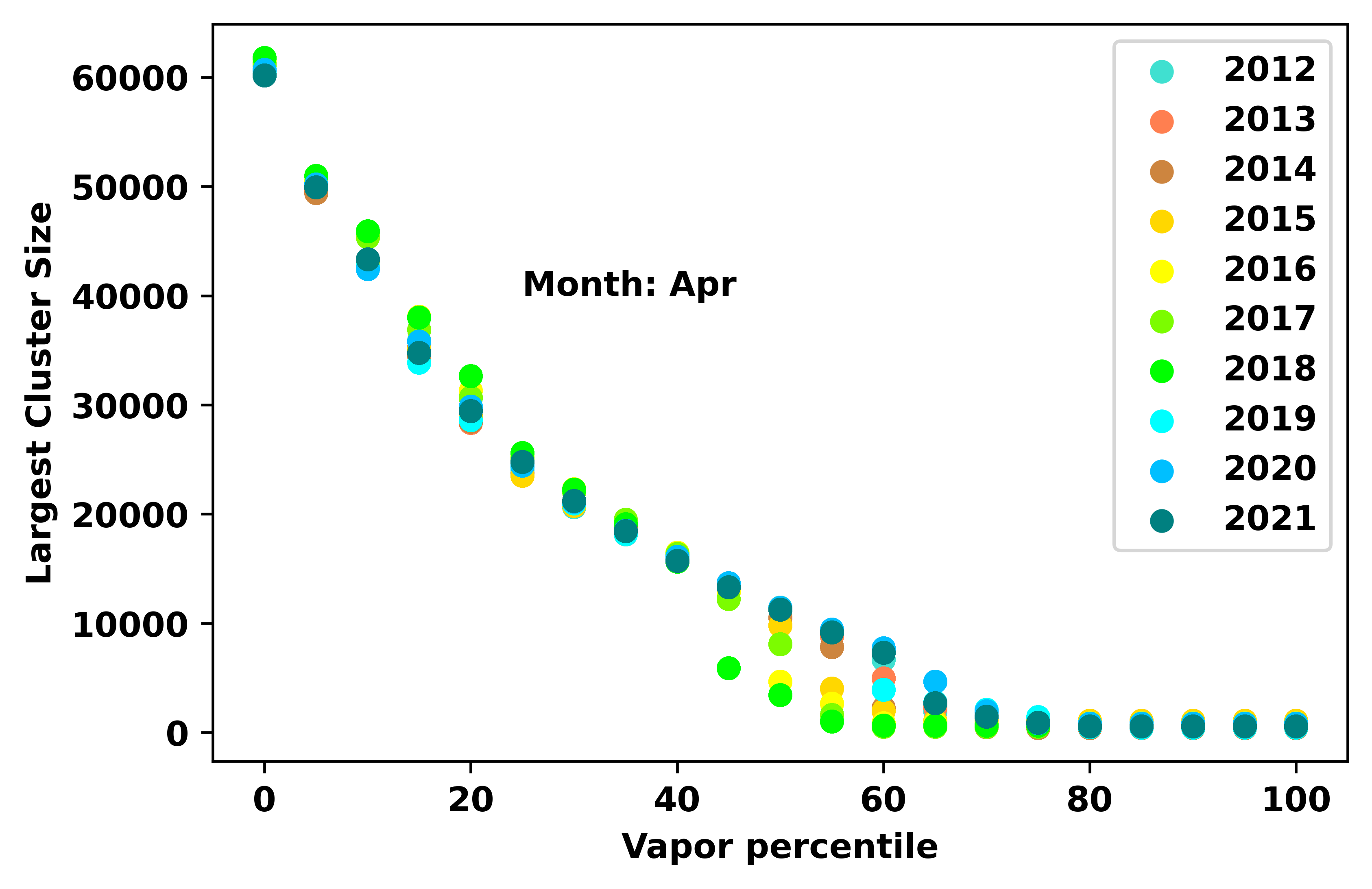} 
	\end{subfigure}
	\begin{subfigure}[t]{0.3\textwidth}
		\centering
		\includegraphics[width=\linewidth]{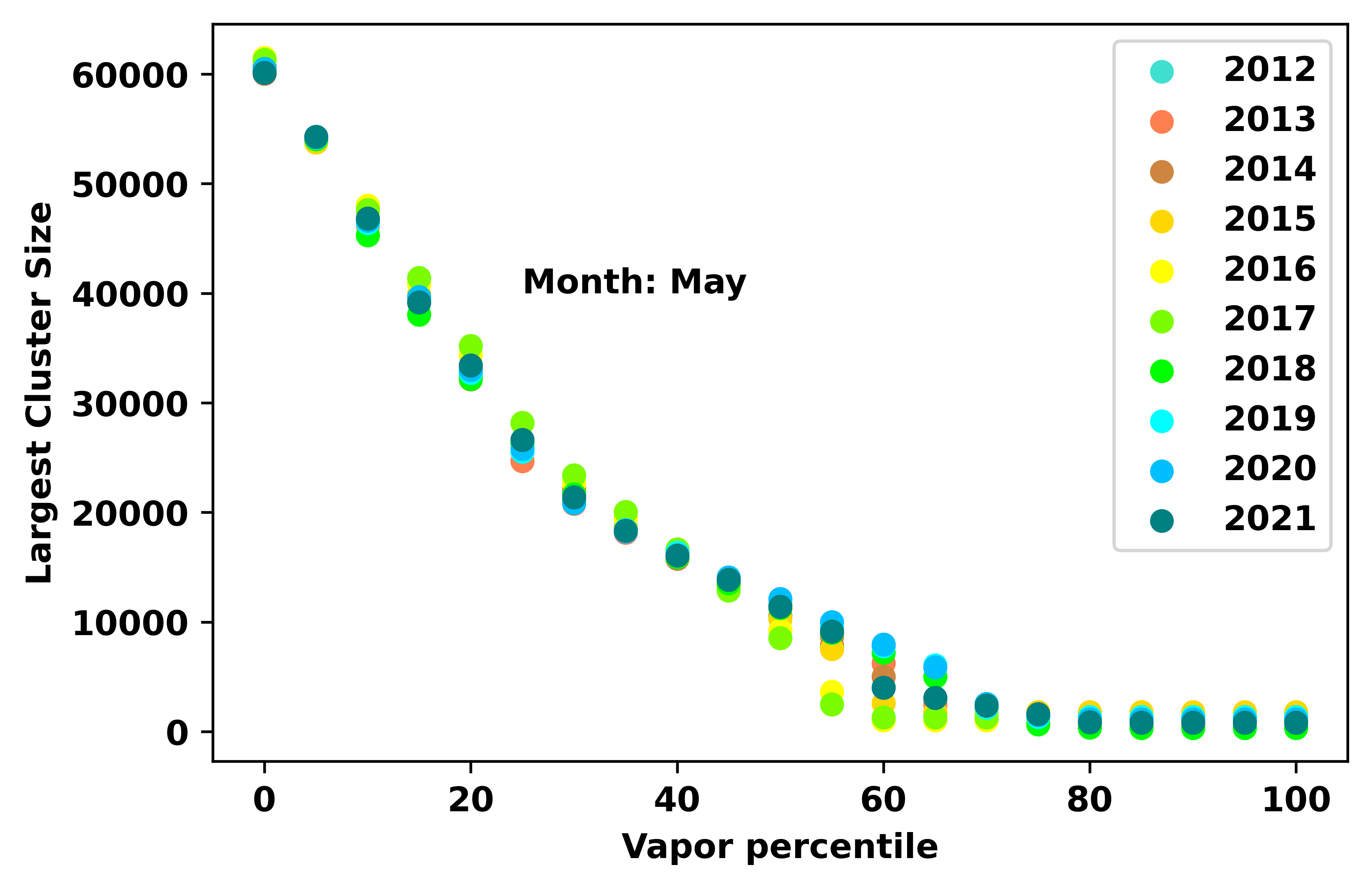} 
	\end{subfigure}
	\begin{subfigure}[t]{0.3\textwidth}
		\centering
		\includegraphics[width=\linewidth]{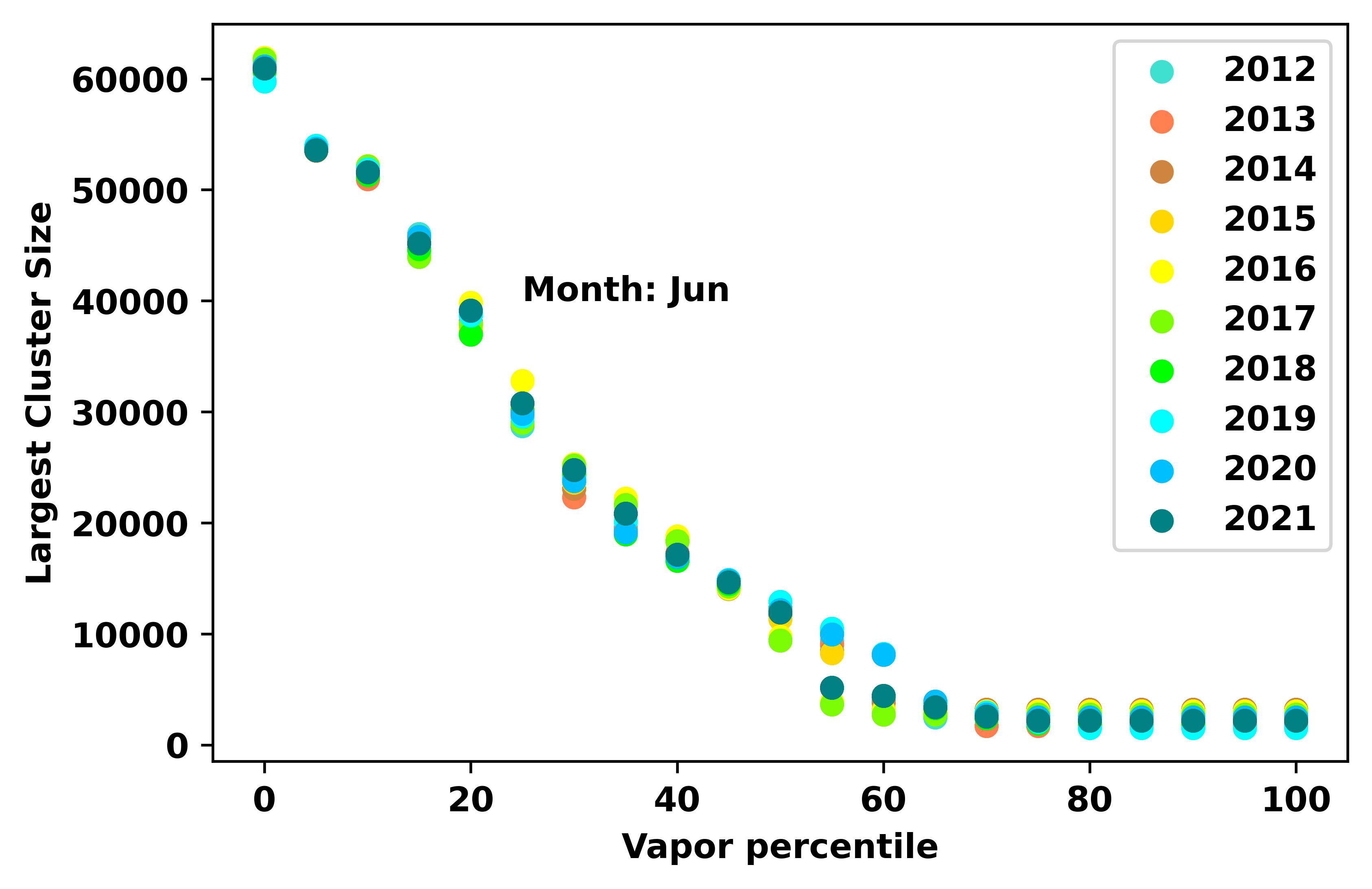} 
	\end{subfigure}		
	\begin{subfigure}[t]{0.3\textwidth}
		\centering
		\includegraphics[width=\linewidth]{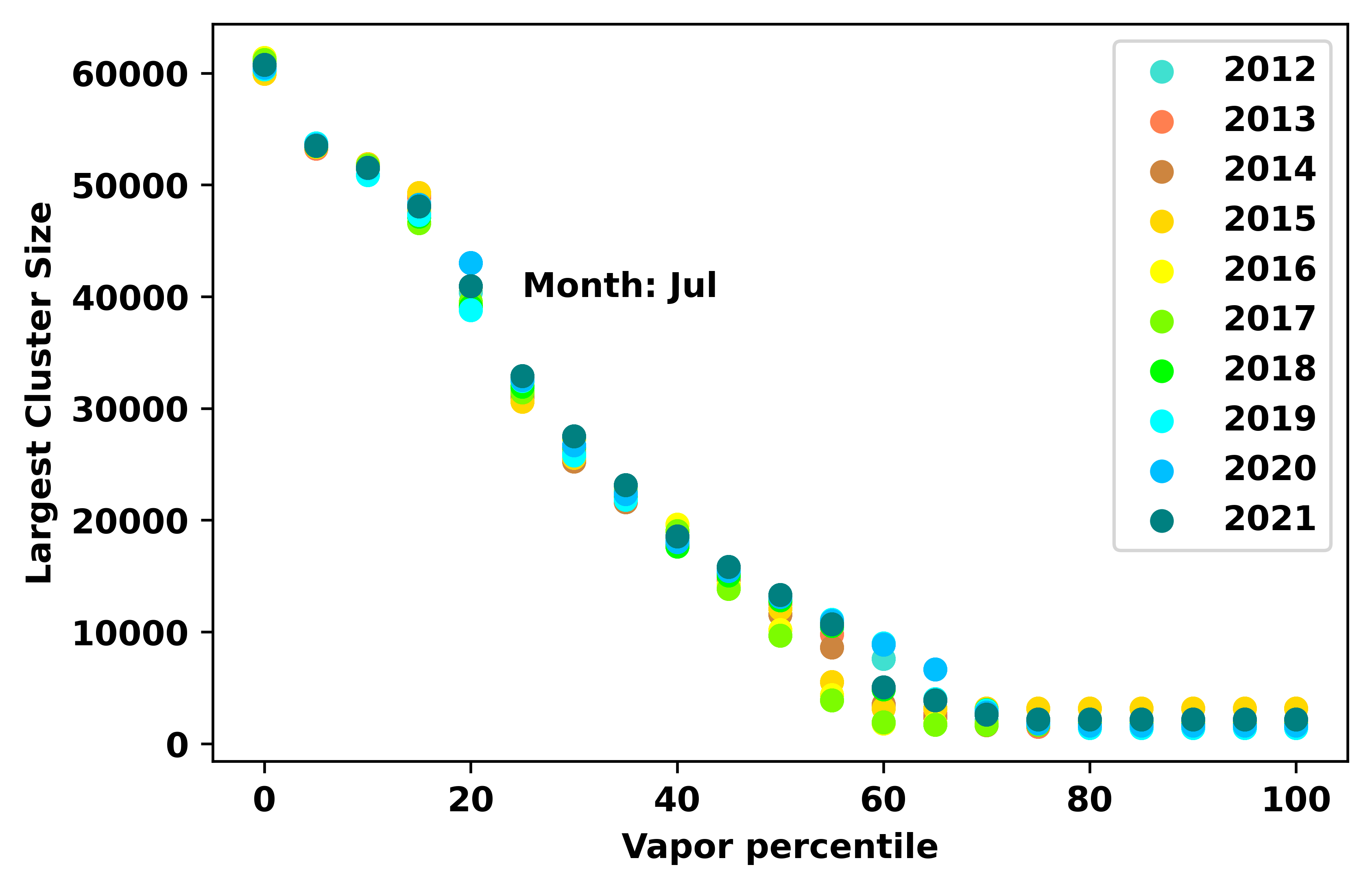} 
	\end{subfigure}
	\begin{subfigure}[t]{0.3\textwidth}
		\centering
		\includegraphics[width=\linewidth]{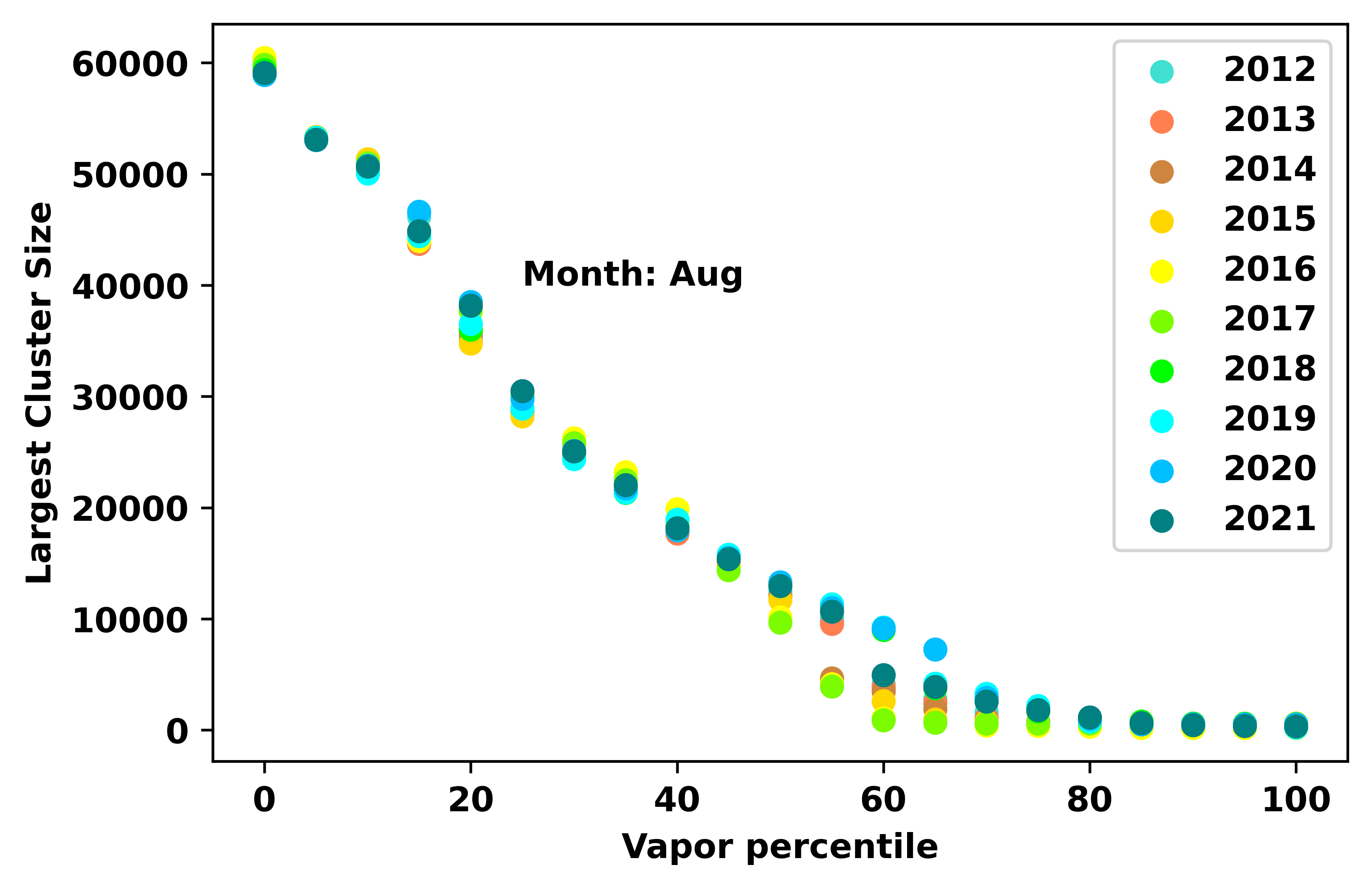} 
	\end{subfigure}
	\begin{subfigure}[t]{0.3\textwidth}
		\centering
		\includegraphics[width=\linewidth]{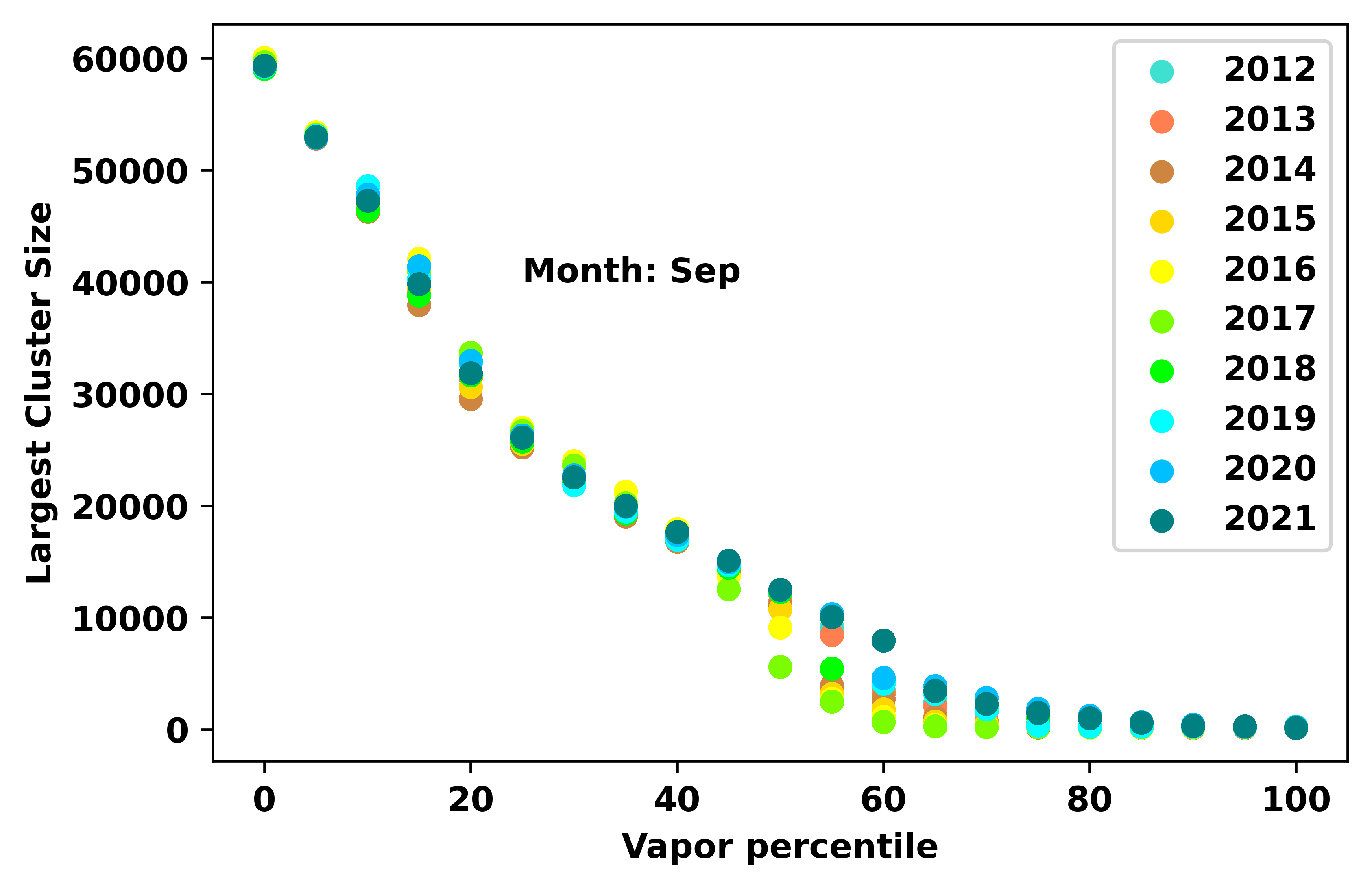} 
	\end{subfigure}		
	\begin{subfigure}[t]{0.3\textwidth}
		\centering
		\includegraphics[width=\linewidth]{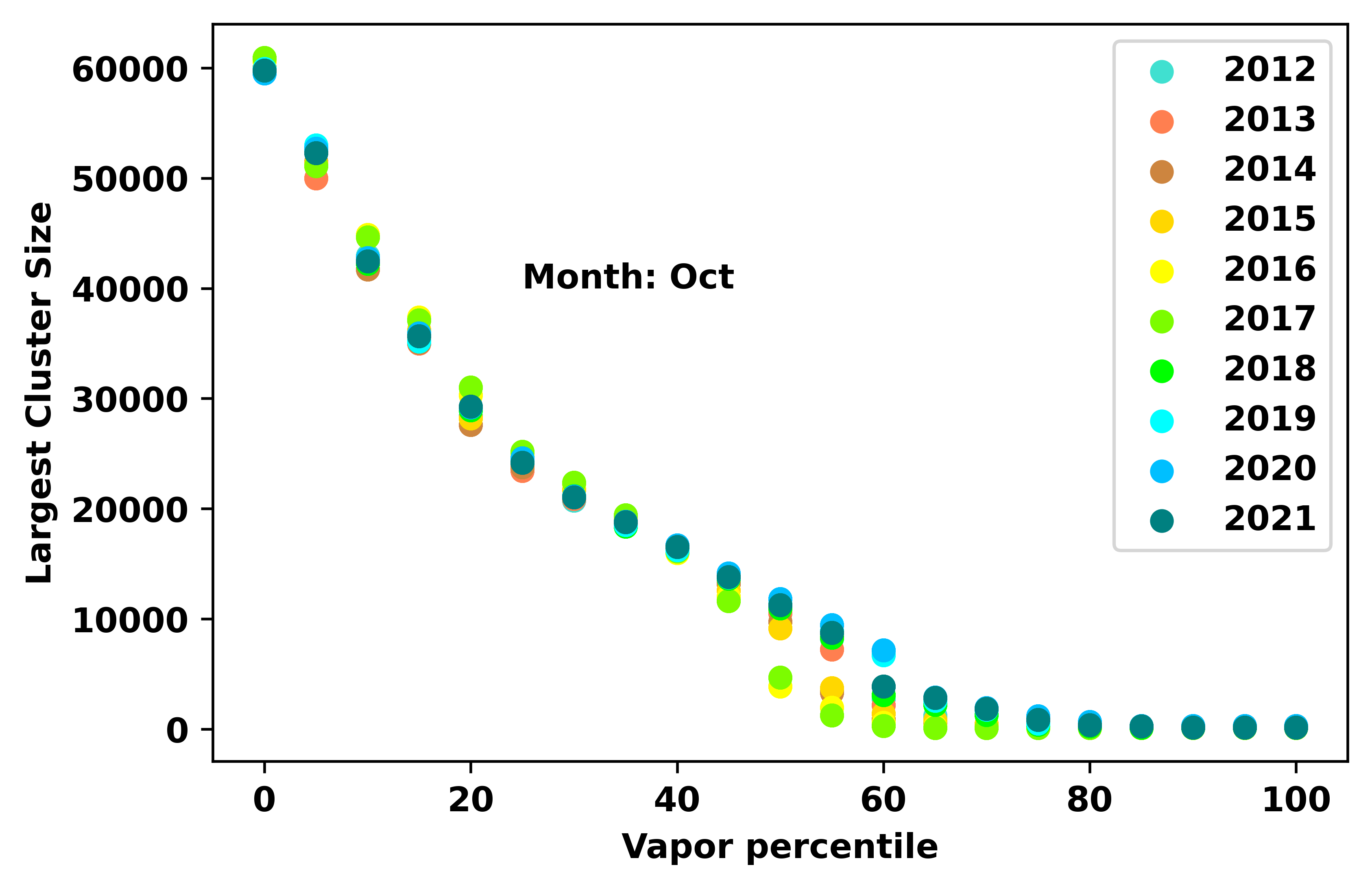} 
	\end{subfigure}
	\begin{subfigure}[t]{0.3\textwidth}
		\centering
		\includegraphics[width=\linewidth]{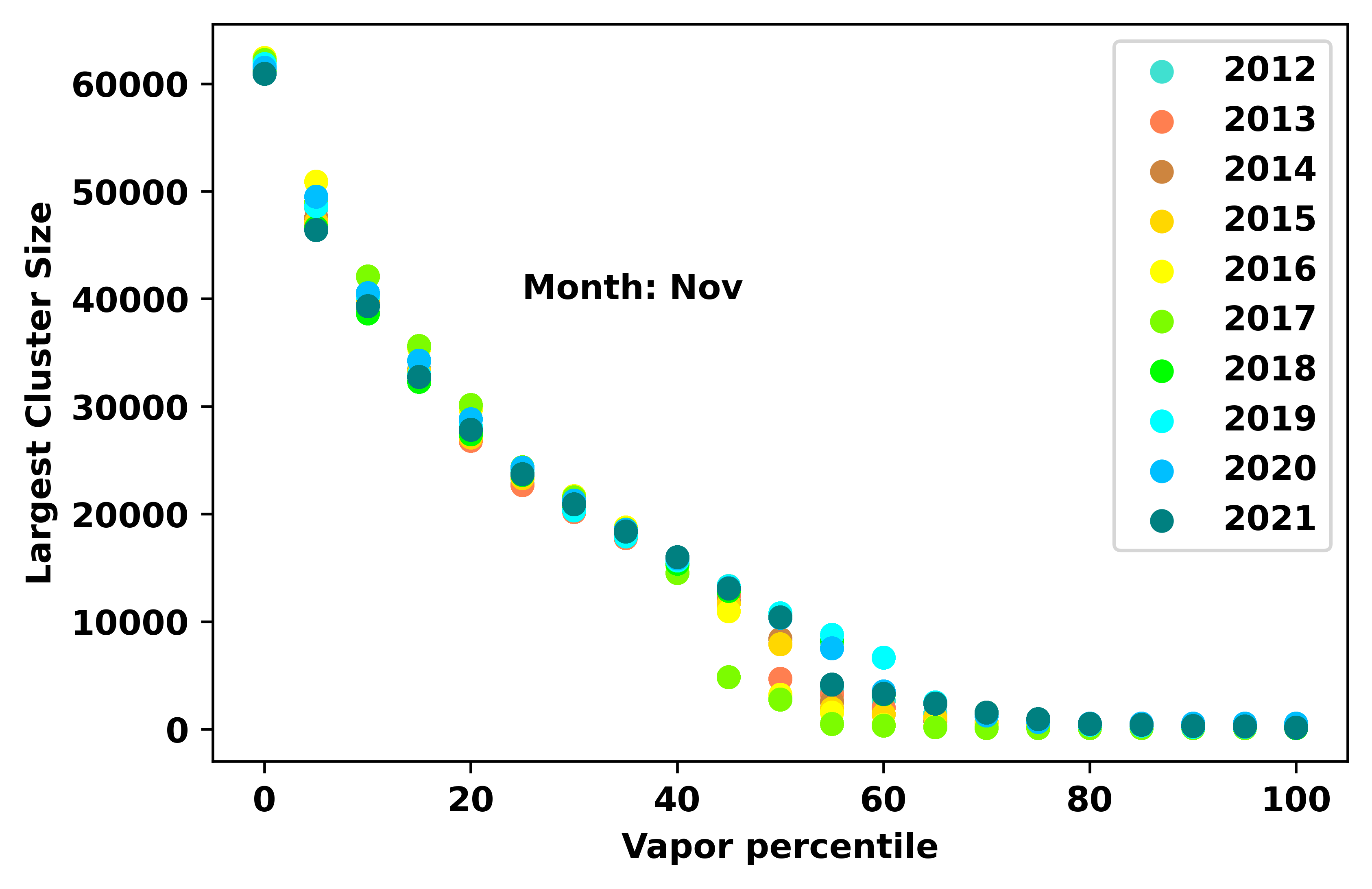} 
	\end{subfigure}
	\begin{subfigure}[t]{0.3\textwidth}
		\centering
		\includegraphics[width=\linewidth]{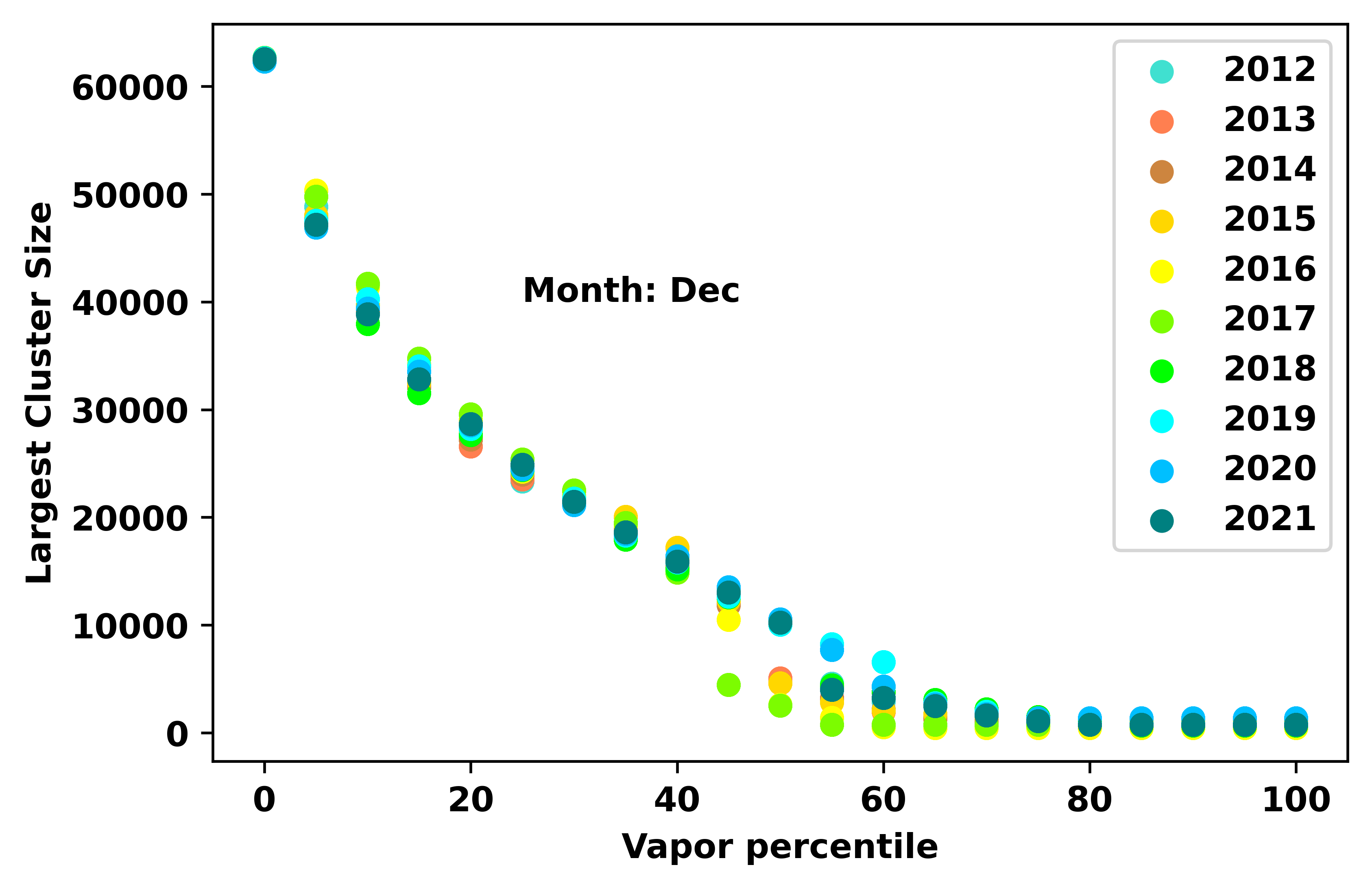} 
	\end{subfigure}		
	\caption{Maximum Size of Cluster vs Vapor Percentile for each month for 10 years, $2012-2021$, at a resolution of $180\times 360$ pixels.}
	\label{10}
\end{figure}

As before, we also plot the size of the largest cluster as a function of vapor percentile for each month for all $10$ years at a resolution of $180\times 360$ pixels. Here too, the data is not scaled as in the previous section since we always consider the same resolution. These plots are shown in Fig. \ref{10}. The distributions show an exponential decay for all months and years, with significant overlap between the data for different years. Some exceptions to this observations occur on and around the $60$th percentile which is very close to $V_{thres} = 55\% $. The distributions agree well with the percolation hypothesis, since at low vapor percentiles there exits one single massive cluster while at very large percentiles, the average size of clusters is small and approximately constant. While we observe that the decays are likely exponential, we make no attempt at fitting them to some fit function. We find that the percolation is not in the Bernoulli universality class ($p_c = 0.59$), since for us the critical probabilty is $p_c \approx 0.10$.

\begin{figure}
	\centering
	\begin{subfigure}[t]{0.45\textwidth}
		\centering
		\includegraphics[width=\linewidth]{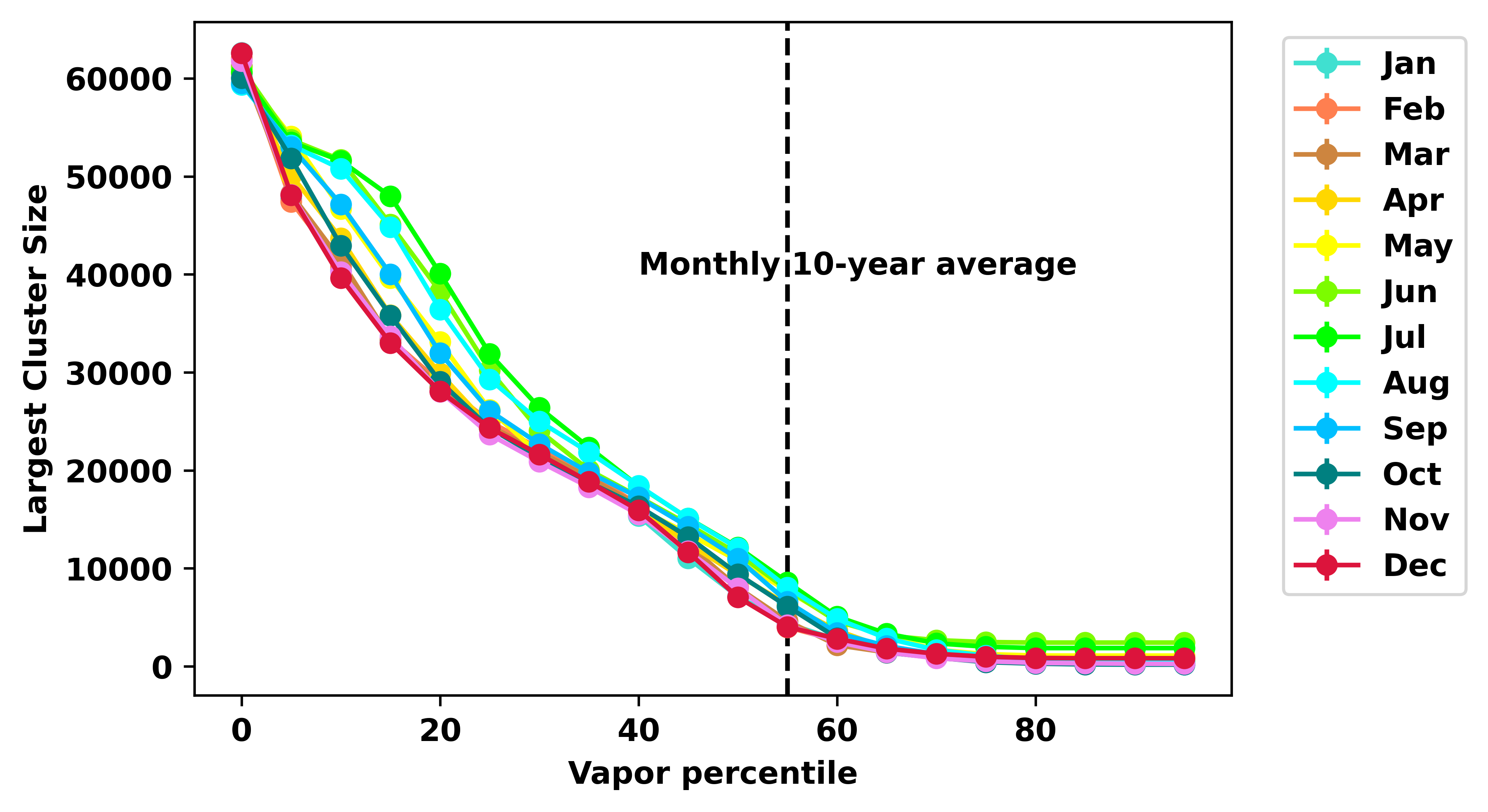} 
	\end{subfigure}
	\begin{subfigure}[t]{0.45\textwidth}
		\centering
		\includegraphics[width=\linewidth]{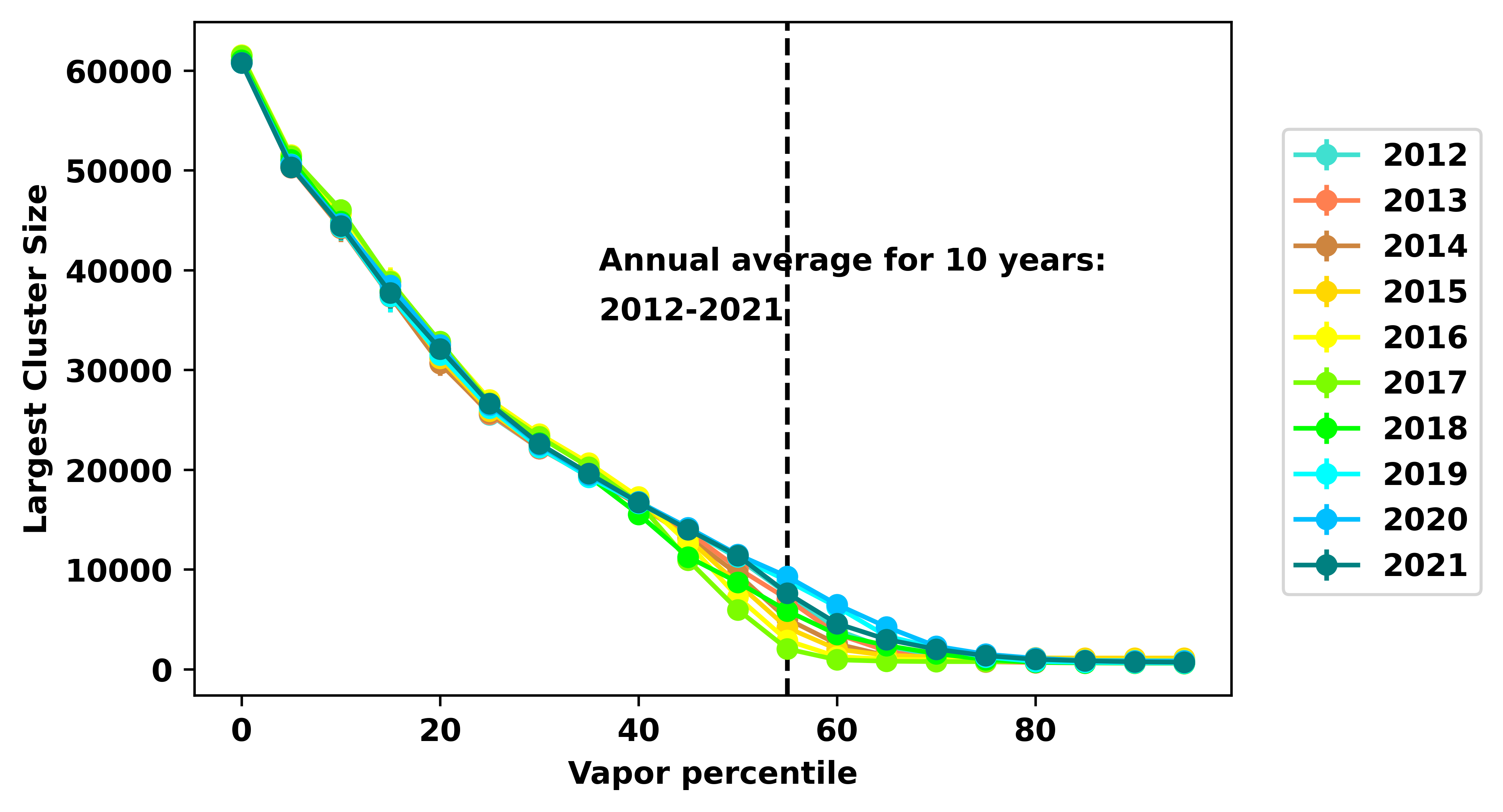} 
	\end{subfigure}
	\caption{Maximum Size of Cluster vs Vapor Percentile averaged over months and years for 10 years, $2012-2021$, at a resolution of $180\times 360$ pixels.}
	\label{11}
\end{figure}

We also plot the above data averaged over years and months in Fig. \ref{11} on the left and right plots, respectivelty. The plot on the left shows a relatively broad distribution for percentiles less than $30\%$, and a narrow or convergent distribution for larger percentiles. The overall tendency of the distribution is one of an exponential decay, although there exist some inflection points in the interim. The plot on the right, which represents data averaged anually, shows a large overlap for all years except around the $60$th percentile. While this is an interesting observation, we have no hypotheses as to appearance of the broadening (other than it being close to the threshold vapor percentile) and the relavance of the $60$th percentile to it.

\subsection{Korcak's Law and Scaling Exponents}

\begin{figure}[H]
	\centering
	\begin{subfigure}[t]{0.45\textwidth}
		\centering
		\includegraphics[width=\linewidth]{Power_law_55.png} 
	\end{subfigure}
	\begin{subfigure}[t]{0.45\textwidth}
		\centering
		\includegraphics[width=\linewidth]{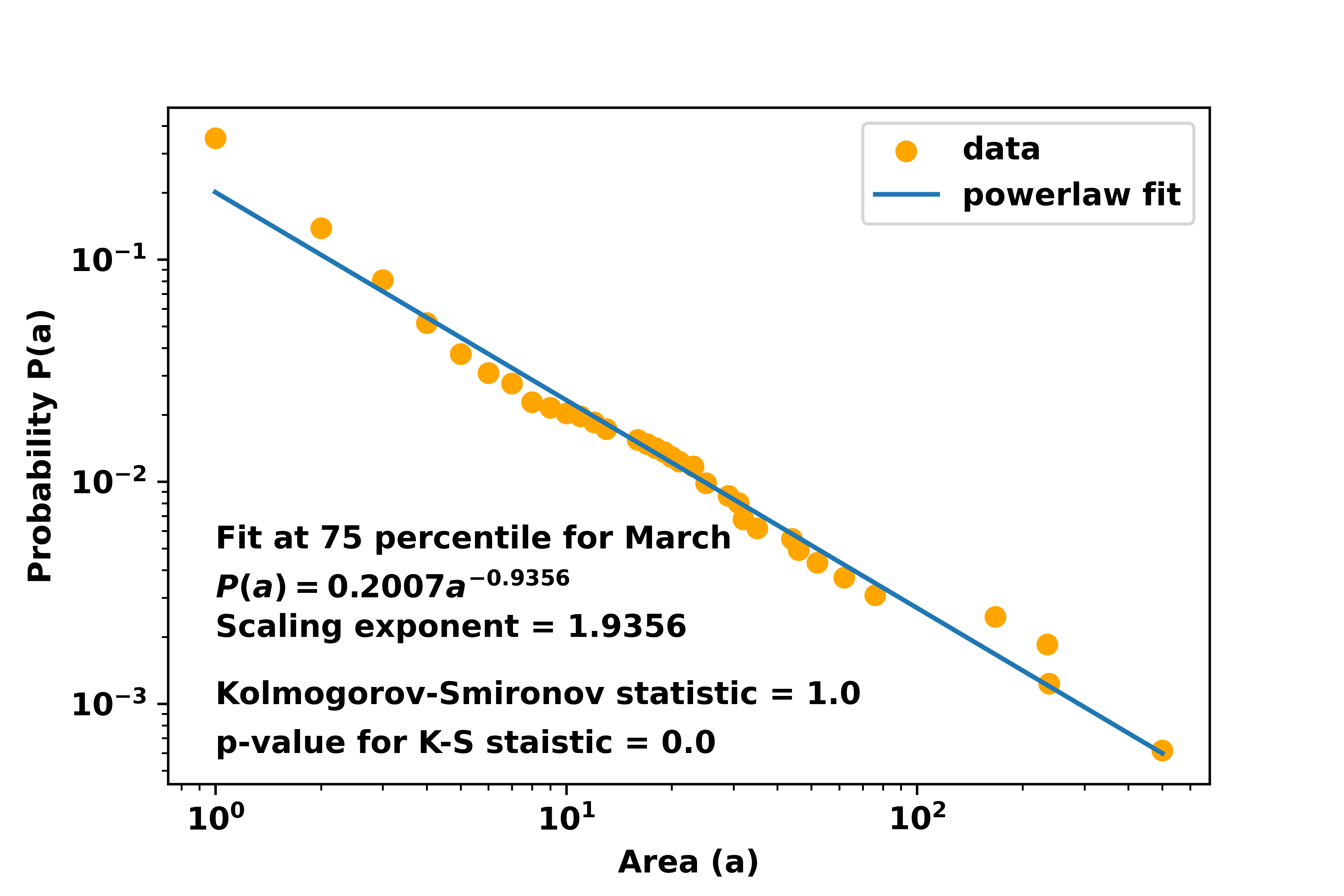} 
	\end{subfigure}
	\caption{Power law fit}
	\label{12}
\end{figure}

 As discussed previously, we are interested in testing the validity of Korcak's law for our distributions. We proceed as before and calculate the probabilities $P(a)$ for all areas of cluster size $a$. We plot the probabilities for the March 2021 data as a function of the cluster areas in Fig. \ref{12} at the percolation threshold (left plot) and at the $75$th percentile (right plot). The data is plotted as the orange scatter points, and the power-law fit is represented by the blue line. We test for the quality of the fit by using the Kolmogorov-Smirnoff (KS) statistic. These values are stated in the plots below. As we can see from the plots below, the fit to the data is close to perfect, with the KS statistic being approximately $1$ and the p-value being $0$. This verifies that our data is in agreement with Korcak's law, with little deviation from the powerlaw behaviour even above the percolation threshold. This suggests that criticality might not even exist in our dataset, although this will require further probing, a task which is relegated to a future work.

\begin{figure}[H]
	\centering
	\begin{subfigure}[t]{0.45\textwidth}
		\centering
		\includegraphics[width=\linewidth]{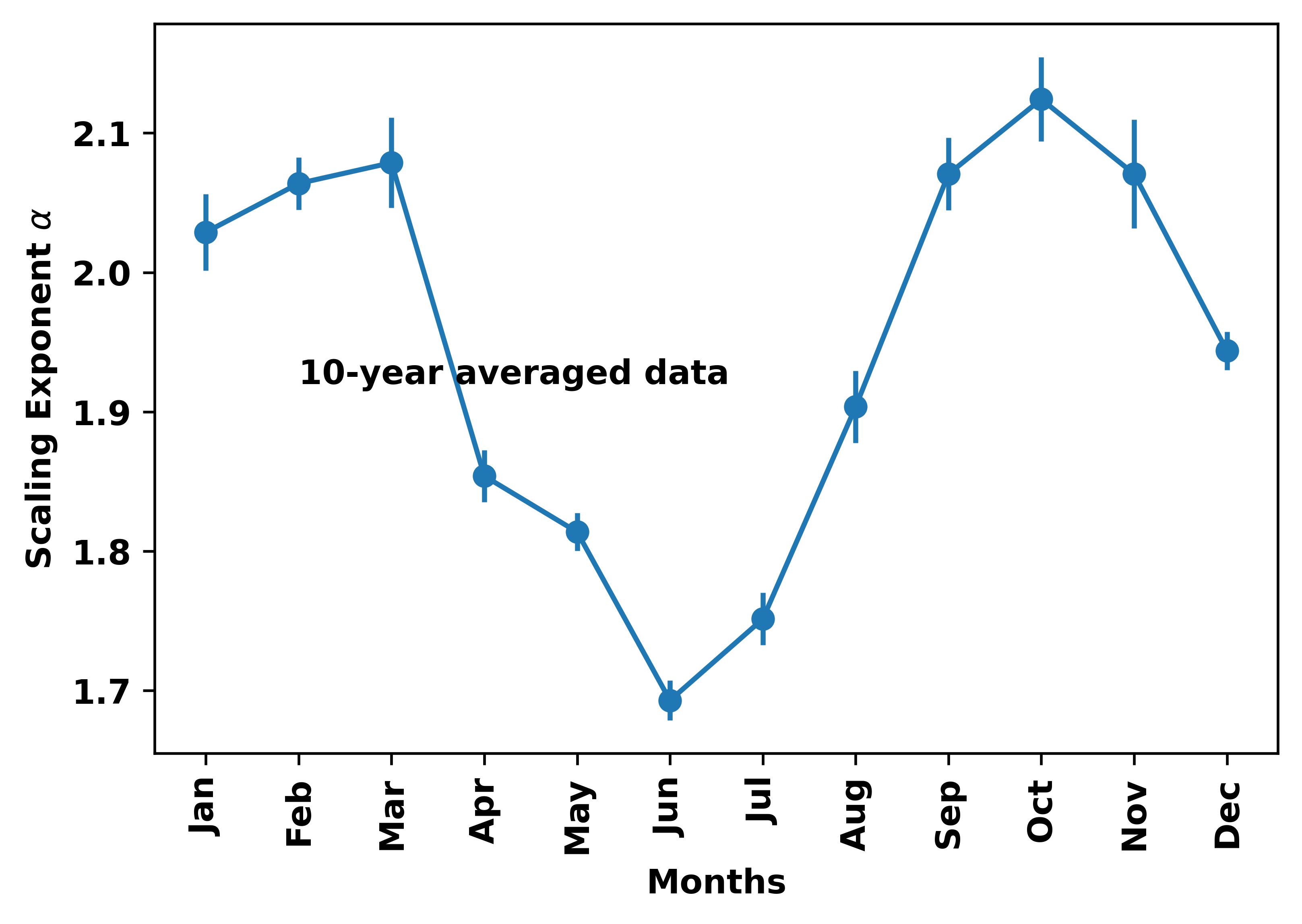} 
	\end{subfigure}
	\begin{subfigure}[t]{0.45\textwidth}
		\centering
		\includegraphics[width=\linewidth]{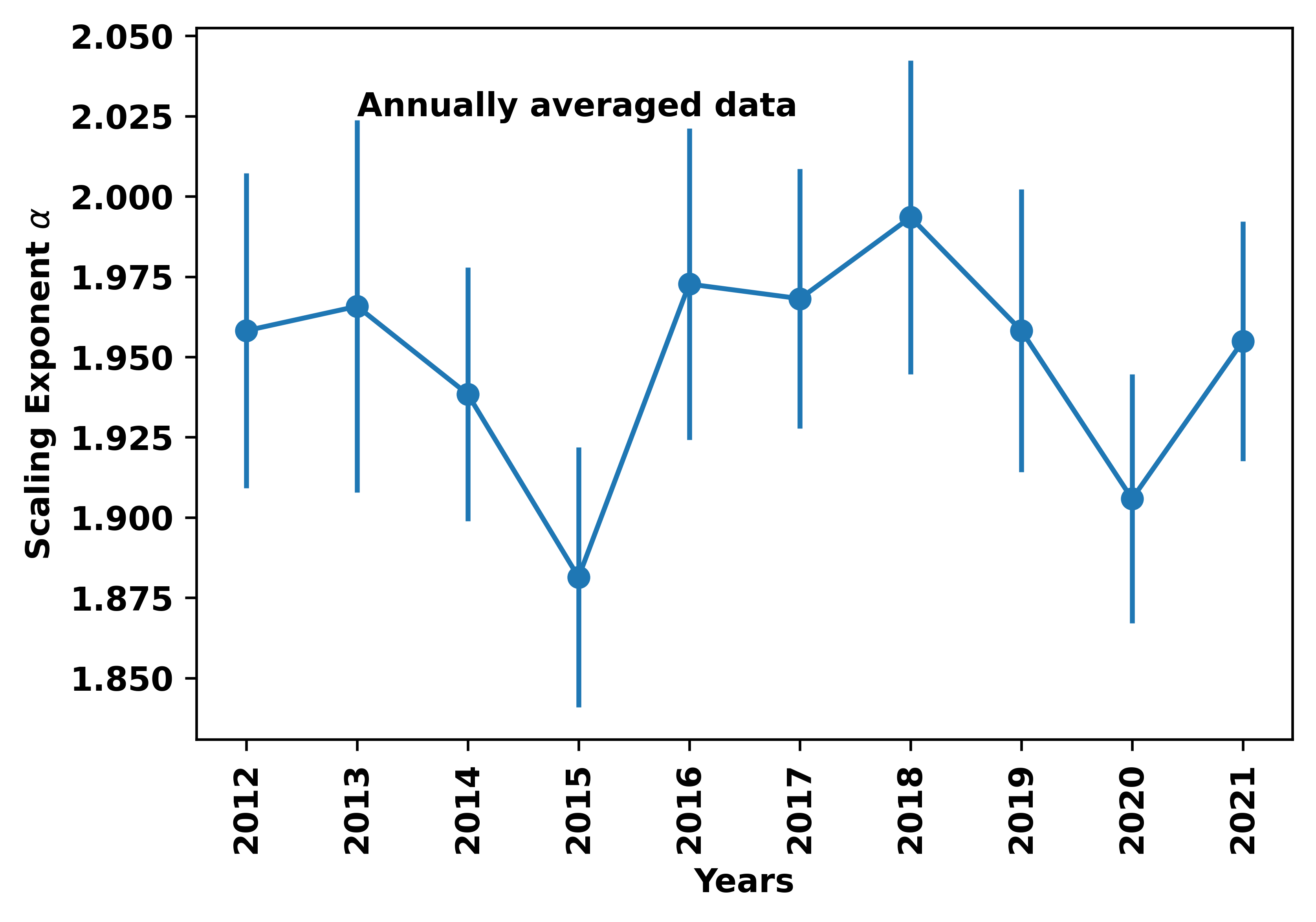} 
	\end{subfigure}
	\caption{Scaling Exponents: the exponents were extracted from the 10-year data and averaged by month and by year as shown above. The month-averaged data shows a greater range with more fluctuations than the year-averaged data.}
	\label{13}
\end{figure}

We plot the scaling exponents for the data at the percolation threshold averaged over years and months in Fig. \ref{13} on the left and right plots respectively. The scatter plots contain the mean scaling exponents and the respective standard deviations. The range of the scaling exponents in both cases lie between $1.70$ and $2.20$ with relatively small deviations from the mean only in the right plot. We have verified that in each case the KS statistic is very close to $1$ and the p-value is roughly $0$.

\section{Conclusions}

In this work we have sought to unravel the underlying fractal nature of the distribution of water vapor above the Earth. We have obtained data from the TERRA MODIS dataset and processed it using python v3.9. Clustering and numerical calculations were performed using author-designed algorithms. 

We first investigated the variation of fractal dimension of the dataset as a function of the resolution of the data at resolutions of $180\times 360$ pixels and $360\times 720$ pixels for the year 2021. We found that the fractal dimension of the vapor distribution as a function of vapor percentile remained universal with values between $1$ and $2$ for both resolutions for all months except January. We examined the potential multi-fractal behavior of our data using the box-counting method to estimate the generalized dimension. Our results indicate that our data is indeed representative of a multi-fractal which satisfies scale invariance at the resolutions examined. The anomalous result for January 2021 was attributed to some inconsitency in constructing the dataset. Further, the scaled number of clusters as a function of vapor percentile was plotted, and we found percolation to occur at $V_{thres} = 55\%$ and at $p_c=0.1$, the latter of which ruled out the 2D Bernoulli class of transitions. This observation was further supported by the results of plotting the scaled maximum cluster size versus vapor percentile. Both sets of representations showed an identical overlap between the data for both resolutions, further confirming the universality of the fractal structure as a function of resolution. We then determined the scaling exponent for this data distribution based on Korcak's law and found our numerical analysis to be in agreement with it. The quality of our fit was tested using the Kolmogorov-Smirnov statistic which approached unity, with a p-value of zero. This confirmed the excellent quality of our fit function and parameters.

We then examined the variation of fractal dimension of the dataset as a function of vapor percentile at a resolution of $180\times 360$ pixels for the years 2012-2021. The fractal dimension of the vapor distribution as a function of vapor percentile remained constant (as a function of vapor percentile) in certain ranges with values between $1$ and $2$ for all months and years. We therefore concluded that the data exhibited approximate fractal behavior in these ranges of vapor percentiles. Additionally, the multi-fractal character of the distribution was analysed using the box-counting technique. The generalized dimensions found were plotted as a function of the moments and vapor percentile, leading to the conclusion that our data indeed shows multi-fractality. Next, we proceeded to quantify the possible percolation transition in the data as a function of vapor percentile. The number of clusters as a function of vapor percentile was plotted, and  we found percolation to occur at $V_{thres} = 55\%$. This observation was further supported by the results of plotting the maximum cluster size versus vapor percentile. We plotted the monthly and yearly averages of all three quantities discussed above and found some interesting trends in the distributions. The possibility of the common 2D Bernoulli transition was ruled out. The data was then fitted to Korcak's law and the scaling exponents were determined. We tested the quality of our fit using the Kolmogorov-Smirnov statistic which approached unity, with a p-value of zero. This affirmed the effectiveness of our fit function and parameters.

In conclusion, we find that the water vapor distribution above the surface of the Earth exhibits a fractal character as a function of data resolution and vapor percentile, the former of which is exact while the latter is approximate. The water vapour clusters  display properties analogous to that of a percolating system as we vary the vapour percentile. At the percolation threshold, the size distribution of these water vapour clusters follows a power law, namely Korkac's law,  with a scaling exponent of $\beta \in (1.6, 2.2)$. To the best of our knowledge, the examination of the global distribution of water vapor from a fractal perspetctive is novel, which underlies the importance of this work.

{\bf Acknowledgements.} AM would like to thank the Science and Engineering Research Board (SERB), Government of India, for financial support through the fellowship Grant No. CRG/2019/001461. The research of BB is partially supported by research grant No. CRG/2019/001461 from the Science and Engineering Research Board (SERB), Government of India. The authors would also like to thank Dr. Anamika Shreevastava (NASA Jet Propulsion Lab, Caltech) for extensive and highly engaging discussions on the research associated with this manuscript. The authors would also like to thank the anonymous referee whose comments provided useful insights into the improvement of the current work.

\end{document}